%% file: PR-April-10.tex
\documentstyle[12pt,graphicx,axodraw]{article}
\newcommand{\s}{{supersymmetry~}}
\newcommand{\co}{{\cal O}}
\newcommand{\beq}{\begin{equation}}

\newcommand{\eeq}{\end{equation}}
\newcommand\beqn{\begin{eqnarray}}
\newcommand\eeqn{\end{eqnarray}}
\newcommand{\bea}{\begin{eqnarray}}
\newcommand{\eea}{\end{eqnarray}}

\newcommand{\nn}{\nonumber\\}
\newcommand{\x}{x^{\mu}}
\newcommand{\y}{x^{5}}
\newcommand{\yp}{x^{5'}}
\newcommand{\bl}{baryon and lepton number~}

\def\fr#1.#2.{{#1\over #2}}

\newcommand{\ps  }{$SU(4)_C\times SU(2)_L\times SU(2)_R$ }

\begin{document}
\pagestyle{empty}
\begin{center}
{\huge \bf \it{Proton Stability }}
\end{center}
\begin{center}
{\huge \bf \it {in} }
\end{center}
\begin{center}
{\huge \bf \it{ Grand Unified Theories,}} 
\end{center}
\begin{center}
{\huge \bf \it{in Strings and in Branes}}
\end{center}
\vspace{0.5cm}

\begin{center}
Pran Nath$^{1}$, and Pavel Fileviez P\'erez$^{2}$ \\  
\end{center}

\begin{center}
{\small $^{1}$ Department of Physics. Northeastern University\\
Boston, MA 02115, USA.\\
email:nath@lepton.neu.edu} 
\end{center}
\begin{center}
{\small $^{2}$ Departamento de F\'{\i}sica. 
Centro de F\'{\i}sica Te\'orica de Part\'{\i}culas,\\ 
Instituto Superior T\'{e}cnico, Av. Rovisco Pais 1, 1049-001. \\ 
Lisboa, Portugal.\\
email:fileviez@cftp.ist.utl.pt}
\end{center}

\vfill
\date{\today}
\begin{abstract}

\skip0.6cm

A broad overview of the current status of proton stability in unified 
models of particle interactions is given which includes non - supersymmetric 
unification, SUSY and SUGRA unified models, unification 
based on extra dimensions, and string-M-theory models. 
 The extra dimensional unification includes  5D and 6D and universal extra dimensional
 (UED) models, and models based on warped geometry. 
 Proton stability in a wide array of string theory and M theory models is
 reviewed. These include Calabi-Yau models, grand unified models with 
Kac-Moody  levels $k>1$, a new class of heterotic string models, 
 models based on intersecting D branes, and  string landscape models.
  The destabilizing effect of quantum gravity on the proton
 is discussed. The possibility of  testing grand unified models, models 
based on extra dimensions and string-M-theory models via 
their distinctive modes is investigated.
The proposed next generation proton decay experiments, HyperK,
UNO, MEMPHYS, ICARUS, LANNDD (DUSEL), and LENA would shed significant 
light on the nature of unification complementary to the 
physics at the LHC.
Mathematical tools for the computation of proton 
lifetime are  given in the appendices. Prospects for the future are
discussed.

\end{abstract}


\newpage
{\small \tableofcontents}

\pagestyle{plain}

\newpage

\section{Introduction}
	The Standard Model of strong, and the electro-weak interactions, 
given by the gauge group $SU(3)_C\times SU(2)_L\times U(1)_Y$,   
is  a highly successful model of particle 
interactions~\cite{Glashow:1961tr,Gross:1973id} 
which has been tested  with great accuracy by the LEP, SLC and Tevatron 
data. The electro-weak sector of this theory~\cite{Glashow:1961tr}, i.e., 
the $SU(2)_L\times U(1)_Y$ sector, provides a fundamental 
explanation of the Fermi constant and the scale
	\beqn
	G_F^{-\frac{1}{2}}\simeq 292.8 {\rm GeV}
	\label{fermi}
	\eeqn
    	has its origin in the spontaneous breaking of the $SU(2)_L\times U(1)_Y$ gauge
	group   and can be understood as arising from the vacuum expectation value ($v$) of the
	Higgs  boson field  ($H^0$) so that  $G_F^{-\frac{1}{2}} =2^{1/4} v$. Thus the scale
	$G_F$ is associate with new physics, i.e., the unification of the electro-weak interactions.
	There are  at least  two  more scales which are  associated with new physics.  First, from
	the high precision LEP data, one finds  that  the gauge coupling constants
	$g_3, g_2,  g_1(=\sqrt{\frac{5}{3}} g_Y)$, where $g_3, g_2, g_Y$ are the gauge
	coupling constants for the gauge groups $SU(3)_C$,  $SU(2)_L$,  $U(1)_Y$,
	 appear to unify within the minimal supersymmetric
	standard model at a scale $M_G$ so that

	\beqn
	M_G\simeq 2\times  10^{16} ~ {\rm GeV}
        \label{gutscale}
	\eeqn
    This scale which is presented here as empirical must also be associated  with new physics.
    A candidate theory here  is grand  unification.     Finally,  one has  the Planck scale  defined by
    \beqn
    M_{\rm Pl} = (8\pi G_N)^{-\frac{1}{2}}\simeq 2.4\times 10^{18} ~{\rm GeV}
    \eeqn
  where one expects  physics to be  described by quantum gravity,
  of which string-M-theory are  possible  candidates. 
Quite remarkable is the fact that the scale $M_G$ where the gauge couping
unification occurs is smaller than the Planck scale by about two orders of magnitude.
This fact has important implications in that one can build a field theoretic description
of unification of  particle interactions without necessarily having a full solution to
the problem of quantum gravity which operates at  the scale $M_{\rm Pl}$.
  Since grand unified theories
 and  models based on strings typically put quarks and  leptons in common multiplets
their unification
 in general leads to proton decay, and thus  proton stability becomes
 one of the crucial tests of such models.  Recent experiments have made such
 limits very stringent, and one expects  that the next generation of experiments will
 improve the lower limits by a factor of ten or more.  Such an improvement may
 lead to confirmation of proton decay which would  then provide us with
 an important window to the nature of the underlying unified structure of matter.
Even if no proton decay signal is seen, we will have much stronger lower limits
than what the current experiment gives, which would constrain the unified
models even more  stringently. This report is  timely  since  many new developments
have occurred since the early  eighties.
On the theoretical side there have been developments such as supersymmetry and
supergravity grand unification, and model building
in  string, in D branes, and in extra dimensional framework.  On the
experimental side Super-Kamiokande  has put the most stringent lower limits thus far
on the proton decay partial life times. Further, we stand at the point where new
proton decay experiments are being planned.  Thus it appears appropriate at this
 time to present a broad view of the current status of unification with proton stability
as its focus.  This is precisely the purpose of this report.
\\

We give now a brief description of  the content of the report.
In Sec.(2) we review the current status of proton decay lower limits from
recent experiments. The most stringent  limits come from the Super-Kamiokande
experiment.
We also describe briefly the proposed  future experiments. These  new generation of
experiments are expected to increase  the lower  limits  roughly by a factor  of ten.
In Sec.(3)  we discuss proton stability in non-supersymmetric scenarios.  In Sec.3.1  we
estimate  the proton lifetime where the B -violating effective  operators  are  induced by
instantons.  In Sec.(3.2) we  discuss the baryon and  lepton number violating dimension
six operators induced by gauge  interactions  which are $SU(3)_C\times SU(2)_L\times U(1)_Y$
invariant.  Proton decay modes from these  $B-L$ preserving interactions are  also discussed.
In Sec.(3.3), we  discuss the general set  of dimension six operators  induced by scalar
lepto-quarks consistent with $SU(3)_C\times SU(2)_L\times U(1)_Y$ interactions.
\\

In Sec.(4) nucleon decay in supersymmetric gauge theories is discussed.
In Sec.(4.1) the constraint on R parity violating interactions to suppress rapid proton decay
from \bl ~violating dimension four operators is analyzed.  However, in general proton decay
from \bl ~violating dimension five operators will occur and in this case it is the most
dominant contribution to proton decay in most of the supersymmetric 
grand unified theories. The analysis
of proton decay dimension five operators requires that one convert the \bl ~violating
dimension five  operators by chargino, gluino and neutralino exchanges to convert
them to \bl ~violating dimension six operators.  The dressing loop diagrams 
depend sensitively on  soft breaking.  Thus in Sec.(4.2) a  brief  review
of  supersymmetry breaking is given.  As is  well known, the soft  breaking  sector  of
supersymmetric theories depends  on CP  phases and thus  the dressing loop
diagrams and  proton decay can be affected by the presence  of  such phases. 
A discussion of this phenomenon is given in Sec.(4.3). Typically in grand unified
 theories the Higgs iso-doublets with quantum numbers  of the MSSM Higgs fields  and the
 Higgs  color- triplets are  unified in a single  representation.  Since we  need  a  pair
 of light Higgs  iso-doublets to break the electro-weak symmetry, while we
 need  the Higgs  triplets to be  heavy to avoid  too fast a  proton decay,  a
 doublet-triplet  splitting is  essential for any viable unified model.   Sec.(4.4) is
 devoted  to this  important topic. The remainder  of Sec.(4) is devoted to a
 discussion of proton decay in specific models.   A discussion of proton decay
 in $SU(5)$ grand unification is  given in Sec.(4.5), while a discussion of
 proton decay in $SO(10)$ models is given in Sec.(4.6). In Sec.(4.7)  a new
 $SO(10)$ framework is given
 where a single constrained  vector-spinor - a 144- multiplet
 is used to break $SO(10)$ down to the residual gauge group symmetry
 $SU(3)_C\times U(1)_{em}$. \\

Sec.(5) is devoted to tests of grand unification through proton decay
and a number of items that impinge on it are discussed.   One of these
concerns the implication of  Yukawa textures on the  proton lifetime.
It is generally believed  that the fermion mass hierarchy may be more easily
understood in terms  of  Yukawa textures  at a  high  scale and there
are  many proposals  for the nature of such textures.  It turns out that
the  Higgs  triplet textures are  not the  same  as  the Higgs doublet
textures,  and a unified framework  allows  for the calculation  of
such textures.  This topic is discussed in Sec.(5.1).   Supergravity
grand unification involves three arbitrary functions: the superpotential,
the Kahler  potential, and the gauge kinetic energy function.
Non-universalities  in gauge kinetic  energy function can affect
both the gauge  coupling unification and proton lifetime. This topic
is discussed in Sec.(5.2).  In grand unified models, the gauge
coupling unification receives threshold  corrections from the low mass
(sparticle) spectrum  as well from  the high scale  (GUT) masses.
Consequently the GUT scale  masses, and  specifically the
Higgs  triplet mass,  are  constrained by the high precision LEP data.
These  constraints are  discussed  in Sec.(5.3).
Model independent tests  of distinguishing GUT models using
meson and anti-neutrino final state are discussed in Sec.(5.4)
where three different models, $SU(5)$, flipped $SU(5)$ and
$SO(10)$  are  considered.  In Sec(5.5) the important issue
of the constraints necessary to rotate  away or  eliminate  the
 \bl ~violating dimension  six operators  induced by gauge interactions
is discussed.
It is shown that it  is possible to satisfy  such constraints  for  the
flipped $SU(5)$ case.  Finally,  an analysis of the upper limits on
the proton lifetime on  \bl ~violating dimension six
operators  induced by gauge interactions is given in Sec.(5.6).
\\

Sec.(6) is devoted to grand unified models in extra dimensions
and the status of proton stability in such models.  In Sec.(6.1)
a discussion of proton stability in grand unified models in
dimension five (i.e.,  with one extra dimension) is given
and various possibilities  where the matter could reside
either on the branes or  in the bulk are discussed.   In these models
it is possible to get a natural doublet-triplet splitting in the
Higgs sector with no Higgs triplets and anti-triples with zero
modes. A review of $SO(10)$ models in 5D is given
in Sec.(6.2) while 5D trinification models are discussed in Sec.(6.3).
6D grand unification models in dimension six, i.e.,  on $R\times T^2$,
are discussed in Sec.(6.4).  Various grand unification possibilities on
the branes, i.e.,  $SO(10)$, $SU(5)\times U(1)$, flipped $SU(5)\times U(1)$,
and  $SU(4)_C\times SU(2)_L\times SU(2)_R$ exist in this case. 
Another class of models  which are closely related to the models
 above are those with gauge-Higgs unification.  Here the Higgs fields
 arise  as part of the  gauge multiplet and hence  gauge and Higgs
 couplings are unified. Various possibilities for the suppression of proton
 decay exist in these models  since proton decay is sensitive  to how  matter
 is located in extra  dimensions.  In Sec.(6.6) a discussion of proton decay
 in models with  universal extra dimensions (UED) is given. In these models
 extra symmetries arise  which can be  used to control proton decay.
 In Sec.(6.7)  proton stability in models with warped geometry is
 discussed.  Such models  lead to a solution to the hierarchy
 problem via a warp factor which depends on the extra dimension.
 Proton decay can be  suppressed through a  symmetry
 which conserves baryon number. Finally, in Sec.(6.8) proton stability in 
kink backgrounds is discussed.
 \\

In Sec.(7) we discuss proton stability in string and brane models.  There  are
 currently five different types of string theories: Type I, Type IIA, Type IIB,  SO(32)
 heterotic and $E_8\times E_8$ heterotic.  These are  all connected by a web
 of dualities and conjectured to be subsumed in a more fundamental M-theory.
 Realistic and semi-realistic model building has been carried out in many of
 them and  most extensive investigations exist for the case of the
 $E_8\times E_8$  heterotic string within the so called Calabi-Yau compactifications
 where the effective  group structure after Wilson line breaking  is
 $SU(3)_C\times SU(3)_L\times SU(3)_R$ and further breaking through the
 Higgs  mechanism is needed  to break the group down to the Standard Model
 gauge group.   Proton stability in Calabi-Yau models is discussed in Sec.(7.1).
 In Sec.(7.2) we discuss  grand  unification in Kac-Moody levels $k>1$.
  It is known  that in weakly coupled heterotic strings one cannot
 realize massless scalars in the adjoint representation at level 1, and one
 needs to go to levels $k>1$ to realize massless scalars in the adjoint
 representations necessary to break the GUT symmetry. However, at level 2
 it is difficult to obtain 3 massless generations while this problem is overcome
 at level 3.   In these models
 baryon and lepton number violating dimension four operators are
 absent due to an underlying gauge and discrete symmetry. However,  \bl
 ~violating dimension five operators  are  present and  one needs to
 suppress them by  heavy Higgs triplets.  A detailed analysis of proton lifetime
 in these models is currently difficult due the problem of generating
 proper quark-lepton masses. In Sec.(7.3) a new class of heterotic string models are  discussed which have the interesting feature that they have the spectrum of 
 MSSM,  while  proton decay is absolutely forbidden
 in these  models, aside  from the proton decay  induced by quantum gravity effects.
Other attempts at realistic model building in 4D models in the heterotic string
framework are also briefly discussed in Sec.(7.3).
\\

Proton decay in M-theory compactifications are  discussed in Sec.(7.4).
The low energy limit of this theory is the 11 dimensional supergravity
theory and one can preserve  $N=1$ supersymmetry if one compactifies
the 11 dimensional supergravity on a seven-compact manifold X  of $G_2$
holonomy.  The manifold  X can be  chosen to give  non-abelian gauge symmetry
and chiral fermion.  Currently quantitative  predictions of proton lifetime  do not exist
due  to an unknown overall  normalization factor which requires  an M theory
calculation for its computation.  However,  it is  still possible to make  qualitative
predictions in this theory.  Thus for a class of X-manifolds,
\bl ~violating dimension five operators  are absent but \bl ~violating dimension
six  operators  do exist  and here one can make the interesting prediction that
the decay mode $p\to e^+_R\pi^0$  is  suppressed relative  to the mode
$p\to e^+_L\pi^0$.  In Sec(7.5) proton decay in intersecting  D brane models
is discussed.  Here  we  consider  proton decay in $SU(5)$ like GUT models in
Type IIA orientifolds with D-6 branes.  It is assumed that  the \bl ~violating
dimension 4  and dimension 5 operators are absent and  that the
observable proton decay arises  from dimension six operators.  The predictions
of the model here may lie within reach of the next generation of proton
decay experiment.  In Sec.(7.6) we discuss proton stability in string landscape
models.  There are a variety of scenarios in this class of models where
the squarks and sleptons  can be  very heavy and thus proton decay via
dimension five  operators  will be suppressed.  Such is the situation on the
so called Hyperbolic Branch of radiative  breaking of the electro-weak symmetry.
A brief review is given in Sec.(7.6) of the possible scenarios within string models
where a hierarchical breaking of supersymmetry can occur.  In Sec.(7.7) a review
of proton decay from quantum gravity effects is  given.  It is conjectured that quantum
gravity does not  conserve  baryon number and thus can catalyze proton decay.
Thus, for example,  quantum gravity effects could induce baryon number violating
processes of the type $qq\to \bar q l$.  Proton decay via  quantum gravity effects
in the context of large extra dimensions are also discussed in Sec.(7.7).
In Sec.(7.8) a discussion of  $U(1)$ string symmetries is given which
allow  the suppression of proton decay  from  dimension four and dimension
five  operators.  In Sec.(7.9)  discrete symmetries for the  suppression of proton
decay are discussed.  However, if the discrete symmetries are global they are
not respected by quantum gravity specifically, for example, in virtual black  hole
exchange and in wormhole tunneling. However, gauged discrete symmetries
allows one to overcome this hurdle. A brief discussion of the classification of such
symmetries is also given in Sec.(7.9).
\\

A number of other topics related  to proton stability in GUTs, strings  and branes  are
discussed in Sec.(8).  Thus an interesting issue concerns the connection between
proton stability and neutrino masses. This connection is especially  relevant in the
context of grand unified models based on $SO(10)$ and the discussion of Sec.(8.1)
is devoted to this case.  Supersymmetric  models with R parity invariance  lead to
the  lowest supersymmetric particle (LSP) being absolutely stable. In supergravity
GUT models the LSP over much of the parameter  space  turns  out to be  the lightest
neutralino. Thus supersymmetry/supergravity models  provide  a  candidate  for
cold dark matter. The recent WMAP data puts  stringent constraints on the amount
of dark matter. The dark matter constraints have  a direct  bearing on predictions
of the proton lifetime in unified models.  This topic is discussed in Sec.(8.2).
In Sec.(8.3) exotic \bl  ~violation is discussed. These include processes
involving $\Delta B=3$ such as $^3H\to e^+\pi^0$, \bl
 ~violation involving higher
generations, e.g., $p\to  \tau^*\to \bar \nu_{\tau} \pi^+$, and proton decay via
monopole catalysis where $M+p\to M+e^+ + {\rm mesons}$.
Finally, Sec.(8.4)  contains speculations on proton decay and the ultimate
fate of the universe.  Sec.(9) contains a summary of the report  highlighting
some of the important elements of the report and outlook for the future.
\\

Many of the mathematical details of the report are relegated to the Appendices.
Thus in Appendix  A mathematical aspects of the grand unification groups $SU(5)$ and
$SO(10)$ necessary for understanding the discussion in the main text are  given.
In Appendix B,  the allowed  contributions arising from dimension five  operators
to proton decay are listed.  In Appendix C a  glossary of dressings of dimension
five operators  by chargino, gluino, and neutralino exchanges is  given.  The dressing loop
diagrams involve sparticle masses, and in Appendix D an analysis of the sparticle
spectra at low energy using renormalization group is  given. Appendix E is
devoted to a discussion of the renormalization group factors of the dimension 5 and
dimension 6 operators.  A detailed discussion of the effective  Lagrangian which
allows one to convert  \bl ~violating  quark-lepton dimension six operator to
interactions involving  baryons and  mesons is  given in Appendix F.
Appendix G gives details of the analysis of testing models, and Appendix H gives 
the details on the analysis of upper bounds.
Appendix I gives a discussion of how one may relate the 4D parameters  to the
parameters of M theory.  Finally, Appendix J is devoted  to a  discussion of the
gauge coupling unification in string and  D brane  models.

\section{Experimental bounds and future searches}
The issue of proton stability has attracted attention over three quarters of
a century. Thus in the period 1929-1949 the law of baryon number conservation was
formulated by Weyl, Stueckelberg and  Wigner~\cite{weyl}, and the first experimental test 
of the idea was proposed by Maurice Goldhaber in 1954~\cite{Goldhaber:1988gb,Reines:1980by}. The 
basic idea of Goldhaber was that  nucleon decay 
could leave $Th^{232}$ in an excited and fissionable state, and thus comparison 
of the measured lifetime to that for spontaneous fission 
could be used to search for nucleon decay. This technique produced a lower limit on
the proton lifetime of $\tau>1.4\times 10^{18}$ years. 
The first direct search for 
proton decay was made by F. Reines, C. Cowan and M. Goldhaber~\cite{Reines:1954pg} 
using a 300 liter liquid scintillation detector, and they set a limit on the  
lifetime of free protons of  $\tau > 1\times 10^{21}$ years  and a 
lifetime for bound nucleons of $\tau > 1\times 10^{22}$ years. 
From a theoretical view point the idea
that proton may be unstable originates in the work on Sakharov in 1967~\cite{sakharov}
 who postulated
that an explanation of baryon asymmetry in the universe requires CP violation  and
baryon number non-conservation. Further, impetus for proton decay came  with 
the work of Pati and Salam in 1973~\cite{Pati:1973rp} and later with
 non-supersymmetric~\cite{Georgi:1974sy,SO(10)}, 
supersymmetric~\cite{Dimopoulos:1981zb}, and supergravity~\cite{Chamseddine:1982jx,Nath:1983aw}
grand unification, 
and from quantum gravity where black hole and worm hole  effects
can catalyze proton decay~\cite{Zeldovich:1976vq,Hawking:1979hw,Page:1980qm,Ellis:1983qm,Gilbert:1989nq}.

Thus spurred by theoretical developments in the nineteen seventies and the 
eighties there were large scale experiments for the detection of proton decay. 
Chief among these are the Kolar Gold Field\cite{Krishnaswamy:1986ig}, 
NUSEX~\cite{Battistoni:1985na}, FREJUS~\cite{Berger:1987ke}, SOUDAN~\cite{Thron:1989cd}, 
Irvine-Michigan-Brookhaven (IMB)\cite{Becker-Szendy:1992hr} 
and Kamiokande~\cite{Hirata:1989kn}. These experiments use either tracking 
calorimeters (e.g. SOUDAN) or Cherenkov effect (IMB, Kamiokande). These experiments yielded
null results but produced improved lower bounds on various proton decay modes.  
In the nineteen nineties the largest proton water Cherenkov detector, Super-Kamiokande, 
came on line for the purpose of searching for proton decay and for the study of 
the solar and atmospheric neutrino properties.
Super-Kamiokande~\cite{SK} is a ring imaging water Cherenkov detector 
containing 50 ktons of ultra pure water held in a cylindrical 
stainless steel tank 1 km underground in a mine in the Japanese Alps. 
The sensitive volume of water is split into two parts. The 2 m thick 
outer detector is viewed with 1885 20 cm diameter 
photomultiplier tubes. When relativistic particles pass through the water 
they emit Cherenkov light at an angle of about $42^0$  from the particle 
direction of travel. By measuring 
the charge produced in each photo multiplier tube and the time at which it is collected,
 it is possible to reconstruct the position and energy of the event as well as 
the number, identity and momenta of the individual charged particles in the event. 

The progress in the last 50 years of proton decay searches 
is shown in Figure~\ref{fig2.4}, where the experimental lower bounds 
for the partial proton decay lifetimes are exhibited. The plot exhibits the power of the 
water Cerenkov detectors in improving the proton decay lower bounds. 
Since Super-kamiokande is currently the most sensitive proton decay experiment,
it is instructive  to examine briefly the signatures of proton decay signals in
this experiment. We focus on the decay mode $p \to e^+ \pi^0$. Since it is one of the
simplest modes it serves well as a general example of proton decay
searches.

\begin{figure}[h]
  \centerline{\epsfxsize=2in \epsfbox{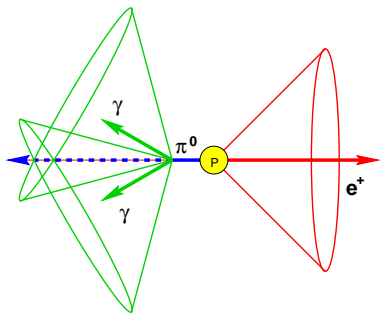}}
  \caption{Idealized $p \to e^+ \pi^0$ decay in Super-Kamiokande~\cite{Viren:1999pk}.}
\label{pepi}
\end{figure}
\begin{figure}[h]
  \centerline{\epsfxsize=2in \epsfbox{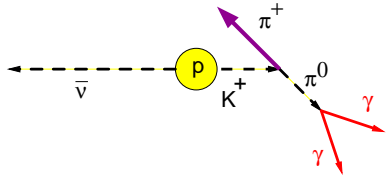}}
  \caption{Idealized $p \to K^+ \nu$ decay in Super-Kamiokande, 
$K^+ \to \pi^+ \pi^0$ case~\cite{Viren:1999pk}}
\label{pkplus}
\end{figure}
\begin{figure}[h]
  \centerline{\epsfxsize=2in \epsfbox{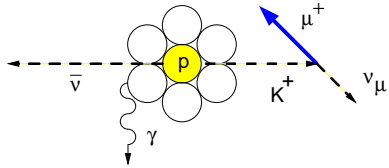}}
  \caption{Idealized $p \to K^+ \nu$ decay in Super-Kamiokande, 
$K^+ \to \mu^+\nu_{\mu}$ case~\cite{Viren:1999pk}}
\label{fig2.3}
\end{figure}
\begin{figure}[h]
  \centerline{\epsfxsize=7.0in \epsfbox{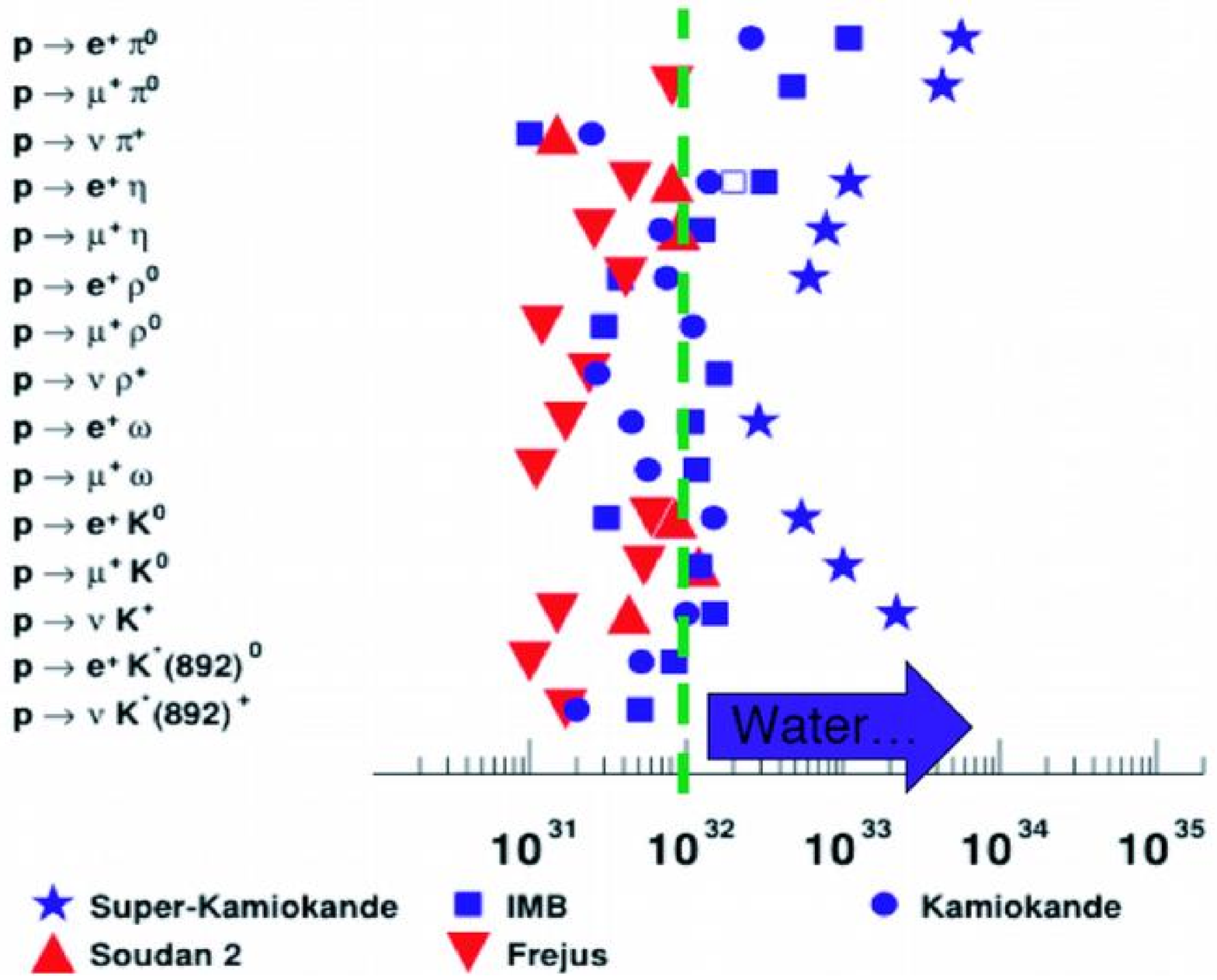}}
  \caption{Experimental lower bounds on proton decay partial lifetimes~\cite{Rubbia:2004yq}}
\label{fig2.4}
\end{figure}

Fig.(\ref{pepi}) gives a schematic presentation  of an ideal $p \to e^+ \pi^0$ 
decay.  Here, the positron, $e^+$ and neutral pion $\pi^0$, exit 
the decay region in opposite directions. The positron initiates an
electromagnetic shower leading to a single isolated ring.  The
$\pi^0$ will almost immediately decay to two photons which will go
on to initiate showers creating two, usually overlapping, rings.
In general, real $p \to e^+ \pi^0$ events will differ from this 
ideal picture  because the pion can scatter or be absorbed 
entirely before it exits the nucleus. In addition the 
proton in the nucleus can have some momentum due to Fermi motion. These two effects, 
i.e., the
pion-nucleon interaction and Fermi motion, serve to spoil 
the balance of the reconstructed momentum.  Further, the pion can decay
asymmetrically where one photon takes more than half of the pion's
energy leaving the second photon to create a faint or even completely
invisible ring. All these effects are taken into account in search for
proton decay signals. Super-Kamiokande experiment 
also searches for the $p \to K^+ \bar{\nu}$ mode 
by looking for the products from the two primary branches 
of the $K^+$ decay (see Figure~\ref{pkplus}). 
In the $K^+ \to \mu^+\nu_{\mu}$ case, 
when the decaying proton is in the ${}^{16}O$, the nucleus will be 
left as an excited ${}^{15}$N.  Upon de-excitation, 
a prompt 6.3 MeV photon will be emitted (See Figure~\ref{fig2.3}). 

An important question for proton decay searches concerns the issue of 
backgrounds.
There are three classes of atmospheric neutrino 
background events that are directly relevant for proton decay searches.
The first is the inelastic charged current events, $ \nu N \to N^´ {e,\mu} + n\pi$, where a 
neutrino interacts with a nucleon in the water and produces a visible
 lepton and a number of pion's. This can mimic proton decay modes 
such as $p \to e^+ \pi^0$. The second class is neutral current pion production, 
$\nu N \to \nu N^´ n \pi$, the only visible products of which are pion's. 
This is the background to, for example, $n \to \nu \eta$. Finally, 
there are quasi elastic charged current events $\nu N \to N^´ {\mu, e}$, events 
which can look like, $p \to K^+ \bar{\nu}$.    
 The current experimental lower bounds 
on proton decay lifetimes are listed in Table 1.

We note that presently the largest lower bound is for the mode
$p \to e^+ \pi^0$. Interestingly the radiative decay
modes $p \to e^+ \gamma$ and $p \to \mu^+ \gamma$ also have 
very strong constraints.

\begin{table}
\begin{center}
\begin{tabular}{|r|r|}
\hline
\hline
 & \\
\textbf{Channel}&\textbf{$\tau_p$ ($10^{30}$ years)}\\ 
 & \\
\hline
\hline
 & \\ 
$p \to invisible$ & 0.21 \\
$p \to e^+ \pi^0$  &  1600  \\
$p \to \mu^+ \pi^0$ &  473  \\
$p \to \nu \pi^+ $   & 25   \\
$p \to e^+ \eta^0 $ & 313   \\
$p \to \mu^+ \eta^0 $ & 126 \\
$p \to e^+ \rho^0 $ & 75   \\
$p \to \mu^+ \rho^0 $ & 110 \\
$p \to \nu \rho^+ $ & 162 \\
$p \to e^+ \omega^0 $ & 107 \\
$p \to \mu^+ \omega^0 $ & 117 \\
$p \to e^+ K^0 $ & 150  \\
$p \to e^+ K^0_S $ & 120 \\
$p \to e^+ K^0_L $ & 51 \\
$p \to \mu^+ K^0 $ & 120 \\
$p \to \mu^+ K^0_S $ & 150 \\
$p \to \mu^+ K^0_L $ & 83 \\
$p \to \nu K^+ $ & 670 \\
$p \to e^+ K^{*}(892) $ & 84 \\
$p \to \nu K^{*}(892) $ & 51 \\
$p \to e^+ \gamma $ & 670 \\
$p \to \mu^+ \gamma $ & 478\\
& \\
\hline
\hline
\end{tabular}
\caption{
\label{current}
Experimental lower bounds on proton lifetimes
~\cite{Eidelman:2004wy}. 
The limits listed are on $\tau/B_i$, where $\tau$ is 
the total mean life and $B_i$ is 
the branching fraction for the relevant mode.  
}
\end{center}
\label{current}
\end{table}  
Recently the Super-Kamiokande collaboration has reported new experimental lower bounds 
on proton decay lifetimes. The improved limits for some of the channels are as 
follows~\cite{Kobayashi:2005pe}:

\begin{eqnarray}
\tau( p \to K^+ \bar{\nu}) \ &>& \ 2.3 \times 10^{33} \ \textrm{years} \\
\tau( p \to K^0 \mu^+) \ &>& \ 1.3 \times 10^{33} \ \textrm{years} \\
\tau( p \to K^0 e^+) \ &>& \ 1.0 \times 10^{33} \ \textrm{years} 
\end{eqnarray}

As will be discussed later in this report, proton decay is a probe of fundamental
interactions at extremely short distances and as such it is an instrument for
the exploration of grand unifications, of Planck scale physics and of  quantum
gravity and more specifically of string theory and M theory. 
For this reason it is crucial to have new experiments to search for proton decay or 
improve the current bounds. Fortunately, there are several proposals currently under
discussion. Thus several new experiments have been 
proposed based mainly on two techniques: the usual water Cherenkov 
detector and the use of noble gases, the Liquid 
Argon Time Projection Chamber (LAr TPC). 
The proposed future experiments based on the water Cherenkov detector 
are: the one-megaton HYPERK~\cite{Itow:2001ee,Nakamura:2003hk}, 
the UNO experiment~\cite{Jung:1999jq} with a 650 kt of 
water, while the experiment 3M~\cite{Diwan:2003uw} is proposed with a 1000 kt 
and the European megaton project MEMPHYS at Frejus~\cite{Mosca:2005mi}.

On the other hand the  ICARUS experiment~\cite{Rubbia:2004yq} is 
based on the Liquid Argon Time Projection Chamber (LAr TPC) 
technique.   
A  more ambitious proposal along similar lines  for proton decay and neutrino
oscillation study (LANNDD) is a 100 kT   liquid Argon TPC which is proposed
for the Deep Underground  Science and Engineering Laboratory (DUSEL) in
 USA~\cite{Cline:2005dm}.
Yet another proposal is of a Low Energy Neutrino Astronomy (LENA) detector consisting
of a 50 kt of liquid scintillator~\cite{Oberaurer:2005}. The LENA detector is suitable for 
SUSY favored decay channel $p\to \bar \nu K^+$ where the kaon will cause a prompt 
mono-energetic signal while the neutrino escapes without producing any detectable signal.
It is  estimated that within ten years of measuring time a lower limit of  
$\tau > 4 \times 10^{34}$ years can be reached~\cite{Oberaurer:2005}.
Basically all those proposals together with Super-Kamiokande 
define the next generation of proton decay experiments. 
These  experiments will either find proton decay or at the
very least  improve significantly the lower bounds and
eliminate many models. Thus, for example, the goal of 
Hyper-Kamiokande is to explore the proton lifetime at least up to 
$\tau_p/B(p \to e^+ \pi^0) > 10^{35}$ years 
and $\tau_p/B(p \to K^+ \bar{\nu}) > 10^{34}$ years 
in a period of about 10 years~\cite{Nakamura:2003hk}.   
Thus the next generation of proton decay experiments mark an important
step to probe the structure of matter at distances which fall
outside the realm of any current or future accelerator.

\section{Nucleon decay in non-supersymmetric scenarios}
As mentioned in Sec.(2) proton decay is a generic prediction of grand 
unified theories. 
There are different operators contributing to the nucleon decay 
in such theories. In supersymmetric scenarios the $d=4$ and 
$d=5$ contributions are the most important, but quite model dependent. 
They depend on the whole SUSY spectrum, on the structure of the Higgs
sector and on fermion masses. The so-called gauge $d=6$ contributions 
for proton decay are the most important in non-supersymmetric grand 
unified theories, which basically depend only on fermion mixing. 
The remaining Higgs $d=6$ operators coming from the Higgs sector 
are less important and they are quite model dependent, since we can have 
different structures in the Higgs sector. 
In this section we will study the stability of the proton in the
Standard Model, and the nucleon decay induced 
by the super-heavy gauge and Higgs bosons. The outline of the 
rest of this section is as follows: 
 In Sec. (3.1)  we  discuss the  B -violating effective  operators   induced by
instantons and estimate the proton lifetime arising from them. 
An analysis of $SU(3)_C\times SU(2)_L\times U(1)_Y$
invariant and $B-L$ preserving 
baryon and  lepton number violating dimension
six operators  induced by gauge  interactions is given in Sec.(3.2). 
Also discussed are the proton decay modes from these  interactions. 
$SU(3)_C\times SU(2)_L\times U(1)_Y$
\bl-violating  dimension six operators can also be induced by scalar
lepto-quark exchange and an analysis of these is given in
Sec.(3.3). We give below the details of these analyzes.

\subsection{Baryon number violation  in the Standard Model}

The Standard Model with gauge symmetry $SU(3)_C \ \times SU(2)_L \ \times U(1)_Y$
has a $U(1)_B$ global symmetry at the classical level, where $B$ is the baryon number,
which implies stability of the lightest baryon, i.e., the proton,  in the universe.
 However, this global symmetry is broken at the 
quantum level by anomalies~\cite{'tHooft:1976up}, i.e. 
the baryonic current $J^{\mu}_B$ is not conserved:  
\begin{equation}
\partial_{\mu} J^{\mu}_B = \frac{ n_f g^2}{16 \pi^2} Tr F_{\mu \nu} 
\tilde{F}^{\mu \nu}  
\end{equation}   
where $n_f$ is the number of generations and 

\begin{equation}
F_{\mu \nu}= \partial_{\mu} A_{\nu} - \partial_{\nu} A_{\mu} 
- i g [ A_{\mu}, A_{\nu}]
\end{equation} 
while 
\begin{equation}
\tilde{F}_{\mu \nu}= \frac{1}{2} \epsilon_{\mu \nu \alpha \beta} \ 
F^{\alpha \beta}
\end{equation}
With the above anomaly  baryon number violation can arise  from instanton transitions 
between degenerate $SU(2)_L$ gauge vacua. 
The B-violating effective operator induced by the  instanton processes is given by
(for details see, for example,~\cite{Espinosa:1989qn}):
\begin{equation}
O_{eff} = c  (\frac{1}{M_W})^{14}  e^{- \frac{2 \pi}{\alpha_2}} 
\prod_{i=1}^{3}(\ \epsilon_{\alpha \beta \gamma} \
 Q_{\alpha L}^i \ Q_{\beta L}^i \ Q_{\gamma L}^i \ L_L^i)
\end{equation}
where i is the generation index. The above interaction leads to violations of
baryon and lepton number so that $\Delta B=\Delta L= 3$.  We note, however, 
the front factor would give a rate  so  that 

\beqn
{\rm Rate} \sim  |e^{-\frac{2\pi}{\alpha_2}}|^2 \sim 10^{-173}
\eeqn
Clearly this is a highly  suppressed  rate irrespective  of other particulars. 
However,  baryon and lepton number violating dimension six and higher
operators can be written consistent with the Standard  Model gauge 
invariance. ~\cite{Weinberg:1979sa,Wilczek:1979hc,Sakai:1981pk}. 
This is the subject of discussion in the remainder of this
section.       

\subsection{Grand unification and gauge contributions to the decay of the proton}
We discuss now a  unifying  framework beyond that  of the Standard Model. There
 are many reasons for doing so.  One of the major ones is the presence  of 
 far  too many arbitrary parameters  in the  Standard Model
  and it is difficult to accept that a fundamental theory should be that arbitrary.
  One example of this is the presence of three independent gauge couplings:
  $\alpha_s$ for the color interactions, $\alpha_2$  for
  $SU(2)_L$,  and $\alpha_Y$ for  the gauge group $U(1)_Y$.
   This arbitrariness could be removed if one had  a 
  semi-simple gauge group, i.e., a grand unified group, with a single 
gauge coupling constant. Thus the  three  gauge coupling constants will be 
unified in such a scheme at a high  scale, but  would be split at low energy due to their different  renormalization group evolution from the grand unification scale to low 
  scales.   Of course,  the correctness of a  specific assumption of grand 
  unification must be tested by a detailed comparison of the predictions of 
  the unified model with the precision LEP data on the couplings. 
  Another virtue   of grand unification is that it leads to an understanding of the 
  quantization of charge, e.g.,  $|Q_e|=|Q_p|$, while such an explanation is missing
  in the Standard Model.  Additionally, grand unification reduces arbitrariness  
  in the Yukawa coupling sector, by relating Yukawa couplings for particles 
  that reside  in the common multiplets.  However, one important consequence of grand unification
  as  noted earlier is  that it leads generically  to proton decay. This arises from the fact that
   in  grand unified models  quarks and leptons fall in common multiplets  and thus
   interactions lead to processes involving  violations of baryon and lepton number.

In this subsection we focus on the non-supersymmetric contributions 
to the decay of the proton  (For an early review  of proton decay in non-supersymmetric
grand unification see Ref.~\cite{Langacker:1980js}).
In particular  we  study the gauge $d=6$ operators. 
Using the properties of the Standard Model fields we can write down 
the possible $d=6$ operators contributing to the decay of the proton, 
which are $SU(3)_C \times SU(2)_L \times U(1)_Y$ 
invariant~\cite{Weinberg:1979sa,Wilczek:1979hc,Sakai:1981pk}:
\begin{eqnarray}
\label{O1}
\textit{O}^{B-L}_I&=& k^2_1
\ \epsilon_{ijk} \ \epsilon_{\alpha \beta} 
\ \overline{u_{i a}^C}_L \ \gamma^{\mu} \ {Q_{j \alpha a}}_L   \
\overline{e_b^C}_L \ \gamma_{\mu} \ {Q_{k \beta b}}_L\\ 
\label{O2}
\textit{O}^{B-L}_{II}&=& k^2_1
\ \epsilon_{ijk} \ \epsilon_{\alpha \beta}
\ \overline{u_{i a}^C}_L \ \gamma^{\mu} \ {Q_{j \alpha a}}_L   \
\overline{d^C_{k b}}_L \ \gamma_{\mu} \ {L_{\beta b}}_L \\
\label{O3}
\textit{O}^{B-L}_{III}&=& k^2_2
\ \epsilon_{ijk} \ \epsilon_{\alpha \beta}
\ \overline{d_{i a}^C}_L \ \gamma^{\mu} \ {Q_{j \beta a}}_L   \
\overline{u_{k b}^C}_L \ \gamma_{\mu} \ {L_{\alpha b}}_L \\
\label{O4}
\textit{O}^{B-L}_{IV}&=& k^2_2
\ \epsilon_{ijk} \ \epsilon_{\alpha \beta}
\ \overline{d^C_{i a}}_L \ \gamma^{\mu} \ {Q_{j \beta a}}_L   \
\overline{\nu_b^C}_L \ \gamma_{\mu} \ {Q_{k \alpha b}}_L
\end{eqnarray}
In the above expressions $k_1= g_{GUT}/ {\sqrt 2} M_{(X,Y)}$, 
and $k_2= g_{GUT}/ {\sqrt 2} M_{(X^{'},Y^{'})}$, where 
$M_{(X,Y)}$, $M_{(X^{'},Y^{'})}$ $\approx M_{GUT}$ 
and $g_{GUT}$ are the masses of the superheavy gauge bosons
and the coupling at the GUT scale. The fields 
$Q_L= ( u_L,d_L)$, and $L_L= ( \nu_L, e_L)$. The indices $i$, $j$ 
and $k$ are the color indices, $a$ and $b$ are the family
indices, and $\alpha, \beta =1,2$. 
The effective operators $\textit{O}^{B-L}_I$ and
$\textit{O}^{B-L}_{II}$ (Eqs.~\ref{O1} and~\ref{O2}) 
appear when we integrate out the superheavy gauge fields 
$(X, Y)=({\bf 3},{\bf 2},5/3)$, where the $X$ and $Y$ fields have 
electric charge $4/3$ and $1/3$, respectively. 
This is the case in theories based on the gauge group $SU(5)$.
Integrating out $(X^{'}, Y^{'})=({\bf 3},{\bf 2},-1/3)$ we obtain the 
operators $\textit{O}^{B-L}_{III}$ and $\textit{O}^{B-L}_{IV}$ 
(Eqs.~\ref{O3} and~\ref{O4}), the electric charge of $Y^{'}$ is $-2/3$, 
while $X^{'}$ has the same charge as $Y$. This is the case of flipped 
$SU(5)$ theories~\cite{DeRujula:1980qc,Barr:1981qv,Derendinger:1983aj,Antoniadis:1989zy}, while 
in $SO(10)$ models all these superheavy fields are present. 
One may observe that all these operators conserve $B-L$, i.e. 
the proton always decays into an antilepton. 
A second selection rule $\Delta S/ \Delta B= -1,0$ is satisfied 
for those operators.

\begin{figure}[h]
  \centerline{\epsfxsize=5in \epsfbox{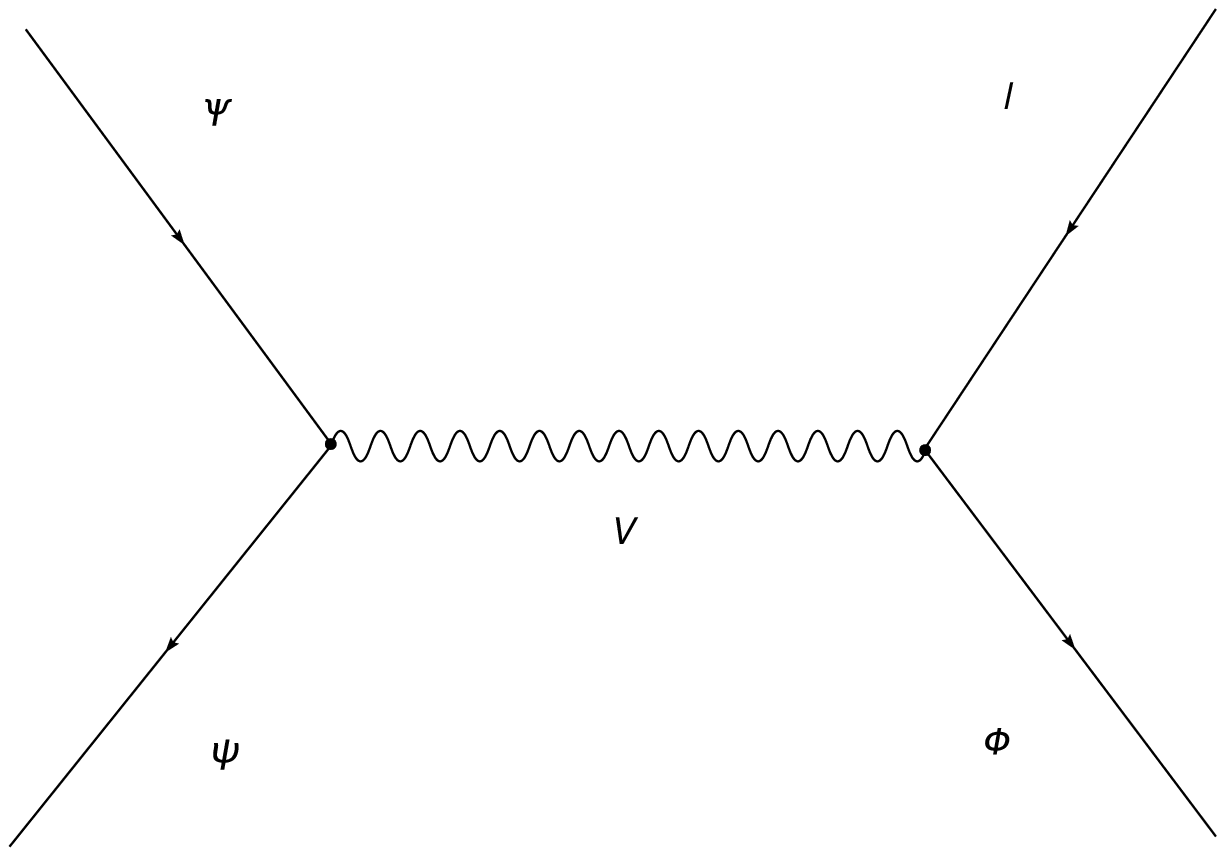}}
\caption{Gauge contributions to the decay of the proton. In this case 
the decay of the proton is mediated by vector leptoquarks. 
The fields $\Psi$, $\psi$ and $\Phi$ are quark fields and $l$ corresponds to the leptons. 
A possible contribution is: $\Psi=u_L$, $\psi=(u^C)_L$, $\Phi=(d^C_{\alpha})_L$ and $l=e_{\beta L}$. 
{\small $\alpha, \beta=1,2$}.}
\label{gauge}
\end{figure}

Using the operators listed above, we can write the effective operators
for each decay channel in the physical basis~\cite{FileviezPerez:2004hn}:

\begin{eqnarray}
\label{Oec}
\textit{O}(e_{\alpha}^C, d_{\beta})&=& c(e^C_{\alpha},
d_{\beta}) \ \epsilon_{ijk} 
\ \overline{u^C_i}_L \ \gamma^{\mu} \ {u_j}_L \ \overline{e^C_{\alpha}}_L \
\gamma_{\mu} \ {d_{k \beta}}_L \\
\label{Oe}
\textit{O}(e_{\alpha}, d^C_{\beta})&=& c(e_{\alpha}, d^C_{\beta}) \ \epsilon_{ijk} \ 
\overline{u^C_i}_L \ \gamma^{\mu} \ {u_j}_L \ \overline{d^C_{k \beta}}_L \
\gamma_{\mu} \ {e_{\alpha}}_L\\
\label{On}
\textit{O}(\nu_l, d_{\alpha}, d^C_{\beta} )&=& c(\nu_l, d_{\alpha}, d^C_{\beta}) \
\epsilon_{ijk} \ \overline{u^C_i}_L \ \gamma^{\mu} \ {d_{j \alpha}}_L
\ \overline{d^C_{k \beta}}_L \ \gamma_{\mu} \ {\nu_l}_L \\
\label{OnC}
\textit{O}(\nu_l^C, d_{\alpha}, d^C_{\beta} )&=& 
c(\nu_l^C, d_{\alpha}, d^C_{\beta}) \
\epsilon_{ijk} \
\overline{d_{i \beta}^C}_L \ \gamma^{\mu} \ {u_j}_L \ \overline{\nu_l^C}_L \
\gamma_{\mu} \ {d_{k \alpha}}_L
\end{eqnarray}
where:
\begin{eqnarray}
\label{cec}
c(e^C_{\alpha}, d_{\beta})&=& k_1^2 [V^{11}_1 V^{\alpha \beta}_2 + ( V_1 V_{UD})^{1
\beta}( V_2 V^{\dagger}_{UD})^{\alpha 1}] \\
\label{ce}
c(e_{\alpha}, d_{\beta}^C) &=& k^2_1  \ V^{11}_1 V^{\beta \alpha}_3 +  \ k_2^2 \
(V_4 V^{\dagger}_{UD} )^{\beta 1} ( V_1 V_{UD} V_4^{\dagger} V_3)^{1 \alpha}\\
\label{cnu} 
c(\nu_l, d_{\alpha}, d^C_{\beta})&=& k_1^2 \ ( V_1 V_{UD} )^{1 \alpha}
( V_3 V_{EN})^{\beta l} \nonumber \\
&+& \ k_2^2 \ V_4^{\beta \alpha}( V_1 V_{UD}
V^{\dagger}_4 V_3 V_{EN})^{1l} \\
\label{cnuc}
c(\nu_l^C, d_{\alpha}, d^C_{\beta})&=& k_2^2 [( V_4 V^{\dagger}_{UD} )^{\beta
 1} ( U^{\dagger}_{EN} V_2)^{l \alpha }+ V^{\beta \alpha}_4
 (U^{\dagger}_{EN} V_2 V^{\dagger}_{UD})^{l1}];  \nonumber \\ 
\alpha = \beta \neq 2.
\end{eqnarray}
In the above $V_1, V_2$ etc are mixing matrices defined so that 
 $V_1= U_C^{\dagger} U$, $V_2=E_C^{\dagger}D$,
$V_3=D_C^{\dagger}E$, $V_4=D_C^{\dagger} D$, $V_{UD}=U^{\dagger}D$,
$V_{EN}=E^{\dagger}N$ and $U_{EN}= {E^C}^{\dagger} N^C$, 
where $U,D,E$ define the Yukawa coupling diagonalization so that 

\begin{eqnarray}
U^T_C \ Y_U \ U &=& Y_U^{diag} \\
D^T_C \ Y_D \ D &=& Y_D^{diag} \\
E^T_C \ Y_E \ E &=& Y_E^{diag} \\
N^T \ Y_N \ N &=& Y_N^{diag}  
\end{eqnarray}    
Further, on may  write $V_{UD}=U^{\dagger}D=K_1 V_{CKM} K_2$,
where $K_1$ and $K_2$ are diagonal matrices containing three and two
phases, respectively. Similarly, 
 leptonic mixing $V_{EN}=K_3 V^D_l K_4$ in case
of Dirac neutrino, or $V_{EN}=K_3 V^M_l$ in the Majorana case, where 
$V^D_l$ and $V^M_l$ are the leptonic mixing at low energy in the Dirac and
Majorana case, respectively. The above analysis points up that the 
theoretical predictions of the proton lifetime from the gauge $d=6$ operators 
require a knowledge of the quantities
$k_1$, $k_2$, $V^{1b}_1$, $V_2$, $V_3$, $V_4$ and $U_{EN}$. In addition
we have three diagonal matrices containing phases, 
$K_1$, $K_2$ and $K_3$, in the case that the neutrino is Majorana.
In the Dirac case there is an extra matrix with two more phases. 
An example of the Feynman graphs for those contributions is 
given in Figure.~(\ref{gauge}). 
Since the gauge $d=6$ operators conserve $B-L$, 
 the nucleon decays into a meson and an antilepton.  
Let us write the decay rates for the different channels. 
Assuming that in the proton decay experiments one can not 
distinguish the flavor of the neutrino and the 
chirality of charged leptons in the exit
channel, and using the chiral Lagrangian techniques 
(see appendices), the decay rate of 
the different channels due to the presence 
of the gauge $d=6$ operators are given by:

\begin{displaymath}
\label{A1}
\Gamma(p \to K^+\bar{\nu})
		= \frac{(m_p^2-m_K^2)^2}{8\pi m_p^3 f_{\pi}^2} A_L^2 
\left|\alpha\right|^2 \times 
\end{displaymath}

\begin{equation}
\times \sum_{i=1}^3 \left|\frac{2m_p}{3m_B}D \ c(\nu_i, d, s^C) 
+ [1+\frac{m_p}{3m_B}(D+3F)] c(\nu_i,s, d^C)\right|^2 
\end{equation}

\begin{eqnarray}
\label{A2}
\Gamma(p \to \pi^+\bar{\nu})
		&=&\frac{m_p}{8\pi f_{\pi}^2}  A_L^2 \left|\alpha
		\right|^2 (1+D+F)^2 
\sum_{i=1}^3 \left| c(\nu_i, d, d^C) \right|^2\\
\label{A3}
\Gamma(p \to \eta e_{\beta}^+) 
		&=& {(m_p^2-m_\eta^2)^2\over 48 \pi f_\pi^2 m_p^3}
A_L^2 \left|\alpha \right|^2 (1+D-3 F)^2 \nonumber \\
&\times& \{ \left| c(e_{\beta},d^C)\right|^2 + \left|c(e^C_{\beta}, d)\right|^2 \} \nonumber \\ 
\\
\label{A4}
\Gamma (p \to K^0 e_{\beta}^+) 
		&=& {(m_p^2-m_K^2)^2\over 8 \pi f_\pi^2 m_p^3}  A_L^2
		\left|\alpha\right|^2 [1+{m_p\over m_B} (D-F)]^2 \nonumber \\
&\times& 
\{ \left|c(e_{\beta},s^C)\right|^2 +  \left|c(e^C_{\beta},s)\right|^2\} \nonumber \\ 
\\
\label{A5}
\Gamma(p \rightarrow \pi^0 e_{\beta}^+)
           &=& \frac{m_p}{16\pi f_{\pi}^2} A_L^2 \left|\alpha\right|^2
		(1+D+F)^2 \{ \left|c(e_{\beta},d^C)\right|^2 + 
		\left|c(e^C_{\beta},d)\right|^2 \} \nonumber \\ 
\end{eqnarray}
where $\nu_i= \nu_e, \nu_{\mu}, \nu_{\tau}$ and $e_{\beta}= e, \mu$.
In the above equations $m_B$ is an average Baryon mass satisfying $m_B \approx
m_\Sigma \approx m_\Lambda$, $D$, $F$ and $\alpha$ are the parameters
of the Chiral Lagrangian. $A_L$ takes into account renormalization 
from $M_Z$ to 1 GeV. (See the appendices  for details of the chiral 
lagrangian technique and the renormalization group effects). 
The analysis above indicates that it is possible to check on different
proton decay scenarios with sufficient data on proton decay modes if
indeed such a situation materializes in future proton decay experiment.

As we explained above the gauge $d=6$ contributions are quite model dependent. 
However, we can make a naive model-independent estimation for the 
mass of the superheavy gauge bosons using the experimental lower 
bound on the proton lifetime. Using 

\begin{equation}
\Gamma_p \approx \alpha_{GUT}^2 \ \frac{m_p^5}{M_V^4}
\end{equation}     
and $\tau (p \to \pi^0 e^+) \ > \ 1.6 \times 10^{33}$ years 
we find a naive lower bound on the superheavy gauge boson masses

\begin{equation}
M_V \ > \ (2.57 - 3.23) \times 10^{15} \ \ \textrm{GeV}
\end{equation}     
for $\alpha_{GUT}= 1/40 - 1/25$. Notice that this value 
tell us that usually the unification scale has to be 
very large in order to satisfy the experimental bounds.
\subsection{Proton decay induced by scalar leptoquarks}
\begin{figure}[h]
  \centerline{\epsfxsize=5in \epsfbox{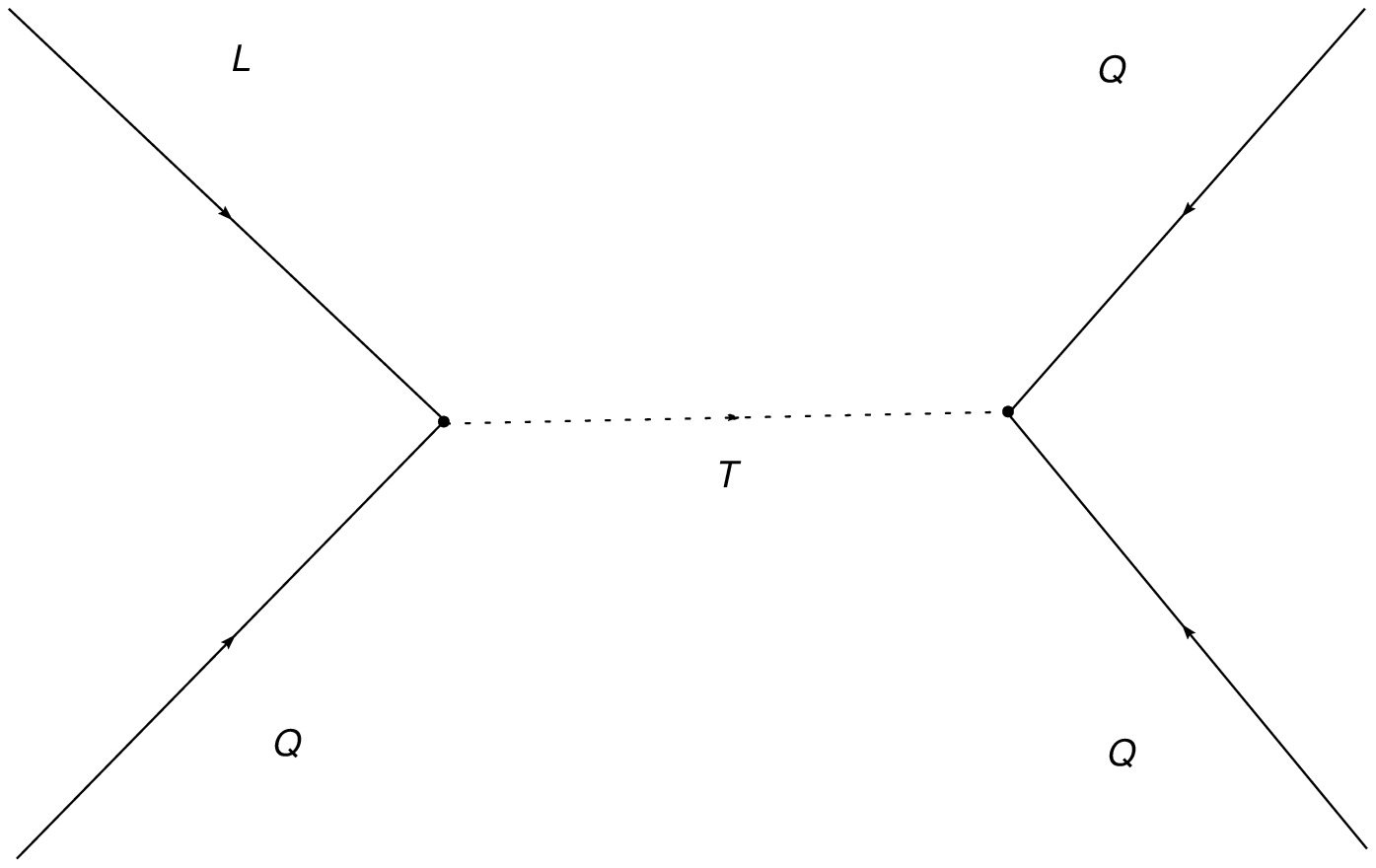}}
  \caption{Higgs contributions to the decay of the proton. 
In this case proton decay is mediated by scalar leptoquarks $T$. 
The fields $Q=(u_L, d_L)$ and $L=(\nu_L, e_L)$.}
\label{higgs}
\end{figure}

In non-supersymmetric scenarios the second most important contributions 
to the decay of the proton are the Higgs $d=6$ contributions. 
In this case proton decay is mediated by scalar leptoquarks $T=({\bf 3},{\bf 1},-2/3)$. 
Here, we will study those contributions in detail.
For simplicity, let us study the case when we have just one scalar 
leptoquark (See Figure.~(\ref{higgs}) for the Feynman graphs.). 
This is the case of minimal $SU(5)$. In this model the scalar 
leptoquark lives in the $5_H$ representation together 
with the Standard Model Higgs. The relevant interactions for 
proton decay are the following:

\begin{eqnarray}
V_T \ & = & \ \epsilon_{ijk} \ \epsilon_{\alpha \beta} \ Q_{i \alpha L}^T \ C^{-1} \ \underline{A} \ Q_{j \beta L} \ T_k  \ 
+ \ {u^C_{i L}}^T \ C^{-1} \ \underline{B} \ e^C_L \ T_i \nonumber \\ 
& + & \ \epsilon_{\alpha \beta} \ Q_{i \alpha L}^T \ C^{-1} \ \underline{C} \ L_{\beta} \ T_i^* 
\ + \ \epsilon_{ijk} \ {u^C_{iL}}^T \ C^{-1} \underline{D} \ d_{j L}^C \ T_i^* \ + \ h.c. \nonumber \\
\end{eqnarray}
In the above equation we have used the same notation as in the previous section.
The matrices $\underline{A}$, $\underline{B}$, $\underline{C}$ 
and $\underline{D}$ are a linear combination of the Yukawa 
couplings in the theory and the different contributions 
coming from higher-dimensional operators. In the minimal $SU(5)$, 
the have the following relations: $\underline{A}=\underline{B}=Y_U$, 
and   $\underline{C}=\underline{D}= Y_D=Y_E^T$.

Now, using the above interactions we can write the Higgs $d=6$ 
effective operators for proton decay

\begin{eqnarray}
\textit{O}_H (d_{\alpha}, e_{\beta})& = & a(d_{\alpha}, e_{\beta}) \ u^T \ L \ C^{-1} \ d_{\alpha} \ u^T \ L \ C^{-1} e_{\beta} \\
\textit{O}_H (d_{\alpha}, e_{\beta}^C)& = & a(d_{\alpha}, e_{\beta}^C)\ u^T \ L \ C^{-1} \ d_{\alpha} \ 
{e^C_{\beta}}^{\dagger} \ L \ C^{-1} {u^C}^* \\
\textit{O}_H (d_{\alpha}^C, e_{\beta})& = & a(d_{\alpha}^C, e_{\beta}) \ {d^C_{\alpha}}^{\dagger} \ L \ C^{-1} \ {u^C}^* 
\ u^T \ L \ C^{-1} e_{\beta} \\
\textit{O}_H (d_{\alpha}^C, e_{\beta}^C)& = & a(d_{\alpha}^C, e_{\beta}^C) \ {d^C_{\alpha}}^{\dagger} \ L \ C^{-1} \ {u^C}^* 
\ {e^C_{\beta}}^{\dagger} \ L \ C^{-1} {u^C}^*\\
\textit{O}_H (d_{\alpha}, d_{\beta}, \nu_i) & = & a(d_{\alpha}, d_{\beta}, \nu_i) \ u^T \ L \ C^{-1} \ d_{\alpha} \ d_{\beta}^T \ L \ C^{-1}\ \nu_i \\
\textit{O}_H (d_{\alpha}, d_{\beta}^C, \nu_i) & = & a(d_{\alpha}, d_{\beta}^C, \nu_i) \ {d^C_{\beta}}^{\dagger} \ L \ C^{-1} \ {u^C}^* 
\ d_{\alpha}^T \ L \ C^{-1}\ \nu_i  
\end{eqnarray}
where
\begin{eqnarray}
a(d_{\alpha}, e_{\beta}) &=& \frac{1}{M_T^2} \ (U^T ( \underline{A} + \underline{A}^T) D)_{1 \alpha} 
\ (U^T \underline{C} E)_{1 \beta}\\
a(d_{\alpha}, e_{\beta}^C) &=& \frac{1}{M_T^2} \ (U^T (\underline{A}+ \underline{A}^T) D)_{1 \alpha} 
\ (E_C^{\dagger} \underline{B}^{\dagger} U_C^*)_{\beta 1}\\
a(d_{\alpha}^C, e_{\beta}) &=& \frac{1}{M_T^2} \ (D_C^{\dagger} \underline{D}^{\dagger} U_C^*)_{\alpha 1} 
\ (U^T \underline{C} E)_{1 \beta}\\
a(d_{\alpha}^C, e_{\beta}^C) &=& \frac{1}{M_T^2} \ (D_C^{\dagger} \underline{D}^{\dagger} U_C^*)_{\alpha 1} 
\ (E_C^{\dagger} \underline{B}^{\dagger} U_C^*)_{\beta 1}\\
a(d_{\alpha}, d_{\beta}, \nu_i)&=& \frac{1}{M_T^2} (U^T ( \underline{A} + \underline{A}^T) D)_{1 \alpha} 
\ (D^T \underline{C} N)_{\beta i}\\
a (d_{\alpha}, d_{\beta}^C, \nu_i) &=&\frac{1}{M_T^2} \ (D_C^{\dagger} \underline{D}^{\dagger} U_C^*)_{\beta 1} \ 
(D^T \underline{C} N)_{\alpha i}
\end{eqnarray}
Here $L=(1- \gamma_5)/2$, $M_T$ is the triplet mass, $\alpha=\beta=1,2$ are $SU(2)$ and $i=1,2,3$ are $SU(3)$  indices. 
The above are the effective operators for the case of one Higgs triplet.  Often unified models have more than one
pair of Higgs  triplets as, for example, for the  case of  $SO(10)$ theories. In these cases we  need to go the
mass diagonal basis to derive  the baryon and lepton number violating dimension six operators by eliminating the heavy fields. 
The above analysis exhibits that the Higgs $d=6$ contributions are quite model dependent, and because of this 
it is possible to suppress them in specific models of fermion masses. For instance, we can 
set to zero these contributions  by the  constraints  
$\underline{A}_{ij}=-\underline{A}_{ji}$ and $\underline{D}_{ij}=0$, 
except for $i=j=3$. 

As we explained above the Higgs $d=6$ contributions to the decay 
of the proton are quite model dependent. However, we can make 
a naive model-independent estimation for the mass of the 
superheavy Higgs bosons using the experimental lower 
bound on the proton lifetime. Using 

\begin{equation}
\Gamma_p \approx |Y_u Y_d|^2 \ \frac{m_p^5}{M_T^4}
\end{equation}     
and $\tau (p \to \pi^0 e^+) \ > \ 1.6 \times 10^{33}$ years 
we find a naive lower bound on the superheavy Higgs boson masses

\begin{equation}
M_T \ > \ 3 \times 10^{11} \ \ \textrm{GeV}
\end{equation}     
Notice that this naive bound tell us that usually the 
Triplet Higgs has to be heavy. Therefore since the 
Triplet Higgs lives with the SM Higgs in the same multiplet 
we have to look for a Doublet-triplet mechanism.     
\section{Nucleon decay in SUSY and SUGRA unified theories}
  Supersymmetry in four space-time dimensions~\cite{Wess:1974tw,wb} arises algebraically from the 
"graded algebra" involving the spinor charge $Q_{\alpha}$ along with the
generators of the Lorentz algebra $P_{\mu}$ and $M_{\mu\nu}$.
Among the remarkable features of supersymmetry is the property that aside from some simple generalization,
the only graded algebra for an S-matrix one can construct from a local relativistic
field theory is the supersymmetric algebra~\cite{Haag:1974qh}. 
The above implies that supersymmetry 
appears as the only unique graded extension of a Lorentz covariant field theory. 
At the level of model building supersymmetric  models enjoy the advantage of a 
no-renormalization theorem~\cite{Iliopoulos:1974zv,Grisaru:1979wc} 
making the theory  technically natural. However, one
apparent disadvantage of supersymmetric theories is that proton stability is 
a priori more difficult relative to  case for non-supersymmetric theories 
since dangerous proton decay  arises from
dimension four and dimension five operators in addition 
to the proton decay induced by gauge bosons as in non-supersymmetric
theories. We  will first discuss proton decay from dimension
four operators which is considered the most dangerous as it can
decay the proton very rapidly. Later we  will discuss proton decay
from dimension five operators specifically in the context of GUT 
models based on $SU(5)$ and $SO(10)$~\cite{Mohapatra:1999vv}. 

In the following we assume that the reader has familiarity  with the basics of 
supersymmetry  and of the minimal supersymmetric  standard model (MSSM) 
which can be found in  a number of modern texts and reviews (see, e.g.,
~\cite{wb,Mohapatra:1999vv,Nilles:1983ge,Haber:1984rc,applied,recent_susy}).
Here, for completeness, we mention some salient features of MSSM as this model
is central to the discussion of low energy supersymmetry.  MSSM is based on the
gauge group $SU(3)_C\times SU(2)_L\times U(1)_Y$ with three generations of matter, 
and two pairs of Higgs  multiplets which are $SU(2)_L$ doublets ($H_1$ and $H_2$) 
where $H_1$ gives  mass to the  down quark and the lepton, and $H_2$ gives mass to the up quark.
Thus the gauge sector in addition to the gauge bosons of the Standard Model consists of
eight gluinos $\lambda_a$ (a=1,..,8), four  $SU(2)_L\times U(1)_Y$ electro-weak 
gauginos $\lambda^{\alpha}$ ($\alpha$=1,2,3) and $\lambda_Y$ which are all Majorana
spinors. Similarly,  in the matter sector MSSM consists in addition to the three generations 
of quarks and leptons, also their superpartners, i.e., three generations of squarks and sleptons.
In the Higgs sector one has in addition to the two pairs of $SU(2)_L$ Higgs doublets, also two
pairs of $SU(2)_L$  Higgsino multiplets.  The renormalizable superpotential in MSSM is  given by
\beq
W = \hat{U}^C Y_u \hat{Q} \hat{H}_u + \hat{D}^C Y_d \hat{Q} \hat{H}_d + \hat{E}^C Y_e \hat{L} \hat{H}_d + \mu \hat{H}_u \hat{H}_d + W_R
\eeq
where $Y_{u,d,e}$ are matrices in generation space and $W_R$ contains the R-parity violating terms 
which are given by 

\begin{equation}
W_{R} = \alpha_{ijk} \ {\hat Q}_i \ { \hat L}_j \ {\hat D}^C_k \ + \
\beta_{ijk} \ {\hat U}^C_i \
{\hat D}^C_j \ {\hat D}_k^C \ + \ \gamma_{ijk} \ {\hat L}_i \ {\hat
  L}_j \ 
{\hat E}_k^C \ + \ a_i \ {\hat L}_i \ {\hat H}_u
\label{rparityviolating}
\end{equation}
where  the coefficient $\beta_{ijk}$ and $\gamma_{ijk}$ obey the symmetry 
constraints  $\beta_{ijk}=- \beta_{ikj}$ and $\gamma_{ijk}= - \gamma_{jik}$.
In the above  we use the usual notation for the MSSM
superfields (see for example~\cite{Haber:1984rc}). The couplings of Eq.(\ref{rparityviolating})
violate $R$-parity 
 where $R$-parity is defined by $R=(-1)^{2S} M$, where $S$ is the spin
and $M=(-1)^{3(B-L)}$ is the matter parity, 
which is $-1$ for all matter superfields and $+1$ for Higgs and 
gauge superfields~\cite{Ibanez:1991pr}. In addition to $R$-parity violation, 
the second term of Eq.(\ref{rparityviolating}) 
violates the baryon number, while the rest of 
the interactions violate the leptonic number. These terms can be eliminated by 
the imposition of R-parity conservation, which requires that the overall
$R$-parity of each term is $+1$. 

The  outline of the rest of this section is as follows: 
In Sec.(4.1) we discuss the constraint on R parity violating interactions to suppress rapid proton 
decay from \bl  ~(B\&L) ~violating dimension four operators.  In addition to  B\&L ~violating dimension
four operators most  supersymmetric grand unified theories also have  B\&L  violating dimension
five operators which typically dominate over the B\&L  ~violating dimension six operators
which arise from gauge  interactions.
A computation of proton decay from dimension five operators  involves dressing of these
operators by  chargino, gluino and neutralino exchanges to convert
them to \bl ~violating dimension six operators. Such dressings depend on the sparticle
spectrum and thus on the nature of soft breaking.  With this in mind we give a brief 
discussion of supersymmetry breaking in Sec.(4.2).  Soft breaking is also affected by the
 CP  phases and thus proton decay is affected by the CP phases. This phenomenon is
discussed in Sec.(4.3).  In Sec.(4.4) a discussion of  Higgs doublet-Higgs triplet problem
is given. Since typically Higgs doublets and Higgs triplets  appear in common multiplets
a splitting to make Higgs doublets light and Higgs triplets heavy is essential to stabilize
the proton. 
Secs(4.5), (4.6) and (4.7) concern discussion of  specific grand unified models. Thus 
in Sec.(4.5) a discussion of $SU(5)$ grand unification is given, and a discussion of
$SO(10)$ grand unification is given in Sec.(4.6). In Sec.(4.7) we discuss a new class
of $SO(10)$ grand unified models based on a unified Higgs sector where a single
pair of $144+\overline{144}$ of Higgs can break the $SO(10)$ gauge symmetry all
the  way down to $SU(3)_C\times U(1)_{em}$. 
\subsection{R-parity violation and the decay of the proton}
It is interesting to ask what the constraints on the coupling structures are if
one does not impose R-parity invariance. Such constraints for the R-parity 
violating couplings from proton decay in low energy supersymmetry have been investigated for some 
time~\cite{Hall:1983id,Barbieri:1985ty,Ben-Hamo:1994bq,Hinchliffe:1992ad,Smirnov:1996bg,Smirnov:1995ey,Bhattacharyya:1998dt,Bhattacharyya:1998bx}. 
However, only recently the bounds coming 
from proton decay have been achieved taking into account  
flavor mixing and using the chiral lagrangian techniques~\cite{Perez:2004th} 
(For several phenomenological aspects of $R$ parity violating interactions see references
~\cite{Allanach:1999bf, Barbier:2004ez, Dreiner:1997uz}).  
Thus the first and the second terms in Eq.(\ref{rparityviolating})
give rise to tree level contributions to proton decay mediated by the 
${\tilde d}^C_k$ squarks. These are the most important contributions, which can be 
 used  to constrain the $R$-parity violating couplings. To extract these we
 write all interactions in the physical basis and  exhibit the proton decay 
 widths into charged leptons  using the chiral lagrangian method. 
 The rates for proton decay into charged anti-leptons are given by
 
\begin{eqnarray}
\Gamma(p \to \pi^0 e^{+}_{\beta}) &=& \frac{m_p}{64 \pi f_{\pi}^2} 
\ A^2_L \ |\alpha|^2 \ (1 + D + F)^2  \ |c(e_{\beta}^+, d^C)|^2 \\
\Gamma(p \to K^0 e^{+}_{\beta}) &=& \frac{(m_p^2 - m_K^2)^2}{32\pi
  f_{\pi}^2 m_p^3} \ A^2_L \ |\alpha|^2 \ [ 1 + \frac{m_p}{m_B} (D -F)]^2 
\ |c(e_{\beta}^+,s^C)|^2 \nonumber \\
\end{eqnarray}
where
\begin{eqnarray}
c(e_{\beta}^+, d^C_{\alpha})&=& \sum_{m=1}^3 \frac{(\Lambda_3^{\alpha m})^* 
\Lambda_1^{\beta m}}{m_{\tilde{d}^C_m}^2}  
\end{eqnarray}
Here $D$ and $F$ are the parameters of 
the chiral lagrangian, $\alpha$ is the matrix element, and $A_L$ 
takes into account the renormalization effects from $M_Z$ to 
$1$ GeV. In the case of the decay channels into 
antineutrinos, the decay rates are as follows~\cite{Perez:2004th}:
\begin{eqnarray}
\Gamma(p \to K^+\bar{\nu})
&=& \frac{(m_p^2-m_K^2)^2}{32\pi m_p^3 f_{\pi}^2} A_L^2
\left|\alpha\right|^2 \nonumber\\
&\times& \sum_{i=1}^3 \left|\frac{2m_p}{3m_B}D \ \tilde{c}(\nu_i, d, s^C)
+ [1+\frac{m_p}{3m_B}(D+3F)] \tilde{c}(\nu_i,s, d^C)\right|^2 \nonumber \\ \\
\Gamma(p \to \pi^+\bar{\nu})
&=&\frac{m_p}{32\pi f_{\pi}^2}  A_L^2 \left|\alpha
\right|^2 (1+D+F)^2
\sum_{i=1}^3 \left| \tilde{c}(\nu_i, d, d^C) \right|^2 \nonumber \\
\end{eqnarray}
where:
\begin{eqnarray}
\tilde{c}(\nu_l, d_{\alpha}, d^C_{\beta})&=& \sum_{m=1}^3 
\frac{(\Lambda_3^{\beta m})^* 
\Lambda_2^{\alpha l m}}{m_{\tilde{d}^C_m}^2}  
\end{eqnarray}
In the above equations the 
couplings $\Lambda_1$, $\Lambda_2$ and $\Lambda_3$ are given by~\cite{Perez:2004th}:
\begin{eqnarray}
\Lambda_1^{\alpha m} &=& \alpha_{ijk} \ U^{1i} \ E^{j \alpha} \ {\tilde
  D}_C^{km}\\
\Lambda_2^{\alpha l m} &=& \alpha_{ijk} \ D^{\alpha i} \ N^{jl} 
\ {\tilde  D}_C^{km}\\
\Lambda_3^{\alpha m} &=& 2 \beta_{ijk} \ U_C^{i1} \ D_C^{j \alpha} 
\ {\tilde  D}_C^{km} 
\end{eqnarray}

The most stringent constraints on $R$-parity violating couplings are obtained 
from the decays into charged leptons and mesons. 
Using $m_p=938.3$ MeV, $D=0.81$, $F=0.44$, $M_B=1150$ MeV,
$f_{\pi}=139$ MeV, $\alpha=0.003$ GeV$^3$, $A_L=1.43$ and 
the experimental constraints~\cite{Eidelman:2004wy} one finds 
\begin{eqnarray}
|c(e^{+}, d^C)|&<& 7.6 \times 10^{-31}\\
|c(\mu^{+}, d^C)|&<& 1.4 \times 10^{-30}\\
|c(e^{+}, s^C)|&<& 4.2 \times 10^{-30}\\
|c(\mu^{+}, s^C)|&<& 4.7 \times 10^{-30} 
\end{eqnarray}
Now, for simplicity assuming that all squarks have the same mass ${\tilde m}$, the
quantity $(\lambda_3^{\alpha m})^* \lambda_1^{\beta m}$ have to satisfy
the following relations~\cite{Perez:2004th}
\begin{eqnarray}
|(\lambda_3^{1m})^* \lambda_1^{1m}|&<& 3.8 \times 10^{-31} \ {\tilde m}^2 \\
|(\lambda_3^{1m})^* \lambda_1^{2m}|&<& 7.0 \times 10^{-31}  \ {\tilde m}^2 \\
|(\lambda_3^{2m})^* \lambda_1^{1m}|&<& 2.1 \times 10^{-30} \ {\tilde m}^2\\
|(\lambda_3^{2m})^* \lambda_1^{2m}|&<& 2.3 \times 10^{-30} \ {\tilde m}^2
\end{eqnarray}
where
\begin{eqnarray}
(\lambda_3^{\alpha m})^* \lambda_1^{\beta m} &=& 
\beta_{ijk}^* \ \alpha_{lpk} \ (U^{1i}_C)^* \ (D_C^{j \alpha})^* \
  U^{1l} \ E^{p\beta}  
\end{eqnarray}
It is easily seen that the constraints on $\alpha_{ijk}$ and
$\beta_{ijk}$ are quite model dependent i.e., they depend on the model for
the fermion masses that we choose. We can choose, for example, the
basis where the charged leptons and down quarks are diagonal, however
still $U_C$ will remain, and $U=K_1 V_{CKM}^{\dagger} K_2$.  $K_1$ and
$K_2$ are diagonal matrices containing three and two CP-violating
phases, respectively.  
In Table I we exhibit the different constraints for two
supersymmetric scenarios, i.e., in the low energy supersymmetry 
${\tilde m} = 10^3$ GeV and in scenarios with large scalar masses 
(split supersymmetry~\cite{Arkani-Hamed:2004fb,Giudice:2004tc} 
or hierarchical supersymmetry breaking~\cite{lands2})
the case ${\tilde m} = 10^{14}$ GeV. 

\begin{table}[h]
\begin{center}
\begin{tabular}{|r|r|r|}
\hline
\hline
 & & \\
\textbf{Couplings}&\textbf{Low energy SUSY}& ${\tilde m}=10^{14}$ GeV\\
 &  & \\ 
\hline
\hline
 & & \\ 
$|(\lambda_3^{1m})^* \lambda_1^{1m}|$ & $ 3.8 \times 10^{-25}$ & 0.0038\\
$|(\lambda_3^{1m})^* \lambda_1^{2m}|$ & $ 7.0 \times 10^{-25}$ & 0.0070\\
$|(\lambda_3^{2m})^* \lambda_1^{1m}|$ & $ 2.1 \times 10^{-24}$ & 0.0210\\
$|(\lambda_3^{2m})^* \lambda_1^{2m}|$ & $ 2.3 \times 10^{-24}$ & 0.0234\\ 
 & & \\
\hline
\hline
\end{tabular}
\end{center}
\caption{Upper bounds for the R-parity violating couplings.}
\label{tab:table2}
\end{table}
The analysis above shows that the $R$-parity 
violating couplings could be large in supersymmetric scenarios with
large susy breaking scale. In the case of SUSY breaking with low scale,
the $R$ parity violating couplings are  small, and this smallness  can
be construed as  a  hint  that $R$-parity is  an exact symmetry of 
a physical theory [See, for example,~\cite{Martin:1992mq,Aulakh:2000sn} 
for the possibility of  an $R$-parity as an exact symmetry 
arising from realistic grand unified theories.].

In the above  we have investigated the constraints from proton stability
 with explicit $R$-parity violation in the minimal supersymmetric 
 version of the Standard Model. One may now investigate  similar 
 constraints in unified  models such as in the simplest supersymmetric 
unified $SU(5)$ model~\cite{Dimopoulos:1981zb}. 
Here the $R$-parity violating interactions are
 $\Lambda^{ijk} \ \hat{10}_i \ \hat{\bar{5}}_j \ \hat{\bar{5}}_k$, $ b_i \ \hat{\bar{5}}_i
\ \hat{5}_H$ and $c_i \ \hat{\bar{5}}_i \ \hat{24}_H \ \hat{5}_H$. In this case at the GUT scale the
couplings satisfy the relations $\frac{\alpha_{ijk}}{2} = \beta_{ijk} = \gamma_{ijk} =
\Lambda_{ijk}= -\Lambda_{ikj}$.  These relations reduce the number of free parameters, 
and lead to a more constrained parameter space.
\subsection{Supersymmetry breaking and SUGRA unification}
Supersymmetric proton decay involves dressing of the \bl ~violating dimension 
five operators by gluino, chargino and neutralino exchanges which convert the dimension
five into dimension six operators.  The dressing loops depend on the masses of the exchanged
sparticles. Thus the prediction of proton lifetime depends in a central way on the soft parameters
which break   supersymmetry. One could in principle add soft parameters by hand to break 
supersymmetry at low energy.  In MSSM the number of such terms is rather 
large~\cite{Dimopoulos:1995ju}  consisting of
30 masses, 39 real mixing angles, and 41 phases,  a total of 110,  making the model
unpredictive. It is thus desirable to generate soft breaking via spontaneous breaking of the 
supersymmetric GUT model for a predictive theory much the same way one generates spontaneous
breaking of a non-supersymmetric GUT model. However, it is well known 
that the spontaneous breaking of global supersymmetry leads to patterns of 
sparticle masses which are typically in contradiction
with current experiment. Further, such a breaking leads to a vacuum energy which is in 
gross  violation of the observed value. For these  reasons a  globally supersymmetric
 grand unification is not a  grand unified theory that has any chance of consistency with
 experiment. These problems are closely associated with global supersymmetry and one needs
 to go to the framework of local-supersymmetry/supergravity~\cite{Nath:1975nj,Freedman:1976xh}
 to resolve them. Thus both of the hurdles mentioned above are  overcome within supergravity grand 
 unification~\cite{Chamseddine:1982jx}.  
 In  order to build models based on  supergravity one needs to use the techniques of applied 
 supergravity  where one couples N=1 supergravity with N=1 chiral multiplets
  and N=1 gauge multiplet belonging to the adjoint representation of the 
 gauge group~\cite{Chamseddine:1982jx,applied,Cremmer:1982en,Bagger:1983tt}.
 The effective $N=1$ applied supergravity lagrangian depends on three arbitrary functions:
 the superpotential $W(z_i)$, the Kahler potential $d(z_i,z_i^{\dagger})$, and the gauge 
kinetic energy function $f_{\alpha\beta}(z_i,z_i^{\dagger})$ where $\alpha, \beta$ are the 
adjoint representation indices, and where $W, d$ are gauge singlets, 
$f_{\alpha\beta}(z_i,z_i^{\dagger})$  is a gauge tensor, and 
$W, d, f_{\alpha\beta}(z_i,z_i^{\dagger})$ are hermitian. 
 The potential that results
 from such a theory is given by~\cite{Chamseddine:1982jx,Cremmer:1982en}
 \beqn
 V=e^{\kappa d}\left [ (d^{-1})^i_j (\frac{\partial W}{\partial z_i} +\kappa^2d_i W) 
 (\frac{\partial W}{\partial z_j} +\kappa^2d_j W)^{\dagger} 
  -3\kappa^2 |W|^2\right] + V_D
 \label{scalarpot}
 \eeqn
 where $\kappa=1/M_{\rm Pl}$ and $V_D$ is the D term contribution to the 
potential.  As may be seen from Eq.(\ref{scalarpot}) the scalar potential is
 no longer  positive definite. As a consequence it is possible to fine tune the vacuum energy to 
 zero after spontaneous breaking of supersymmetry. A remarkable  aspect of supergravity 
 formulation is that it is now possible to break supersymmetry spontaneously and still recover
  soft  parameters  which are phenomenologically viable.  To achieve this one postulates two
  sectors:  a hidden sector where supersymmetry is broken and a visible sector  where 
 fields of the physical sector reside. The only communication between  the two sectors occurs
 via gravity. 
 
 The simplest way to achieve the breaking of supersymmetry is through  a singlet scalar field
 with a superpotential of the form $W_h=m^2( z +B)$. Assuming a flat Kahler potential, i.e., 
 $d=zz^{\dagger}$,  a minimization of the potential then leads to the result 
 $<z>= \kappa^{-1} a(\sqrt 2- \sqrt 6), a=\pm 1 $. It is now seen that $<z>= O(M_{Pl})$.
 For this reason no direct interactions between the visible and the hidden sector are 
 allowed since they will lead to  sparticle masses   $O(M_{\rm Pl})$ in the 
 visible sector~\cite{Chamseddine:1982jx,Barbieri:1982eh}. 
 With communication
 between the two sectors arising only from gravitational interactions, the problem of
 large masses is avoided. Further, in the above example one can fine tune the vacuum
 energy to zero by setting $B=-\kappa^{-1} a(2\sqrt 2- \sqrt 6) $.  The above phenomenon 
 is in fact  a super Higgs effect where after spontaneous breaking the fermionic  partner of
 the graviton becomes massive by absorbing the fermionic partner of the chiral field z.
 It has a mass which is given by
 \beqn
 m_{{3}/{2}}=\frac{1}{2}  |<W(z)>|  exp({\frac{\kappa^2}{4} <Z>^2})
 \eeqn
 The above leads to a  gravitino mass of $m_{\frac{3}{2}} \sim \kappa m^2$ and implies
 that an $m\sim 10^{10-11}$ GeV will lead to   $m_{\frac{3}{2}}$
  in the electroweak region~\cite{Chamseddine:1982jx,Barbieri:1982eh}. 
 A realistic model building involves a decomposition of the superpotential so that 
 $ W=W_h(z)+ W_v(z_i)$ so that the hidden sector superpotential $W_h$ depends
 only on the gauge singlet field z while the visible sector superpotential $W_v$
 depends only the visible sector fields $z_i$ and has  no dependence 
 on z~\cite{Chamseddine:1982jx,Barbieri:1982eh}. 
Integrating out the hidden sector then leads to soft parameters in the visible sector. 
For the case of supergravity grand unification an extra complexity arises because 
of the presence of the grand unification scale $M_G$.  The  appearance of such
 a scale in the soft parameters would throw the sparticle spectrum out of the electroweak
 region.  Quite  remarkably it is shown that the grand unification scale cancels out of 
 the soft parameters~\cite{Chamseddine:1982jx}.
 
 We now summarize the conditions under which the soft breaking in the minimal 
supergravity model are derived. These consist of  
 (i) The hidden sector is assumed  a gauge singlet which breaks  super-symmetry
 by a super Higgs  effect;
(ii) There is no direct interaction between the hidden sector and the visible sector 
except for gravity so the communication of breaking to the visible sector occurs
only via gravitational interactions; 
(iii) The Kahler potential is assumed to have no generational dependence;
(iv) The cubic and higher field dependent parts of the  gauge kinetic energy function $f_{\alpha\beta}$ 
are  assumed negligible. Thus effectively  $f_{\alpha\beta}\sim \delta_{\alpha\beta}$. 
Under  these assumptions it is then found that the scalar  potential is of the 
form~\cite{Chamseddine:1982jx,Hall:1983iz,Nath:1983aw}
\beqn
-{\cal L_{SB}} =m_{\frac{1}{2}} \bar \lambda^{\alpha} \lambda^{\alpha} 
+ m_0^2 z_az_a^{\dagger} + (A_0W^{(3)}+B_0W^{(2)} +h.c.)
\label{softlag}
\eeqn
 where for the case of MSSM one has
 \beqn
 W^{(2)}= \mu_0 H_1 H_2;~~
 W^{(3)}= \tilde Q Y_{U}  H_2 \tilde u^c + \tilde Q Y_{D}  H_1 \tilde d^c +
 \tilde L Y_{E}  H_1 \tilde e^c
 \eeqn
(We note that in the appendices we use $H_1=H_d$, and $H_2=H_u$.)  
Now a remarkable aspect  of soft breaking is that  it leads to spontaneous breaking of 
 the electroweak symmetry~\cite{Chamseddine:1982jx}. 
 This is most efficiently achieved by radiative breaking of the
 electroweak symmetry by renormalization group
  effects~\cite{Inoue:1982pi,Ibanez:1982fr,Alvarez-Gaume:1983gj,Ellis:1983bp,Ibanez:1983di,Ibanez:1984vq}. To exhibit this  consider the
 effective scalar potential. The renormalization group improved scalar potential for
 the Higgs fields is  given by
 \begin{eqnarray}
 V&=&m_1^2|H_1^2| +m_2^2 |H_2|^2 -m_3^2 (H_1H_2+h.c.) \nonumber\\ 
&+& \frac{(g_2^2+g_Y^2)}{8} (|H_1|^2-|H_2|^2)^2 + \Delta V_1,\nonumber\\
 \Delta V_1&=& (64\pi^2)^{-1} 
 \sum_a (-1)^{2s_a}(2s_a+1) M_a^4 \left [ ln \frac{M_a^2}{Q^2}- \frac{3}{2}\right ] 
 \end{eqnarray}
where $s_a$ is the spin of the particle a, $\Delta V_1$ is  
the one loop correction~\cite{Coleman:1973jx,Weinberg:1973ua}
to the
effective potential, and all parameters, i.e., $g_2, g_Y, m_i$ etc are  running parameters
evaluated  at the scale $t=ln(M_G^2/Q^2)$ where   Q  is taken to be  in the electro-weak region. 
The boundary conditions on these  parameters are~\cite{9309277} 
$\alpha_2(0)= \alpha_G= \frac{5}{3} \alpha_Y(0);$
$m_i^2(0)=m_0^2+\mu_0^2, ~i=1,2;$ and $m_3^2(0)= - B_0\mu_0$.
Now  $SU(2)_L\times U(1)_Y$ electro-weak symmetry breaks  when the determinant  
of the Higgs mass$^2$ matrix turns negative and further one requires  that the potential
be bounded from below for a valid minimum to exist. Thus one requires the constraints
on the  Higgs  parameters  so that
$(i) ~~m_1^2 m_2^2 -2m_3^4<0,$  and $(ii)  ~m_1^2+m_2^2 -2|m_3^2|>0$, 
where the first constraint indicates that the determinant  of the Higgs  mass$^2$ matrix
turns negative while the second constraint  requires the potential to  be  bounded  from
below.  Minimization of the potential, i.e.,  $\partial V/\partial v_i=0$  where 
$v_i =<H_i>$ is the VEV of the neutral component of the Higgs $H_i$, gives two
constraints
  \begin{eqnarray}
(a)~~ M_Z^2&=&2(\mu_1^2-\mu_2^2\tan^2\beta)(\tan^2\beta -1)^{-1},\nonumber\\ 
(b)~~\sin 2\beta &=& 2m_3^2(\mu_1^2 +\mu_2^2)^{-1}
 \end{eqnarray}  
Here $\mu_i^2=m_1^2+\Sigma_i$ where $\Sigma_i$ is the loop 
correction~\cite{Gamberini:1989jw,Arnowitt:1992qp} 
 and $\tan\beta= v_2/v_1$.
The electroweak symmetry breaking constraint (a) can be used to fix $\mu$
using the experimental value of  the Z  boson mass $M_Z$, and the
 constraint (b) can be utilized to eliminates  $B_0$ in favor of $\tan\beta$.  
  Thus the supergravity model at low energy can be parametrized by  
 \beqn
 m_0, m_{1/2}, A_0, \tan\beta, sign (\mu)
\label{para1}
 \eeqn
 The number of soft parameters in the minimal supersymmetric standard
 model allowed by  the ultra-violet behavior of the
 theory~\cite{Girardello:1981wz}  is quite large and thus the result of Eq.(\ref{para1})
  is a big improvement.
    While the assumption of a super Higgs effect using a scalar field  is the simplest 
  way to break supersymmetry, there 
  are other ways such as gaugino condensation~\cite{nilles,taylortr}
   where one can accomplish a similar 
  breaking. Non-perturbative  effects are needed to produce such a condensate 
  which makes the condensate analysis more  difficult. However,  if the gaugino 
condensate~\cite{nilles} does occur the gravitino mass generated by such a condensate will be of 
  size  $m_{\frac{3}{2}} \sim \kappa^2  <\lambda \gamma^0 \lambda>$. 
  In this case the condensate  $ |< \lambda \gamma^0 \lambda >| \sim (10^{12-13})$ GeV 
  will lead  to an  $m_{\frac{3}{2}} $  again in the electro-weak region. 
  Further,   the result of Eq.(\ref{para1})
 arises from certain simple assumptions about the nature  of the Kahler potential
 and on the gauge kinetic energy function that were stated in the paragraph preceding
 Eq.(\ref{softlag}). 
 On the other hand, the nature of the Kahler potential in supergravity is 
 determined by the physics at the Planck scale of which we have as yet not
 a firm grasp. For this reason it is reasonable to explore deformations of the
 Kahler potential from the flat Kahler potential limit, i.e.,
  consider non-universalities~\cite{Soni:1983rm,Kaplunovsky:1993rd}.
 One possibility is  to consider non-universalities in the  Higgs sector, 
 and in the third generation sector and also allow for non-universalities  in the
 gaugino sector  by allowing for field dependent gauge kinetic energy function
 $f_{\alpha\beta}$. For instance, non-universalities for the 
  Higgs boson masses at the GUT scale arising from deformations of the
  Kahler potential will lead 
  to~\cite{Matalliotakis:1994ft,Olechowski:1994gm,Polonsky:1994rz,Nath:1997qm}
  \beqn
  m_{H_i}(0)=m_0(1+\delta_i),  ~i=1,2
  \eeqn
  For the case  of non-universalities an additional correction  term   arises at low 
  energies in the renormalization group evolution~\cite{Martin:1993zk}, i.e., 
  \beqn
  \Delta m_{H_1}^2 =-\frac{3}{5} S_0p, ~~ \Delta m_{H_2}^2 =- \frac{3}{5} S_0p 
    \eeqn
  where $S_0$ is given by 
  \beqn
  S_0=Tr(Ym^2) = m_{H_2}^2-m_{H_1}^2 
  +\sum_{i=1}^{n_g} (m_{\tilde Q}^2-2m_{\tilde u}^2+ m_{\tilde d}^2  
 -  m_{\tilde L}^2  + m_{\tilde e}^2 )\nonumber\\
 \eeqn 
 Here all the masses are taken at the GUT scale, and $n_g$  is the number of 
  generations and p is defined by 
 $ p=\frac{5}{66} [1-  (\frac{\alpha_1(t)}{\alpha_1(0)})]$, 
 where $\alpha_1(0)=g_1^2(0)/4\pi$ is the $U(1)$ gauge coupling constant
 at the GUT scale.  The $Tr(Ym^2)$ term vanishes for the universal case but
 contributes in the presence of non-universalities~\cite{Martin:1993zk}.  
 Similarly, non-universalities can be introduced in the third generation sector.  
 \\
 An important aspect of SUGRA models is the possibility of  realizing radiative breaking
 of the electroweak symmetry on the so called hyperbolic branch (HB)~\cite{Chan:1997bi}. 
  To see how this
 comes about we consider the radiative symmetry breaking constraint expressed in 
 terms of the soft parameters only 
 \beqn
C_1m_0^2+C_3m'^2_{1/2}+C_2'A_0^2+\Delta \mu^2_{loop}=
	\mu^2+ M_Z^2/2 
\label{hb1}
\eeqn
where $m_{1/2}'=m_{1/2}+\frac{1}{2}A_0C_4/C_3$, and $C_1$ etc
are determined purely in terms of gauge and Yukawa couplings,
and $\Delta\mu^2_{loop}$ is the loop correction~\cite{Arnowitt:1992qp}.    
The correction $\Delta\mu^2_{loop}$  plays an important role as it 
controls  the behavior of radiative breaking specially for moderate to 
large values of $\tan\beta$. 
To see this phenomenon we note that the coefficients $C_2'$, $C_3$ are positive
and the loop corrections are typically small for small $\tan\beta$ when
 $Q=M_Z$. In this case one finds that $C_1>0$ and thus Eq.(\ref{hb1})
implies that the soft parameters lie on the surface of an ellipsoid.  
However, as $\tan\beta>5$ the loop correction
 $\Delta\mu^2$ becomes sizable  and also
  $C_1(Q)$ develops a significant Q dependence. 
  One may then choose a  Q value where  $\Delta\mu^2$ is 
  minimized. Quite remarkably then one finds that 
  $C_1(Q_0)$ turns negative. The implications of this switch in sign
  means that the soft  parameters can get large while $\mu$ 
  remains fixed.  Thus if one thinks of $\mu/M_Z$ as  the fine tuning
  parameter, then in this case the switch in sign implies that for 
  a  fixed fine tuning, the soft parameters  lie on the surface of 
   a hyperboloid. This is the hyperbolic branch of radiative breaking
   of the electroweak symmetry  and this branch does not  limit the
   soft parameters stringently the way the ellipsoidal branch did~\cite{Chan:1997bi}.
  The  so called focus point region~\cite{Feng:1999zg} is included in
  the hyperbolic branch~\cite{Chan:1997bi,Baer:2003ru}.

There are several novel phenomena  that occur  on the  hyperbolic 
branch. Thus as  $m_0$ and $m_{\frac{1}{2}}$  get large  with $\mu$ 
remaining relatively small,  the  light chargino becomes higgsino like  while the
lightest neutralino  and the next to the lightest
neutralino become  degenerate and also
essentially higgsino like.  Typically the following
pattern of masses  emerges when  $m_0$ and $m_{1/2}$ 
get large on  HB~\cite{ccnwmap}:
$m_{{\tilde \chi}_1^0}< m_{{\tilde \chi}_1^{\pm}}<m_{{\tilde \chi}_2^{0}}$.
This relation holds  at the tree level and there 
could be important loop corrections to this relation. 
 The mass differences 
 $\Delta M^{\pm} = m_{{\tilde \chi}_1^{\pm}}- m_{{\tilde \chi}_1^0}$
 and $\Delta M^{0} = m_{{\tilde \chi}_2^{0}}- m_{{\tilde \chi}_1^0}$ 
 depend significantly on the location on HB. For the
 deep HB region with large $m_0$ and $m_{1/2}$ and small $\mu$
  these mass differences will be 
 typically  small, ie., O(10) GeV.  The implications for
 such a scenario are many. Thus   
 the usual missing energy signals 
 in the decay of the chargino and in other sparticle decays 
 would not work as in the usual SUGRA scenario which implies
 that one must look for alternative signals to search for supersymmetry
 on the hyperbolic branch in this region.  
 Quite 
 remarkably dark matter constraints can be satisfied on HB. 
 Since  $m_0$ is typically large on  HB,  with $m_0$ becoming
 as large as 10 TeV, in the deep HB region, proton lifetime
 is enhanced  significantly especially in the deep HB region.
The HB region is essentially like the split SUSY scenario which 
is discussed  elsewhere in this report  in greater depth. 
 There are also a variety of other approaches to supersymmetry breaking. 
 Chief among these is the gauge mediated breaking.  The reader is directed
  to recent reports for reviews~\cite{Giudice:1998bp,Luty:2005sn}.
  
  An interesting issue concerns the origin of  $\mu$. For phenomenological reasons
   we expect  $\mu$  to be of  electroweak size.  The challenge to achieve  a
   $\mu$ of  electroweak size 
   while the other scales  appearing in  the theory are $M_G$ and
   $M_{\rm Pl}$ is the so called $\mu$ problem. One possibility is that such a term in
   absent in the theory for the case of unbroken supersymmetry and arises only as 
   a consequence of breaking of supersymmetry. In this circumstance a term appearing
   in the Kahler potential of the form $H_1H_2$ can be transferred by a Kahler transformation
   into the superpotential and a $\mu$ term  naturally appears in the superpotential  
  which is of size the  weak supersymmetry breaking 
  scale~\cite{Chamseddine:1982jx,Soni:1983rm,Giudice:1988yz}.  There are indications
  that a term of the form $H_1H_2$ can arise in 
  string theory~\cite{Antoniadis:1994hg, Nath:2002nb}.
  Another issue of theoretical interest concerns the stability of the weak- scale hierarchy.
A potential danger arises from non-renormalizable couplings in supergravity models
since they can lead to power law divergences which can destabilize the hierarchy.
This  problem has been  investigated at one loop~\cite{Bagger:1993ji,Gaillard:1994sf}
and at two loops~\cite{Bagger:1995ay}. At the one loop  level the 
minimal supersymmetric standard model appears to be  safe  from divergences~\cite{Bagger:1993ji}.
At the two loop level divergences can appear when the visible sector is directly coupled
to the hidden sector where supersymmetry breaking occurs~\cite{Bagger:1995ay}. 
 We end this section by directing the reader to Appendix D where the mass matrices 
 for the sparticles are discussed
 since these matrices enter in the computation of the  dressing diagrams
 for the dimension five operators. 
\subsection{Effect of CP violating phases on proton lifetime}   
CP  phases  affect proton lifetime. As is well known the CP phase that appears in the SM via the 
CKM matrix is not sufficient to generate the desired amount of baryon asymmetry in the universe. 
 Here supersymmetry is helpful.
  The soft breaking sector of \s ~provides a new source of CP violation.  
    This new source of CP violation arises from the soft breaking sector of  supergravity
   and  string theory models.  
 Usually  this type of CP violation is called explicit CP violation.  
   If we allow for explicit CP violation, then the parameter space of 
 mSUGRA allows for two phases which can be chosen to be the phase of
 $\mu_0$  and the phase of the trilinear coupling parameter $A_0$.
Including these  the parameter space  of mSUGRA for the complex case  is

\beqn
 m_0, m_{1/2}, A_0, \tan\beta, \theta_{\mu_0}, \alpha_0
\label{para2}
 \eeqn
 where $\mu_0=|\mu_0|exp(i\theta_{\mu_0})$, and 
 $A_0=|A_0|exp(i\alpha_0)$.  
  For the case of non-universal sugra model  one also has more
 CP violating phases.  These phases can arise in the trilinear 
parameters and in the gaugino sector. 
 Thus more generally we will have phases in addition to $\theta_{\mu}$ 
  so that
 \beqn
 m_i=|m_i| e^{i\xi_i} ~(i=1,2,3); ~~A_f=|A_f|e^{i\alpha_{A_f}}, ~f=flavor
 \eeqn
  where $m_i$ (i=1,2,3) are the gaugino masses for $SU(3)_C\times SU(2)_L\times U(1)_Y$
 gauge sectors. 
 Not all the phases are independent and only certain combination of them appear
 after field redefinitions.  As indicated already in the context of CP phases  in 
 the Standard Model  one  needs to  make certain that the  
 constraints from experiment on the  electric dipole moments (edm)   of elementary 
 particles are satisfied.  Currently the most sensitive  experimental limits  are for
 the   edm of the  electron, of the neutron and of the $^{199}Hg$ atom. 
The  current limits on these are~\cite{Regan:2002ta,Harris:1999jx,Romalis:2000mg}
\beqn
|d_e| < 2\times 10^{-27} \ \textrm{ecm}, ~~|d_n| < 6\times 10^{-26} \ \textrm{ecm}, ~~|d_{Hg}| < 2\times 10^{-28} \ \textrm{ecm}
\eeqn  
Now one approach to satisfy these  constraints in supersymmetric theories  is  to simply  assume the
CP phases  to be small~\cite{ellis}.  In this circumstance the CP phases play no role in the 
supersymmetry phenomenology and  have no effect on the proton lifetime.  
However, as pointed out already one needs  a new source of CP violation for generating baryon asymmetry
in the universe and from that view  point it  is  useful to  have the possibility   that at least 
one or more of the SUSY phases are large.
 Now it turns out that  there are a variety of ways
in which one  can have large CP phases in \s ~and consistency with experiment on the  
edm~\cite{Nath:1991dn,dimo,bagger,Chang:1998uc}.
One such possibility is  mass suppression where one may have  large sparticle masses 
especially for the first  two generations.  In this case some of the sparticles but not all 
would have to   be  massive with masses lying  in the TeV range. For instance the 
heaviness of the sfermions   for the first two generations  will guarantee 
 the  satisfaction of the edms while the gluino, the chargino and the neutralino 
could be light enough to be accessible at the LHC. This is precisely the situation 
that arises  also on the hyperbolic  branch (HB) of radiative  breaking of the electroweak
symmetry. 

Another is the intriguing possibility for the suppression of the edmss~\cite{incancel}.
In \s ~ there are 
three different types  of contributions to the edm of the elementary particles. These
  arise  from the  electric-dipole operator, the  chromoelectric dipole 
operator and from the purely gluonic dimension six operator of 
Weinberg~\cite{Weinberg:1989dx}.
 In general these  operators receive contributions from the gluino, 
 from the chargino,  and from the neutralino  exchanges.   Now in certain 
 arrangement of phases there are cancellations among the contributions from the
 gluino, from the chargino and from the neutralino exchanges, as well as 
 among the contributions from the electric dipole,
 from the chromoelectric dipole and from the purely gluonic dimension six  
 operators.   These allow the reduction of the edms of the electron, of the neutron and 
 of the $^{199}Hg$ atom below their current experimental limits (for further 
 developments see 
 Refs.~\cite{Brhlik:1998zn, Falk:1998pu,Bartl:1999bc,Pokorski:1999hz,Accomando:1999uj,Brhlik:1999ub,Abel:2001mc}).
  Additionally,  it turns out
 that there is a scaling which approximately preserves  the smallness of the edms as one  
 scales  in $m_0$ and $m_{1/2}$ by a common factor.   Thus with the help of 
 scaling, given a point in the parameter space where the edm is small  one 
 can generate  a trajectory where the edms remain small ~\cite{Ibrahim:1999af}. 
 Using this procedure one can generate a region in the moduli space where the phases
 are large and the  edms are within the current  experimental bounds. 
 
The presence of large CP  phases affect all the supersymmetric phenomena.
As an example the phases will lead to a mixing of the CP even and the  CP odd 
 Higgs bosons~\cite{Pilaftsis:1998dd}  which makes 
 the Higgs boson and dark matter searches  more interesting and more intricate. 
 The inclusion of CP phases also  has  an effect on the proton lifetime. To see
 this we note that the inclusion of phases in the gaugino masses and in 
 the parameter $\mu$ affect the chargino, the neutralino, the  squark and the
 slepton  mass matrices.  Thus, for example, with the  inclusion of phases the
 chargino mass matrix is 
\beq
M_C=\left(\matrix{|M_2|e^{i\xi_2} & \sqrt 2 m_W \sin\beta \cr
	\sqrt 2 m_W \cos\beta &| \mu|e^{i\theta_{\mu}} }
            \right)
\label{chmatrix}
\eeq
which  can be diagonalized  by the following  biunitary  transformation 
\beq
U^* M_C S^{-1}={\rm diag}(m_{{\tilde \chi}_1^+},  m_{{\tilde \chi}_2^+} )
\label{usmatrices}
\eeq
where U and S are unitary matrices.  To exhibit the sensitivity of the 
 chargino masses   on the phases we note that 
\beqn 
 m_{{\tilde \chi}_1^+}^2  m_{{\tilde \chi}_2^+}^2 = |\mu M_2|^2 + M_W^4\sin^2 (2\beta) 
 -2 |\mu M_2| M_W^2 \sin (2\beta) \cos(\theta_{\mu}+\xi_3)
 \label{chamass}
 \eeqn
The last term in Eq.(\ref{chamass})  
changes sign as $(\theta_{\mu}+\xi_3)$ varies from 
0 to $\pi$ which exhibits the sharp phase dependence of the chargino masses. 
 Consequently the chargino 
propagators that enter in the dressing of the \bl ~violating dimension five operators
 are sensitive to the CP phases. 
  A similar situation holds for other sparticle exchanges in the
 dressing loops, e.g., the  neutralino and  the squark exchanges etc. 
Thus, for example, the up-squark mass matrix  in the presence of phases 
 becomes  
\begin{displaymath}
M_{\tilde{u}}^2=\left(\matrix{M_{\tilde{Q}}^2+m{_u}^2 + M_{Z}^2(\frac{1}{2}-Q_u
s^2_W)\cos2\beta & m_u(A_{u}^{*} - \mu \cot\beta) \cr
   	          	m_u(A_{u} - \mu^{*} \cot\beta) & 
m_{\tilde{u}}^2+m{_u}^2 + M_{Z}^2 Q_u s^2_W \cos2\beta}
		\right) \nonumber
\end{displaymath}
 where $\mu$ and $A_{u}$ are complex. Consequently,  the  squark masses 
dependent on the phases. The phase  dependence can be  quite  significant similar
 to the phase dependence for the chargino  case discussed above. 
 CP phases also enter in the fermion-sfermion- gaugino vertices. The dependence
 there arises from the diagonalizing matrices, i.e.,  from 
 U and S matrices   that appear in Eq.(\ref{usmatrices}) and similar matrices arising from the 
 diagonalization of the  squark sector.
The above are the  two main avenues by which the CP phases enter  proton decay,
i.e., via modifications of the sparticle masses and via the vertices. 
The effects of these modifications can be included by following the 
standard  procedure where one expresses  the squark and slepton fields in terms 
of their sources. Thus, for example, one can write 
\beqn
\tilde u_{iL}=2\int [\Delta_{ui}^L \frac{\delta \it L_I}{\delta \tilde u_{iL}^{\dagger}},
+\Delta_i^{LR}\frac{\delta \it L_I}{\delta \tilde u_{iR}^{\dagger}}]
\nonumber\\
\tilde u_{iR}=2\int [\Delta_{ui}^R\frac{\delta \it L_I}{\delta \tilde u_{iR}^{\dagger}}
 +\Delta_i^{RL} \frac{\delta \it L_I}{\delta \tilde u_{iL}^{\dagger}}].
\eeqn
where $L_I$  contains all the  fermion-sfermion-chargino,  fermion-sfermion-neutralino,
 and fermion-sfermion-gluino  interactions.  In the above $\Delta_{ui}^L$,
$ \Delta_{ui}^R$,  $\Delta_{ui}^{LR}$,  $\Delta_{ui}^{RL}$ 
 are linear combinations of the propagators for the mass eigen states. For the 
 CP conserving case  one has $\Delta_{ui}^{LR} =\Delta_{ui}^{RL}$, but 
 is no longer the case when CP violating phases are  present, and  in the presence  
 of CP phases  $\Delta_{ui}^{LR} \neq\Delta_{ui}^{RL}$.
  This is yet another way  in which CP violating effects enter in the dressing loop function. 
 Of  course as pointed out above  the propagators  for the mass  eigen states 
as well as the vertices are also  dependent on the phases.   

In addition to the above, CP  phases  can modify drastically the nature  of interference
involving different generations in the dressing loops.
Specifically,  for supersymmetric proton decay  the major  contributions arise from
the dressing loops involving the second and the third generations.  The phases
define the relative strength with which they interfere, and with appropriate choice
of phases  a constructive interference  can become  destructive interference suppressing
the dressing loop. This is one of the  ways  in which  the proton lifetime can be enhanced. 
The above  analysis shows that phases do affect  proton lifetime and
the effects can be  quite significant.  An analysis of proton lifetime 
with the inclusion of phases is given 
in Ref.~\cite{Ibrahim:2000tx} where it is found that the CP phases 
that enter via the dressing loops can affect the proton 
lifetime estimates  by much as a factor of 2 or even more.
\subsection{Doublet-triplet splitting problem}
One of the main issues in GUT  model building is the doublet-triplet splitting. 
Thus in the simplest $SU(5)$ model one has two  Higgs multiplets
$5_H$ and $\bar 5_H$ and the simplest  scheme to affect doublet-triplet splitting
is via fine tuning where  one takes  the following combination 
\beqn
W_{G} =\lambda_1[\frac{1}{3} \Sigma^3 +\frac{1}{2} M \Sigma^2] 
+\lambda_2 H_{2}[\Sigma + 3M']H_1
\eeqn
where $\Sigma$ is a 24-plet of Higgs whose VEV formation breaks $SU(5)$ and  
where M is of  size  $M_G$.  Now minimization of the effective potential generates  
a VEV for the  $\Sigma$ field and assuming that  the VEV formation breaks 
$SU(5)\to SU(3)_C\times SU(2)_L\times U(1)_Y$ one has
\beqn
<\Sigma^i_j>= M ~diag(2,2,2,-3,-3) 
\eeqn
A fine tuning $M'=M$ then makes the Higgs doublets light while Higgs triplets  are 
supermassive with masses of order the GUT scale if $M$ is of size $M_G$.  
 There  are  alternate  possibilities where one can  avoid a fine tuning in order to recover
  light Higgs doublets. One  well known mechanism for this is the 
  missing partner mechanism~\cite{Grinstein:1982um,Masiero:1982fe} 
  where one replaces the 24-plet of Higgs  with $50,\overline{50}, 75$ Higgs representations.
  Consider for instance  a Higgs sector  of the form 
  \beqn
  W_G'= \lambda_1 50^{ijk}_{Hlm} 75^{lm}_{Hij}H_{2k} + \lambda_2 \overline{50}^{ij}_{Hklm}
  75^{lm}_{Hij}H_{1}^k + W_G"(75_H) 
  \label{wgprime}
  \eeqn
  Let us assume that the scalar potential generated by $W_G"(75_H)$ supports a VEV formation for
  the  $75$ plet field with  $<75>\sim M$. Inserting this VEV growth in the rest of $W_G'$ one
  finds  that the Higgs triplets become supermassive while the Higgs doublets remain light. 
  To see  this  more  clearly let  us look at the  $SU(3)_C\times SU(2)\times U(1)$ content
  of $50$ plet  representation 
  
\begin{displaymath}
50_H=(1,1,-12)+ (3,1,-2)+ (\bar 3,2, -7) + (\bar 6, 3,-2) + 
\end{displaymath}
\begin{equation}
(6,2, -7) + ( 8, 2, 3)  + (15, 1,-2). 
 \label{50h}
\end{equation}
    Quite  remarkably one finds that  there is no $SU(2)$ -doublet-color-singlet  in the  above
    and similar is the case for  $\overline{50}_H$. Thus the VEV formation of $75$  plet and 
    breaking of the $SU(5)$ symmetry leave a pair of light  Higgs  doublets coming  from
    $5_H$ and $\bar 5_H$. On the other hand one finds that Eq.(\ref{50h}) contains  a Higgs  
    color triplet $(3,1,-2)$ which can tie  up with the color anti-triplet from $H_{2}$ making
    them supermassive.   Thus in this fashion the color triplets and  anti-triplets  from
    $H_1^i$ and $H_{2i}$ become  superheavy while the Higgs doublets  remain light.
   There are a variety of other avenues for splitting the doublets from the  triplets.

  An interesting possibility for realizing light Higgs iso-doublets without the
  necessity of fine tuning   arises in 
   $SU(6)$ ~\cite{Berezhiani:1989bd}.
Thus consider an $SU(6)$ grand unification where the Higgs  sector of the theory  
consists of a $35$-plet field  $\Sigma$ and a pair of $6(H)$ and $\bar 6(\bar H)$ multiplets.
In particular consider the superpotential in the Higgs sector so that: 
\begin{equation}
W =  M  Tr \Sigma^2 \ + \ h Tr \Sigma^3 + \rho Y ( \overline{H} H - \Lambda^2)
\label{supersu6}
\end{equation}
where $Y$ is an auxiliary $SU(6)$ singlet field. This model has a 
global $SU(6)_{\Sigma} \times U(6)_H$ symmetry. The superpotential of
Eq.(\ref{supersu6}) can lead to spontaneous breaking of this symmetry
with VEV formation of the  $\Sigma$, 
$H$, and  $\overline{H}$ fields such that
\begin{equation}
<\Sigma> = V_{\Sigma} \ diag(1, 1, 1, 1, -2, -2)
\end{equation}
and
\begin{equation}
<H >^T = <\overline{H}>^T= V_H \ (1, 0, 0, 0, 0, 0)
\end{equation}
where $V_{\Sigma}= M / h$, and $V_H= \Lambda$.  Here $<H >$, and 
$<\overline{H}>$ break $SU(6)$ down to $SU(5)$, while $<\Sigma>$ breaks
$SU(6)$ down to $SU(4) \times SU(2) \times U(1)$, 
which together lead to the breaking of the local $SU(6)$ symmetry 
down to  residual gauge group symmetry $SU(3)_C \times SU(2)_L \times U(1)_Y$. 
At the same time the global symmetry $SU(6)_{\Sigma} \times U(6)_H$ 
is broken down to $[SU(4) \times SU(2) \times U(1)]_{\Sigma} \times U(5)_H$. 
All the Goldstones bosons are eaten by the $SU(6)/SU(3)_C \times SU(2)_L \times U(1)_Y$
coset gauge bosons which become super-heavy, and only a pair of Higgs doublets remain massless. 
These are the pseudo-Goldstone bosons which are identified as the MSSM Higgs doublets.  
The matter sector of $SU(6)$ consists of three families each containing  
($\overline{6} + \overline{6}^´ + 15$), and one $20$-plet.of matter. These
have the following $SU(5)$ decompositions
\begin{center}
\begin{eqnarray}
20 = 10 + \overline{10}= ( q \ + \ u^C \ + \ e^C )_{10} \ + \ 
( Q^C \ + \ U \ + \ E )_{\overline{10}} \\
15 = 10 \ + \ 5 = ( q \ + \ u^C \ + \ e^C )_{10} \ + \ ( D \ + \ L^C )_{5} \\
\overline{6} = \overline{5} \ + \ 1 = ( d^C \ + \ l )_{\overline{5}} \ + \ n \\
\overline{6}^´ = \overline{5}^{'} \ + \ 1^{'} 
= ( D^C \ + \ L )_{\overline{5}^{'}} \ + \ n^{'} 
\end{eqnarray}
\end{center} 
As in $SU(5)$  supersymmetric grand unified model this model also contains
baryon and lepton number violating dimension five operators and one needs
a mechanism to suppress them. An investigation of proton decay in this class
of models is  given in  Ref.~\cite{Shafi:1999tn}. 
  The doublet-triplet  splitting in the context of  SO(10) will be discussed in 
   Sec.(\ref{sub_so10}), and for the
   case of models with extra dimensions in Sec.(\ref{sec_extradim}). 
\subsection{Proton decay in $SU(5)$ supersymmetric grand unification}
The decay of the proton in the minimal SU(5) model is governed by

\begin{equation}
W_Y=-\frac{1}{8}f_{1ij}\epsilon_{uvwxy}H_1^u10_i^{vw}10_j^{xy}+
f_{2ij}\bar H_{2u}\bar 5_{iv} 10_j^{uv}
\end{equation}
where $\bar 5_{ix}$ and $10_i^{xy}$ (i=1,2,3) are the
$\bar 5$ and 10 of SU(5) which contain the three
generations of quarks and leptons, and
$H_1, H_2$ are the $\bar 5$,5 of Higgs, and $f's$ are the Yukawa couplings.
After the breakdown of the GUT symmetry there is a splitting of the
Higgs multiplets where the Higgs triplets become super-heavy and
the Higgs doublets remain light by one of the mechanisms discussed in Sec.(4.4).
 One can now integrate on the Higgs triplet field and
obtain an effective interaction at low energy which contains baryon and lepton
number  violating dimension five operators with chirality LLLL and RRRR such
that
\begin{eqnarray}
\it W(LLLL)&=& \frac{1}{M} \epsilon_{abc}(Pf_1^uV)_{ij}(f_2^d)_{kl}
( \tilde u_{Lbi}\tilde d_{Lcj}(\bar e^c_{Lk}(Vu_L)_{al}-
\nu^c_kd_{Lal})+...)\nonumber\\  &+& H.c. \nonumber\\
W(RRRR)&=& -\frac{1}{M} \epsilon_{abc}(V^{\dagger} f^u)_{ij}(PVf^d)_{kl}
(\bar e^c_{Ri}u_{Raj}\tilde u_{Rck}\tilde d_{Rbl}+...) \nonumber\\
&+& H.c. 
\label{LLLLRRRRR}
\end{eqnarray}
where  V is the CKM matrix and $f_i$, $P_i$ are generational phases
\begin{equation}
P_i=(e^{i\gamma_i}), ~\sum_i \gamma_i=0; ~i=1,2,3
\label{genphases}
\end{equation}
 Both LLLL and RRRR interactions must be taken into account in a full
 analysis and their relative strength depends on the part of the parameter
 space where their effects are computed.
The operators  of Eq.(\ref{LLLLRRRRR}) are dimension five operators which
must be dressed via the  exchange of
gluinos, charginos and neutralinos. The dressings give
rise to  dimension six operators. A partial analysis of the 
dressing loops was given in Refs.~\cite{Dimopoulos:1981dw,Ellis:1981tv}, 
and a full analysis was first given in Refs.~\cite{Nath:1985ub,Arnowitt:1985iy}
and worked on further in Refs.~\cite{Hisano:1992jj,Goto:1998qg,Fukuyama:2004xs}.
These dimension six operators are
then used in the computation of proton decay.
In the dressings one takes into account the L-R mixings, 
where, the mass diagonal states for sfermions are related to the
chiral left and right states by a unitary transformation. After
dressing of the dimension 5 by the gluino, the chargino 
and the neutralino
exchanges one finds baryon and lepton number violating  
dimension six operators with chiral structures  LLLL, LLRR, RRLL and RRRR
in the Lagrangian.  In the minimal SU(5) model the 
 dominant decay modes of the proton involve pseudo-scalar bosons and
anti-leptons, i.e.,  
\begin{eqnarray}
\bar\nu_iK^+,\bar\nu_i\pi^+,                                               
e^+K^0,\mu^+K^0,e^+\pi^0,\mu^+\pi^0,
e^+\eta,\mu^+\eta;  ~ i=e,\mu,\tau
\end{eqnarray}
The  relative strengths of these decay modes depend on various
factors, such as quark masses, CKM  factors,
and the third generation effects in the loop diagrams which are parametrized by $y^{tk}_1$
etc.  The various decay modes and some of the factors that control
these decays  modes are summarized in table below. 
$$  $$
\begin{center} \begin{tabular}{|c|c|c|}
\multicolumn{3}{c}{Table:~Leptonic decay modes of the proton } \\
\hline
Mode & quark factors  & CKM factors \\
\hline
$\bar \nu_eK$  &~~~~$V_{11}^{\dagger}V_{21}V_{22} $ ~~~~& $m_dm_c$  \\
\hline
$\bar \nu_\mu K $ &$V_{21}^{\dagger}V_{21}V_{22}$ & $m_sm_c$   \\
\hline
$\bar \nu_\tau K $  &$V_{31}^{\dagger}V_{21}V_{22} $ & $m_bm_c$  \\
\hline
~~~~$\bar \nu_e \pi,\bar \nu_e \eta $ ~~~~&$V_{11}^{\dagger}V_{21}^2 $& $m_d m_c$    \\
\hline
$\bar \nu_\mu \pi,\bar \nu_\mu \eta$ &$V_{21}^{\dagger}V_{21}^2$ & $m_sm_c$ \\
\hline
$\bar \nu_\tau \pi,\bar \nu_\tau\eta $ &$V_{31}^{\dagger}V_{21}^2 $& $m_bm_c$  \\
\hline
$eK $ & $V_{11}^{\dagger}V_{12} $&  $m_dm_u$  \\
\hline
$\mu \pi,\mu \eta $ & $V_{11}^{\dagger}V_{21}^{\dagger}$&$m_sm_d$   \\
\hline
\hline
\end{tabular}
\label{leptonicmodes}
\end{center}
$$  $$
The  order of magnitude estimates can be gotten by keeping in mind
$m_uV_{11}<<m_cV_{21}<<m_t V_{31}$.
In general the   most dominant mode is $\bar\nu K$  in the minimal
 supersymmetric $SU(5)$ model. 
In the analysis below we will ignore the mixings among the neutrinos, a good
approximation for a detector with size much smaller than the neutrino oscillation
length. In this approximation the chargino exchange contributions involving the
second generation to this decay is~\cite{Nath:1985ub} 
\begin{eqnarray}
\Gamma(p\rightarrow\bar\nu_iK^+)&=&\frac{\beta_p^2m_N}{M_{T}^232\pi f_{\pi}^2}
(1-\frac{m_K^2}{m_N^2})^2|{\cal A}_{\nu_iK}|^2 A_L^2(A_S^L)^2
|(1+\frac{m_N(D+F)}{m_B})|^2  \nonumber \\ &&
\end{eqnarray}
where $\beta_p$ is defined by Eq.(\ref{matrixelements}) and where we have  used a 
subscript p to distinguish it from the $\beta$ in $\tan\beta$ and where
\beq 
{\cal A}_{\nu_iK}=(\sin 2\beta M_W^2)^{-1}\alpha_2^2P_2m_cm_i^dV_{i1}^{\dagger}
V_{21}V_{22}[{I}(\tilde c;\tilde d_i;\tilde W)+
 { I}(\tilde c;\tilde e_i;\tilde W)]
\eeq
Here $I(\tilde c;\tilde d_i;\tilde W)$ are dressing loop functions as defined in Ref.~\cite{Nath:1985ub}.
 Further, one can take into account 
the contribution of the third generation exchange via corrections parametrized by
  $y_i^{tk}$ where~\cite{Nath:1985ub}
\begin{equation}
y_i^{tK}=\frac{P_2}{P_3}(\frac{m_tV_{31}V_{32}}{m_c V_{21}V_{22}})
(\frac{I(\tilde t,\tilde d_i,\tilde W)+I(\tilde t,\tilde e_i,\tilde W)}
{I(\tilde c,\tilde d_i,\tilde W)+I(\tilde c,\tilde e_i,\tilde W)})
\end{equation}
Here  $P_2$ and $P_3$ are the relative intrinsic
parities of the second and the third generation as defined by Eq.(\ref{genphases}). 
The ratio $P_2/P_3$ is a relative
phase factor which can generate a constructive or a destructive interference between
the second generation and the third generation contributions. An enhancement of the
proton lifetime can occur by a destructive interference and the  maximum destructive
interference occurs when $P_2/P_3=-1$.  Similarly one can take into account the
gluino and the neutralino exchange contributions  to the dressing loops.  Thus, for  example,
the gluino  exchange contribution can be parametrized by $y_{\tilde g}$ where~\cite{Nath:1985ub}

\beqn
y_{\tilde g}= \frac{P_1}{P_2} \frac{\alpha_3}{\alpha_2} \frac{m_u V_{11}}{m_c V_{21} V_{21}^{\dagger} V_{22}}
\frac{ H(\tilde u; \tilde d:\tilde g)-H(\tilde  d:\tilde d;\tilde g)}{ I(\tilde c;\tilde s;\tilde W) +I(\tilde c;\tilde  \mu;\tilde W)}
\label{gluinoexchange}
\eeqn
where I and H are  loop functions
as defined in Ref.~\cite{Nath:1985ub}. It is now easily seen that 
 the gluino contribution given by Eq.(\ref{gluinoexchange})
 vanishes when the $\tilde u$ and $\tilde d$
squarks are  degenerate.  

In general the contributions of both the LLLL and the RRRR dimension five operators
to the proton decay amplitudes  are important 
and their relative contributions  vary  depending on the part of the parameter space one
is in.  Specifically, for example, the RRRR dimension five operators can make a
significant contribution to the $\bar \nu_{\tau}K$ mode. The important  contribution of
the RRRR operators was first observed in Ref.~\cite{Nath:1985ub} and later also
noted in Ref.~\cite{Goto:1998qg,Lucas:1996bc,Babu:1998wi,Fukuyama:2004pb}.
 Further, the relative  contributions of the dressing loop can modify the
relative  strength of the partial decay widths. Thus consider the situation
where the third generation contribution cancels approximately the
second generation contribution in the $\bar \nu K^+$ mode.
In this case the subdominant mode  $\bar \nu \pi^+$  will be relatively  enhanced
and become comparable to the $\bar \nu K^+$ 
mode~\cite{Nath:1985ub,Arnowitt:1985iy}.
 In addition to the nucleon decay modes involving pseudo-scalar
bosons and anti-leptons, one also has in general decay modes
involving vector bosons and anti-leptons. The source of these
modes are the same baryon number violating dimension six quark operators
that  give rise to the decay modes that give rise to
pseudo-scalar and anti-lepton modes.
The vector decay modes of the proton are
\begin{eqnarray}
\bar\nu_iK^*,\bar\nu_i\rho,\bar\nu_i\omega, 
e K^*,\mu K^*,e\rho,\mu\rho,e\omega,\mu\omega; ~i=e,\mu,\tau
\end{eqnarray}
A chiral lagrangian analysis of these  modes 
is carried out in Ref.~\cite{Kaymakcalan:1983uc}.
However, the vector meson decay modes have generally smaller branching 
ratios than  the corresponding pseudo-scalar decay modes.  An analysis of these  vector
boson decay modes for the supergravity $SU(5)$ model is given 
in Ref.~\cite{Yuan:1986km}. Another interesting mode is $p\to e^+\gamma$. While
this mode would be suppressed by a factor of $\alpha$, it has some interesting features
in that it is a relatively clean mode  free of strong final state interactions
and nuclear absorption. An estimate of the decay rate is given in Ref.~\cite{Silverman:1980ha}. 
A more recent analysis of  this decay mode is  given in Ref.~\cite{ngamma}. A closely related 
process is the decay of the bound  neutron so that~\cite{ngamma}.
\beqn
n\to \gamma\bar \nu
\eeqn
This decay is interesting since the anti-neutrino will escape detection 
in the detector and the only visible signal will be just a photon of energy 
about  half  a  GeV~\cite{ngamma}. A estimate of the  lifetime here 
gives $10^{38\pm 1}$ yr.  
\\
The issue of viability of the supersymmetric grand 
unification and specifically of the 
minimal supersymmetric  $SU(5)$ has recently been 
analyzed~\cite{Dermisek:2000hr,Murayama:2001ur}. 
The work of Ref.~\cite{Murayama:2001ur} which is focused 
on the minimal $SU(5)$ model analyzed the dual  constraints
arising from gauge coupling unification and 
proton partial lifetime limits for the
 $\bar \nu K^+$  mode and found them to  be incompatible. Thus according to
the work of  Ref.~\cite{Murayama:2001ur}
 gauge coupling unification in the minimal supersymmetric
$SU(5)$ constrains the Higgs triplet mass to lie in the range
\beqn
3.5\times 10^{14}\leq M_{T}\leq 3.6\times 10^{15} ~~{\rm GeV}
\label{triplet1}
\eeqn
at  the $90\%$ confidence level. Using the partial lifetime 
lower limit on  the  $\bar \nu K^+$ mode of $6.7\times 10^{32}$ 
yr (the  current limit for this mode is $>2.3\times 10^{33}$ yr) they find a lower
limit on the Higgs triplet mass of~\cite{Murayama:2001ur}
\beqn
M_{T} \geq 7.6\times 10^{16} ~~{\rm GeV}
\label{triplet2}
\eeqn
The above led the authors of Ref.~\cite{Murayama:2001ur} to conclude that 
the minimal supersymmetric $SU(5)$ is ruled out. However, as is well-known 
the minimal supersymmetric $SU(5)$ is not a realistic model since the relation 
between fermion masses are not in agreement with experiment. 

There are a number of ways in which the incompatibility 
of Eq.(\ref{triplet1}) and Eq.(\ref{triplet2}) can be overcome. 
Thus for example, the addition of Planck scale corrections can drastically alter 
the picture~\cite{Ellis:1979fg,Nath:1996qs}. 
An analysis along these lines with inclusion of higher dimensional 
operators~\cite{Bajc:2002bv,Bajc:2002pg,Emmanuel-Costa:2003pu}, 
crucial for fermion masses, and inclusion of mixings between 
fermion and sfermions is carried out in 
references~\cite{Bajc:2002bv,Bajc:2002pg}. The work of
Refs.~\cite{Bajc:2002bv,Bajc:2002pg} concludes that the 
uncertainty in the theoretical predictions is as 
much as  $10^3$ or even larger for the minimal model to be ruled out 
when modifications of the above type are included 
(For an earlier analysis of uncertainties in the prediction of proton decay
lifetime in the context of non-supersymmetric grand unification 
see Ref.~\cite{Ellis:1980jm}.). The constraint of Eq.(\ref{triplet2}) on the 
SU(5) model can be significantly softened if the Higgs sector at the GUT scale
contains higher dimensional operators. Thus, for example, if the superpotential
in the Higgs sector contains operators of the  $Tr(\Sigma^2)^2/M_{\rm Pl}$ and 
$Tr(\Sigma^4)/M_{\rm Pl}$,  then the gauge coupling unification and the Higgs
triplet constraints can be reconciled more easily in certain regions  of the 
parameter space of the Higgs potential. One consequence  of the 
addition of higher dimensional operators  is to  generate a splitting in
the GUT masses of $\Sigma_3$ and $\Sigma_8$.  This splitting turns
out to be rather useful in softening the constraints on the SU(5) GUT model.
Specifically, in reference~\cite{Bajc:2002pg}, an explicit analysis shows that it is 
possible  to satisfy the bound on $M_T$  from proton decay once the splitting 
between the masses of the fields $\Sigma_3$ and $\Sigma_8$ is taken into 
account. As  pointed out above  such a splitting is quite natural when 
higher-dimensional operators are included in the Higgs sector.

There are additional ways in which one can find compatibility of gauge coupling
unification and  the KamioKande lower limits on the proton lifetime.  
For example, presence of additional matter in the desert between $M_Z$ and $M_G$ 
could increase  the Higgs triplet mass removing the constraint. Another 
possibility is to enhance the proton lifetime by fine tuning
or by a discrete symmetry if there are additional Higgs triplet
fields present~\cite{Arnowitt:1993pd}.
Thus, for example, with many Higgs  triplet fields the
proton decay inducing  dimension five operators are governed by the interaction
\begin{equation}
\bar T_1J+\bar K T_1+\bar T_iM_{ij}T_j
\end{equation}
In the above we have made a redefinition of fields so that the Higgs  triplet and
anti-triplet that couple with matter are labeled  
$T_1, \bar T_1$,  while J and $\bar K$ are matter currents,
 and $M_{ij}$ is the Higgs triplet mass matrix.  A suppression of
proton   decay in these theories can be engineered
if~\cite{Arnowitt:1993pd}

\begin{equation}
(M^{-1})_{11}=0
\label{m11suppression}
\end{equation}
A suppression of this type can occur in the presence of many Higgs triplet 
fields  by a discrete symmetry, or
by a non-standard embedding~\cite{Arnowitt:1993pd,Gomez:1998zf}.
Another possibility for the suppression of proton decay is via gravitational smearing effects
discussed in Sec.(5.2).
\subsection{Nucleon decay in SO(10) theories
\label{sub_so10}}
The SO(10) is an interesting group in that  a single spinor representation of SO(10) can
accommodate  a full generation of quarks and leptons.  Thus the 16-plet of SO(10) has
the following decomposition in terms of SU(5)  
\begin{equation}
16=10+\bar 5+ 1
\end{equation}
where the $\bar 5$ and $10$ plets accommodates the full set of one generation of quarks
and leptons and in addition on has the singlet field  which is a right handed neutrino 
  needed for  generating See-Saw masses for the neutrinos.  One, of course, must 
break the SO(10) gauge symmetry down to $SU(3)_C\times SU(2)_L\times U(1)_Y$ and
further break $SU(2)_L\times U(1)_Y$ down to $U(1)_{em}$. Now a combination of 
$45_H$ and a $16_H+\overline{16}_H$ can break the symmetry down to the Standard  Model
gauge group symmetry. Further, a 10-plet of Higgs gives the two $SU(2)_L$ doublets of 
Higgs that are needed to break $SU(2)_L\times U(1)_Y$ down to $U(1)_{em}$. 
Thus a $45$, $16_H+\overline{16}_H$ and a $10$ plet of Higgs are a minimal set that
is needed to break $SO(10)$ down to $SU(3)_C\times U(1)_{em}$. Now the Higgs content 
of a model is determined not only by the requirement that the $SO(10)$ gauge group 
completely breaks down to $SU(3)_C\times U(1)_{em}$, but also by the constraint
that one produce Yukawa couplings, quark-lepton mass matrices, and neutrino textures
consistent with the current experiment. Further, the stringent proton decay limits
put further constraints on the Higgs content of a model. Attempts  to satisfy 
partially or in whole these constraints has led to a huge number of $SO(10)$ models
with a variety of Higgs structures. Following is a list of the some of the most
commonly employed Higgs representations:
\beqn
10_H, 16_H+\overline{16}_H, 45_H, 54_H, 120_H, 126_H+\overline{126}_H, 210_H
\eeqn
More recently the following Higgs structure  has  been used 
\beqn
144_H+\overline{144}_H
\eeqn
to accomplish a one step breaking of SO(10) down to the Standard Model gauge group. 
We will discuss this possibility in greater detail later. In most
models the Higgs contents of the model do contain the 45-plet representation. This 
representation is also interesting as it enters  in accomplishing doublet-triplet  
splitting. There are many ways in which the VEV formation can take place in 
the 45-plet consistent with the Standard  Model gauge  group 
$SU(3)_C\times SU(2)\times U(1)_Y$. Some of the possible  directions
for the $<45>$ plet  VEVs are  
\begin{displaymath}
 v_1 i\sigma_2 (1,1,1,1,1), ~v_2 i\sigma_2 (0,0,0,-1,-1),
~v_3 i\sigma_2 (1,1,1,0,0),
\end{displaymath}
\beqn
~v_4 i\sigma_2 (\frac{2}{3},\frac{2}{3},\frac{2}{3},-1,-1)
\eeqn 
Here the VEV formation $v_1$ breaks SO(10) down to $SU(5)\times U(1)$, 
$v_2$ is along the third component $T_{3R}$ of $SU(2)_R$ and breaks the
SO(10) symmetry down to $SU(3)_C\times SU(2)_L\times U(1)_{T_{3R}}\times U(1)_{B-L}$,
$v_3$ is along
the $B-L$ direction and breaks the SO(10) symmetry down to 
$SU(3)_C\times SU(2)_L\times SU(2)_{R}\times U(1)_{B-L}$,
while $v_4$ is along the hypercharge $Y$  direction and breaks the 
 SO(10) symmetry down to  $SU(3)_C\times SU(2)_L\times U(1)_Y\times U(1)$.
 Thus  the VEV formations for the cases  $v_2,v_3,v_4$  all break $SU(5)$.
\\

The Yukawa couplings for the 16-plets at the cubic level can be generated
by 10, $\overline{126}$ and 120 plets of Higgs. The coupling of the 
10-plet to the 16-plet of matter  in the superpotential is the
following

 \begin{equation}
       f_{ab}\tilde \psi_a B \Gamma_{\mu}\psi_b\phi_{\mu}
\end{equation}
where a, b are the generation indices. The coupling of the 120 plet to
matter is

 \begin{equation}
 \frac{1}{3!}
f_{ab}\tilde \psi_a B \Gamma_{\mu}\Gamma_{\nu}\Gamma_{\lambda}
\psi_b\phi_{\mu\nu\lambda}
\end{equation}
and the coupling of the $\overline{126}$ to matter is given by 

\begin{equation}
 \frac{1}{5!}f_{ab}\tilde \psi_a B 
\Gamma_{\mu}\Gamma_{\nu}\Gamma_{\lambda}
\Gamma_{\rho}\Gamma_{\sigma}\psi_b\Delta_{\mu\nu\lambda\rho\sigma}
\end{equation}
The couplings of these can be explicitly computed using the
so called Basic Theorem derived in Ref.~\cite{Nath:2001uw}.
The decomposition of these  in terms of $SU(5)\times U(1)$ representations
is  discussed in the Appendix A.
\\

An interesting phenomenon in SO(10) is the possibility of a natural doublet-triplet splitting
in SO(10). Consider, for  example, two 10 plets of SO(10) Higgs fields $10_1$, $10_2$, and
a $45$ plet of Higgs and consider a  superpotential for the Higgs  fields of the form

\beqn
W_H=  M10_2^2 + \lambda 10_1.45.10_2 
\eeqn
Consider now that a VEV formation takes place for the  45-plet field  so that
\beqn
<45>=diag(v,v,v,0,0)\times i\sigma_2
\label{45vev}
\eeqn
We may decompose the 10-plet of Higgs in SU(5) representations so that $10=5+\bar 5$.
The above leads to the following mass matrices  for the doublets and the triplets. 
Thus for the Higgs doublets one finds
\beqn
\left(\overline{5}_1^d \ \overline{5}_2^d \right)
\left(\begin{array}{cc}0&0\\
                  0&M_2\end{array}\right)                
	   \left(\begin{array}{c}5_1^d\\
                  5_2^d\end{array}\right)               
\label{sp1}
\eeqn
Here one finds  that  one pair of Higgs doublets is light while the second pair
is supermassive. For the case of the Higgs triplet one finds  the following mass
matrix 
\beqn
       	\left( \overline{5}_1^t \ \overline{5}_2^t \right)
  	 \left(\begin{array}{cc}0&\lambda v\\
                  \lambda v& M_2\end{array}\right)                
	   \left(\begin{array}{c}5_1^t\\
                  5_2^t\end{array}\right)               
\label{sp2}
\eeqn
Here both pairs of Higgs triplets  are superheavy.  Further, the Higgs  triplet
  combination which enters in the Higgsino mediated proton decay have an
  effective mass which is  given by~\cite{Lucas:1996bc} 
 \beqn
 M_{eff}^t=\frac{\lambda^2 v^2}{M_2}
 \label{split3}
 \eeqn 
  The  above  allows the possibility of raising $M_{eff}$ by adjustment of $\lambda v$
  and $M_2$. Of  course one must check that the unification of  gauge couplings  
  is maintained~\cite{Lucas:1996bc,Babu:1993we}. 
 It  is also possible to get a strong suppression of \bl violating dimension five  operators 
 as we now discuss. For this  purpose  we consider  a bit more  elaborate  Higgs 
 structure.  Thus consider the case when the  Higgs potential and the Higgs  interactions
 with matter have the form~\cite{Babu:1993we}
 
 \beqn
 W_{MH}= M10_{3H}10_{3H} +  \lambda_110_{1H}45_H10_{2H}  
 +\lambda_2 10_{2H}45_H10_{3H} + J^M_i 10_{iH}
 \label{j1h}
 \eeqn 
 where the $45$ plet of Higgs develops a VEV as in Eq.(\ref{45vev}) and 
 the $45'$ -plet develops a VEV as follows 
 \beqn
<45''>=diag(0,0,0,v',v')\times i\sigma_2
\eeqn
Here one has  three  color triplets  and anti-triplets coming from the $10_{i}$ (i=1,2,3)
and also three  iso-doublet pairs. The mass matrix in the Higgs doublet and in 
the Higgs triplet  sectors are  
\beqn
M^t= \left(\begin{array}{ccc} 0&\lambda_1 v &0 \\
                - \lambda_1 v&0 &  0\\
                0 & 0& M\end{array}\right),                
     M^d= \left(\begin{array}{ccc} 0& 0 &0 \\
                0&0 &  \lambda_2 v'\\
                0 & -\lambda_2 v'& M \end{array}\right)
      \eeqn
Here one has one pair of light Higgs doublets while all the Higgs  triplets  are heavy. 
If we define the fields so that the Higgs multiplet that couples with matter is 
$10_{1H}$ of Higgs, then only the coupling $J^M_110_{1H}$ appears in Eq.(\ref{j1h}) and
one finds that the $(M^t)^{-1}_{11}=0$ (see Eq.(\ref{m11suppression})) 
and thus there are no dimension five operators
arising from the exchange of the Higgs  triplets and we have a strong suppression of 
proton decay.

In Ref.~\cite{Anderson:1993fe} an attempt is  made at the analysis  of fermion masses  in a class of
SO(10) models and a more detailed analysis  of one model was given in Ref.~\cite{Lucas:1996bc} where
an investigation of proton decay rates along with quark-lepton textures was carried out. 
A mechanism of the type  Eq.(\ref{sp1}) and Eq.(\ref{sp2}) is  used in the analysis of  Ref.~\cite{Lucas:1996bc}
to get a doublet-triplet splitting. The Higgs sector  of the model 
consists of two 10-plets of Higgs  $10_{1H}, 10_{2H}$ and three  45-plets of Higgs
$45_{1H}, 45_{2H}, \tilde{45}_H$ which develop VEV's in the $B-L$, hypercharge and in 
the $SU(5)$ invariant direction, and in addition one has  an $SO(10)$ singlet field
S which develops a VEV of Planck size. Only the third generation of matter  has  
cubic couplings, i.e., $O_{33}= 16_310_116_3$  while couplings where the first or second  
generation of matter enter are quartic or higher suppressed  by appropriate  mass factors, 
i.e., the  effective operators are  of the form 
\beqn
O_{ij}= (\prod_{k=1}^{n} M_k^{-1})  16_i 45_1..45_m 10 45_{m+1} ...45_n 16_j
\label{oij}
\eeqn
Here $M_k$ could  be  order  the Planck scale  or the GUT scale as  needed
to get  the right textures. 
For the model discussed in Ref.~\cite{Lucas:1996bc} the branching ratios 
of proton decay into different modes differ significantly from the predictions 
of a generic SU(5) model. The analysis of neutrino
 masses is not included  in this work.  
 
A somewhat different scheme is adopted for doublet-triplet splitting in the work
of Ref.~\cite{Babu:1998wi}. Here a 45-plet of SO(10) is used to break the SO(10)
symmetry in the B-L direction, a pair of $16_H+\overline {16}_H$ is used to break
the $B-L$ symmetry, and 10-plets of Higgs  are used to break the electroweak symmetry.
Specifically one considers  two 10 plets  of Higgs  $10_{1H}$ and $10_{2H}$, one 
$45_H$ adjoint Higgs  and a  pair of $16_H+\overline{16_H}$ of Higgs. The superpotential is of the
form
\begin{eqnarray}
W_H&=&M_{10}10_{2H}^2  + M_{16}16_H.\overline{16}_H + \lambda_1 10_{1H}.45_H.10_{2H} 
+ \lambda_2 \overline{16}_H.\overline{16}_H.10_{1H} \nonumber \\ 
\end{eqnarray}
Assuming that the  $45_H$ and   $\overline{16}_H$  develop VEVs we have the following mass matrix 
\beqn
       	\left(\overline{5}_{10_1} \  \overline{5}_{10_2} \ \overline{5}_{16} \right)
  	 \left(\begin{array}{ccc}0 & \lambda_1<45> &   \lambda_2 <\overline{16}_H>   \\
       -\lambda_1<45>  &M_{10}& 0\\
       0 & 0 & M_{16}\end{array}  \right)                
	   \left(\begin{array}{c}5_{10_1}\\
                  5_{10_2}\\
		  5_{\overline{16}}		  
		  \end{array}\right)               		  
\eeqn
Here one finds again that  with the VEV of $45$ in the $B-L$ direction that one has one pair of light Higgs
doublets  while the  Higgs triplets all become heavy. Here  the light Higgs  doublet that couples to 
the down quarks is a linear combination of the Higgs doublets from the $10_{1H}$ and from 
${16}_H$. Thus the  two Higgs doublets  of MSSM are  
  \beqn
 H_u=10_{1H}, ~~H_d=\cos\alpha 10_{1H}+\sin\alpha 16_{H}
 \eeqn
where  $\tan\alpha =\lambda_2 <\overline{16}_H>/M_{16}$. 
In the  model of  Ref.~\cite{Babu:1998wi} the matter-Higgs  interaction is  taken to be of the form
\begin{eqnarray}
W_{MH}&=& h_{33} 16_3.16_3.10_H + h_{23} 16_216_3 10_H + \nonumber \\
&+&\frac{1}{M} (\lambda_{23} 16_216_310_H45_H 
+ \lambda_{23}'16_216_316_H16_H)+
\nonumber\\
&+& \frac{1}{M} (\lambda_{12} 16_116_210_H45_H 
+ \lambda_{12}'16_116_216_H16_H + f_{ij} 16_i16_j\overline{16}_H \overline{16}_H) \nonumber \\
\label{yuktexture}
\end{eqnarray}
In the above the cubic couplings  are the typical Yukawa couplings which contribute only to the 
  quark-lepton textures in the generations  2 and 3 sectors. The quartic interactions with coefficients
  $\lambda_{ij}$ contribute to  textures  in all three generations while the term with coefficient 
  $f_{ij}$ contributes to Majorana mass  matrix for the neutrinos.   A detailed analysis of quark-lepton
  textures, of neutrino oscillations  and of proton decay modes in given in Ref.~\cite{Babu:1998wi}.
  An interesting aspect of this  analysis is that  the corrections  to $\alpha_3(M_Z)$ from heavy thresholds
  is rather small and thus unification of gauge coupling constants  is well preserved. 
  Further, update of this work can be found in Ref.(\cite{jcpati}).

The work of Ref.~\cite{Dutta:2004zh} gives an analysis of proton decay in SO(10) model where the Yukawa couplings
arise from a Higgs structure consisting of 10, 120 and $\overline{126}$ plet representations. Additionally 
a 210 multiplet is  used  to break SO(10). There are six pairs of higgs doublets which arise from the 
10-plet (H), the $\overline{126}$- plet ($\bar \Delta$), the $120$- plet ($D$), 
 and from the 210-plet ($\Phi$). Thus one has  the following set of Higgs doublets 
 $h_u=(H_u^{10}, D_u^1, D_u^2, \bar \Delta_u, \Delta_u,\Phi_u)$ and 
 $h_d=(H_d^{10}, D_d^1, D_d^2, \bar \Delta_d, \Delta_d,\Phi_d)$. 
 Each of the sets produce a $6\times 6$ Higgs doublet mass matrix and 
a fine tuning is needed to get to the MSSM Higgs doublets which are 
now linear  combinations of the above six Higgs doublets for each $H_u$ and
$H_d$. A similar situation holds in the Higgs triplet sector. 
Here one has the following sets of fields  for the Higgs triplets ($h_T$) and Higgs anti-triplets
($h_{\bar T}$): $h_T=(H_T^{10}, D_T^1, D_T^2, \bar \Delta_T, \Delta_T,   \Delta_T', \Phi_T)$ and 
 $h_{\bar T}=(H_{\bar T}^{10}, D_{\bar T}^1, D_{\bar T}^2, \bar \Delta_{\bar T}, \Delta_{\bar T},\Delta_{\bar T}
 \Phi_{\bar T})$ and the Higgs triplet mass matrix is a $7\times 7$ matrix. 
We note that the dimension five operators are  only mediated  by interactions arising
from 10-plet and 120-plet mediations but these interactions are modified as  a 
consequence of the mixings in the  Higgs  triplet sector. Thus the rigid relationship
between the Higgs doublet and the Higgs  triplet couplings no longer  exist. Using this
flexibility the analysis of~\cite{Dutta:2004zh} shows that it is possible  to fine tune 
parameter in the textures to suppress  both LLLL and RRRR dimension five proton decay operators.
Another $SO(10)$ model where the Higgs sector is composed of $10_H$, $126_H$, $\overline{126}_H$,
and $210_H$ is discussed in Ref.~\cite{Aulakh:2003kg}.
\subsection{Proton decay in models with unified symmetry breaking}
In all the models  discussed above the symmetry breaking is carried out with more than
one multiplets  of Higgs. However, it is tempting to think that in a truly
grand  unified scheme only a single representation of the  Higgs  multiplet might accomplish
the breaking to the Standard Model gauge group and even all the way down to the residual gauge 
group $SU(3)_C\times U(1)_{em}$. We will discuss this idea within the context of SO(10)~\cite{bgns} 
although the idea could have  a more general  validity. For the case of SO(10) model building 
typically the  Higgs multiplets  used are  $45_H$ -plets  and
$16_H+\overline{16}_H$ of Higgs and for getting the light higgs doublets one uses  in addition
$10$ plet of Higgs. Thus we see three different Higgs representations that are used to break
 SO(10)  down to  $SU(3)_C\times U(1)_{em}$.  It is possible,  however, to achieve the
breaking of $SO(10)$ to $SU(3)_{3}\times U(1)_{em}$  with a single irreducible representation, i.e., with
a single $144$ plet of Higgs and its conjugate which is a very economical way to break the gauge
symmetry~\cite{bgns}.  The 144  plet of Higgs can be decomposed  under $SU(5)\times U(1)$  as follows
  \beqn\label{144plet}
 144= \bar 5 (3) +5(7) +10(-1) +15(-1) + 24(-5) +40(-1) + \overline{45} (3)
 \label{144}
 \eeqn
 The decomposition contains the $24-plet$ of Higgs which is in the adjoint  representation of
 $SU(5)$ and further it carries a $U(1)$ charge of $-5$. Thus  once the Standard Model singlet
 in it acquires a VEV one will have a change in the rank of the gauge group and the $SO(10)$
 symmetry will break down to the Standard Model gauge group $SU(3)_C\times SU(2)_L\times U(1)_Y$.
  The $SU(5)$ multiplets $\bar 5(3), 5(7)$ and $\overline{45}(3)$  all contain fields which
 have the same identical quantum numbers as the Standard Model Higgs doublet. Thus in addition
 to two doublets arising from $\bar 5(3), 5(7)$ one has one more doublet arising from the 
 $45$-plet which can be seen from the following $SU(2)\times SU(3)\times U(1)_Y$ decomposition 
 \beqn\label{45}
45=(2,1)(3) + (1,3)(-2) + (3,3)(-2) + (1,\bar 3) (8)\nonumber\\
 + (2,\bar 3) (-7) +(1,\bar 6)(-2)
+(2,8)(3)
\eeqn 
Thus we find that one has three pairs of Higgs doublets arising from $144+\overline{144}$ leading
to a $3\times 3$  Higgs doublet mass matrix and a fine tuning is required to get  a light Higgs
doublets~\cite{bgns}.  Such a fine tuning can be justified within the framework of  recent 
ideas of string landscapes~\cite{landscape,landscape1,lands2,lands3,lands4}.
Since one has  a light pair of Higgs doublets one can break the 
$SU(2)\times U(1)_Y$ gauge  symmetry down to $U(1)_{em}$. Thus one  finds  that with a
single pair of $144+\overline{144}$ one can break the $SO(10)$ symmetry down to the
residual gauge group $SU(3)_C\times U(1)_{em}$
\beqn
SO(10)\to SU(3)_C\times U(1)_{em} ~~~{\rm :144~~breaking}
\label{144breaking}
\eeqn 
In the Higgs triplet sector one finds that there are four Higgs triplets
and anti-triplets two of which arise from $\bar 5(3), 5(7)$ and two from $45+\overline{45}$ 
leading to a  $4\times 4$ Higgs triplet mass  matrix which factorizes further into 
$3\times 3$ and $1\times 1$ block diagonal forms. Further, all the Higgs  triplets are heavy. 
The interactions of the $144$-plet Higgs are quartic. Thus the superpotential that 
 accomplishes the symmetry breaking of Eq.(\ref{144breaking}) has the form 

\begin{eqnarray}\label{generalsuperpotential}
 {\mathsf W_H}= M(\overline{144}_H\times 144_H)+
\sum_{i=1,45,210} \frac{\lambda_1}{M'} (\overline{144}_H\times 144_H)_i
(\overline{144}_H\times 144_H)_i  +\cdot\nonumber \\
\end{eqnarray}

Of course, many additional self-interactions can be included on the right hand side of 
Eq.(\ref{generalsuperpotential}) but the terms exhibited are  sufficient to accomplish
the desired breaking. There  are no cubic interactions of the $144$ with the $16$-plet of 
matter and the lowest such interaction is quartic. Thus the matter-Higgs interactions are
\beqn
{\mathsf W_Y}=
\sum_{j=10,120,\overline{126}}\frac{\lambda_j}{M'} (16\times16)_j(144\times 144)_j
\eeqn
and terms with $144$ replaced by $\overline{144}$ can also be added. 
We note that $<144>/M'$ is typically O(1) and thus 
the above interactions give baryon and lepton number violating dimension five 
operators  when one of the $144$ or $\overline{144}$ is replaced with a VEV. 
As already noted above the Higgs triplets arise from the $5$ and $\bar 5$ and
also from the $45$ -plet  in the $144$. Thus there are now more than one 
sources of \bl violation. Because of this there is the possibility of internal suppression
of the \bl violating interactions. One can thus easily enhance the proton lifetime
by this internal cancellation procedure still allowing for the possibility of observation 
of proton decay in the next generation of 
proton decay experiment.  
\\
Analyzes of higher gauge groups also exist such as, for example, $SU(15)$~\cite{su15a,su15b,su15c}.
Proton decay for this case is  discussed in Ref.~\cite{su15c}.

\section{Testing grand unification}
In this section we investigate the possibility 
of making tests of grand unified theories through the decay of the proton. 
A variety of phenomena can influence  such tests and we investigate  them 
here. In Sec.(5.1) we give a discussion of the effects of Yukawa  textures on the
proton lifetime.  The Yukawa textures at a high scale play the important role of providing a
possible explanation for fermion masses. However, the textures in the Higgs triplet 
sector can be very different than in the Higgs triplet sector and this phenomenon
has an important bearing on the proton lifetime.  In Sec.(5.2) we discuss the possible
effects of gravity on predictions of grand unification. Specifically such effects arise
in supergravity
grand unification which involves three arbitrary functions: the superpotential,
the Kahler  potential, and the gauge kinetic energy function. The 
non-universalities  in gauge kinetic  energy function are  known to affect 
gauge coupling unification. But they can also affect proton lifetimes.
In Sec.(5.3) we discuss the effects on proton lifetime from gauge coupling unification.
This is so because the gauge
coupling unification receives threshold  corrections from the low mass
(sparticle) spectrum  as well from  the high scale  (GUT) masses in grand unified models.
Since the gauge couplings are given to  a high precision by the LEP data, the gauge
coupling unification leads to constraints on the GUT scale masses, including the Higgs triplet
mass,  
 and hence on the proton lifetime. 
In Sec.(5.4), a model independent analysis of distinguishing various GUT models using
meson and anti-neutrino final state is given. Specifically 
three different models, $SU(5)$, flipped $SU(5)$ and
$SO(10)$  are  analyzed.  In Sec(5.5) we discuss the
 constraints necessary to   eliminate  the
 \bl ~violating dimension  six operators  induced by gauge interactions.
Specifically it is shown that such constraints can be  satisfied  
 for  the case of  flipped $SU(5)$.  In Sec.(5.6)  we discuss the upper limits on
the proton lifetime on  \bl ~violating dimension six
operators  which arise from  gauge interactions.
The upper bound is helpful in determining  if 
a given GUT model is allowed or disallowed by experimental lower limits.
\\

\subsection{Textures, Planck scale effects and proton decay}

The quark-lepton masses and mixing angles  pose a challenge in understanding their 
hierarchical structure.  It  is suggested that  perhaps such structure  may be 
understood from simple hypotheses at high scale, i.e.,  the grand unification scale or the 
string  scale~\cite{Georgi:1979df,Ramond:1993kv,Froggatt:1978nt}. 
Thus, for example, in grand unification where the b quark and the $\tau$ lepton fall in the same multiplet 
the experimental ratio $m_b/m_{\tau}\sim 3$ at low energy can be understood by the equality of the 
$b-\tau$ Yukawa couplings at the grand unification scale. This occurs  in supergravity  grand unification
but not in ordinary (non-supersymmetric) grand unification giving further support to the validity 
 of supersymmetry.  However, the same  does not hold for
 $m_s/m_{\mu}$ and $m_d/m_e$. This discrepancy   is attributed to the possibility that the Yukawa
 couplings  at the high scale  have   textures. That  is the couplings have a matrix form
 in the flavor space.   Thus in MSSM the Yukawa interactions at the high scale will have the form 

	\begin{equation}
	W_{d}=  H_2u^c Y^uq  + H_1d^c Y^dq+ H_1 l Y^e e^c
	\end{equation}
 where $Y^u, Y^d,  Y^e$ are the texture matrices.  A simple choice for these are the ones
	by Georgi-Jarlskog (GJ)~\cite{Georgi:1979df}   which  (assuming no CP phases) are
	\begin{equation}	
	Y^u=\left(\matrix{0&c&0\cr
                  c&0&b\cr
                  0&b&a\cr}\right),                  
  	 ~~Y^{d,e}=\left(\matrix{0&f&0\cr
                  f&e(1,-3)&0\cr
                  0&0&d\cr}\right)             
	  \end{equation}                
where  a-f  have a hierarchy  of  scales so that  
$a\sim ~O(1)$  and the quantities b-f are appropriate powers of $\epsilon$  where $\epsilon~<~1$.
In addition to the GJ textures there are also a variety of other suggestions.
  Chief among these  are those of   Ref.~\cite{Ramond:1993kv} which classify many
possibilities.   There are various approaches to generating textures~\cite{q-textures,texture-string}:
 grand unification,   Planck scale corrections,  models
based on an Abelian  $U(1)_X$ symmetry, and string based  models. 
A possible origin of the parameter $\epsilon$ is from the ratio of mass scales, e.g., 
$\epsilon={M_{GUT}}/{M_{str}}$~\cite{Jain:1994hd,Nath:1996qs}.
Thus  in the context of supergravity unified models this ratio can arise
from higher dimensional operators.
In the energy domain below the string scale after integration over
the heavy modes of the string one has an effective theory of the type
$ W=W_3+\sum_{n>3}W_n.$ where $W_n (n>3)$ are suppressed by the string (Planck) scale  and  
in general contain the adjoints which develop 
VEVs$\sim$ $O(M_{GUT})$. After VEV formation of the  heavy fields 
$W_n\sim O({M_{\rm GUT}}/{M_{\rm string}})^{n-3}\times $~(operators  in ~${\rm W_3}$).
With the above one can generate the necessary hierarchies in the textures.

 A  technique similar to the  addition of Planck scale corrections 
 to generate  textures is due to  Froggatt and Nielsen~\cite{Froggatt:1978nt} 
who observed  that a way to generate hierarchy of
mass scales is through non-renormalizable interactions involving a flavon field which carries 
some non-trivial quantum numbers under a  $U(1)_X$ symmetry.  If the Standard Model fields
possess quantum numbers under this $U(1)_X$ symmetry which are flavor dependent, then
a hierarchy could be generated when the  flavon field develops a vacuum expectation value.
Thus, for example, a  term in the  superpotential involving the up quarks would have the form
\beqn
 Y^u_{Nij} q_iH_2 u^c_j (\frac{\theta}{M})^{n_{ij}}
\eeqn
where $\theta$ is the flavon field with a  $U(1)_X$ charge of $-1$  and the subscript N  on
$Y^u_{Nij}$ refers to the non-renormalizable nature of the interaction.  Invariance under
$U(1)_X$ requires  
\beqn
n_{ij}=n_{q_i} +n_{H_2}+n_{u^c_j} 
\eeqn
where $n_{q_i}$ is the $U(1)_X$ charge of the field $q_i$ etc. 
VEV formation for the flavon field will lead to a Yukawa interaction for the up quarks 
of the form
	
\beqn
Y^u_{ij} q_iH_2 u^c_j;  ~~
Y^u_{ij} =  Y^u_{Nij}(\epsilon)^{n_{ij}},~~\epsilon\equiv(\frac{<\theta>}{M})
\eeqn
If the VEV formation for the flavon field occurs below the scale M (so that $\epsilon <1$)
then desirable fermion mass hierarchies can occur with appropriate choices  of $\epsilon$
and of  the $U(1)_X$  charges. This is essentially the Froggatt-Nielsen approach which has been
examined in a variety of scenarios.

Typically string models  lead to an  anomalous $U(1)_A$ symmetry and this case has been 
examined  quite extensively. The cancellation of anomalies impose many constraints 
limiting the choices for the generation dependent $U(1)_X$  charges.  However, that
still leaves one  with many possibilities~\cite{u1-textures}.  However, more severe restrictions arise when
one includes as  a constraint the size of allowed baryon and lepton number violating
interaction such as $QQQL$.  The number of allowed models is then drastically 
reduced~\cite{Dreiner:2003hw,Dreiner:2003yr,Harnik:2004yp}.
 In a variant of the same  approach the analysis  of
Ref.~\cite{Jack:2003pb} has considered  an anomaly-free $U(1)$ along with some
simple ansatz regarding the origin of Yukawas. The analysis leads to an automatic  
conservation of baryon number~\cite{Jack:2003pb}. 

Proton decay involves  textures not only in the quark-lepton
Yukawa coupling sector, but also involves textures in the Higgs triplet sector~\cite{Nath:1996qs,Babu:1995cw}.
In general the  Higgs  triplet textures  are not the same as  the Higgs doublet
textures so that
\begin{equation}
W_t= T u^c Y^u_t e^c + 
\epsilon_{\alpha \beta \gamma} ( \bar{T}_{\alpha} d^{C}_{\beta} Y^d_t u^C_{\gamma} 
+ T_{\alpha} u_{\beta} \tilde{Y}^u_t d_{\gamma}) + \bar{T} l Y^e_t q 
\end{equation}
where $Y^d, Y^u, \tilde Y^u,Y^e$  are the Higgs triplet textures. 
In Ref.~\cite{Nath:1996qs} Higgs triplet textures for the
case of SU(5) corresponding to the Georgi-Jarlskog textures were  classified and their form
found to be  significantly different from the  textures in the up and down quark sectors in the
Higgs doublet  sectors.   An example of such textures based on Planck scale operators in 
 $SU(5)$ is~\cite{Nath:1996qs}
 \begin{equation}
               Y^u_t=\left(\matrix{0&{4\over 9}c &0\cr
                  {4\over 9}c &0&-{2\over 3}b\cr
                  0&-{2\over 3}b&a\cr}\right),     
~Y^{d,e}_t=\left(\matrix{0&{8\over 27}F(-1,1) &0\cr
       {8\over 27}F(-1,1) &{4\over 3}e(-1,4) &0\cr
         0&0&{2\over 3}d(-1,1)\cr}\right).~    \\ \\     
\end{equation}
and   $\tilde Y^u_t=Y^u_t$.
As already stated   proton decay is affected by  textures both in the
doublet sector and in  the Higgs triplet  sector. For the $SU(5)$ case 
the $\bar \nu K^+$  mode is enhanced roughly by a factor of $\sim (\frac{9}{8}\frac{m_s}{m_{\mu}})^2$ 
by the inclusion of Higgs triplet textures. In general textures affect differentially 
the different decay modes. Thus proton decay modes hold important information on 
GUT physics and this includes also textures both in the doublet as well as in the triplet sectors. 
More recent analysis of textures in GUT models can be found in 
Refs.~\cite{Babu:1998wi,Bajc:2004xe,Dutta:2004hp,Berezhiani:1998hg}.
		  
\subsection{Gravitational smearing effects} 
Gravitational smearing effects can modify the unification of gauge coupling constants as well
as affect analysis of proton decay. Consider, for example, the gauge kinetic energy function
for gauge fields for a gauge  group $G$. Here the conventional kinetic energy term $-(1/2)Tr (FF)$, where 
F is the Lie valued  field strength in the adjoint representation of the 
gauge  group can be
modified by the  addition of the non-renormalizable  operator~\cite{Hill:1983xh,Shafi:1983gz}  
\beqn
{\cal L}_n =\frac{c}{2M_{\rm Pl}} Tr(FF \Phi) 
\eeqn
where $\Phi$ is a  scalar field in a representation of the gauge group such that $Tr(FF\Phi)$ is a
gauge group scalar which develops a  VEV and enters  in the  spontaneous breaking of the gauge 
group symmetry. Thus after spontaneous 
breaking the  gauge kinetic  energies in the $SU(3)_C\times SU(2)_L\times U(1)_Y$ will be modified
and a proper normalization will lead to splitting of the $SU(3)_C\times SU(2)_L\times U(1)_Y$
fine structure constants for these so that~\cite{Hall:1992kq}
\beqn
\alpha_G^{-1}(M_G)\to \alpha_G^{-1}(M_G)
(1+r_1 \frac{cM}{M_{\rm Pl}}, 1+r_2 \frac{cM}{M_{\rm Pl}},1+r_3 \frac{cM}{M_{\rm Pl}})
\eeqn
where $r_i$ depend on the nature of the gauge group. These splittings affect the analysis of 
gauge coupling 
unification~\cite{Hall:1992kq,Langacker:1994bc,Dasgupta:1995js,Ring:1995wc,Huitu:1999eh}.
Further, the GUT breaking will bring in
heavy thresholds.
With inclusion of the splittings due to quantum gravity effects  and of heavy thresholds 
the renormalization group evolution in the vicinity of the unification scale can be written
as follows 
\begin{equation}
\alpha_i^{-1}(Q)=\alpha_G^{-1} + \frac{cM}{2M_P}\alpha_G^{-1}r_i
+ C_{ia}\log\frac{M_a}{Q}
\end{equation}
where $M_a$ are the heavy thresholds, $C_{ia}$ are one loop renormalization group beta function
 and $Q$ is the renormalization group scale. 
Now by a transformation $M_a=M_a^{eff}e^{\chi_a}$ one can absorb
 the quantum gravity correction by defining effective heavy thresholds so that
$
\alpha_i^{-1}(Q)={\alpha^{eff}_G}^{-1}+C_{ia}\log(M_a^{eff}/Q)
$
where  $\alpha^{eff}_G$ is $\alpha_G$ evaluated at $M^{eff}_G$ where  $M^{eff}_G$=$M_G$
exp(-5$\Delta_g$), and $\Delta_g=(\frac{\pi cM}{M_P}\alpha_G^{-1})$ so that 
$
 (\alpha^{eff}_G)^{-1}=\alpha_G^{-1}-(15/2\pi)\Delta_g, ~M_a^{eff}=M_a e^{-k_a\Delta_g}$,
 where $k_a$ are pure numerics that depend on the specifics of the gauge group,  on the  
 representations $\Phi$ and on the heavy thresholds. 
The main point of the above  illustration is that quantum gravity effects warp the 
 heavy thresholds and it is these warped thresholds that enter in the renormalization group
 analysis. On the other hand, proton decay is  controlled by the unwarped heavy fields.
 This means that the masses of the lepto-quarks $M_V$ that enters in proton decay from
 heavy gauge boson exchange  and of the Higgs triplet field $M_{H_3}$  that enters  in the 
 proton decay from dimension five operators  can be significantly  different from the
 values one obtains from the renormalization group analysis. Indeed prediction of proton
 lifetime will depend sensitively on the gravitational effects and conversely the observation
 of a proton decay mode can be  utilized  along with renormalization group analysis to 
 estimate the amount of Planck scale effects. \\

 Consider, for specificity $SU(5)$ and $\Phi$  a  $24$-plet of scalar field
 in the adjoint representation of $SU(5)$. The VEV formation 
 $<\Phi>=  M diag (2,2,2, -3, -3)$ gives the  heavy thresholds 
 as follows:
 $(3, 2, 5/3) + (\bar{3}, 2, -5/3)$ massive vector bosons of mass
$M_V$, $(1,3,0) + (1,\bar{3},0)$ massive color Higgs triplets of
mass  $M_{T}$, $(1,8,0)+(1,3,0)$ massive
$\Sigma$--fields of mass $M_{\Sigma}$ and a massive singlet $\Sigma$
field. Here $r_i=(-1,-3,2)$ for i in $U(1)$, $SU(2)_L$ and $SU(3)_C$ and
the gravitational warping generates  
an effective scaling of the heavy masses so that 
$k_a=(-\frac{3}{5},\frac{3}{10},5)$, 
where a=1,2,3 refer to $\Sigma,V,M_{T}$ masses. As noted above the heavy masses
that enter in proton decay are  the unwarped ones. 
Thus, for example, an experimental determination of $p\rightarrow \bar\nu K^+$
would provide a determination of $M_{T}$ while the renormalization group 
analysis provides a determination of $M_{T}^{eff}$ allowing for a determination
of c~\cite{Ring:1995wc,Dasgupta:1995js,tobewells}. 
To see these effects more clearly we look at  the experimental constraints on the 
current data. The RG analysis of  Ref.~\cite{Murayama:2001ur} gives
$3.5\times 10^{14}\leq M_{T}\leq 3.6 \times 10^{15} GeV$,
while Super-Kamiokande data demands  
$ M_{T}\geq  2\times 10^{17} GeV$.
This appears to eliminate  the  $SU(5)$ model. However, inclusion of the Planck scale  
effects requires only that  
\beqn
3.5\times 10^{14}\leq M_{T} e^{-5\Delta_g}\leq 3.6 \times 10^{15} GeV
\label{pknumodi}
\label{tripmodi}
\eeqn
The  above  implies that with $c\sim 1$ one can achieve consistency with the SuperK data.
However, we add a note of caution. In Eq.(\ref{pknumodi})  we have
not taken into account the corrections to the gaugino masses that arise as a consequence
of quantum gravity effects~\cite{Drees:1985bx,Drees:1985jx,Ellis:1985jn,Dasgupta:1995js}. 
Inclusion of these affects involve an overlap of the Planck scale and GUT scale effects
and bring in a new parameter $c'$ generally distinct from $c$.  The gluino, the chargino 
and the neutralino masses are thus modified and since they enter in the dressing loop 
integrals for proton decay in the mode $p\to \bar\nu K^+$, Eq.(~\ref{pknumodi}) is 
affected. Because of this the effects of gravitational smearing in this sector are more
model dependent. However, $c'$  does not enter in the analysis of $p\to \pi^0e^+$ 
which is thus a cleaner channel to observe the gravitational smearing effects.  
Similar modification will also arise in SO(10) analysis. However, here there are
many more possibilities for Planck scale corrections since the Higgs  structure of 
SO(10)  models  is more complex. Thus  Higgs fields that enter
at the GUT scale to accomplish $SO(10)$ breaking include  large representations
such as  $45$, $54$, $210$ etc. which can give rise to higher dimensional operators
\beqn
Tr(FF \Phi_{45}),  ~~Tr(FF \Phi_{54}), ~~Tr(FF \Phi_{210})
\eeqn
where, however, the first term is zero due to anti-symmetry. 
After VEV formation for these scalars, one would find gravitational corrections
to the renormalization group evolution which also indirectly affects proton decay estimates
as discussed above. An RG analysis including gravitational corrections in SO(10)
is given recently in Ref.~\cite{Parida:2005hu}. 
 
\subsection{Constraints from gauge coupling  unification}
The analysis of  the previous sections exhibits that the proton
lifetime from dimension five operators depends critically on the mass of the 
Higgs triplet while that from dimension six operators depends on the 
mass of the superheavy gauge boson. It turns out that these masses 
are also strongly constrained  by the condition that gauge couplings
unify at high scale~\cite{Georgi:1974yf}.  Thus consider 
the renormalization group equations for the gauge couplings
~\cite{Gell-Mann:1954fq,Callan:1970yg,Symanzik:1970rt}:
\begin{equation}
\mu \frac{d}{d \mu} g_i(\mu)=\beta_i(g_i(\mu))
\end{equation}  
where the functions $\beta_i$ at one-loop level are given by:
\begin{equation}
\beta_i(g_i(\mu))= \frac{g_i^3}{16 \pi^2} [\frac{2}{3} T(F) d(F) + \frac{1}{3} T(S)d(S) -\frac{11}{3} C_2(G_i)]
\end{equation}
with $i=1,2,3$ for $U(1)_Y$, $SU(2)_L$ and $SU(3)_C$. In the 
above expression the fermion representations are assumed to transform 
according to the representation $F$ with dimension $d(F)$, while the scalars 
transform in the representation $S$ with dimension $d(S)$. For an irreducible 
representation $R$ we have,
\begin{eqnarray}
R^a \ R^b &=& C_2(R) I \\
Tr ( R^a R^b ) &=& T(R) \delta^{ab}
\end{eqnarray}
where $R^a$ is a matrix representation of the generators of the group. 
$T(R)$ and $C_2(R)$ are related by the identity,
\begin{equation}
C_2(R) d(R)= T(R) r
\end{equation}
with $r$ the number of generators of the group and $d(R)$ is the dimension 
of the representation. $C_2(R)$ is the quadratic Casimir 
operator of the representation R. For the group $SU(N)$ 
$T(N)=1/2$ and $T(Adj)=N$. In the case of the $U(1)_Y$ group 
we can use the above formula for $\beta_1$, with $C_2(G)=0$ and $T(R)=Y^2$ 
(See for example~\cite{Jones:1981we}), 
where the electric charge is defined by $Q=T_3 + Y$. In the above expression 
we have taken the scalar representation to be complex, and the fermion 
representation to be chiral.

\begin{figure}[h]
\begin{center}
\includegraphics[width=6in]{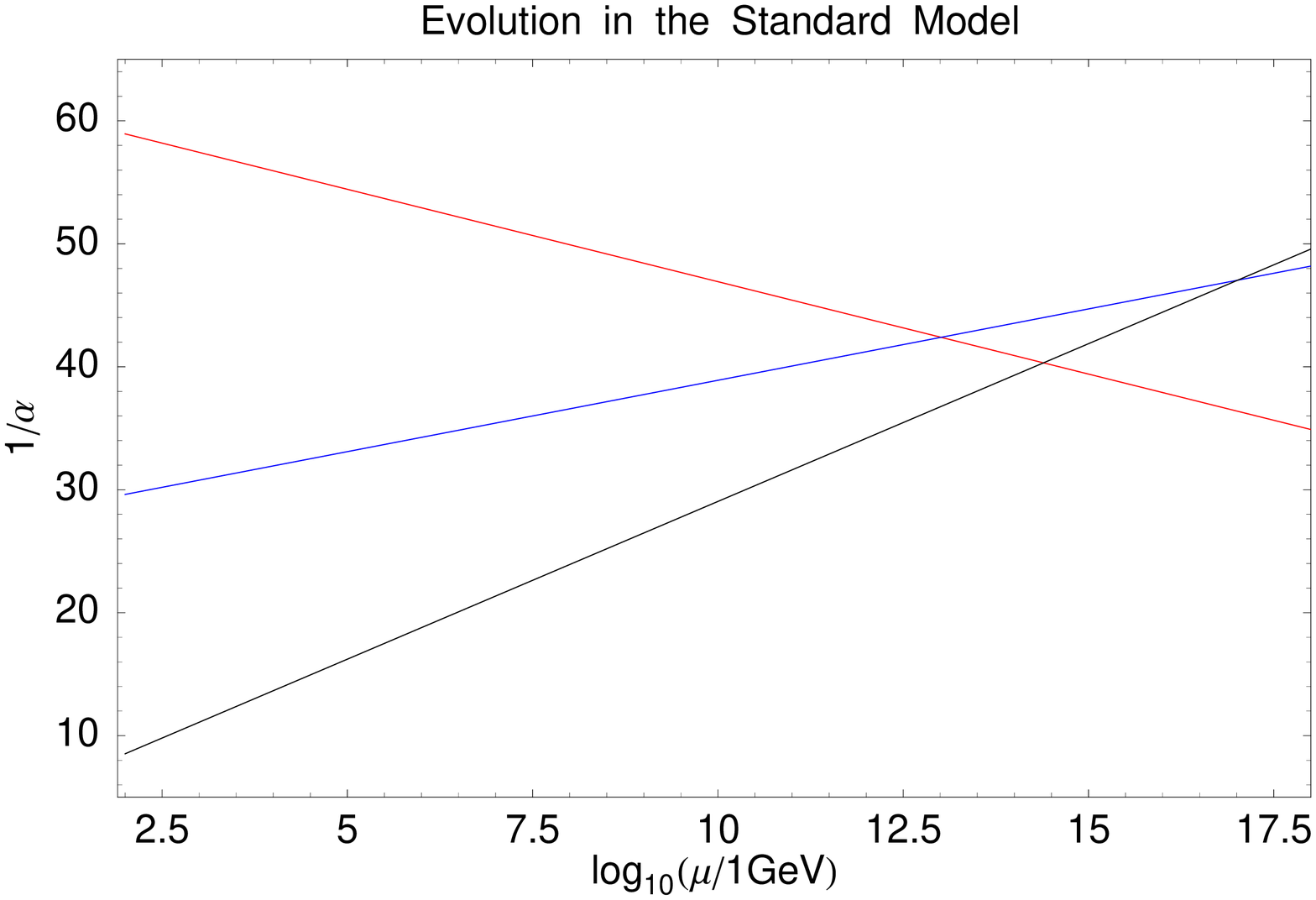}
\caption{\label{runningsm} Values of the gauge couplings of the Standard Model at different scales. 
As input parameters we take $\alpha_s (M_Z)_{\overline{MS}} = 0.1187$, $\alpha (M_Z)_{\overline{MS}}=1/127.906$, 
and $\sin^2 \theta_W (M_Z)_{\overline{MS}}= 0.2312$.  Here the three couplings do not have
a common intersection.}
\end{center}
\end{figure}

The equation for the running of the gauge couplings at one-loop 
level is 

\begin{equation}
\alpha_i(M_Z)^{-1}= \alpha_{GUT}^{-1} + \frac{b_i}{2 \pi} ln \frac{M_{GUT}}{M_Z}
\end{equation}
where $\alpha_{i}=g_{i}^2 / (4 \pi)$. Using the general expression $\beta_i$ one
finds for the Standard Model

\begin{equation}
b_1^{SM}=41/10, \ \ b_2^{SM}=-19/6, \ \ b_3^{SM}=-7 
\end{equation}
As is well known the above beta functions do not allow the unification of 
gauge couplings. See Figure.~(\ref{runningsm}) for details.

Next  we consider  the minimal non-supersymmetric $SU(5)$, where the matter is unified in 
${\bar 5}$ and ${10}$, the Higgs sector is composed by ${5_H} = ( H, T)$ and 
${24_H} = (\Sigma_8, \Sigma_3, \Sigma_{(3,2)}, \Sigma_{(\bar{3},2)},\Sigma_{24})$, 
while the gauge fields live in ${24_V}$. Using the SM decomposition one gets the 
following equations for the $B_i$,

\begin{eqnarray}
B_1^{SU(5)} &= & b_1^{SM} \ - \ \frac{105}{6} r_V \ + \ \frac{1}{15}r_T \\
B_2^{SU(5)} &=& b_2^{SM} \ + \ \frac{1}{3} r_{\Sigma_3} \ - \ \frac{21}{2} r_V \\ 
B_3^{SU(5)} &=& b_3^{SM} \ + \ \frac{1}{2} r_{\Sigma_8} \ - \ 7 \ r_V \ + \ \frac{1}{6} r_T
\end{eqnarray} 
where:

\begin{equation}
\label{r}
r_I=\frac{\ln M_{GUT}/M_{I}}{\ln M_{GUT}/M_{Z}}
\end{equation}
and where
$M_I$ is the mass of the additional particle $I$ ($M_Z \leq M_I \leq M_{GUT}$).
Now, following Giveon {\it et al}~\cite{Giveon:1991zm}, 
The equations for the running 
of the gauge couplings (replacing $b_i$ by $B_i$) can be put in a more 
suitable form  in terms of 
differences in the coefficients $B_{ij}(=B_i-B_j)$ and low 
energy observables~\cite{Giveon:1991zm}.
 One finds two relations that hold at $M_Z$~\cite{Giveon:1991zm}

\begin{eqnarray}
\frac{B_{23}}{B_{12}}&=&\frac{5}{8} \frac{\sin^2
\theta_w-\alpha_{em}/\alpha_s}{3/8-\sin^2 \theta_w},\\
\nonumber\\
\ln
\frac{M_{GUT}}{M_Z}&=&\frac{16 \pi}{5 \alpha_{em}}
\frac{3/8-\sin^2 \theta_w}{B_{12}}.
\end{eqnarray}
Using the  experimental values at $M_Z$ in the
$\overline{MS}$ scheme~\cite{Eidelman:2004wy} of 
$\sin^2 \theta_w=0.23120 \pm 0.00015$, $\alpha_{em}^{-1}=127.906 \pm 0.019$ and
$\alpha_{s}=0.1187 \pm 0.002$, one obtains

\begin{eqnarray}
\label{condition1}
\frac{B_{23}}{B_{12}}&=&0.719\pm0.005,\\
\nonumber\\
\label{condition2}
\ln \frac{M_{GUT}}{M_Z}&=&\frac{184.9 \pm 0.2}{B_{12}}.
\end{eqnarray}

\begin{figure}[h]
\begin{center}
\includegraphics[width=4in]{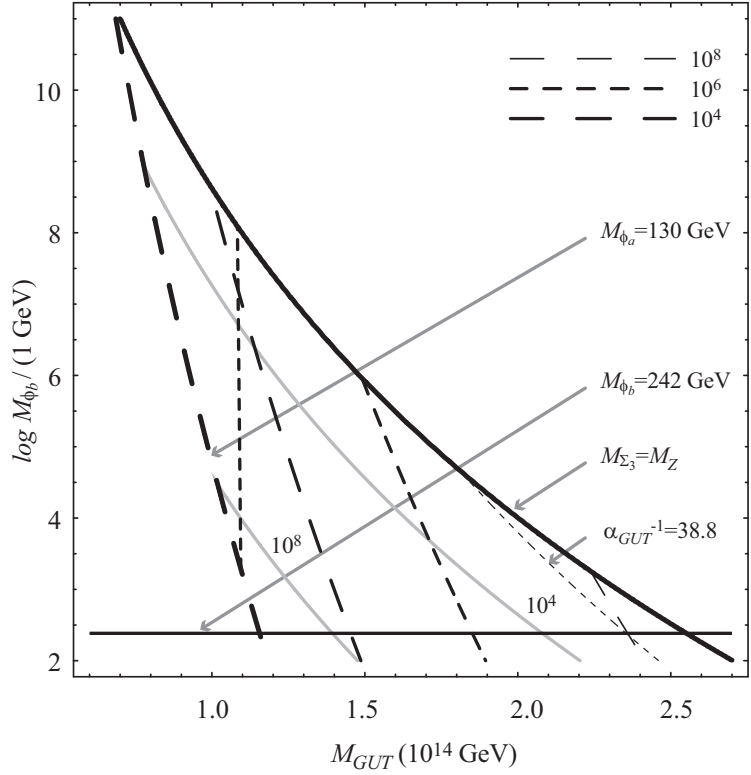}
\end{center}
\caption{\label{oneloop} Plot of lines of constant $M_{\Sigma_3}$ and
$M_{\Phi_a}$ in the $M_{GUT}$--$\textit{log}\, (M_{\Phi_b}/1\,\textrm{GeV})$
plane, assuming exact one-loop unification.  The central values for the
gauge couplings as given in the text are used. All the masses are given in GeV units.
The triangular region is bounded from the left (below) by the experimental
limit on $M_{\Phi_a}$ ($M_{\Phi_b}$). The right bound is $M_{\Sigma_3} \geq
M_Z$. The two grey solid (thick dashed) lines are the lines of constant
$M_{\Sigma_3}$ ($M_{\Phi_a}$). The line of constant $\alpha^{-1}_{GUT}\,$ is
also shown. The region to the left of the vertical dashed line is excluded by
the proton decay experiments if 
$\alpha=0.015$\,GeV$^3$~\cite{Dorsner-Ricardo}.}
\end{figure}

The above two relations constrain the mass spectrum of the extra 
particles that leads to an exact unification at
$M_{GUT}$ and this approach offers a simple way to test 
 unification for a given model. 
The fact that the SM with one Higgs doublet cannot yield
unification is now more transparent in light of
Eq.(~\ref{condition1}). Namely, the resulting SM ratio is simply
too small ($B_{23}^{SM}/B_{12}^{SM}=0.53$) to satisfy equality in
Eq.~(\ref{condition1}). In minimal non-supersymmetric $SU(5)$ we have the same problem, 
since the colored triplet and superheavy gauge bosons 
have to be very heavy to avoid problem with proton decay $(
B_{23}^{SU(5)}/ B_{12}^{SU(5)} \leq 0.60)$. 
Now, in a minimal realistic non-supersymmetric 
grand unified theory based on $SU(5)$~\cite{Dorsner:2005fq}, 
 the Higgs sector is extended by 
${15_H} = (\Phi_a, \Phi_b, \Phi_c)$, where 
the fields $\Phi_a$, $\Phi_b$, and $\Phi_c$ transform as 
$(1,3,1)$, $(3,2,1/6)$ and $(6,1, -2/3)$, respectively. 
Here it is possible to generate neutrino masses, satisfy 
all experimental bounds on proton lifetimes and 
achieve unification. In this case we have additional 
contributions to the parameters 
$B_{12}$ and $B_{23}$ (see Table 3):

\begin{table}[h]
\begin{center}
\begin{tabular}{|r|r|r|r|r|}
\hline
\hline
& & & & \\
     & Minimal $SU(5)$ & $\Phi_a$ & $\Phi_b$ & $\Phi_c$\\
& & & & \\
\hline 
\hline
& & & & \\
$B_{23}$ & $B_{23}^{SU(5)}$ & $\frac{2}{3}r_{\Phi_a}$
& $\frac{1}{6} r_{\Phi_b}$ &$-\frac{5}{6} r_{\Phi_c}$\\
$B_{12}$ & $B_{12}^{SU(5)}$ & $-\frac{1}{15}r_{\Phi_a}$ 
& $-\frac{7}{15} r_{\Phi_b}$ &$\frac{8}{15} r_{\Phi_c}$\\
& & & & \\
\hline
\hline
\end{tabular}
\caption{Contributions to the $B_{ij}$ coefficients in a 
realistic minimal non-SUSY $SU(5)$~\cite{Dorsner:2005fq}.}
\label{tab:table2}
\end{center}
\end{table}
A knowledge of $B_{12}$ and $B_{23}$ allows one to exhibit 
the entire parameter space where it is
possible to achieve exact unification (See figure~\ref{oneloop}). 
The triangular region in Fig.~\ref{oneloop} represents the
available parameter space under the assumption that $\Psi_T$,
$\Sigma_8$ and $\Phi_c$ reside at or above the GUT scale. The
region is bounded from the left and below by experimental limits
on $M_{\Phi_a}$ and $M_{\Phi_b}$. The 
right bound stems from a requirement that $M_{\Sigma_3} \geq M_Z$. 
We note that in this scenario it is possible to predict the maximal value
for the GUT scale, which allows one to define the upper bound on the 
proton decay lifetime (See Section~5.6 
for details.). In this minimal non-supersymmetric scenario 
light leptoquarks $\Phi_b$ are predicted in order 
to achieve unification. Therefore it is
a possibility to test the idea of grand unification at the next
generation of collider experiments~\cite{Dorsner:2005fq}. 
For studies in a different extension of the Georgi Glashow 
model see Ref.~\cite{FGlashow}. 
  
\begin{figure}[h]
\begin{center}
\includegraphics[width=6in]{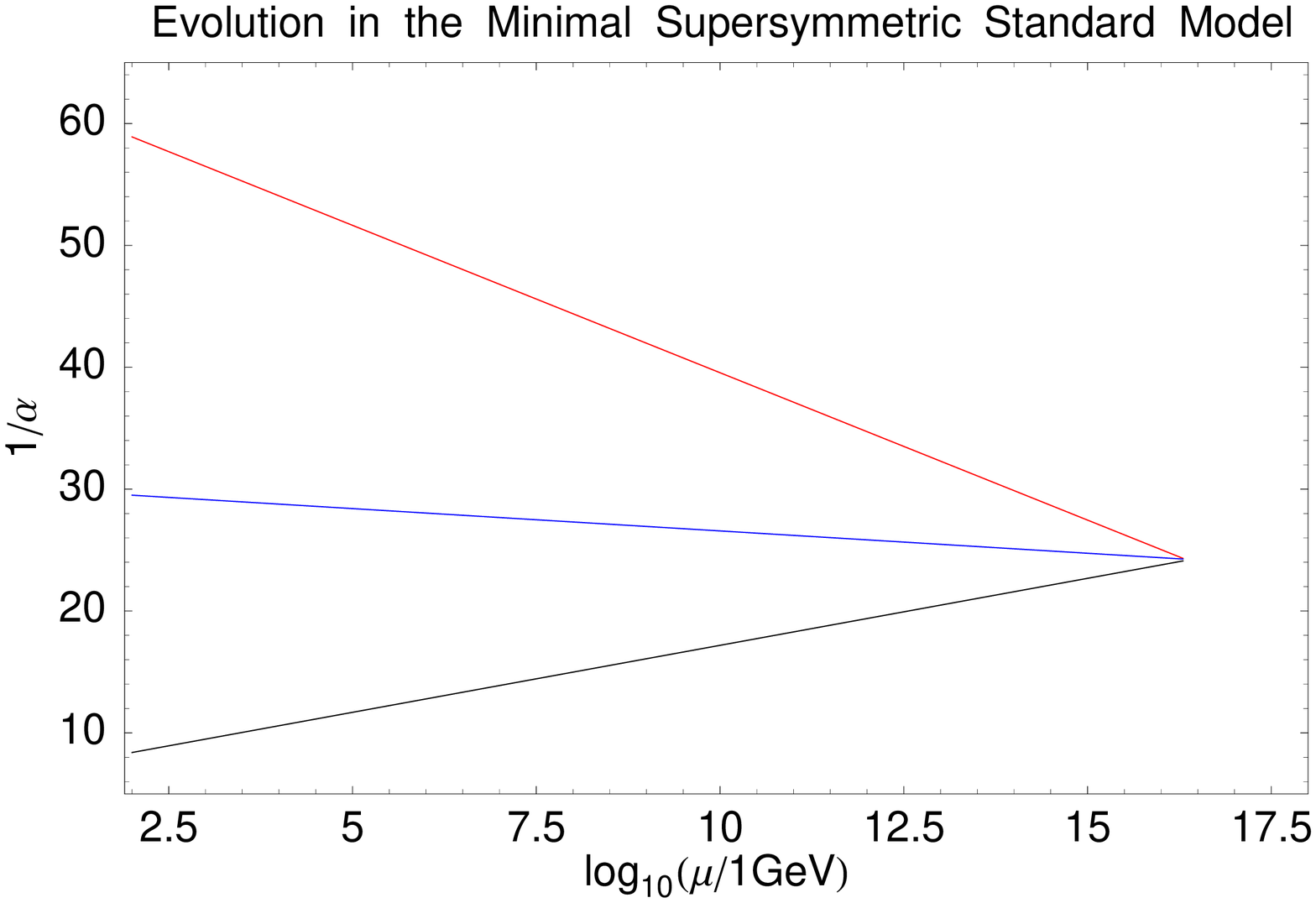}
\end{center}
\caption{\label{runningmssm} Values of the gauge couplings at different scales, 
in the $\overline{DR}$ scheme, in the context of the minimal supersymmetric standard model. 
For simplicity all superpartner masses are taken at $M_Z$ scale. The input parameters 
in the $\overline{MS}$ scheme are listed in Figure.~(\ref{runningsm}). 
Here the gauge couplings unify at  a high scale of $M_G\sim 2\times 10^{16}$ GeV.}
\end{figure}

Let us investigate the constraints in supersymmetric scenarios. 
In the minimal supersymmetric standard model the equations 
for the running are given by:  

\begin{eqnarray}
B_1^{MSSM} &=& b_1^{SM} \ + \ \frac{21}{10} \ r_{\tilde q} \ + \ 
\frac{2}{5} \ r_{\tilde G}\\
B_2^{MSSM} &=& b_2^{SM} \ + \ 2 \ r_{\tilde G} \ + \ \frac{13}{6} \ r_{\tilde q}\\
B_3^{MSSM} &=& b_3^{SM} \ + \ 2 \ r_{\tilde q} \ 
+ \ 2 \ r_{\tilde G}  
\end{eqnarray}
assuming the same mass $M_{\tilde q}$ for all scalars and the same 
mass for Higgsinos and gauginos $M_{\tilde G}$. In this case as is well-known 
it is possible to get unification at the scale $M_{GUT} \approx 10^{16}$ GeV, 
if the supersymmetric particles are around 1 TeV, or if one has only 
the gauginos and higgsinos at the $10^2$-$10^3$ GeV 
scale~\cite{Dimopoulos:1981yj,Marciano:1981un,Amaldi:1991zx}. 
See Figure.~(\ref{runningmssm}) where we show the values of the gauge 
couplings at different scales in the context of the MSSM. 

To discuss the constraint on the Higgs triplet mass we list the equations 
for the running in the case of the minimal supersymmetric $SU(5)$:

\begin{eqnarray} 
B_1^{SSU(5)}&=& B_1^{MSSM} \ + \ \frac{2}{5} \ r_T \ 
- \ 10 \ r_V \\
B_2^{SSU(5)}&=& B_2^{MSSM} \ + \ 2 \ r_{\Sigma_3} \ - \ 6 \ r_V \\
B_3^{SSU(5)} &=& B_3^{MSSM} \ + \ r_T \ - \ 4 r_V \ + \ 3 \ r_{\Sigma_8}
\end{eqnarray} 

Assuming that $\Sigma_3$ and $\Sigma_8$ have the same mass and 
using the equations above one finds~\cite{Hisano:1992mh}:

\begin{equation}
( 3 \alpha_2^{-1} \ - \ 2 \alpha_3^{-1} - \alpha_1^{-1})(M_Z) 
= \frac{1}{2 \pi} \ \left( \frac{12}{5} \ \ln \frac{M_T}{M_Z} - 2 \ \ln \frac{M_{SUSY}}{M_Z} \right)
\label{higgstripletconst}
\end{equation}
Eq.(\ref{higgstripletconst}) is a very useful in constraining the Higgs  triplet mass.
In reference~\cite{Murayama:2001ur} the authors concluded that the 
triplet mass $M_T \leq 3.6 \times 10^{15}$ GeV, in order to satisfy 
the above constraint in the context of the minimal supersymmetric SU(5).
However, when the fields $\Sigma_3$ and $\Sigma_8$ have 
different masses~\cite{Bajc:2002pg} the bound on $M_T$ is quite different. 
This is a possible solution, which implies that in the context of 
the minimal supersymmetric SU(5) it is still possible to satisfy the 
experimental bounds on proton decay lifetimes.

\subsection{Testing GUTs through proton decay }

As shown in the previous section the proton decay predictions arising 
 from the gauge $d=6$ operators depend on the fermion mixing, 
i.e. the predictions are different in each model for 
fermion masses~\cite{DeRujula:1980qc}. 
Let us analyze the possibility to test the realistic grand unified
models, the $SU(5)$, the flipped $SU(5)$ and $SO(10)$ theories, respectively. 
Let us make an analysis of the operators in each theory, and study the physical
parameters entering in the predictions for proton decay.
Here we do not assume any particular model for fermion masses,
in order to be sure that we can test the grand unification idea.

As an example we discuss now the specific case of $SU(5)$ with symmetric up Yukawa couplings.
Here we consider the simplest grand unified theories, 
which are theories based on the gauge group $SU(5)$. 
In these theories the unification of quark and leptons is realized in two
irreducible representations, $10$ and $\overline{5}$. The minimal Higgs
sector is composed of the adjoint representation $\Sigma$, and two Higgses
$5_H$ and $\overline{5}_H$ in the fundamental and anti-fundamental
representations~\cite{Georgi:1974sy,Dimopoulos:1981zb}. If one wants to
keep the minimal Higgs sector and have a realistic $SU(5)$
theory, one needs to introduce non-renormalizable operators, Planck 
suppressed operators, to get the correct quark-lepton mass relations. A
second possibility is introduce a Higgs in the $45_H$ representation. In order 
to generate neutrino mass in these theories we have to add $15_H$ Higgs~
(See for example~\cite{Dorsner:2005fq}) or the right handed neutrinos.   
In this case we have only the operators $\textit{O}^{B-L}_I$ (Eq.~\ref{O1}), 
and $\textit{O}^{B-L}_{II}$ (Eq.~\ref{O2}) contributing to the decay of the
proton. Let us study the prediction for proton decay in a $SU(5)$ theory 
with $Y_U = Y_U^T$. In this case we have $U_C = U K_u$, where $K_u$ is a
diagonal matrix containing three phases which gives
~\cite{FileviezPerez:2004hn}:

\begin{eqnarray}
\label{sumSU(5)1} \sum_{l=1}^3 c(\nu_l, d_{\alpha},
d^C_{\beta})_{SU(5)}^* c(\nu_l, d_{\gamma},
d^C_{\delta})_{SU(5)}&=& k_1^4 (V_{CKM}^*)^{1 \alpha}
(K_2^*)^{\alpha \alpha} (V_{CKM})^{1 \gamma} K_2^{\gamma \gamma}
\delta^{\beta \delta} \nonumber \\
\end{eqnarray}
In this case the clean channels to test the scenario are~\cite{FileviezPerez:2004hn}:

\begin{eqnarray}
\label{xxx1} \Gamma(p \to K^+\bar{\nu})&=& k_1^4 \left[ A^2_1
|V_{CKM}^{11}|^2+A^2_2 |V_{CKM}^{12}|^2\right] C_1
\\
\label{xxx2} \Gamma(p \rightarrow \pi^+ \bar{\nu}) &=& k_1^4
\left|V_{CKM}^{11}\right|^2 C_2
\end{eqnarray}

where 

\begin{eqnarray}\label{a}
C_1&=&\frac{(m_p^2-m_K^2)^2}{8\pi m_p^3 f_{\pi}^2} A_L^2
\left|\alpha\right|^2
\\
\label{b} C_2&=&\frac{m_p}{8\pi f_{\pi}^2}  A_L^2 \left|\alpha
        \right|^2 (1+D+F)^2
\end{eqnarray}
where the notation is as in Appendix G.
Here we have two expressions for $k_1$, which are independent
of the unknown mixing matrices and the phases.
Thus it is possible to test $SU(5)$ grand unified theory 
with symmetric up Yukawa matrices through these two processes
~\cite{FileviezPerez:2004hn}. These results are valid 
for any unified model based on $SU(5)$ with $Y_U= Y_U^T$. 
Similar tests can be investigated for other gauge groups.
Specifically a discussion of the tests for the gauge groups $SO(10)$ and
flipped $SU(5)$ is given in Appendix G.

\subsection{Proton decay in flipped $SU(5)$}

In the previous section we have shown the possibility to 
make a clear test of realistic grand unified theories with symmetric 
Yukawa couplings through the proton decay into a meson and antineutrinos.
It is thus interesting to investigate how these conclusions change if
one departs from the flavor structure of the minimal renormalizable theories.
It is well known that the gauge $d=6$ proton decay cannot be
rotated away, i.e., set to zero via particular choice of
parameters entering in a grand unified theory, in the framework of
conventional $SU(5)$ theory with the Standard Model particle 
content~\cite{Mohapatra:1979yj,Nandi:1982ew}. So, it 
would appear 
 that the gauge $d=6$ operators and proton decay induced by 
 them are genuine features of matter unification. Now 
this conclusion has some caveats as we now discuss.
To understand the issues more clearly it is useful to investigate
the constraints that might allow one to rotate away the baryon and
lepton number violating dimension six operators induced by gauge
interactions. Thus  consider the model based on conventional $SU(5)$. 
Setting $k_2=0$ in Eqs.~(9)-(10) the relevant coefficients that enter 
in the decay rate formulas are:

\begin{eqnarray}
\label{5a} c(e^C_{\alpha}, d_{\beta})_{SU(5)}&=& k_1^2 \left[
V^{11}_1 V^{\alpha \beta}_2 + ( V_1 V_{UD})^{1
\beta}( V_2 V^{\dagger}_{UD})^{\alpha 1}\right] \\
\label{5b} c(e_{\alpha}, d_{\beta}^C)_{SU(5)} &=& k_1^2 V^{11}_1
V^{\beta \alpha}_3 \\
\label{5c} c(\nu_l, d_{\alpha}, d^C_{\beta})_{SU(5)}&=& k_1^2 (
V_1 V_{UD} )^{1 \alpha} ( V_3 V_{EN})^{\beta l}, \
\textrm{$\alpha=1$ or $\beta = 1$}\\
\label{5d} c(\nu_l^C, d_{\alpha}, d^C_{\beta})_{SU(5)}&=&0
\end{eqnarray}

It is now easy to see that the demand to rotate away proton decay leads
to conflict with experiment. In order to
set Eq.~(\ref{5b}) to zero, the only possible choice is
$V_1^{11}=0$. [Setting $(V_3)^{\beta\alpha}$ to zero would violate
unitarity.] If we now look at Eq.~(\ref{5c}), there is only one
way to set to zero the coefficient entering in the decay channel
into antineutrinos. Namely, we have to choose $(V_1 V_{UD})^{1
\alpha}=0$. This, however, is not possible since it would imply
that, at least, $V_{CKM}^{13}$ is zero in conflict with experiment.

Next we investigate the same issue in flipped
$SU(5)$.  However, before doing so we give a brief discussion of  it.
  The gauge group in this case is  $SU(5)\times U(1)$ and the hypercharge 
is a linear combination of generators in $SU(5)$ and in $U(1)$, and so strictly speaking one does not
have  a unified gauge group.  The particles  reside in the multiplets $\bar 5$, $10$ 
and in an SU(5)  singlet and the assignments differ from those  of the usual SU(5) as
given in Appendix A. Thus in the flipped $SU(5)\times U(1)$ case the particle
content of the multiplets  is as below: For the $\bar 5$ of  SU(5) we have  

\begin{eqnarray}
{\overline {5}}= \left(\begin{array}{c}u_{ L a}^c\\
e^-_L\\  -\nu_{eL}
  \end{array}\right)
\end{eqnarray}
where  subscript  a is the color  index. For the 10 plet of $SU(5)$ and for the singlet we have

\begin{eqnarray}
10 =\left(\begin{array}{ccccc}0 & d_3^c &-d_2^c  &-u^1 &-d^1\\ -d_3^c &0
&d_1^c  &-u^2 &-d^2\\ d_2^c &-d_1^c &0  &-u^3 &-d^3 \\
 u^1 &u^2 &u^3  &0 &\nu^C \\ d^1 &d^2 &d^3 &-\nu^C
&0
  \end{array}
 \right)_L
, ~~1=e^+
\end{eqnarray}
The $d=6$ proton decay  from gauge interactions  is again mediated by lepto quarks but
their quantum number assignments are different so we label them with a prime: 
 $V'=(X', Y')$. This time the relevant $d=6$ coefficients are: 
\begin{eqnarray}
\label{7a} c(e^C_{\alpha}, d_{\beta})_{SU(5)'}&=&0\\
 \label{7b}
c(e_{\alpha}, d_{\beta}^C)_{SU(5)'} &=&
k_2^2 (V_4 V^{\dagger}_{UD} )^{\beta 1} ( V_1 V_{UD} V_4^{\dagger} V_3)^{1 \alpha}\\
\label{7c} c(\nu_l, d_{\alpha}, d^C_{\beta})_{SU(5)'}&=&  k_2^2
V_4^{\beta \alpha}( V_1 V_{UD} V^{\dagger}_4 V_3 V_{EN})^{1l}, \
\textrm{$\alpha=1$ or $\beta = 1$}\\
\label{7d} c(\nu_l^C, d_{\alpha}, d^C_{\beta})_{SU(5)'}&=& k_2^2
\left[ ( V_4 V^{\dagger}_{UD} )^{\beta
 1} ( U^{\dagger}_{EN} V_2)^{l \alpha }+ V^{\beta \alpha}_4
 (U^{\dagger}_{EN} V_2 V^{\dagger}_{UD})^{l1}\right], \nonumber \\
\textrm{$\alpha=1$ or $\beta = 1$}
\end{eqnarray}
where the subscripts $SU(5)'$ stands for flipped $SU(5)$. 
Let us see if it is possible to rotate away the proton decay in
flipped $SU(5)$. To set Eq.~(\ref{7c}) to zero, we can only choose
$V_4^{\beta\alpha}= (D_C^{\dagger} D)^{\beta \alpha}=0$, where
$\alpha=1$ or $\beta=1$. We could think about the possibility of
making both Eqs.~(\ref{7b}) and (\ref{7d}) zero, choosing $(V_4
V_{UD}^\dagger)^{\beta 1}=0$, however, this is in contradiction
with the measurements of the CKM angles. Since in flipped $SU(5)$
the neutrino is Majorana, we only have to suppress Eq.~(\ref{7b}).
This can be accomplished by setting
$(V_1 V_{UD} V_4^{\dagger} V_3)^{1 \alpha}=(U_C^{\dagger} E)^{1
  \alpha}=0$~\cite{Dorsner:2004jj}.
We note that this constraint is unrelated to the  constraint on
$V_4$. Thus, there is no contradiction with the unitarity constrains
nor conflict with any experimental measurements of mixing angles.
Consequently in the context of flipped $SU(5)$, it is
possible to \textit{completely} eliminate or rotate away the gauge $d=6$
contributions in a consistent way, by imposing the necessary conditions 
at 1\,GeV~\cite{Dorsner:2004jj}.

In contrast in the minimal renormalizable flipped $SU(5)$ it is not possible
to satisfy the first condition, since $Y_D=Y_D^T$ implies
$V_4=K^*_d$, where $K_d$ is a diagonal matrix containing three 
phases. However, as discussed already we have to take
into account the nonrenormalizable operators, which are 
important for fermion masses and which invariably lead to
modification of naive predictions. Thus in general, in the
context of flipped $SU(5)$, one is allowed to impose the
necessary constraints and remove the gauge operators for proton decay.
In summary the main difference between $SU(5)$  and
flipped $SU(5)$ is that the unitary constraint that prevents
one to eliminate proton decay in conventional $SU(5)$ does not
operate in the latter case. In other words, the coefficients which
depend on $\alpha$ and $\beta$ with $\alpha=1$ or $\beta =1$ have
different consequences in those two scenarios [see Eqs.~(\ref{5b}) and
(\ref{7c})].

\subsection{Upper bound on the proton lifetime in GUTs}

In the previous section we have discussed the different ways to
 test grand unified theories through the decay of the proton. 
In this section we discuss the possibility of finding an upper bound 
on the total proton decay lifetime~\cite{Dorsner:2004xa}.
In order to establish an upper bound on the total proton lifetime one
may focus on the gauge $d=6$ contributions since all other 
contributions can be set to zero in searching for  upper limits.  
Proton lifetime induced by superheavy gauge boson exchange can be
written as follows
\begin{equation}
\tau_p = C \ M_X^4 \ {\alpha^{-2}_{GUT}} \ m_p^{-5}
\end{equation}
Here $C$ is a coefficient which contains all information about
the flavor structure of the theory,  $M_X$ is the mass of the
superheavy gauge bosons, and  $\alpha_{GUT}=g^2_{GUT}/4 \pi$, where
$g_{GUT}$ is the coupling defined at the GUT scale (the scale of
gauge unification). To find a true upper bound on the total
lifetime the maximal value of $C$
 is needed. Then, for a given value of $M_X$ and $\alpha_{GUT}$ 
it is possible to bound the GUT scenario prediction for the nucleon lifetime.
However, minimization of the total decay rate is very difficult since 
in principle 42 unknown parameters enter in the decay. 
The upper bound on the proton lifetime in the case of
Majorana neutrinos reads as:
\begin{equation}
\tau_p\leq 6.0^{+0.5}_{-0.3} \times 10^{39} \
\frac{(M_X/10^{16}\,\textrm{GeV})^4}{\alpha_{GUT}^2} \
(0.003\,\textrm{GeV}^3 / \alpha)^2\,\textrm{years}
\end{equation}
where the gauge boson mass is given in units of $10^{16}$\,GeV.
Details of the
analysis is given in Appendix H and here we present only the results~\cite{Dorsner:2004xa}.

\begin{figure}[h]
\begin{center}
\includegraphics[width=4.5in]{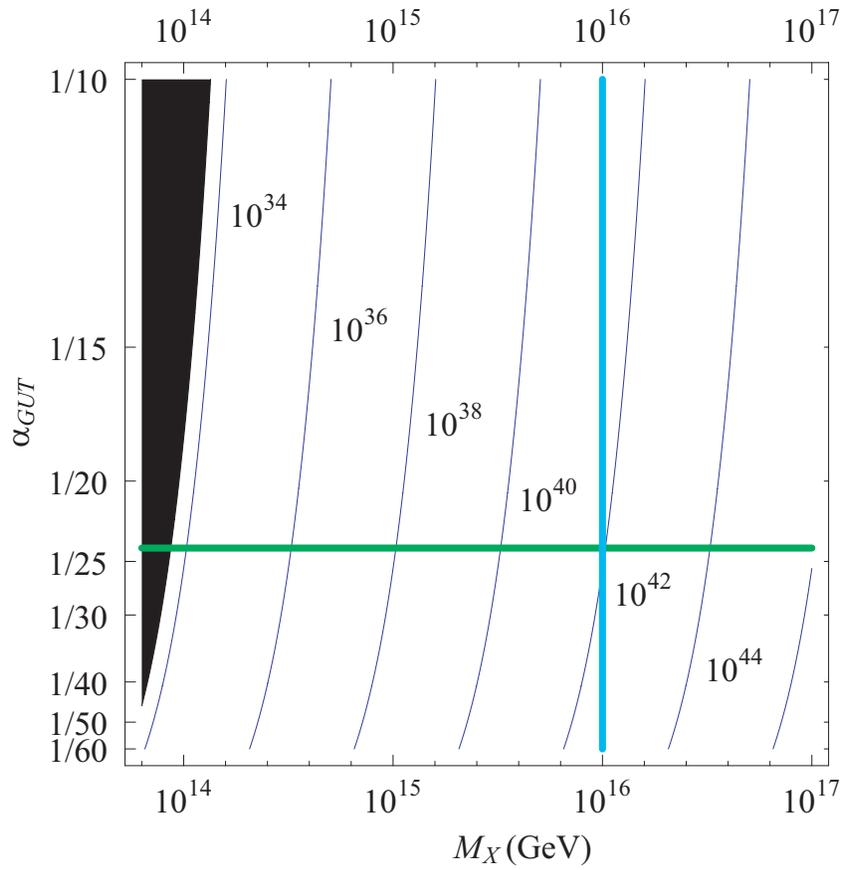}
\end{center}
\caption{\label{figure1} Isoplot for the upper bounds on the total
proton lifetime in years in the Majorana neutrino case in the
$M_X$--$\alpha_{GUT}$ plane. The value of the unifying coupling
constant is varied from $1/60$ to $1/10$. The conventional values
for $M_X$ and $\alpha_{GUT}$ in SUSY GUTs are marked in thick
lines. The experimentally excluded region is given in black~\cite{Dorsner:2004xa}}.
\end{figure}

\begin{figure}[h]
\begin{center}
\includegraphics[width=4.5in]{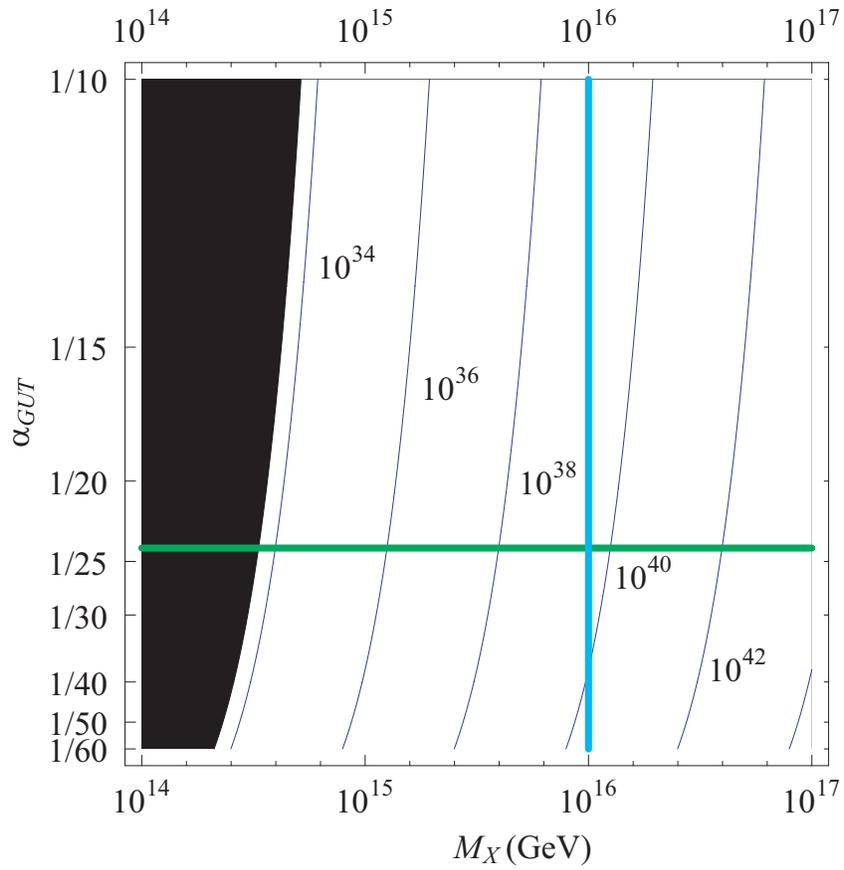}
\end{center}
\caption{\label{diracnew} Isoplot for the upper bounds on the total
proton lifetime in years in the Dirac neutrino case in the
$M_X$--$\alpha_{GUT}$ plane. The value of the unifying coupling
constant is varied from $1/60$ to $1/10$. The conventional values
for $M_X$ and $\alpha_{GUT}$ in SUSY GUTs are marked in thick
lines.  The experimentally excluded region is given in black~\cite{Dorsner:2004xa}}.
\end{figure}

The proton decay bounds in the $M_X$--$\alpha_{GUT}$ plane for
the Majorana (Dirac) neutrino case are in Fig.~\ref{figure1}
(\ref{diracnew}). These results, in conjunction with the experimental
limits on nucleon lifetime, set an absolute lower bound on mass of
superheavy gauge bosons. Since their mass is identified with the
unification scale after the threshold corrections are incorporated
in the running this also sets the lower bound on the
unification scale. Using the most stringent limit on partial
proton lifetime ($\tau_p \geq 50 \times 10^{32}$\,years) for the
$p \rightarrow \pi^0 e^+$ channel~\cite{Eidelman:2004wy} and setting
$\alpha=0.003$\,GeV$^3$, the bound on $M_X$ reads:

\begin{equation}
M_X \geq 3.04^{+0.3}_{-0.3} \times 10^{14}
\sqrt{\alpha_{GUT}}\,\textrm{GeV}
\end{equation}
where $\alpha_{GUT}$ \textit{usually}\/ varies from $1/40$ for
non-supersymmetric theories to $1/24$ for supersymmetric theories.
For example, if we take a non-supersymmetric value
$\alpha_{GUT}=1/39$, one obtains

\begin{equation}
\label{GUT} M_X \geq 4.9 \times 10^{13}\,\textrm{GeV}
\end{equation}

We note that the above result implies that any non-supersymmetric 
theory with $\alpha_{GUT}=1/39$ is eliminated if its 
unifying scale is bellow $4.9 \times 10^{13}$\,GeV
regardless of the exact form of the Yukawa sector of the theory.
Further, a majority of non-supersymmetric extensions of the
Georgi-Glashow $SU(5)$ model yield a GUT scale which is slightly
above $10^{14}$\,GeV. Hence, as far as the experimental limits on
proton decay are concerned, these extensions still represent
viable scenarios of models beyond the SM. Region of $M_X$ excluded
by the experimental result is also shown in Figs.~\ref{figure1}
and~\ref{diracnew}. 
 The plots of Fig.(\ref{figure1},\ref{diracnew}) exhibit that it is possible to
satisfy all experimental bounds on proton decay in the context of
non-supersymmetric grand unified theories. For example in a minimal
non-supersymmetric GUT~\cite{Dorsner:2005fq} based on 
$SU(5)$ the upper bound on the total proton decay 
lifetime is $\tau_p \ \leq \ 1.4 \times 10^{36}$ years~\cite{Dorsner-Ricardo}.

\section{Unification in  Extra Dimensions and
Proton Decay}
\label{sec_extradim}
Over the recent past models based  on large extra dimensions have received 
considerable attention.  The largeness of the extra dimension implies that the
 compactification scale  is small
compared to the Planck scale, and guided by  a desire for new physics at accelerators
this scale is often chosen to lie in the TeV region,  limited only by the constraints of
the precision data.  The extreme smallness of the compactification scale compared to
the GUT scale or Planck scale implies that baryon and lepton number violating 
dimension six operators would only be suppressed by the inverse of the TeV scale
and thus lead to unacceptable rate for proton decay.  This is an important hurdle for
the large extra dimension models. In this section we discuss various scenarios
where proton stability can  be achieved in such models with the help of discrete 
symmetries.  We briefly outline the main items discussed in this section. 

 In Sec.(6.1) we consider grand unified models based on one extra space-time 
dimension, and discuss proton stability within such models. It is shown that with 
discrete  symmetries it is possible to get a natural doublet-triplet splitting in the Higgs sector.
In Sec.(6.2) we give a review of SO(10) models based in 5D, and give a discussion of 5D
trinification models  in Sec.(6.3).  Sec.(6.4) is devoted to a discussion
of grand unification models in 6D, where several  unification scenarios
are analyzed. These include  $SO(10)$, $SU(5)\times U(1)$, flipped $SU(5)\times U(1)$,
and  $SU(4)_C\times SU(2)_L\times SU(2)_R$.
In Sec.(6.5) we discuss  gauge-Higgs unification.  Here the Higgs fields
 arise  as part of the  gauge multiplet and thus  gauge and Higgs
 couplings are unified. In these models proton decay is sensitive to how
matter is located in extra dimensions. A discussion of proton decay
 in models with  universal extra dimensions (UED) is given in Sec.(6.6).
Proton decay is suppressed in these models due to the existence of extra
symmetries. In Sec.(6.7) we give a discussion of proton stability in models with
 warped geometry. In this class of models proton stability arises via a symmetry
which conserves the baryon number.
 Sec.(6.8) is devoted to a discussion of proton stability in 
kink backgrounds.

\subsection{Proton decay in models  with 5D} 
In this subsection we  discuss proton decay in theories with one extra dimension. 
Theories with extra
dimensions have a long history beginning with the work of Kaluza and Klein 
in the nineteen twenties~\cite{kaluza,klein, O'Raifeartaigh:1998pk, freund}.  
More recently interest in 
theories with extra dimensions emerged with the realization that string 
theories could allow for low scale compactifications which removes the 
rigid  relationship that exists between the string scale and the Planck
scale  in the weakly coupled heterotic strings~\cite{Kaplunovsky:1987rp}.
Thus, in the context of the weakly coupled Type I string compactifications
the string scale  can be quite low~\cite{Witten:1996mz,Lykken:1996fj}
 and there has
been much work in model building along these
 lines~\cite{Antoniadis:1990ew,Antoniadis:1996hk,Arkani-Hamed:1998rs,Shiu:1998pa}
and important constraints have been placed on the size of such dimensions
from experiment~\cite{Nath:1999fs,Giudice:1998ck,Han:1998sg}.
An interesting phenomena in such theories is the power law evolution of the
gauge coupling 
constants~\cite{Taylor:1988vt,Dienes:1998vg,Ghilencea:1998st, Dienes:1996du}
 which allows for a meeting of the 
coupling constants at a low scale although  in such a scheme
the unification of the gauge couplings is not a prediction of the model but
rather an accident. The second more serious issue concerns stability of the 
proton. This is so because  if one wishes  to formulate  
 unified models with low scale extra dimensions then 
 dimension  five and dimension six baryon and lepton number violating
operators are suppressed only by the inverse powers of a mass  order
a  TeV which would lead to disastrous proton decay. 
An early suggestion to  achieve proton stability is to have quarks - leptons
in the bulk~\cite{Kobakhidze:2001yk}. In the  model of Ref.~\cite{Kobakhidze:2001yk}
$B$ and $L$ are  separately conserved  and the proton is stable with a unification
scale  in the TeV region. In this model TeV scale  mirror  particles  could be
produced at colliders~\cite{Kobakhidze:2001yk}. Another way to suppress proton
decay is to assume that the baryon number is gauged in the bulk and the symmetry
is broken on a brane different from the physical brane~\cite{Arkani-Hamed:1998sj}.
Other suggestions to suppress proton decay require imposition of discrete 
symmetries~\cite{Dienes:1998vg,Shiu:1998pa,Kakushadze:1998wa,Ellis:1997ec}.
 Such discrete symmetries are discussed 
in detail in Ref.~\cite{Kakushadze:1998wa}
where a generalized matter parity of the type $Z_3\times Z_3$ 
is proposed in an extended MSSM type model where proton decay operators are
suppressed to high orders.
However, suppression of proton decay may require an exact or  almost exact 
baryon number conservation, since otherwise proton decay may be induced  by
quantum gravity effects~\cite{Adams:2000za}. It is argued that in order  to suppress this 
type  of proton decay one would  need a high scale similar to what one has in
grand unified theories~\cite{Adams:2000za}.\\

We would not pursue further the analysis of proton decay in extra dimension theories
with low scale. Rather, we turn our attention now to the more realistic scenarios with
high scale  extra dimensions.  Typically this is the situation in heterotic string 
models where the size  of the extra dimension is of order the inverse of the compactification
scale $M_C$ which one expects  is  close  to the string scale.
It turns out that the study of  such  models do have important benefits, 
the most prominent being that
they provide a natural solution to the doublet-triplet splitting in the
Higgs sector. Often they also lead to a reduction of the gauge symmetries 
without the necessity of invoking the Higgs mechanism.
Thus, we consider grand unified theories in higher dimensions where reduction to 4 dimensions 
is accomplished by orbifold  compactification. It has been known for some time that an orbifold
compactification can reduce symmetries beginning with the work of 
Scherk and Schwarz~\cite{Scherk:1979zr,Scherk:1978ta}
and  follow  up works~\cite{Fayet:1985ua,Fayet:1985kt,Hosotani:1983xw,Hosotani:1988bm}.
(For a  discussion of  
generalized symmetry breaking on orbifolds see Refs.~\cite{Bagger:2001qi,Bagger:2001ep}).
Orbifold compactifications have  played a major role in recent works in the exploration of low scale 
extra dimensions putting lower limits of a few TeV on such 
dimension~\cite{Nath:1999fs,Giudice:1998ck,Han:1998sg}.
More recently interest has focused on grand unified models with extra  dimensions  and
here an interesting development is the  reduction of the gauge symmetry by orbifold 
compactification~\cite{Kawamura:2000ev, Kawamura:2000ir,Kawamoto:2001wm,Hebecker:2001jb,Bagger:2001qi,Hall:2001pg} 
which has in addition some very interesting features  such as  automatic doublet-triplet splitting.
The simplest possibility is a   GUT theory formulated in 5 dimensions. 
  Thus  let us  consider a  5D  space with coordinates  $x^M =(x^{\mu}, x^5)$ 
where $\mu=0,1,2,3$.  We  assume that the fifth  dimension $x^5$ is compacted
on $S^1/(Z_2\times Z_2')$ where the $Z_2$ and  $Z_2'$ are  defined  as  follows:
$Z_2$ corresponds to the  transformation 

\beqn
x^5\to -x^5
\label{d5-1}
\eeqn 
while $Z_2'$ corresponds to the transformation  
\beqn
x^{5'}\to -x^{5'}
\label{d5-2}
\eeqn
where  $x^{5'}=x^5+\pi R/2$.
 We focus on the $Z_2$ orbifolding first and return to the $Z_2'$ orbifolding later. We begin by
considering a super Yang-Mills field in the  bulk. The $N=1$ super Yang-Mills in 5D consists of 
the multiplet $(V^M, \Sigma, \lambda^i,f^a)$, where $V^{M}$ is the vector field with
M=0,1,2,3,5, $\Sigma$ is  a  real scalar field, $\lambda^i$ are simplectic 
Majorana spinors and $f^a$  (a=1,2,3)  are a triplet of 
auxiliary real fields [$V_M$ is a Lie valued quantity so that
$V_M= gV_M^{\alpha} T^{\alpha}$ where $tr(T^{\alpha}T^{\beta})=\frac{1}{2}\delta_{\alpha\beta}$,
and $\lambda$ and  $\Sigma$ are  similarly defined.].
 For specificity we consider
first the unified gauge group $SU(5)$  and assume that the super Yang Mills multiplet
belongs to the adjoint representation of $SU(5)$. The 5D super Yang-Mills Lagrangian is 
given by~\cite{Pomarol:1998sd,Delgado:1998qr,Arkani-Hamed:2001tb}

\begin{eqnarray}
{\cal L}_5^g &=&\frac{1}{g^2}\{ -\frac{1}{2} tr(V_{MN})^2 + tr(D_M\Sigma)^2
+tr(\bar\lambda i\gamma^M D_M\lambda) -tr(\bar\lambda [\Sigma,\lambda]) 
\nonumber \\ 
&+&  tr(f^a)^2\}
\label{d5-4} 
\end{eqnarray}
where $D_M\sigma =\partial_M \Sigma -i[V_M, \Sigma]$.  The action is invariant under 
the following supersymmetry transformations 

\begin{eqnarray}
\delta_{\xi}V^M &=& i\bar\xi^i\gamma^M\lambda^i \nonumber\\ 
\delta_{\xi} \Sigma &=& i\bar\xi^i\lambda^i \nonumber\\
\delta_{\xi}\lambda^i &=& (\sigma^{MN}V_{MN} -\gamma^M D_M \Sigma)\xi^i 
-(f^a\sigma^a)^{ij}\xi^j \nonumber \\
\delta_{\xi}f^a &=& \bar  \xi^i (\sigma^a)^{ij} \gamma^M D_M \lambda^j 
-i [\Sigma, \bar \xi^i (\sigma^a)^{ij} \lambda^j], 
\label{d5-5}
\end{eqnarray}
where $\xi^i$ are the transformation parameters and 
$\sigma^{MN}=[\gamma^M,\gamma^N]/4$. From the 4D view point, the 5D N=1  vector multiplet is an
N=2, 4D multiplet. We would like to reduce this multiplet to an N=1 multiplet on the  $x^5=0$ brane
which we  consider to be the physical brane. To achieve this we consider the $Z_2$ transformation which
acts on the bulk fields  so that 
\beqn
f(x^{\mu},y)\to f(x^{\mu},-x^5)=P f(x^{\mu},x^5)
\label{d5-7}
\eeqn
where $P=\pm 1$.   We take the fields
$V_{\mu}$, $\lambda_L^1$, $f^3$ to have even parity, and the fields $V_5, \Sigma, \lambda_L^2, f^{1,2}$
to have  odd parity.  Further, we assign to $\xi_L^1$ an even parity and to $\xi_L^2$ an  odd parity.
Now the fields with odd parity vanish on the $x^5=0$ boundary, and the transformations on the
$x^5=0$ brane reduce to the following~\cite{Mirabelli:1997aj}
\beqn
\delta_{\xi}V^{\mu} =i\bar\xi^{1\dagger}_L\bar \sigma^{\mu}\lambda_L^1 -i\lambda^{1\dagger}_L\bar\sigma^{\mu}\xi^1_L
\nn
\delta_{\xi} \lambda_L'= \sigma^{\mu\nu}V_{\mu\nu} \xi^1_L -i D\xi_L^1\nn
\delta_{\xi} D= i\bar\xi^{1\dagger}_L\bar \sigma^{\mu} D_{\mu}\lambda_L^1 + h.c.
\label{d5-8}
\eeqn
where $D\equiv (f^3-\partial_5 \Sigma)$. Eqs.(\ref{d5-8}) constitute the transformations of an N=1 gauge 
multiplet with components
\beqn
V_{\mu}, \lambda_L^1, D\equiv (f^3-\partial_5 \Sigma )
\label{d5-9}
\eeqn
on the  $x^5=0$ brane. We note the appearance of $\partial_5\Sigma$ in the auxiliary field D. 
While $\Sigma$
has odd  $Z_2$ parity and vanishes on the $x^5=0$ brane,  $\partial_5\Sigma$  has even $Z_2$ parity and 
is non-vanishing on the $x^5=0$ boundary. \\

 Analogous to the vector multiplet we assume that the Higgs 
multiplets reside also in the bulk and for model building we consider two hypermultiplets
consisting of two complex scalar fields and two
 Dirac fermions  $(H_i^s,\psi^s)$ (i=1,2)  where $H_i^s$ are complex Higgs doublets and $\psi^s$ 
 are Dirac spinors. 
  We identify
these multiplets as  follows
\beqn
\{(H^1_1, \psi_R^1), (H_2^1,  \psi_L^1)\}\nn
 \{ (H^2_1, \psi_R^2), (H_2^2,  \psi_L^2)\}
\label{d5-10}
\eeqn 
The 5D bulk Lagrangian for the Higgs  multiplet is then given by~\cite{Delgado:1998qr} 

\beqn
{\cal L}_5^H = |D_M H_i^s|^2 +i\bar \psi_s  \gamma^M D_M \psi^s 
-(i\sqrt 2 H_s^{i\dagger}\bar \lambda_i \psi^s + h.c.)\nn
-\bar\psi_s \Sigma \psi^s -H_s^{i\dagger} (\Sigma)^2 H_i^s 
-\frac{g^2}{2} \sum_{m.\alpha} [H_s^{i\dagger}(\sigma^m)^j_i T^{\alpha}H_j^s]^2
\label{d5-11}
\eeqn
However, care is needed in the  reduction of the Higgs bulk Lagrangian to the
boundary. Analogous to the case of the vector multiplet one should  begin
with  off shell hypermultiplets $(H^s_i, \psi^s, F_i^s)$ which break
up into the $Z_2$ parity even multiplets
$(H^1_1, \psi_R^1, F_1^1)$, $(H_2^2, \psi_L^2, F_2^2)$
and the $Z_2$ parity odd multiplets 
$(H_2^1,  \psi_L^1, F_2^1)$,  $(H^2_1, \psi_R^2, F^2_1)$.
 As we go to the boundary  $x^5=0$ only the $Z_2$  even parity multiplets survive
 and the surviving multiplets are~\cite{Pomarol:1998sd,Delgado:1998qr}
 ${\cal H}_1= (H_1^{1\dagger}, \bar \psi_R^1, F_1^{\dagger})$, and 
 ${\cal H}_2= (H_2^2, \psi_L^2, F_2)$ where 
 $F_1= F_1^1-\partial_5H_2^1$ and $F_2= F_2^2-\partial_s H_1^2$. 
 Here  ${\cal H}_2$  is the multiplet that
 couples to the up quark and ${\cal H}_1$ is the multiplet that couples to the 
 down quark and the lepton. We note that on the boundary the auxiliary fields
 are modified and this phenomenon is much similar to the modification of the D term 
 on the boundary discussed above for the  case of  the vector multiplet.\\
 
In the preceding analysis we have  seen that the action of $Z_2$ orbifolding  reduces  N=2 
supersymmetry down to $N=1$ supersymmetry on the boundary. However, the SU(5) gauge  
symmetry is left intact. We consider now  the action of the $Z_2'$ orbifolding which leaves the $N=1$
supersymmetry  intact but reduces the $SU(5)$ gauge  symmetry down to  the Standard Model gauge
group. 
To accomplish this   we consider $Z_2'$ transformation such that the field $f(x^{\mu}, x^5)$ which
belongs to  the  fundamental representation of $SU(5)$ transforms so that

\beqn
f(x^{\mu}, x^{5'})\to f(x^{\mu}, -x^{5'})=P' f(x^{\mu}, x^{5'})
\label{d5-13}
\eeqn
where $x^{5'}=x^5+\pi R/2$ and $P'$ is a $5\times 5$ matrix with 
$P'=diag (-1,-1,-1,1,1)$. Thus the fields  with $SU(3)_C$ color
indices will transform with parity $-$ and  the  fields  with $SU(2)$ indices 
will transform  with $Z_2'$ parity  $+$. 
We  identify $H_5$ with $H_2^2$ as the  one that gives mass to the up quarks, and
$H_{\bar 5}$ with $H_1^1$ which gives mass to the down quarks and the leptons.
Similarly, we define $\hat H_{\bar 5}= H^2_1$ and $\hat H_5= H^1_2$.
One has  then the following transformations for the Higgs multiplets 
under $Z_2^{'}$ transformations  
  
 \beqn
 H_5(x^\mu,\yp) \to H_5(x^{\mu},-\yp)= P'  H_5(x^{\mu},\yp),\nonumber\\ 
H_{\bar 5}(\x, \yp)\to  H_{\bar 5}(\x,-\yp)= P'  H_{\bar 5}(\x,\yp),\nonumber\\
  \hat H_5(\x, \yp)\to \hat H_5(\x,-\yp)= - P' \hat H_5(\x, \yp),\nonumber\\
\hat H_{\bar 5}(\x, \yp)\to \hat H_{\bar 5}(\x,- \yp)= - P' \hat H_{\bar 5}(\x, \yp).
\eeqn
Thus under $Z_2\times Z_2'$ transformations a field can be classified as $f_{\pm\pm}(\x,\y)$.
It is instructive to carry out a normal mode expansion for these. 
\beqn
  f_{++} (x, x^5) &=& 
       \sqrt{1 \over {\pi R}} \sum_{n=0}^{\infty} \frac{1}{\sqrt{2^{\delta_{n,0}}}}
       f^{(2n)}_{++}(x) \cos({2nx^5 \over R}),\nonumber\\
  f_{+-} (x, x^5) &=& 
     \sqrt {1 \over {\pi R}}     \sum_{n=0}^{\infty} f^{(2n+1)}_{+-}(x) \cos({(2n+1)x^5 \over R}), \nonumber\\
  f_{-+} (x, x^5) &=&   \sqrt{1 \over {\pi R}}
      \sum_{n=0}^{\infty} f^{(2n+1)}_{-+}(x) \sin({(2n+1)x^5 \over R}), \nonumber\\
 f_{--} (x, x^5) &=&  \sqrt{1 \over {\pi R}}  \sum_{n=0}^{\infty} f^{(2n+2)}_{--}(x) \sin ({(2n+2)x^5 \over R}).
\eeqn
The  above implies that the modes  $f^{(2n)}_{++},  f^{(2n+1)}_{+-}, f^{(2n+1)}_{-+},
 f^{(2n+2)}_{--}$ have masses  $2n/R$, $(2n+1)/R$, $(2n+1)/R$ and $(2n+2)/R$. 
 One notices  that  only $f_{++}$ contains massless  modes corresponding to the
 case when $n=0$. The other  modes all acquire masses scaled by the inverse of the 
 compactification radius, i.e., proportional to $1/R$. 
We  exhibit the mode  expansion for the Higgs multiplets in Table (\ref{higgsmode})
where we have decomposed the Higgs $5$  plets  in SU(3) color triplets, and $SU(2)$ doublets
and the Higgs  $\bar 5$ in  the  SU(3) color anti-triplets, and $SU(2)$ doublets, i.e., $H_5=(H_u,  H_T)$, 
$H_{\bar 5}=(H_d, H_{\bar T})$,
$\hat H_5= (\hat H_u, \hat H_T)$, and $ \hat H_{\bar 5}=(\hat H_d, \hat H_{\bar T})$.\\

\begin{table}[h]
\begin{center}
\begin{tabular}{|r|r|r|}
\hline\hline
4D fields  & $Z_2 \times Z_2'$ parity  & Mass \\
\hline
$H^{(2n)}_u$ & (+,+) & 2n/R\\
$\hat  H^{(2n)}_u$ & $(-,-)$ & (2n+2)/R\\
$H^{(2n)}_d$ & $(+,+)$ & 2n/R\\
$\hat H^{(2n+2)}_d$ & $(-,-)$ & (2n+2)/R\\
\hline
\hline
$H_T^{(2n+1)}$ & $(+,-)$ & (2n+1)/R\\
$H_{\bar  T}^{(2n+1)}$ & $(+,-)$ & (2n+1)/R\\
$\hat H_T^{(2n+1)}$ & $(-,+)$ & (2n+1)/R\\
$\hat H_{\bar  T}^{(2n+1)}$ & $(-,+)$ & (2n+1)/R\\
\hline
\end{tabular}
\end{center}
\caption{ P and ${\rm P}'$ parities of the components of bulk Higgs multiplets.} 
\label{higgsmode}
\end{table}

In Table (\ref{higgsmode}) the entries above the double  horizontal line are the  Higgs
 doublet modes. Here  for $n=0$ we  have massless modes  in $H_u$ and $H_d$.
  The entries  below  the double horizontal line  are the  Higgs triplets (denoted  by the subscript $T$) 
  and the color anti-triplets
(denoted by the subscript  $\bar T$). Here  we  see  that none of the  Higgs triplets
and  anti-triplets have massless modes.  Thus we  see  a  natural  doublet-triplet 
splitting by the assignment of the  parities  as  described above. The  Higgs
triplets and anti-triplet produce  a tower of  massive Kaluza-Klein modes whose
masses are  scaled  by the inverse  radius of  the circle $S^1$.\\

We look now at the  transformation properties of the vector  multiplet. These  fields
have transformations  like bi-fundamentals because they carry two $SU(5)$ indices. 
It is easily seen that  the Lagrangian is invariant under the following $Z_2'$
transformations

\beqn
 &~& V_{\mu}(\x, \yp) \to V_{\mu}(\x, -\yp) = P' V_{\mu}(x^\mu, \yp) P^{'-1} , \nonumber \\
 &~& \lambda_L^1(x^\mu, \yp) \to \lambda_L^1(x^\mu, -\yp) =  P' \lambda_L^1(x^\mu, \yp) P^{'-1} , \nonumber \\
  &~& \lambda_L^2(x^\mu, \yp) \to \lambda_L^2(x^\mu, -\yp) = - P' \lambda_L^2(x^\mu, \yp) P^{'-1} , \nonumber \\
 &~& \Sigma(x^\mu, \yp) \to \Sigma(x^\mu, -\yp) = - P' \Sigma(x^\mu, \yp) P^{'-1} \nonumber\\
&~& V_{5}(x^\mu, \yp) \to V_{5}(x^\mu, -\yp) = - P' V_{5}(x^\mu, \yp) P^{'-1} 
\label{1908}
\eeqn
 It is easy to infer that
the transformation of the generators of  $SU(5)$ under $P'$ are

\beqn
P'T^aP^{'-1}= T^a, ~~ P'T^{\hat a}P^{'-1}= -T^{\hat a}
\label{generators}
\eeqn
where  $T^a$ are the generators of the Standard Model gauge group $G_{\rm SM}$ 
 and $T^{\hat a}$ are in the remaining set.
  The mode expansion  of the vector multiplet components is listed  in Table \ref{vectormode}
 where the subscripts $\pm$  on the modes  specify their properties under 
 $Z_2\times Z_2'$  transformations. We find that only the fields  with $(+,+)$ 
 parities have  zero modes and  they  transform under
 $SU(3)_C \times SU(2)_L$  as  $(8,1)+(1,3)+(1,1)$. These zero modes  are precisely
 the   gauge  vector  multiplets of MSSM  which we label $V_{\mu}^a$.  All the  remaining vector
  fields  $V_{\mu}^{\hat a}$, i.e.,  the lepto-quarks,   acquire  masses.
 Specifically, we  note  that the vector multiplet  which transforms like 
 $(3,2)+(\bar 3,2)$ under  $SU(3)_C\times SU(2)$   has only  massive modes. 
Thus the above  orbifolding naturally splits the lepto-quarks from the 
Standard Model gauge  bosons. \\

\begin{table}[h]
\begin{center}
\begin{tabular}{|r|r|r|}
\hline\hline
4D fields  & $SU(3)\times SU(2)$ reps & Mass \\
\hline
$V_{\mu ++}^{a(2n)}$, $\lambda_{++}^{1a(2n)}$ & $(8,1)+(1,3)+(1,1)$ & 2n/R\\
$V_{5--}^{a(2n+2)}$,$\lambda_{--}^{2a(2n+2)}$, $\Sigma_{--}^{a(2n+2)}$ &  $(8,1)+(1,3)+(1,1)$ & (2n+2)/R\\
$V_{\mu +-}^{\hat a(2n+1)}$, $\lambda_{+-}^{1\hat a(2n+1)}$ & $(3,2)+(\bar 3,2)$ & (2n+1)/R\\
$V_{5-+}^{\hat a(2n+1)}$,$\lambda_{-+}^{2\hat a(2n+1)}$, $\Sigma_{-+}^{\hat a(2n+1)}$ &  $(3,2)+(\bar 3, 2)$ & (2n+1)/R\\
\hline
\end{tabular}
\end{center}
\caption{ P and ${\rm P}'$ parities for the components of bulk gauge multiplets} 
\label{vectormode}
\end{table}

 In setting up the Lagrangian in 5D we have to make sure that the  Lagrangian is invariant 
 under the full $Z_2\times Z_2'$  transformations. 
This set up is dependent on how the matter is  located  in the 5D space. 
 One could locate such matter either in the bulk, or on the orbifolds.  There are  two
 invariant orbifold points corresponding to $x^5=0$ and  $x^5=\pi R/2$  which are 
 the end points of the fundamental domain $x^5=(0, \pi)$. 
   When matter, is located at
 the $x^5=0$ brane, one can maintain the full $SU(5)$  symmetry, while
 when matter is located at the $x^5=\pi R/2$ brane, only the standard model symmetry
  can be maintained. 
  In fact, there are three  scenarios for the location of matter and we classify the three possibilities as  
  follows~\cite{Hall:2001pg,Altarelli:2001qj,Hebecker:2001jb}. 
 \begin{enumerate}
 \item
 Matter on the  SU(5) brane
 \item
 Matter  in the bulk
 \item
 Matter on the SM brane
 \end{enumerate}
 Let  us begin by discussing case (1).  We need to assign parities to the quark and  lepton 
 fields.  For  $Z_2$ transformations,    P  is  $+$ for color and  $+$ for SU(2).
 For quarks and leptons, one way to determine the  $P'$ parities is to require that cubic
 SU(5) invariant interactions with matter-matter-Higgs transform with an over all sign when
 one uses the $P'$  parities of Higgs as given in Table (\ref{higgsmode}). 
This gives the following possibilities 
 
 \beqn
 10: ~~P'(Q, U^C, E^C)= \eta_{10} (+, -, -)\nonumber\\
 \bar 5: ~~~~P'(D^C, L) =\eta_{\bar 5} (-,+)
 \label{8.10a}
 \eeqn  
  where  $\eta_{\bar  5,  10}$  are overall  signs of $\bar 5$  and $10$ multiplets, i.e., 
 $\eta_{\bar  5,  10}=\pm 1$.  With the above we have  
  
  \beqn 
 P'(10.10.5_H)= -(10.10.5_H)\nonumber\\
 P'(10.\bar 5.\bar 5_H)= -\eta_5 \eta_{10} (10.\bar 5. \bar 5_H)
 \label{8.10b}
 \eeqn 
Using Eq.(\ref{8.10b}) we  can write  a $Z_2\times Z_2'$ invariant 5D Yukawa interaction  
in the form

\beqn
{\cal{L}}_5=\int d^2\theta \frac{1}{2}(\delta(x^5)-\delta(x^5-\pi R))  f_{5u} 10.10.5_H
\nonumber\\
+ \int d^2\theta \frac{1}{2}(\delta(x^5)-   \eta_{\bar 5}\eta_{10} \delta(x^5-\pi R))  f_{5u} 10.\bar 5.\bar 5_H
+h.c.
\label{8.10c}
\eeqn
The $Z_2\times Z_2'$ invariance of  Eq.(\ref{8.10c}) is easily checked  by using Eq.(\ref{8.10b}). 
On integration over the fifth coordinate  one gets  the following effective Higgs-quark-lepton interaction   in 4D

\begin{equation}
{\cal{L}}_4= {\cal{L}}_0 +{\cal{L}}_{KK}
\end{equation}
\begin{equation}
{\cal{L}}_0 = \int d^2\theta (f_1 QU^cH_u^{(0)} +f_2 QD^cH_d^{(0)} +f_2 LEH_d^{(0)}) + h.c.
\end{equation}
\begin{eqnarray}
{\cal{L}}_{KK}&=&\sum_{n=1}^{\infty}\sqrt 2 \int d^2\theta (f_1 QU^cH_u^{(2n)} +f_2 QD^cH_d^{(2n)} +f_2 LEH_d^{(2n)})
\nonumber \\
&+&
\sum_{n=1}^{\infty}\sqrt 2 \int d^2\theta (f_1 QQH_T^{(2n+1)}  + 
f_1 U^CE^CH_T^{(2n+1)} 
+ f_2 QLH_{\bar T}^{(2n+1)}  \nonumber \\ 
&+& f_2 QLH_{\bar T}^{(2n+1)} ) 
\end{eqnarray}
where  $f_1=f_{5u}/\sqrt{2\pi R}$, $f_2=f_{5d}/\sqrt{2\pi R}$.  One finds that ${\cal{L}}_0$ which contains the
zero Higgs  modes is  precisely  what one has in the  minimal SU(5) theory for the Higgs  doublets.  
However, unlike the minimal SU(5) of 4D theory, here one has a natural  Higgs  doublet-triplet splitting
and one has no zero Higgs  triplet  modes. 
 The ${\cal{L}}_{KK}$ contains the  Kaluza-Klein excitations of the Higgs doublets and the Higgs triplets 
 and  anti-triplets.   \\
 
 There is no dimension five  proton decay in this theory since the  Higgs  triplet mass terms are of 
 the form~\cite{Hall:2001pg}

 \beqn
 \sum_{n=0}^{\infty} R^{-1}\int d^2\theta  (H_T^{(2n+1)} \hat H_T^{(2n+1)} + H_T^{(2n+1)} 
 \hat H_T^{(2n+1)}) +  h.c.
 \label{21.8a}
 \eeqn
Since
 $\hat H$ does not  connect to the quarks and leptons there  is no dimension five
 proton decay mediated by Higgs  triplets in this model.  Further,  as  shown in Ref.~\cite{Hall:2001pg}
  the model has an 
 overall  $U(1)_R$  invariance which kills the proton decay via  dimension  four operators from the 
 term  $10.\bar 5.\bar 5$ where all multiplets  are matter multiplets. 
We pause to contrast  the situation here with that in 4D supersymmetric theories.  As discussed in
 Sec.5, in 4D supersymmetric grand unified theories, even with R parity one typically has \bl ~violating
 dimension five  operators which lead to proton decay, and because of that there exist  overlapping
 constraints on the GUT scale from the current experimental limits on the proton lifetime   and from the 
 gauge coupling unification.  This issue  lead  us to  consider in detail the twin constraints of
  gauge coupling unification in 4D theories and proton  stability
   in Sec.(5.3).  In contrast in  higher dimensional theories  of the type discussed above, 
   one does not have any dimension five induced  proton decay.  However, the gauge coupling
   unification constraint can still affect  proton decay via dimension six operators.   Specifically, 
   here  Kaluza-Klein tower of states can affect  proton decay lifetimes. A detailed  discussion of this topic
   is given  at the end of Sec.(6.4). 
   It needs to be pointed out that the analysis of gauge coupling unification in higher dimensional
 theories in by no means unique, but rather has a  significant model dependence.  However, in a 
class of models the situation is even improved~\cite{Hall:2001xb} over the  supersymmetric $SU(5)$ model  in 4D. A more detailed 
 discussion of this topic is outside the scope of this report, but the reader is refereed  to a number  of
 recent works for an update\cite{Dienes:1996du,Mohapatra:1997sp,Perez-Lorenzana:1999qb, Hall:2001pg,Hall:2001xb}. 
 
Although, there no proton decay from dimension four and five  operators in models of the above
 type, there is, however,   proton decay from  dimension six operators induced by gauge 
interactions. 

Assuming that all  the three generations  are located on the SU(5) brane,  one has 
a dimension six  operator in this case, leading to a proton decay width for 
the mode  $p\rightarrow  e^+\pi^0$ which is \cite{Hebecker:2002rc}
\begin{equation}
\Gamma (p\rightarrow  e^+\pi^0)= (\frac{\pi g_4}{4M_C})^4
\frac{ 5\alpha^2A_R^2  m_p}{4\pi f_{\pi}^2} (1+D+F)^2. 
\label{21.8b}
\end{equation}
With
 $F=0.47$, $ D=0.8$, $ f_{\pi}=0.13$ GeV, $\alpha=0.01 $(GeV)$^3$, 
  $g^2_4/(4\pi)=0.04$ ,   $A_R=2.5$ one  finds 

\begin{equation}
\tau (p\rightarrow  e^+\pi^0)\simeq 1.4\times  10^{34} (\frac{M_C}{10^{16} \rm{GeV}})^4 \rm{yr}.
\label{21.8c}
\end{equation}
The current experiment already puts a lower limits on $M_C$ of $M_C \simeq 8\times 10^{15}$ GeV. 
\\

We consider  now case (2) where one has matter in the bulk.  Here  
 one starts with  complete SU(5) multiplets  involving 
10 and $\bar 5$. However,  $P'$ splits  these so that  only certain components of these 
multiplets have zero modes.  For example, with a specific  choice of  $P'$ parities, only
$U^c$ and $E^c$ in the 10 plet and  only $D^c$  in the  $\bar 5$ plet   have  zero modes.
To complete  the multiplets  one can add a copy of the 10 and $\bar 5$ which have 
an overall opposite   $P'$ parity to the previous multiplets.  
Since in this case the zero modes  arise  from different 
multiplets  there  are no X and Y gauge interactions  which can produce \bl 
violating dimension six operators. There are, however,    Kaluza-Klein excitations of
the  bulk  matter fields and X and Y gauge bosons do connect the  zero modes matter
fields with their KK counterparts. But these lead to operators which are at least  
dimension eight and suppressed by $M_C^4$.  Their contributions to proton decay 
is  far too small to be relevant.  
\\

Next we  consider case  (3) where one  has matter  confined to the SM brane. Here the
X and Y boson wave-functions  vanish at the location of the SM brane and thus
one has no couplings of these gauge  bosons to  the  SM matter  fields and
consequently no  \bl violating dimension six operator. So there  is  no proton 
decay from the usual X and Y boson exchange.  
 However,  we now  show that  non-minimal  couplings  such as  derivative  
couplings  can lead  to proton decay.  One can write  in general  on the SM brane 
a non-minimal operator with  one derivative  as  follows~\cite{Hebecker:2002rc}

\begin{equation}
{\cal L}_{5N}= \frac{\gamma_{ij}}{M_P} \delta (x^{5'}) 
\int d^2\theta d^2\bar\theta \psi^{c\dagger}_i ( D_5 e^{2V}) \psi_j + h.c.
\label{nm1}
\end{equation}
The effective \bl violating dimension 6  operators  can be obtained by 
an integration over  the  X and Y gauge  bosons, and one has
 
\begin{equation}
O_{6}\simeq  \delta \gamma_{ij}\gamma_{kl} \frac{g_4^2}{M_CM_P}
\int d^2\theta d^2\bar\theta   \sum_{\hat a} ( \psi^{c\dagger}_i  T^{\hat a}  \psi_j) 
( \psi^{c\dagger}_k  T^{\hat a}   \psi_l) 
\label{nm2}
\end{equation}
In the  above  $\gamma_{ij}$  and $\delta$  are strong interaction parameters  which
are  typically O(1).   The proton lifetime resulting from above is 

\begin{equation}
\tau (p\rightarrow  e^+\pi^0)= 3.5\times  10^{34} (\delta \gamma_{11})^{-2} 
(\frac{M_C^{1/2}M_P^{1/2} } {10^{16} \rm{GeV}})^4
 \rm{years}
\label{nm3}
\end{equation}
Clearly the result of Eq.(\ref{nm3}) has a significant model  dependence.  If one  assumes that 
$M_P$ is around the Planck scale,  since  such type  couplings  are  expected  to  arise  from Planck
scale  corrections,  one  has $M_P\simeq 10^{18}$ GeV. Then an $M_C$ around  $10^{15}$ GeV or 
larger,  will put this  lifetime out of reach of the next generation of experiments unless a  suppression 
is manufactured from the  front factors   $(\delta \gamma_{11})^{-2}$. 
\\
\subsection{SO(10) models in 5D}

 The SO(10) models in 5D which have been investigated by  a number of 
 authors~\cite{Dermisek:2001hp,Kim:2002im}.
  Here  the gauge  multiplet $\cal {V}$  is 45 dimensional belonging to the adjoint 
 representation of SO(10). In 4D language the 5D vector  multiplet will consist of the N=1  vector  multiplet 
 $V$ and an N=1  chiral multiplet $\Sigma$.  
  We take the Higgs multiplet to lie in  the 10 plet representation of SO(10) 
  so in 5D it is  a 10 dimensional
 hypermultiplet ${\cal H}_{10}$.  In 4D it would correspond to two N=1 chiral  superfields  $H_{10}, \hat  H_{10}$. 
 Similar to the SU(5) case  we have the following transformations  under  $Z_2$

  \beqn
 H_{10}(x^\mu,\y) \to H_{10}(x^{\mu},-\y)= P  H_{10}(x^{\mu},\y)\nonumber\\ 
  \hat H_{10}(\x, \y)\to \hat H_{10}(\x,-\y)= - P^T \hat H_{10}(\x, \y)
\label{10-1}
\eeqn
 with $P^2=I$  where  $P$ is now a $10\times 10$ matrix.  We choose $P$ so that  
 $P=1_{5\times 5}\times 1_{2\times 2}$.  We assume similar transformations under  $Z_2'$, 
 with $\yp$ replacing $\y$ and $P'$ replacing $P$ and for $P'$ we choose~\cite{Dermisek:2001hp,Kim:2002im} 
 
 \beqn
 P'=diag (-1,-1,-1,1,1)\times  (1,1).
 \label{10-2}
 \eeqn 
 As  in the case of SU(5) the $Z_2$ orbifolding breaks the N=2 \s in 4D to an N=1 \s. The $Z_2'$  orbifolding 
 breaks the SO(10)  gauge group to an $SO(6)\times SO(4) $ gauge group.  Since $SO(6)\sim SU(4)$ and
 $SO(4)\sim SU(2)_L\times SU(2)_R$, we classify the  fields  according to their $SU(4)\times SU(2)_L\times SU(2)_R$
 representations. Thus the 45 plet  of vector fields  which belong to the  adjoint representation of SO(10) can be 
 classified  in the $SU(4)_C\times SU(2)_L\times SU(2)_R$ representations  as  follows: V(15,1,1), V(1,3,1), V(1,1,3), V(6,2,2)
 and an identical decomposition holds  for the 45 plet  of  the chiral scalar superfield $\Sigma$.  The 
 Higgs multiplets H and  $\hat H$   which belong to the 10 plet representation of SO(10) decompose as  
 H(6,1,1), H(1,2,2), $\hat H(6,1,1)$,  $\hat H(1,2,2)$.   The $Z_2\times Z_2'$ properties of these  fields  are  exhibited
 in Table. (\ref{so10parity}).   The 16-plet spinor representation of SO(10) can be decomposed under  $SU(4)_C \times SU(2)_L\times SU(2)_R$
 as $(4,2,1) + (\bar 4, 1,2)$.  
 The generalization of a  $Z_2'$  transformation on a spinor  is~\cite{Dermisek:2001hp}
  $P'=e^{-\frac{3\pi}{2} (B-L)}$.
 Now under  the $SU(4)_C$ decomposition  $SU(4)_C\to SU(3)_C\times U(1)_{B-L}$,  one finds
 $(4,2,1)\to 3_{1/3}+ 1_{-1} $ which leads to $P'=-i$ for the $(4,2,1)$ multiplet. Thus 16-pet spinor has  $Z_2'$ parities
 given by $(4,2,1)_{-i}+ (\bar 4, 1,  2)_{i}$.
 
\begin{table}[h]
\begin{center}
\begin{tabular}{|r|r|}
\hline
\hline
$SU(4)\times SU(2)_L\times SU(2)_R$ N=1 multiplets & $Z_2\times Z_2'$ parities\\
\hline
V(15,1,1), V(1,3,1), V(1,1,3), H(1,2,2) & $(+, +)$\\
V(6,2,2), H(6,1,1) & $(+,-)$ \\
$\Sigma(6,2,2)$, $ \hat H(6,1,1)$ & $(-, +)$\\
$\Sigma (15,1,1), \Sigma (1,3,1), \Sigma (1,1,3)$, $\hat H(1,2,2)$ & $(-,-)$\\
\hline 
\hline
\end{tabular}
\end{center}
\caption{ P and ${\rm P}'$ parities of SO(10) vector and chiral multiplets} 
\label{so10parity}
\end{table}
As discussed earlier in the $Z_2\times Z_2'$ compactification there are two in-equivalent orbifold points: 
 $x^5=0$ and $x^5=\pi R/2$.  At $x^5=0$, the wave-functions for all the gauge bosons are non-vanishing
and one has an SO(10) invariance.  On the other hand at  $x^5=\pi R/2$, the $V(6,2,2)$  gauge bosons 
have their wave-functions vanishing, and the gauge symmetry is reduced to   \ps.  Thus we  can classify the 
models at the two orbifold points as  
\begin{enumerate}
\item
SO(10) brane model
\item
G(4,2,2) brane model
\end{enumerate}
Analogous to the SU(5) 5D model there is no proton decay in these models from dimension 4 or dimension 5 operators.     
For the case  of SO(10) brane  proton decay from  dimension six operators can occur. However, this proton decay
is proportional to $M_C^{-4}$ as seen in  Eq.(\ref{21.8c}). 
An estimate of  $M_C$ for the model of Ref.~\cite{Kim:2002im} gives  a value 
too low to be compatible with the current lower bounds  on the proton lifetime.  
 We focus next on the $G(4,2,2)$ brane model.  Here to reduce the gauge symmetry further  and to reduce the rank of 
 the gauge group one needs to invoke the Higgs
 mechanism.  One possibility is to consider addition of $16+\overline{16}$ of Higgs multiplets. Now  under 
 \ps the 16 plet  decomposes  so that $16 =(4,2,1) +(\bar 4, 1, 2)$ and one gives  VEV to $\chi^c +\bar \chi^c$ where
 $\chi^c=(\bar 4, 1,2)$.   A  VEV formation for this  combination then breaks the \ps symmetry down to the
 symmetry of the standard model gauge group.   Since the wave-function for the $V(6,2,2)$ gauge bosons  
 vanishes on the \ps  brane, there  is  no proton decay of the usual  sort  from the mediation of X and Y gauge
 bosons. However, proton decay can occur from derivative  terms on the 
 \ps brane as given in Eq.(\ref{nm3}).
 Analysis of gauge 
 coupling unification in Ref.~\cite{Kim:2002im} gives an estimate of  $M_C\sim 2\times  10^{14}$ GeV and
  $M_P$  is  identified with the unification scale in string models  and taken to be  $\sim 2\times 10^{17}$
  GeV. In this case  the analysis of  Ref.~\cite{Kim:2002im} gives 
 \beqn
 \tau (p\rightarrow e^+\pi^0) \sim  7\times 10^{33\pm 2} {\rm yr}
 \eeqn
 where the $\pm 2$ reflects  the uncertainties  due to $\delta, \gamma_{11}, M_C$ and $M_P$.

Another possible class of $SO(10)$ models in 5D is based on
embedding of a four-dimensional flipped $SU(5)$ model in a
five-dimensional $SO(10)$ model~\cite{Barr:2002fb}. This approach
can preserve the best features of both  the  flipped $SU(5)$ and of $SO(10)$.
Namely, the missing partner mechanism, which naturally achieves both
doublet-triplet splitting and suppression of dimension 5 proton
decay operators, can be realized as in flipped $SU(5)$, while the
gauge couplings unify as in $SO(10)$~\cite{Dorsner:2003yg}.

In this approach orbifold compactification leaves two inequivalent
points. One has an $SO(10)$ invariance while the other has flipped
$SU(5)$ invariance. To break the rest of the way to the Standard
Model one can either use Higgs fields that originate from the
bulk~\cite{Barr:2002fb} or reside on the flipped $SU(5)$
brane~\cite{Dorsner:2003yg}. In both cases the split between the
doublets and the triplets is done through the four-dimensional
flipped-$SU(5)$ missing partner mechanism. As before, there is no
proton decay from dimension 4 or dimension 5 operators. On the other
hand, the strength of dimension 6 gauge contributions depends on
the exact location of matter fields. If they originate from the bulk
then the dimension 6 operators are strongly suppressed; if they are
situated on either the $SO(10)$ or the flipped $SU(5)$ brane some
suppression in the Yukawa sector is needed to avoid experimental
bounds since $M_C \sim 5.5 \times 10^{14}$\,GeV~\cite{Dorsner:2003yg}.

\subsection{5D Trinification}

 5D trinification models have also been considered~\cite{Carone:2005rk,Carone:2005ha}.
  The trinification is based on the gauge   group 
 $SU(3)_C\times SU(3)_L\times SU(3)_R\times Z_3$ where the discrete symmetry permutes 
 the three labels C,L,R which gives a single gauge  coupling constant g at the unification 
 scale.  The gauge fields for the system can be decomposed in representations of 
  $SU(3)_C\times SU(3)_L\times SU(3)_R$ so that they fall into the sets   
 \beqn
 (8,1,1)+ (1,8,1) + (1,1,8)
 \eeqn
The $Z_2\times Z_2'$ parities of the vector multiplet  V are  defined 
as in 
Eq.(\ref{1908}) where  $P, P'$ are given by $P=P_C+P_L+P_R$ and similarly for $P'$.  We make the  following
assignments 
\beqn
(P_C; P_L; P_R) =(1,1,1; 1,1,-1;1,1,-1)\nn
 (P_C'; P_L'; P_R') =(1,1,1; 1,1,-1;1,1,1)
\eeqn
With the above assignments  one has 
\beqn
V(8,1,1) ~= ~\left(
\begin{array}{cc|c} (+,+) & (+,+) & (+,+) \\ (+,+) & (+,+) & (+,+)
\\ \hline (+,+) & (+,+) & (+,+) \end{array} \right), \qquad
  \eeqn

\beqn V(1,8,1)~=~ \left( \begin{array}{cc|c} (+,+) & (+,+) &
(-,-) \\ (+,+) & (+,+) & (-,-) \\ \hline (-,-) & (-,-) & (+,+)
\end{array} \right), \qquad 
\eeqn

\beqn V(1,1,8) ~=~ \left( \begin{array}{cc|c}
(+,+) & (+,+) & (-,+) \\ (+,+) & (+,+) & (-,+) \\ \hline (-,+) &
(-,+) & (+,+) \end{array} \right). \qquad 
 \eeqn

Now as usual in addition to the possibility of putting matter in the bulk one may put matter on the $x^5=0$ brane
or on  the $x^5=\pi R/2$  brane.  Suppose we  consider the last possibility. In this case  the gauge bosons 
odd under $P'$  vanish at $x^5=\pi R/2$ and the gauge symmetry is reduced to $SU(3)_C\times SU(2)_L
\times U(1)_L\times SU(3)_R$.  There are no dimension six operators to produce proton decay in these models.
In the usual triunification models, proton decay can arise from the dimension five operators generated by 
the Higgs  triplets in the 27 plet  representations. Here, however, since  at  the  orbifold point one already 
has a reduced symmetry, a  further reduction of the gauge symmetry involves only 
small representations~\cite{Carone:2005rk}.
Consequently there are no dimension five operators  arising from them and hence there  is no proton decay 
from this sector either.  

\subsection{6D models}

There are a number of works which have explored GUT model 
building in 6D~\cite{Hall:2001zb, Asaka:2001eh, Buchmuller:2004eg,Haba:2004qf,Gogoladze:2005zh}.
 In such models one begins with a
space  $R^4\times T^2$  where $T^2$ is a two torus and  one orbifolds $T^2$ in a way similar to  what 
we discussed in 5D.
One  model studied in detail in the context of proton decay is the specific 
compactification~\cite{Asaka:2001eh,Buchmuller:2004eg}
 $T^2/(Z_2\times Z_2'\times Z_2^{''})$. 
The Lagrangian density for the vector multiplet in this case  is 

\beqn
{\cal L}_6= \frac{1}{g^2}tr(-\frac{1}{2} V_{MN}V^{MN} + i\bar\lambda \Gamma^{M} D_{M} \lambda) 
\label{12.8a}
\eeqn 
where $\Gamma^{M}$ satisfy the Clifford  algebra in 6D. 
Defining $V_M=(V_{\mu},V_{\alpha})$, where  $\mu = 0,1,2,3$ as usual and 
$\alpha =5,6$ the transformation properties of 
$V_M, \lambda_1, \lambda_2$ under $Z_2\times Z_2'\times Z_2^{''}$ are 

\beqn
PV_{\mu}(x^{\mu}, -x^5, -x^6)P^{-1}= V_{\mu}(x^{\mu}, x^5, x^6),\nn
PV_{\alpha}(x^{\mu}, -x^5, -x^6)P^{-1}= -V_{\alpha}(x^{\mu}, x^5, x^6),\nn
P\lambda_1(x^{\mu}, -x^5, -x^6)P^{-1}= \lambda_{1}(x^{\mu}, x^5, x^6),\nn
P\lambda_2(x^{\mu}, -x^5, -x^6)P^{-1}= -\lambda_{2}(x^{\mu}, x^5, x^6),
\label{12.8b}
\eeqn 
and we choose  $P=I$.  Here $(V_{\mu}, \lambda_1)$ form an N=1 vector multiplet and
$(V_{\alpha}, \lambda_2)$ form an N=1 chiral multiplet. The zero modes arise only from the
vector  multiplet. Next under $Z_2^{'}$ 

\beqn
P'V_{\mu}(x^{\mu}, -x^5, -x^6 +\pi R_6/2)  P^{'-1}= V_{\mu}(x^{\mu}, x^5, x^6+\pi R_6/2),\nn
P'V_{\alpha}(x^{\mu}, -x^5, -x^6 +\pi R_6/2)P^{'-1}= -V_{\alpha}(x^{\mu}, x^5, x^6+\pi R_6/2),
\label{12.8c}
\eeqn
where  for $P'$ we choose

\beqn
P'= diag (1,1,1, 1, 1)\times \sigma_2.
\label{12.8d}
\eeqn
Similarly for $Z_2^{''}$ 

\beqn
P^{''}V_{\mu}(x^{\mu}, -x^5 +\pi R_5/2, -x^6)P^{''-1}= V_{\mu}(x^{\mu}, x^5 + \pi R_5/2, x^6),\nn
P^{''}V_{\alpha}(x^{\mu}, -x^5 +\pi R_5/2, -x^6)P^{''-1}= -V_{\alpha}(x^{\mu}, x^5 +\pi R_5/2, x^6),
\label{12.8e}
\eeqn
where  for $P^{''}$ we choose

\beqn
P^{''}= diag (-1, -1, -1, 1, 1)\times\sigma_0. 
\label{12.8f}
\eeqn
Now the  mode expansion of a function on the torus depends on its parities and there are
eight cases corresponding to  the eight  permutations $\pm\pm\pm$. These  have  the 
following mode expansions 
\beqn
f_{+++}(x^{\mu}, x^{\alpha}) =\sum_{m\geq  0}  (\pi^2 R_5R_6)^{-\frac{1}{2}}
 \frac{1}{2^{{\delta_{m,0}},{\delta_{n,0}}} }
f_{+++}^{(2m,2n)}(x^{\mu}) \cos(\frac{2mx^5}{R_5} + \frac{2nx^6}{R_6})\nn
f_{++-}(x^{\mu}, x^{\alpha}) =\sum_{m\geq  0} (\pi^2 R_5R_6)^{-\frac{1}{2}}
f_{++-}^{(2m,2n+1)}(x^{\mu}) \cos(\frac{2mx^5}{R_5} + \frac{(2n+1)x^6}{R_6})\nn
f_{+-+}(x^{\mu}, x^{\alpha}) =\sum_{m\geq  0} (\pi^2 R_5R_6)^{-\frac{1}{2}}
f_{+-+}^{(2m+1,2n)}(x^{\mu}) \cos(\frac{(2m+1)x^5}{R_5} + \frac{2nx^6}{R_6})\nn
f_{+--}(x^{\mu}, x^{\alpha}) =\sum_{m\geq  0} (\pi^2 R_5R_6)^{-\frac{1}{2}}
f_{+--}^{(2m+1,2n+1)}(x^{\mu}) \cos(\frac{(2m+1)x^5}{R_5} + \frac{(2n+1)x^6}{R_6})\nn
f_{-++}(x^{\mu}, x^{\alpha}) =\sum_{m\geq  0} (\pi^2 R_5R_6)^{-\frac{1}{2}}
f_{-++}^{(2m+1,2n+1)}(x^{\mu}) \sin(\frac{(2m+1)x^5}{R_5} + \frac{(2n+1)x^6}{R_6})\nn
f_{-+-}(x^{\mu}, x^{\alpha}) =\sum_{m\geq  0} (\pi^2 R_5R_6)^{-\frac{1}{2}}
f_{-+-}^{(2m+1,2n)}(x^{\mu}) \sin(\frac{(2m+1)x^5}{R_5} + \frac{2nx^6}{R_6})\nn
f_{--+}(x^{\mu}, x^{\alpha}) =\sum_{m\geq  0} (\pi^2 R_5R_6)^{-\frac{1}{2}}
f_{--+}^{(2m,2n+1)}(x^{\mu}) \sin(\frac{2mx^5}{R_5} + \frac{(2n+1)x^6}{R_6})\nn
f_{---}(x^{\mu}, x^{\alpha}) =\sum_{m\geq  0} (\pi^2 R_5R_6)^{-\frac{1}{2}}
f_{---}^{(2m,2n)}(x^{\mu}) \sin(\frac{2mx^5}{R_5} + \frac{2nx^6}{R_6}) \nonumber 
\eeqn
\beqn
\label{12.8g}
\eeqn
where the subscripts label the $P,P',P^{''}$ parities.  
The vector multiplet in its $G_{SM}'=SU(3)_C\times SU(2)_L\times U(1)^2$ decomposition takes on
the following parity assignments 

\beqn
(8,1,0,0)_{+++}, ~(1,3,0,0)_{+++}, ~(1,1,0,0)_{+++}, ~(1,1,0,0)_{+++}\nn
(3,2,-5,0)_{++-}, ~(\bar 3,2,5,0)_{++-}\nn
(3,1,4,-4)_{+-+}, ~(1,1,6,4)_{+-+}, ~(\bar 3,1,-4, 4)_{+-+}, ~(1,1,-6,-4)_{+-+}\nn 
(3,2,1,4)_{+--}, ~(\bar 3,2,-1,-4)_{+--}
\label{12.8h}
\eeqn
Now at the orbifold point
$x^5=\pi R/2, x^6=0$, one finds that the gauge vector  bosons with parities $++-$ and $+--$ vanish
and thus only the first and third lines of Eq.(\ref{12.8h}) survive and these generators can
be  assembled into representations of $SU(4)_C\times SU(2)_L\times SU(2)_R$ so that 

\beqn
(15,1,1)= (8,1,0,0)_{+++}+ (\bar 3, 1, -4, 4)_{+-+} + (3,1,4,-4)_{+-+} + (1,1,0,0)_{+++},\nn
(1,3,1)= (1,3,0,0)_{+++},\nn
(1,1,3)=(1,1,0,0)_{+++} + (1,1,6,4)_{+-+} +  (1,1,-6, -4)_{+-+} \nonumber 
\eeqn
\beqn
\label{12.8i}
\eeqn

We see then that the surviving  gauge fields at this orbifold point consist of the sets 
(15,1,1) +(1,3,1) +(1,1,3) which are  precisely the gauge fields for the group 
 $SU(4)_C\times SU(2)_L\times SU(2)_R$. Thus the orbifold point $x^5=\pi R_5/2, x^6=0$,
 can appropriately be labeled G(4,2,2) orbifold, since G(4,2,2) is the surviving
 gauge  symmetry at  this orbifold point. \\
 
   Next we consider the orbifold point $x^5=0, ~x^6= \pi R_6/2$. Here  the surviving operators are
  those with parities $+++$ and $++-$ and consist of the first two lines  of
  Eq.(\ref{12.8h}). They can be  assembled  into the  (24,0) and (1,0) representations of 
  $SU(5)\times U(1)$ as follows

 \beqn
(24,0)= (8,1,0,0)_{+++} + ~(1,3,0,0)_{+++}+ ~(1,1,0,0)_{+++}+ \nn  
(3,2,-5,0)_{++-} + ~((3,2,5,0)_{++-},\nn
(1,0)= ~(1,1,0,0)_{+++}.
\label{12.8j}
\eeqn
Clearly then it is appropriate  to call this orbifold  point an $SU(5)\times U(1)$ orbifold. 
As in the 5D case  a 10-plet of Higgs multiplet in 6D contains two chiral multiplets $H,\hat H$.
For H the $Z_2\times Z_2'\times Z_2^{''}$ parities can be assigned as follows in $G_{SM'}$ 
decomposition 

\beqn
H(1,2,3,2)_{+++}, H(1,2,-3,-2)_{+-+},
H(3,1,-2,2)_{++-},H(\bar 3,1,2,-2)_{+--}. \nn
\label{13.8a}
\eeqn  
Proceeding as  before  we  consider the orbifold point $x^5=\pi R_5/2, x^6=0$. One finds that
the non-vanishing Higgs multiplets here  fall into the (1,2,2) representation of \ps since

\beqn
H(1,2,2) = H(1,2,3,2)_{+++} +H(1,2,-3,-2)_{+-+}. 
\label{13.8b}
\eeqn
Similarly at the orbifold point $x^5, x^6=\pi R_6/2$,  one finds that the following non-vanishing Higgs 
multiplets fall into the  (5,2) representation of $SU(5)\times U(1)$~\cite{Asaka:2001eh}

\beqn
H(5,2)= H(1,2,3,2)_{+++} +H(3,1,-2,2)_{++-}. 
\label{13.8c}
\eeqn
Thus we can classify the  6D orbifold points as follows

\begin{enumerate}
\item
SO(10) brane
\item
$SU(5)\times  U(1)$ brane
\item
Flipped $SU(5)\times U(1)$ brane
\item
\ps  brane
\end{enumerate}
In the orbifold breaking of the gauge  symmetry the rank of the group is typically not reduced. To reduce the 
rank down to the standard  model gauge group symmetry one needs  to introduce $16 + \overline{16}$ of Higgs.
The choice of the Higgs structure to break the symmetry down to the SM gauge group depends  on the details
of the model. Further, proton decay is very sensitive  to placement of generations  in the compact space
and there are a variety of models each with a different scenario.  We would not discuss 
the specific details  of their constructions. Rather, in the following  we comment on some general 
features common to these constructions.\\

There  is no dimension 4 or dimension 5 proton decay in models of this type for reasons similar to the
case of 5D models. Proton decay from dimension six  operators is very model dependent. For example,
placement of all three generations on the  \ps brane will suppress proton decay  from X and Y exchange.
A similar  situation holds  if  the first  generation is placed on the \ps brane and the second and 
third  generations on the flipped  $SU(5)\times U(1)$ and the $SU(5)\times U(1)$ branes.  When 
dimension six operators  from the  X and Y generations are allowed,  one finds  that there  is a 
modification due to the exchange of the towers of KK states. Thus  the  mass of a (m,n) KK state is

\beqn
M_X^2(m,n)= (2m+1)^2 M_5^2 + (2n)^2 M_6^2,
\label{13.8d}
\eeqn  
where $M_5\equiv R_5^{-1}$ and $M_6\equiv R_6^{-1}$. The effective mass that enters in the dimension six
operator is  $\tilde M_X$ where 

\beqn
(\tilde M_X)^{-2} = 2\sum_{m,n=0}^{\infty} ((2m+1)^2M_5^2+(2n)^2 M_6^2)^{-1}. 
\label{13.8e}
\eeqn 
For the case when $M_6/M_5\to 0$, one finds that    
$(\tilde M_X)^{2}= \frac{4}{\pi^2} M_5^2$ which is correctly the 5D result. 
For the case  of the double  summation the sum actually diverges. However, infinite summation on (n,m) is
not really justified since above an effective scale  $M_*$ the theory becomes strongly interacting. 
Because of this one ought to  use a cutoff  so that one counts KK  states only below $M_*$. This can
be  done by putting a cutoff so that $M_X(m,n)\leq  M_*$. 
One then using $M_5=M_6=M_C$ 

\beqn
{(\tilde M_X)^{-2}} \simeq \frac{\pi}{4}M_C^{-2}(ln \frac{M_*}{M_C} +2.3).
\label{13.8f}
\eeqn 
The above modification leads  to an enhancement of the proton lifetime similar to what happens in the
5D case. Also as in the  case of the 5D analysis  derivative couplings can produce  proton decay. 
Beyond these general observations the details of the proton decay are highly model dependent. 
As an example, we  note that the work of Ref.~\cite{Buchmuller:2004eg} investigates a specific 
model where the three 
generations of 16 plets of matter are located at different branes. Thus generation 1 is placed
on the $SU(5)\times U(1)$ brane, generation 2 is placed  on the flipped $SU(5)\times U(1)$ brane,
and generation 3 is placed on the \ps brane. There are additional assumptions  regarding the
Higgs structure and  flavor sector  of the theory. In this model the dominant proton decay
branching ratios are~\cite{Buchmuller:2004eg}.

\beqn
BR(\pi^0e^+)=(71-75)\%,~~BR(\bar \nu \pi^+)=(19-23)\%,~~ \nn
BR(\mu^+\pi^0) = (4-5)\%,
\label{13.8g}
\eeqn
while the other modes are typically less  than 1\%.
An interesting signature  of Eq.(\ref{13.8g}) is the strong suppression of the mode
$\mu^+K^0$ compared to the predictions of the 4D models.
 The analysis of Ref.~\cite{Buchmuller:2004eg}
calculates the life time for the $e^+\pi^0$ mode  so that

\beqn
\tau(p\to e^+\pi^0)= 5.3\times 10^{33} (\frac{0.01 {\rm GeV}^3} { \alpha})^2  (\frac{M_C}{9\times 10^{15}})^4
yr.  
\eeqn
Using  $\alpha =0.01$ GeV$^3$, and $M_C=2\times 10^{16}$ GeV as 
indicated by the unification of the gauge coupling constants,
one  finds that $\tau(p\to e^+\pi^0)\simeq 1\times 10^{35}$ yr.  This life time lies  within reach of the next generation 
of proton decay experiments.\\

\subsection{Gauge-Higgs unification}  

Another class of model which are closely related  are models with gauge-Higgs
couplings  unification~\cite{Hall:2001zb}. Here the Higgs doublet fields  arise as  a  part of the vector
multiplet and hence there  is a unification of the  gauge and  Higgs couplings. There  are
several variants  of such models. We discuss briefly an SU(6) model in 6D compactified on 
$T^2/(Z_2\times Z_2')$ of  Ref.~\cite{Hall:2001zb}. One introduces  an SU(6) vector  multiplet in the bulk
which can be decomposed under  4D N=1 supersymmetry as  the multiplets $V, V_5, V_6, \Sigma$. 
To construct the  $T^2/(Z_2\times Z_2')$ orbifold one considers  the following operations:
${\cal Z}_5$: $(x^5,x^6)\to (-x^5,  x^6)$;  ${\cal Z}_6$: $(x^5,x^6)\to (x^5, - x^6)$;
${\cal T}_5$: $(x^5,x^6)\to (x^5+l_5,  x^6)$; ${\cal T}_6$: $(x^5,x^6)\to (x^5,  x^6+l_6)$ where
$l_5=2\pi R_5$ and $l_6=2\pi R_6$.  One can choose the  transformations for the fields  under the 
above transformations  
so that the zero modes  correspond to the $SU(3)_C\times SU(2)_L\times U(1)_Y$  components. 
Corresponding to  ${\cal Z}_5$ and  ${\cal Z}_6$ transformations we choose 

\beqn
V(-x^5,x^6) =P_ZV(x^5,x^6)P_Z^{-1},\nonumber\\
V(x^5, -x^6) =P_ZV(x^5,x^6)P_Z^{-1},\nonumber\\
\Sigma(-x^5,x^6) = -P_Z\Sigma(x^5,x^6)P_Z^{-1},\nonumber\\
\Sigma(x^5,-x^6) = -P_Z\Sigma(x^5,x^6)P_Z^{-1}.
\eeqn
and similarly

\beqn
V_5(-x^5,x^6) = -P_ZV_5(x^5,x^6)P_Z^{-1},\nonumber\\
V_5(x^5, -x^6) = P_ZV_5(x^5,x^6)P_Z^{-1},\nonumber\\
V_6(-x^5,x^6) =P_ZV_6(x^5,x^6)P_Z^{-1},\nonumber\\
V_6(x^5,-x^6) = -P_ZV_6(x^5,x^6)P_Z^{-1}.
\eeqn 
where $P_Z$ is chosen so that~\cite{Hall:2001zb}

\beqn
P_Z = diag (1,1,1,1,1,-1).
\eeqn
Under ${\cal T}_5$ and ${\cal T}_6$ the  fields transform as  follows 

\beqn
V(x^5+l_5,x^6) =P_TV(x^5,x^6)P_T^{-1},\nonumber\\
V(x^5, x^6+l_6) =P_TV(x^5,x^6)P_T^{-1}.
\eeqn
and identical relations hold for the  other fields, where $P_T$ is chosen so that~\cite{Hall:2001zb}

\beqn
P_T= diag (1,1,1,-1,-1,-1).
\eeqn
With  the  above assignments, $SU(6)$ breaks down to  
$SU(3)_C\times SU(2)_L\times U(1)_Y\times U(1)_X$.
The $P_Z, P_T$ parities of  the $V$ and $\Sigma$ components  can now be  exhibited.  

\begin{equation}
  V:\: \left( \begin{array}{c|c|c}
   (3\times 3){(+,+)} & (3\times 2){(+,-)} &  (3\times 1){ (-,-)} \\ \hline
  (2\times 3){(+,-)} & (2\times 2){(+,+)} & (2\times 1){ (-,+)} \\ \hline
  (1\times 3){(-,-)} & (1\times 2){(-,+)} &  (1\times 1){ (+,+)} 
      \end{array} \right),
\label{eq:V6trans}
\end{equation}

\begin{equation}
  \Sigma:\: \left( \begin{array}{c|c|c}
   (3\times 3){(-,+)} & (3\times 2){(-,-)} &  (3\times 1){ (+,-)} \\ \hline
  (2\times 3){(-,-)} & (2\times 2){(-,+)} & (2\times 1){ (+,+)} \\ \hline
  (1\times 3){(+,-)} & (1\times 2){(+,+)} &  (1\times 1){ (-,+)} 
      \end{array} \right),
\label{eq:V6trans}
\end{equation}
where $(3\times 3)(+,+)$ means that all elements of a $(3\times 3)$ matrix have $P_Z, P_T$ 
parities $(+,+)$ and $(3\times 2)(+,-)$ etc are similarly defined.
 Looking at the $\Sigma$ fields, one finds that fields with $(+,+)$ 
parities  have the $SU(3)_C\times SU(2)_L\times U(1)_Y$ quantum numbers of 
$(1,2,\frac{1}{2}) + (1,2,-\frac{1}{2})$.  These fields  then qualify as Higgs doublets
of MSSM allowing for the possibility of gauge-Higgs unification since $\Sigma$ is part of
the original vector  multiplet in 5D. Before  proceeding further, it is instructive  to
identify the  residual gauge symmetry at various orbifold points. We label the
orbifolds by $(x^5,x^6)$ values. Thus  the residual symmetries at the various orbifold points
are: (i)(0,0): $SU(5)\times U(1)_X$, (ii) ($\pi$ R,0): $SU(3)_C\times SU(2)_L\times U(1)_Y\times
U(1)_X$, (iii) (0,$\pi$ R):  $SU(3)_C\times SU(2)_L\times U(1)_Y\times U(1)_X$,
(iv) ($\pi$ R, $\pi$ R): $SU(3)_{\tilde C}\times SU(2)_L\times U(1)_{\tilde X}$.  As in previous scenarios,
proton decay is sensitive to how matter is located in the compact extra dimensions. 
If we place matter on the (0,0) orbifold  point, the residual  symmetry is  $SU(5)\times U(1)_X$
and one has dimension six operators from X and Y gauge bosons. On the other hand 
if the quark lepton generations  are  placed at  the other orbifold  points with 
reduced gauge symmetry, e.g., at  the orbifold  point (0, $\pi$ R),  
dimension six  proton decay from the X and Y gauge  bosons will
be absent. However, as discussed earlier  one can have proton  decay from derivative 
couplings although  such decays will be suppressed by volume of the extra  
dimensions. We note in passing that dimension 5 proton decay through Higgs triplet 
mediation is absent  since there are no couplings of the Higgs triplets to quarks and leptons.\\

\subsection{Proton decay in universal extra dimension (UED) models
\label{ued}}

We turn now to a discussion of proton decay in the  universal extra 
dimension (UED) models.
In these models  
it is possible  to control proton decay via  the use of extra symmetries that 
might arise in models  with universal extra 
dimensions~\cite{Appelquist:2000nn,Appelquist:2001mj}. Thus in six dimensions 
with two universal extra 
dimensions the standard model particles  are charged  under the $U(1)$ symmetry which
arises due to the extra dimensions $x_4$ and $x_5$ and thus this symmetry may be labeled
as $U(1)_{45}$.
 Even after  compactification a discrete $Z_8$ symmetry survives. The  
symmetry allows only very high dimension baryon and lepton number violating  operators,
i.e., dimension sixteen or  higher which leads to a suppression of proton decay.
In six dimensions  the Lorentz  symmetry is $SO(1,7)$  and  in six dimensional  space
on can introduce Dirac  matrices $\Gamma^M$ (M=0,1,..,5) which are  $8\times 8$ and
can define a $\Gamma^7$ matrix so that $\Gamma^7=\Gamma^0\Gamma^1..\Gamma^5$. Using 
$\Gamma^7$ one  can define chiral eigenstates  ${\cal  \Psi}_{\pm}$ of chiralities  $\pm$ 
and thus a six  dimensions  $\cal{\psi}$ can be  broken up into two $\cal{\psi}_{\pm}$.
Each of the six dimensional chiralities  states are full four  component Dirac fields 
in four dimensions and  can be  further decomposed in left and right chiral projections
under  the four dimensional chiral projection. An interesting result is that the
Standard  Model gauge  and gravitational anomalies  cancel only for  certain
combinations of  chiral assignments which are one of the following two 
possibilities~\cite{Appelquist:2000nn}

\beqn
(i) ~{\cal Q_+, U_-, D_-, L_+, E_-, N_-}; ~(ii) ~{\cal Q_+, U_-, D_-, L_-, E_+, N_+}
\eeqn
where all the quark-lepton fields are in six dimensions and where $\cal{N}$ is a gauge
singlet that is  needed for  the cancellation of gravitational  anomaly. On compactification
the zero modes of ${\cal Q_+, U_-, D_-}$ etc  fields will  be the standard model fields.
The $U(1)_{45}$ quantum numbers of  the fields are as  follows
\beqn
(u_L,d_L, u_R, d_R)(-\frac{1}{2}), ~(\nu_L, e_L, \nu_R, e_R)(\mp \frac{1}{2})   
\label{6dqn}
\eeqn
Because of Eq.(\ref{6dqn}) one can immediately see that  lepton and  baryon number violating
operators  of  the type $QQQL/M$ are forbidden. Thus Lorentz invariance
in six  dimensions severely constraints the operators and  the allowed lepton and baryon number
violating operators must have at least three quarks and  three leptons. This constraint
leads to interesting new signals for proton decay. Thus consider  the following operator
allowed by the above constraints~\cite{Appelquist:2000nn}                                    
\beqn
{\cal O}_{17} =\frac{C_{17}}{\Lambda^{11}} (\bar {\cal L_+ D_-})^3 \tilde {\cal H}
\label{o17}
\eeqn
where  $\tilde {\cal H}$ is the conjugate Higgs doublet  in six dimensions, and $\Lambda$ 
is the scale up to which the six dimensional effective theory is  valid. On compactification
one can obtain the effective baryon and lepton number violating operator  in four dimensions.
The effective operator in four dimensions contains  the term $(\bar \nu_Ld_R)^2(\bar l_Ld_R)$
which implies proton decay modes of the type, $\pi^+\pi^+e^-\nu\nu$ and   
$\pi^+\pi^+\mu^-\nu\nu$. As estimate of proton decay into these modes is then 

\beqn
\tau(p\to \pi^+\pi^+l^-\nu \nu)\simeq 
\frac{10^{35}yr}{C_{17}^2} (\frac{(2\times 10^{-12}}{P_5 f(\pi)})  
(\frac{M_C}{0.5 TeV})^{12} (\frac{\Lambda}{5M_C})^{22}
\label{pipienunu}
\eeqn
 Here  $P_5$ is the phase space factor  which is estimated to be $\leq 2\times 10^{-12}$, 
$f(\pi)$ is a $\pi\pi$ form factor which is expected to be $O(1)$, and $M_C=1/R$ is the compactification
scale. Setting $C_{17}=1$ and the ratios within the braces to unity one find  that 
$\tau(p\to \pi^+\pi^+l^-\nu \nu)\simeq 10^{35}$ yr. The current experimental limits on 
the mode $p\to \pi^+\pi^+e^-$ is $\tau_p> 3\times 10^{31}$ yr. Thus we  see  that with
the default values of the parameters in Eq.(\ref{pipienunu}) the partial lifetime 
$\tau(p\to \pi^+\pi^+l^-\nu \nu)$
 is much larger by orders of 
magnitude than the current limits of similar type processes. One must, however, keep in mind
the extreme sensitivity of the theoretical predictions because of the high powers on
quantities which are currently unknown. 
The above results have been derived using the six dimensional symmetry. On compactification
of the two extra dimensions, the SO(1,5) symmetry including the $U(1)_{45}$ subgroup symmetry 
is broken and a simple choice is compactification of $T^2/Z_2$ orbifold of equal radii. In this
case the $U(1)_{45}$ symmetry is broken down to a $Z_8$ symmetry. This discrete symmetry is sufficient
to guarantee  that there  are no baryon and  lepton number violating processes with less than
three quarks and three  leptons. Of  course it remains to be  seen if the considerations  of 
Casimir energy indeed lead  to the vacuum state with the desired  symmetry. Some  progress 
along this direction is made in Ref.~\cite{Ponton:2001hq}. 
 Further development of this scheme has been carried out in the analysis of Ref.~\cite{Mohapatra:2005wg}
where issues of neutrino masses and of dark matter are also addressed. The gauge group investigated here
include $SU(2)_L\times U(1)_{I_{3R}}\times U(1)_{B-L}$ and $SU(2)\times SU(2)_R\times U(1)_{B-L}$
and compactifications on a $T^2/Z_2$ or $T^2/Z_2\times Z_2'$ orbifolding is considered. The dominant decay
mode of the neutron in this model  is $n\to 3\nu$.  Aside from the power law 
suppression of proton decay, a similar mechanism for the generation of small neutrino masses 
is also valid. Further, in this model dark matter could consist of two components consisting of 
Kaluza-Klein excitations of the neutrino and of the photon. In summary in UED models  a discrete
subgroup of the Lorentz symmetry in six dimensions continues to forbid dangerous proton decay operators
when reduction to four dimension is carried out.

\subsection{Proton decay in warped geometry}

Warped geometry presents a possible solution to the hierarchy problem without necessarily using
supersymmetry. Thus in Refs.~\cite{Randall:1999ee,Randall:1999vf} 
Randall and Sundrum proposed a metric  of the form 
\beqn
ds^2 =e^{-2k|y|} \eta_{\mu\nu} dx^{\mu}dx^{\nu} +dy^2 
\eeqn
where y is the  coordinate of the extra dimension limited to $0\leq y\leq \pi r_c$ 
where $r_c$ may be considered the compactification radius for the extra  dimension.
The action of the theory consists  of a Planck brane at $y=0$ and a TeV brane  at $y=\pi r_c$
and the geometry is  a slice of $AdS_5$. The AdS geometry creates  a warp factor and 
  mass scales at the two branes are related by an exponential hierarchy. In the original
  formulation of RS  all the standard model particles are located at the TeV brane. 
  Later it was  realized that to solve  the hierarchy problem one needs  only the Higgs
  fields  on the TeV brane and the remaining standard model fields including quarks,
  leptons and the gauge bosons could live in the
   bulk~\cite{Davoudiasl:1999tf,Pomarol:1999ad,Grossman:1999ra,Gherghetta:2000qt}. 
   This procedure leads  naturally to 
  a hierarchy of the Yukawas couplings if different generations of standard model fermions 
  are  located at different points in the bulk~\cite{Grossman:1999ra,Gherghetta:2000qt,Huber:2000ie,Huber:2003tu}.
  One still  has  to address the issue of dangerous proton decay operators  in the theory.
  A possible way to address this  problem is to assume a gauged baryon number 
  symmetry~\cite{Goldberger:2002pc,Agashe:2002pr}.  
  However, to make such a  
  symmetry compatible with grand unification, one needs to break 5D GUT by boundary 
  conditions~\cite{Kawamura:2000ev,Hall:2001pg,Hebecker:2001wq} and extract zero modes
  for a single generation from different multiplets.
 The remaining components of the multiples have  only KK modes.  
Thus in the work of Ref.~\cite{Agashe:2004ci,Agashe:2004bm} a non-supersymmetric extra dimensional
 Randall-Sundrum
 (RS) model~\cite{Randall:1999vf} has been explored. 
The specific model of Ref.~\cite{Agashe:2004ci}  assumed the grand unified group is broken
to the Standard Model gauge group by boundary conditions on the Planck brane and the matter is composed
from different replicas of multiplets~\cite{Agashe:2002pr}. 
For example, for the case of SO(10) one assumes three 16 -plet representations
for each generation as  shown below  

\beqn
\left(\begin{array}{c}
 \left(u_L, d_L, u^{c'}_R, d^{c'}_R, \nu_L', e_L', e^{c'}_R, \nu^{c'}_R \right)_{B=1/3}\\
  \left(u_L', d_L', u^{c}_R, d^{c}_R, \nu_L', e_L', e^{c'}_R, \nu^{c'}_R 
 \right)_{B=-1/3}\\
  \left(u_L', d_L', u^{c'}_R, d^{c'}_R, \nu_L, e_L, e^{c}_R, \nu^{c}_R \right)_{B=0}
  \end{array}\right)
\eeqn
where only the unprimed fields  have  zero modes and the subscript indicates the baryon number of the
multiplet. Thus one finds  that a full  generation of matter
arises from three replicas of 16-plet of matter. The baryon number assignment of the multiplets 
corresponds to the baryon number  of the zero modes. The assumption that baryon number is conserved
leads to a $Z_3$ symmetry
\beqn
\Phi \to ~exp{(2\pi i(B-\frac{n_c-\bar n_c}{3}))} ~\Phi
\eeqn
Here the multiplet  $\Phi$ carries  the baryon number $B$ and $n_c(\bar n_c)$ is  the color (anti-color) index.
The quantum numbers assignments are such that the zero modes which constitute the standard model  particles
are  not charged under  $Z_3$ while the other states  are. This also applies to the gauge vector bosons
of  SO(10) where  the gauge bosons  which enter in the Standard Model are  not charged  under  $Z_3$ 
but the lepto-quarks are charged. Thus exotic particles with non-vanishing baryon number B cannot decay
into the Standard Model particles. 
In this scenario the lightest Kaluza-Klein particle (LKP) will be stable 
and could be a candidate  for dark matter. Of course, the baryon number gauge  symmetry cannot be exact 
as it would lead to an undesirable  massless gauge  boson. The analysis of Ref~\cite{Agashe:2004ci}
 has analyzed the implications
of  such breaking on the Planck brane. It  is shown that  if the 
symmetry is broken such that  $\Delta B \neq \frac{1}{3},\frac{2}{3}$, proton decay will be
suppressed by a Planck mass and the LKP mode could be long lived with as much as  $10^{10}$ times the age of 
the universe~\cite{Agashe:2004ci,Servant:2004ke}.\\ 
 
 In another work which is motivated by RS models~\cite{Agashe:2003zs,Agashe:2004rs} 
 unification of gauge couplings with
 composite Higgs and  a composite  right handed top quark are 
 considered~\cite{Agashe:2005vg}. 
 Thus RS models
 where most or all of the Standard Model fields are in the RS bulk may have a dual to
 a purely 4D composite Higgs scenario  via a ADS/CFT correspondence~\cite{Aharony:1999ti,Arkani-Hamed:2000ds}.
 Motivated by this observation it is then suggested
 that in the running of the gauge unification one should project out the Higgs  above
 a compositeness scale $\Lambda_{comp}$. It is further suggested that the largeness of the
 top Yukawa couplings indicates that either $t_L$ or $t_R$ or both may be composite.
 However, precision electroweak data on $Z\to b\bar b$ indicate the elementarily of 
 $b_L$ and hence of $t_L$ and thus it is argued that $t_R$ should be  composite
~\cite{Agashe:2003zs}.
 In running of the gauge coupling constants above the 
 scale $\Lambda_{comp}$ one should then replace H and $t_R$ by the strong 
 dynamics  so that
 \beqn
 \alpha_i(Q)= \alpha_U +SM-\{H,t_R\} + {\rm strong ~dynamics} + M_U-{\rm corrections}  
 \eeqn
 Now if the  strong dynamics cancels out in the differential running as  would be the case  
 if the SM gauge group is embedded in a simple factor of G then one will have 
 \beqn
 \alpha_i(Q)-\alpha_1 = SM-\{H,t_R\} + M_U-{\rm corrections}  
 \label{acs1}
 \eeqn
 While Eq.(\ref{acs1}) improves the unification relative to the  Standard Model running,
  a variant of the scenario improves it  still further. Here one include $t_R^c$ along with $H, t_R$ 
  on the right hand of Eq.(\ref{acs1}). With this modification and assuming  that the corrections  
 from heavy states at the  unification scale are small as is conventional, 
 one finds a unification scale of $M_U\sim 10^{15}$ GeV. This scale is too low 
 to suppress proton decay from the exchange of states with  masses of this size
 which generate  baryon and lepton number violation such as lepto-quarks.
  Additionally there are also composite states  which can generate  proton 
  decay in this model. However, it is envisioned that the model  arises from
  a string or orbifold compactification where processes of the above type
  are suppressed by symmetries or orbifold projections.
 \subsection{Proton stability in kink backgrounds}
 Another approach to suppression of proton decay operators in extra dimensional models 
 comes from fermion localization mechanism~\cite{Arkani-Hamed:1999pv,Mirabelli:1999ks,Kaplan:2001ga} 
 where chiral fermions are localized in
 solitonic backgrounds~\cite{Jackiw:1975fn}. 
 With this mechanism the quarks and leptons have  Gaussian
 wave functions in the extra dimension under a  kink background. In this scenario
  the Yukawa couplings  will be suppressed since they  involve overlap of 
  two quark or lepton wavefunctions. This mechanism for the suppression of 
  proton decay in extra dimensional models 
is explored in  Ref.~\cite{Kakizaki:2001ue}  where it is  proposed  that the same mechanism that leads to a 
hierarchy of  quark-lepton
masses and couplings is also responsible for the longevity of the proton.    
Specifically,  in the analysis of Ref.~\cite{Kakizaki:2001ue} the quark-lepton chiral multiplets are
localized under a kink background along a spatial extra dimension and the smallness of the 
Yukawa couplings and of the operators that govern proton decay result  from the overlap of 
their wave functions and are exponentially
 suppressed.

In summary, in this section we have investigated proton decay in grand unified models based in extra dimensions.
The most commonly studied  models  are those using compactifications of five and six dimensions to 
four dimensions. While the focus of most model building has been on SU(5) and SO(10)  in extra dimensions,
other possibilities such as SU(6) and SU(3)$^3$ have also been investigated. The main attractive feature
of such model building is a natural  doublet -triplet splitting, which makes the color triplets superheavy
while the  SU(2)$_L$  Higgs doublets remain light. In some models there is a residual $U(1)_R$ invariance 
which kills proton decay from dimension four and five operators leaving the exchange of X and Y gauge bosons
as the main possible source of proton decay. However, proton decay from X and Y exchanges turns out to be 
highly model dependent as it depends critically on how the matter fields  are  located in the extra dimensions.
If the matter fields  are  assumed to propagate in the bulk, then a full generation of quarks  and leptons 
must arise from split multiplets which have no normal X and Y gauge  interactions  among them. In such 
models  proton decay can arise only via higher than six dimensional operators which is far  too small to be
of relevance  for any experiment in the foreseeable  future. The usual dimension six operators  can also 
be forbidden by location of matter on certain brains. For example, for the SO(10) case placing all three
generations on the \ps brane will give vanishing dimension six operators from the normal X and Y exchanges
since the wave functions for  the X  and  Y gauge  bosons vanish on the  \ps brane. However, with other
choices  of locating matter on branes,  one will  have in general proton decay from dimension six operators.
Additionally proton decay can arise from derivative couplings. Consequently, predictions of proton decay 
in higher dimensional models vary over a wide range, from predictions of an essentially absolutely forbidden 
case  to the case  where  it could be  just around the corner. Turning this observation around, whole classes
of models would be eliminated by the observation of proton decay. Thus proton decay is an important 
discriminator of higher dimensional grand unified  models. 

\section{Proton Decay in String Models}
The string theory holds out the hope 
of unifying all the interactions of nature
 including gravity (For a review see~\cite{reviews,polchinski}). 
 There  are  five types  of  known string theories:
 Type I, Type IIA, Type IIB, $SO(32)$ heterotic and $E_8\times E_8$ heterotic. 
 These theories are known to be connected by a web  of dualities.
 Indeed all these five theories  may have  a common origin in a more 
 fundamental theory which is the so called M -theory, and  whose  low
 energy limit is  an 11 dimensional supergravity.
   We will first discuss proton decay in the heterotic string models~\cite{Gross:1984dd}. 
   Historically this is the
 class of models  which were investigated in great detail in the 
 beginning~\cite{Ibanez:1987sn,Greene:1986bm} 
 and there  has been a revival of interest in these  models  more recently. 
 The $E_8\times E_8$ heterotic string model after compactification can generate 
 a large variety of models  since models with rank up to 22 are allowed. Many 
 possibilities  for model building exist and the models investigated  include
 those based  on  free fermionic constructions, on orbifolds~\cite{Dixon:1985jw}
 and on  Calabi-Yau compactifications~\cite{Candelas:1985en}.  
 The number of possibilities is 
 rather large one may use additional principles to reduce the number of
 models. Below we will discuss in some detail models based on some  specific Calabi-Yau 
 manifolds which come close to being realistic.  We will also discuss the situation 
 regarding proton stability in string models based on Kac-Moody levels $k>1$.
  Later  we will discuss proton stability in  the more recent  class  of models,
  based on Type IIA or Type IIB or more generally M  theory models.
 We will also discuss proton decay induced by quantum gravity via wormhole and blackhole
 effects and the role of  $U(1)$ abelian gauge symmetries and 
 discrete symmetries in controlling dangerous proton decay.  
 
A brief outline for the rest of the section is as follows:  In Sec.(7.1) we discuss
proton stability in Calabi-Yau models. A discussion of   grand  unification in 
Kac-Moody levels $k>1$ is given in Sec.(7.2). The  $k>1$ levels are needed
to realize massless scalars in the adjoint representation 
necessary to break the GUT symmetry.  It turns out, however, that at level 2
 it is difficult to  obtain 3 massless generations but it is possible to overcome 
 this problem at level 3.  
Baryon and lepton number violating dimension four operators are
 absent in these models due to an underlying gauge and discrete symmetry. 
There are, however, present the  \bl
 ~violating dimension five operators  and  it is necessary to 
 suppress them by  heavy Higgs triplets.  One problem in such models concerns
the generation of proper quark-lepton masses.  In the absence of such mass
generation it is difficult to carry out a detailed analysis of  proton lifetime.
 A new class of heterotic string models are  discussed in Sec.(7.3).

This class of  model have
an  MSSM  massless  spectrum, and no baryon and lepton number violating
operators exist except for those 
    induced by quantum gravity. Also discussed in Sec.(7.3) are other 
 attempts at realistic 4D model building .
\\

In Sec.(7.4) proton decay in M-theory compactifications are  discussed.
While 	quantitative predictions of proton lifetime do not exist in models
based on such compactifications due to an unknown  overall normalization
factor, still qualitative predictions of proton life time are possible and are
discussed.  A review  of  proton decay in intersecting  D brane models
is given in Sec.(7.5).  The case discussed in some detail is of  $SU(5)$ like GUT models in
Type IIA orientifolds with D-6 branes.  The analysis focusses on the \bl~violating
dimension six operators while it is assumed that   the \bl ~violating
dimension 4  and dimension 5 operators are absent. Quite interestingly the predictions
of proton lifetime lie within reach of the next generation of proton
decay experiment.  
A discussion of proton stability  in string landscape
models is given in Sec.(7.6). There exist a number of scenarios of soft breaking 
of supersymmetry where the squarks and sleptons can become superheavy 
and proton decay from dimension five operators is suppressed.   A discussion
of proton decay arising from quantum gravity effects is given in Sec.(7.7). 
It is widely conjectured that quantum gravity does not conserve baryon number 
and can generate proton decay. Also discussed in Sec.(7.7) is proton decay in
higher dimensional models via quantum gravity effects.  The suppression of
proton decay from U(1) string symmetries is given in Sec.(7.8). Finally a discussion
of discrete symmetries that allow for the suppression of proton decay is given in
Sec.(7.9).  

\subsection{Proton Stability in Calabi-Yau Models}
We begin with a discussion of a class of heterotic string models 
which on compactifications maintain $N=1$ supersymmetry~\cite{Candelas:1985en}.
These compactifications are of the type $M_4\times K$ where $M_4$ is the 
four dimensional Minkowski space and $K$ is a compact six-dimensional 
Calabi-Yau manifold~\cite{yau}.  The fact that  one has residual $N=1$ supersymmetry after compactification is 
attractive for model building. 
A specific interesting case is the manifold $CP^3\times CP^3/Z_3$ with coordinates
$x_i,y_i$ (i=0,1,2,3) [These obey the constraints
  $P_1\equiv \sum x_i^3+a x_0x_1x_2+a_2+x_0x_1x_3=0$,
$P_2=x_0y_0+c_1x_1y_1+c_2x_2y_2+c_3x_3y_3+c_4x_2y_3+c_5x_3y_2=0$, and
$P_3=\sum y_i^3+ b_1y_0y_1y_2+b_2y_0y_1y_3=0$.].  There are nine complex or eighteen real
parameters that  enter in K. The zero modes of  K are given by the Hodge
numbers. For the model above one has~\cite{Greene:1986bm} 

\beqn
h_{2,1}=9, ~~h_{1,1}=6
\eeqn
which imply that there are nine 27-plet generations  and six $\overline{27}$ 
generations which leads to a net three  generations of matter. 
The non-simply connected  nature  of $CP^3\times CP^3/Z_3$  manifold allows 
for the breaking of the $E_6$  gauge symmetry by Wilson loops and one 
has~\cite{Hosotani:1983xw,Witten:1985xc}

\beqn
E_6\to SU(3)_C\times SU(3)_L\times SU(3)_R
\eeqn
In terms of $[SU(3)]^3$ there will be nine families of nonets of leptons  
$L_r^l(1,3,\bar 3)$ from the nine generations of $27$, and six families of mirror
leptons $\bar L_r^l(1,\bar 3,3)$. There would also be seven nonet  of 
quarks $Q_l^a(3,\bar 3, 1)$ and four families of mirror quarks $\bar Q_a^l(\bar 3,3,1)$;
seven nonents of anti-quarks $(Q^c)_a^r(\bar 3, 1,3)$ and four nonets of mirror
anti-quarks $(\bar Q^c)^a_r(3,1,\bar 3)$. Here  (a,l,r)=(1,2,3) label 
(color, left, right) components. In the standard particle notation these nonets 
are given by 

\beqn
L=(l^{\alpha}, H^{\alpha}, H'_{\alpha}, e^C, \nu^C, N), ~~Q=(q^{\alpha}, D),
~~Q^c=(u^C, d^C, D^C)
\eeqn  
where $l^{\alpha}$, $l^{\alpha}$, $H^{\alpha}$, $H'_{\alpha}$, and $q^{\alpha}$ 
are the lepton, Higgs-boson, and quark $SU(2)_L$ doublets, D and $D^C$ are color 
Higgs triplets, and $N, \nu^C$ are $SU(5)$ singlets  while $N$ is also an
$SO(10)$ singlet.\\

An important constraint in model building on Calabi-Yau manifolds is that of matter 
parity $M_2$ which for the three generation models is 
defined by~\cite{Bento:1987mu,Lazarides:1988zd} 

\beqn
M_2=C U_Z; ~~C=(1,1,\sigma)\times (1,1, \sigma),~~
\sigma=\left(
\begin{array}{cc}
 0 & 1 \\ 1 & 0
\end{array}
\right)\nonumber
\\
U_Z=diag(1,1,1)\otimes diag(-1,-1,-1)\otimes diag(-1,-1,1)	    
\eeqn
where C is  a transformation of the Calabi-Yau coordinates 
$(x_0,x_1,x_2,x_3)\times (y_0, y_1,y_2,y_3)$ 
and $U_Z$ is an element of $SU(3)_C\times SU(3)_L\times SU(3)_R$. 
Under  the  constraint of the  discrete  symmetry C the number of parameters on the
Calabi-Yau manifold reduce to five complex parameters [In this case  the
constraints read 
$P_1\equiv \sum x_i^3+a (x_0x_1x_2+x_0x_1x_3)=0$,
$P_2=x_0y_0+c_1x_1y_1+c_2(x_2y_2+x_3y_3)+c_3(x_2y_3+c_5x_3y_2)=0$, and
$P_3=\sum y_i^3+ b_1(y_0y_1y_2+ y_0y_1y_3)=0$. Thus instead of nine complex parameters
 for the general  case, we have here just five complex
parameters for the restricted space.].
\begin{table}[h]
\begin{center}
\begin{tabular}{|r|r|}
\hline\hline
C-even states & C -odd states\\
\hline
$L_{1+}, L_{3+}, L_5, L_7, L_{8+}$ ~~~&  $L_{1-}, L_{3-}, L_6, L_{8-}$~~~\\
$Q_1, Q_2, Q_3, Q_{4+}, Q_{6+}$ ~~~& $Q_{4-}, Q_{6-}$ ~~~\\
$Q_1^C, Q_2^C, Q_3^C, Q_{4+}^C, Q_{6+}^C$ ~~~& $Q_{4-}^C, Q_{6-}^C$~~~\\
$\bar L_1, \bar L-2$ ~~~& $\bar L_3, \bar L_4, \bar L_5, \bar L_6$ ~~~\\
$\bar Q_{1+}, \bar Q_{3+}, \bar Q^C_{1+}, \bar Q^C_{3+}$ ~~~~& 
$\bar Q_{1-}, \bar Q_{3-}, \bar Q^C_{1-}, \bar Q^C_{3-}$ ~~~\\
\hline\hline
$M_2$-even states ~~~& $M_2$ -odd states~~~\\
\hline
$l_r, e^C_r, \nu^C_r$ ~~~& $l_n, e^C_n, \nu^C_n$ ~~~\\
$q_r, u^C_r, d^C_r$ ~~~& $q_n, u^C_n, d^C_n$~~~\\
$D_n, D^C_n, N_n$ ~~~& $D_r D_r^C, N_r$ ~~~\\
$H_n, H_n'$ ~~~& $H_r, H_r'$ ~~~\\ 
\hline
\hline
\end{tabular}
\end{center}
\caption{ C parities and matter parities from Ref.~{\cite{Greene:1986bm}}
 where $L_{1\pm}=(L_1\pm L_2)/\sqrt 2$ etc.}
\label{gkmr}
\end{table}
To distinguish between C even and C odd states we  will adopt the following
convention:
$i=(n,r)$, $n=C~{\rm even}$, ~$r=C~~{\rm odd}$.
From Table (\ref{gkmr}) we find that for the lepton nonet one has
$n=1+, 3+, 5, 7,8+$, and $r=1-, 3-, 6,8-$.  Combining these with the
values of $U_Z$ one gets the  $M_2$ parities
of the particle states  listed in Table (\ref{gkmr}). \\

Now matter parities restrict the interaction structure. To exhibit this we 
first display the superpotential for the Calabi-Yau models without 
any restriction. Here one has

\beqn
W_3= \lambda^1 det Q^C+ \lambda^2 detQ + \lambda^3 detL -\lambda^4 tr(QLQ^C)
\label{w3a}
\eeqn
where we have suppressed the generation indices. The superpotential in explicit 
detail is given by\footnote{The full analysis of the couplings from first principles
for the general case  is  difficult. Part of the problem relates  to the 
computation of the kinetic energy normalizations which require that one  
calculate not just the superpotential but also the Kahler potential. While progress
has been made~\cite{Candelas:1987se}, a complete determination of Yukawa interactions
from first principles is still lacking.}$^,$\footnote{A related topic is the  phenomenology of string 
inspired E(6) models.  See, e.g.,\cite{Drees:1987eq,Hewett:1988xc} and references therein}
\beqn
W_3=\lambda^1_{ijk} d_iU_jD_k
+ \lambda^2_{ijk} u^Cd^CD^C + \lambda^3_{ijk} (-H_iH'_jN_k -H_i\nu^Cl_k +H'e^Cl_k)\nonumber\\
 -\lambda^4_{ijk} (D_iN_jD^C_k-D_ie^Cu^C_k +D_i \nu^Cd^C_k 
   + q_il_jD^C_k -q_i H_j u^C_k -q_iH'_jd^C_k) 
\label{w3b}
\eeqn
Matter parity restricts the couplings\footnote{The couplings satisfy the restrictions 
$\lambda^{1,2,3}_{rst} =\lambda^{1,2,3}_{mnr},
~~\lambda^4_{rst}=0=\lambda^4_{mnr}=\lambda^4_{mrn}=\lambda^4_{rmn}.
$}. 
Interactions of Eq.(\ref{w3b}) contain two SU(5) singlets:
the C even  $N$  and  the C odd $\nu^c$, and a VEV growth for these leads to a  
spontaneous breaking of the $[SU(3)]^3$ symmetry down to the Standard Model gauge group symmetry.
The breaking occurs in two steps where first N develops a VEV which breaks the
$[SU(3)]^3$ symmetry as  follows 
\begin{eqnarray}
SU(3)_C\times SU(3)_L\times SU(3)_R \buildrel <N_{C+}> \over \longrightarrow
SU(3)_C\times SU(2)_L\times  SU(2)_R\times U(1)_{ B-L} \nonumber 
\end{eqnarray}
while  the  C odd $\nu^C$ VEV breaks it down further to the  SM gauge group

\begin{displaymath}
SU(3)_C\times SU(2)_L\times  SU(2)_R\times U(1)_{ B-L} \buildrel
<\nu^c_{C-}>\over  \longrightarrow  
SU(3)_C\times SU(2)_L\times U(1)_Y
\end{displaymath}
 Quite remarkably the lowest minimum after  spontaneous breaking is the one that
 preserves matter parity~\cite{Nath:1988xn}. 
 After spontaneous breaking there  will be  mass growth for the matter fields.
  One finds  that only three generation remain and the remaining (exotic)
 states become massive. There  is also a mixing among D and  d  states. Here including
 symmetry breaking at the electro-weak scale one finds
 
\beqn
W^{D-d}_3=DMD^C+ D M' d^C + d\mu d^c   
  \eeqn
where   $M, M', \mu$ are matrices.
Only the combinations that preserve matter parity enter so that 
$M_{mn}=-\lambda^4_{mjn}<N_j>$, $M_{mr}'=-\lambda^4_{mjr}<\nu^C_j>$, 
$\mu_{rs}=\lambda^4_{rjs}<H_j'>$ etc. Diagonalization by a bi-unitary 
transformation leads to eigenstates $\hat D, \hat d, \hat{D^C}, \hat{d^C}$. 
One has~\cite{Campbell:1987va}
\beqn
\left(\begin{array}{c} 
        D^c \\
	d^c
        \end{array} 
\right)=\left( \begin{array}{cc} C^1 & S^1 \\
	S^1 & C^1 
\end{array} \right)
\left(\begin{array}{c}
\hat D^c \\
\hat d^c
\end{array}
\right)
\label{mixings}
\eeqn
where $S^1, C^1$ etc are mixing matrices and only states with the same matter parity 
mix but states of different C parities get  mixed. Similarly one can define a relation between
$D, d$ and $\hat D, \hat d$, by replacing $C^1, S^1$ by $C^2, S^2$. 
The sizes of $S^1$ and
$S^2$ are very different

\beqn
S^1\sim \frac{M'}{(M^2+M^{'2})^{1/2}}\sim 1, ~~S^2\sim \frac{\mu M'}{M^2+M^{'2}}\sim
10^{-13}
\eeqn 
Thus $S^2$ is much suppressed compared to $S^1$.
 There are  two types of exchanges that can mediate  proton decay through dimension
five operators. These are from~\cite{Arnowitt:1988ax,Arnowitt:1989ud}

 \begin{enumerate}
 \item
 $\hat D$ exchange
 \item 
 $\hat  d $ exchange
 \end{enumerate}
The $\hat D$ exchange gives the dominant contribution to proton decay and the 
contribution from this exchange is~\cite{Arnowitt:1988ax}

\beqn
\Gamma (p\to \bar  \nu_{\mu} +K^+) =\frac{f^2\alpha^2}{M_{\hat  D}^2} \frac{M_N}{32\pi f_{\pi}^2} 
[1-\frac{M_K^2}{M_N^2}] |A_{\nu_{\mu}K}|^2   A_L^2 (A_S^L)^2 |1+y^{tK}|^2
\label{Dexchange}
\eeqn
where $M_D$ is the D quark mass, $A_S^L(A_L)$ are the  short-range (long-range) RG factors, 
$\alpha$ is the three-quark matrix element of the proton, $y^{tk}$ is  the correction 
from the third  generation exchange, and  $A_{\nu_{\mu}K}$ is the dressing loop function.
In the above we have included a fudge factor f which is put there to 
 account for the fact that the couplings in Calabi-Yau manifolds are  not fully
 known (The normalization f=1 corresponds to the SU(5) GUT model). Using the current data
 on the $\bar \nu_{\mu}K^+$ mode  one finds  the following limit on $M_D$ 
 \beqn
 M_D\geq (\frac{Bf}{10^{-5}}) (\frac{\alpha}{0.01 {\rm GeV^3}}) \times 10^{16} GeV
 \eeqn  
 where B depends on the dressing loops that convert dimension five to dimension six 
 operators.
Next we  consider the p decay that can arise from the exchange of $\hat d$. One finds
that because of mixings of Eq.(\ref{mixings}), there are interactions of the 
type 
$\lambda^2 S^1u^c_n\hat d^c_n\hat d^c_s, ~\lambda^4 S^2 \hat d_s e^c_n u^c_s$, 
where n mean C parity plus and s means C parity minus. The proton lifetime via
exchange of the C odd $d_s$ can be estimated~\cite{Arnowitt:1989ud}

\beqn
\tau_p\sim (\frac{\tilde m_{\tilde d_s}} {10^9{\rm GeV}})^4 
(\frac{\alpha_{em}}{\lambda^2\lambda^4})^2 (\frac{10^{-13}}{S_1S_2})^2\times (10^{34} 
{\rm yr}) 
\eeqn
For the superstring models being considered  on has $\tilde m_{\tilde d_s}\sim 10^{15}$ GeV.
Thus proton decay via $d_s$ exchange is totally negligible and the dominant 
decay comes from the D exchange as  discussed above. 
An alternative  approach is to suppress proton decay from the isosinglet  D exchange
by use of discrete symmetries, specifically by extension of the  so called  $Z_3$
baryon parity of Ref.~\cite{Ibanez:1991pr,Ibanez:1991hv}  to include the  isosinglet  
quarks~\cite{Castano:1994ec}.
\subsection{Kac-Moody level $k>1$ string models and proton decay}
As discussed above there is a large number of possibilities for models building
 in string theory and one way to limit such constructions is to use the 
 constraint of grand unification. Such constructions depend on the nature of the gauge
 symmetry  which is turn depend on the Kac-Moody level which enters in 
 the operator product expansion of  world sheet currents [The product of two currents 
 can be expanded so that
 $j_a(z) J_b(w) \sim  i  f_{abc} (z-w)^{-1} j_c(w) + $ $(k/2)\delta_{ab}$ $(z-w)^{-2 }+..$
 where k is the Kac-Moody level. k  is a positive integer for the case
 of non-abelian gauge groups but for abelian case k is not constrained.].
 The level 1 is the most  widely studied case.  In these models 
  grand unified groups such as $SU(5)$, $SO(10)$, and $E_6$ can be obtained~\cite{Lewellen:1989qe}.
  One problem encountered here  is the absence of massless scalar fields in 
  the adjoint representation of the gauge group which can be used to break 
  the unified gauge symmetry. In grand unified theories based on the weakly 
  coupled heterotic string massless scalars  in the adjoint representation 
  along with N=1 supersymmetry and chiral fermions can only be  realized 
  for $k>1$~\cite{Lewellen:1989qe}. 
    At level 2, while it is possible to get massless 
  scalars  in the adjoint representation, it is difficult to get
  three massless generations of  quarks and leptons in this case. 
  Although there is no firm theorem to this effect, all analyzes to
  achieve k=2 models  with three generations have been unsuccessful. 
  Perhaps a simple way to understand this result is that  the orbifold group
  for level 2 is  $Z_2$. Since the numbers of  chiral families are  related
  to the fixed points  in the twisted  sectors, this number will then
  be even~\cite{Kakushadze:1997mc}. 
  At level three it is possible  to get  the massless scalars in the adjoint 
  representation as well as  get three massless  generations of quarks and
  leptons~\cite{Kakushadze:1997mc}.\\ 
  
 Thus there has been considerable work over the past few years on the level 3 
  models~\cite{Kakushadze:1997bk,Kakushadze:1996tm,Kakushadze:1996jm,Kakushadze:1997ne,Kakushadze:1997mc}.
   The construction of the models requires realizing a $Z_3$ outer automorphism symmetry
   not  present in 10 dimensions and one needs rules  for model building which have been
   realized  within the framework of asymmetric orbifolds.
    Thus models building 
   at level 3 requires special techniques and is significantly more difficult than level
   1 constructions. Using these techniques, 
    models with gauge groups $SU(5)$, $SO(10)$,
  and $E_6$ have been constructed which have $N=1$ space-time supersymmetry, 
  three  chiral families and massless scalars in the adjoint representation
 of the gauge groups. 
  Specifically the number  of adjoint scalars is just one.
  Additionally these models have a non-abelian hidden sector.
 The phenomenology of the $E_6$ model as well as of the related  SO(10) 
  model has been worked out in some detail~\cite{Kakushadze:1997bk}.
   Here with the assumption of dilaton
  stabilization by a non-perturbative  mechanism, the gaugino condensation scale
  in this model is found to be around $10^{13}$ GeV which gives
  a weak  SUSY breaking scale of $\sim$ TeV. However, there are some  
  undesirable features as well. Thus although one has massless scalars in the 
  adjoint representation, the adjoint Higgs is  flat modulus. Further, the
  Higgs doublet mass matrix is rank six and all the  Higgs  are in general
   superheavy. If one uses the Dimopoulos-Wilczek mechanism
  then  one gets two pairs of light Higgs doublets which is undesirable.
  Thus typically one needs a fine tuning to get a pair of  light Higgs 
  doublets. Lepton-number violating dimension four operator $LLE^C$ and 
  $LQD^C$, and the baryon-number violating dimension four operator $U^CD^CD^C$ are
  absent due to the underlying gauge  and discrete symmetries of the model.
 However, \bl violating dimension  five operators are present and one needs to 
 use heavy Higgs triplets to suppress proton decay rates from these operators.
 A detailed analysis of proton decay life time would require computation of 
 the quark-lepton textures. But these  are problematic since  the leptons and 
 down quarks have the same mass  matrices. Thus while many of the features  of
 the models investigated have the right flavor, on the whole the models appear not to be 
 phenomenologically viable rendering a detailed investigation of 
 proton stability in these  models not compelling.    
\subsection{A new class of heterotic string models}
Recently a new class of heterotic string ~\cite{Gross:1984dd}   have been proposed
 which lead to 
some  remarkably attractive features from the point of view of phenomenology and  these
models are  worthy of attention. The models have the remarkable feature that the spectrum
is exactly that of MSSM. 
  Specifically in the work of   Ref.~\cite{Bouchard:2005ag}  a compactification
 of the heterotic string on a Calabi-Yau threefold with ${\cal Z}_2$ fundamental 
 group coupled with an invariant $SU(5)$  bundle is achieved.  The spectrum of this
 model consists of three generation of matter and in addition 0, 1, or 2 Higgs doublet
 conjugate pairs depending on the part of the  moduli space one is in.  Specifically it is 
 possible  to get a heterotic string model with precisely the MSSM spectrum with a 
 single pair of Higgs. 
 The gauge group in the visible sector is 
 $SU(3)_C\times SU(2)_L\times U(1)_Y$. In this model proton decay from dimension 4, 5
 and 6 operators is absent. 
 Another recent work which finds an exact MSSM spectrum from string 
 theory is that of Ref.~\cite{Braun:2005nv}.  Here one finds three families of 
 quarks and leptons, each family with a right-handed  neutrino  and one pair of 
 Higgs doublets while the gauge group in the visible  sector is 
 $SU(3)_C\times SU(2)_L\times U(1)_Y\times U(1)_{B-L}$. The proton 
 is again stable in this model with no dimension four, five, or  six lepton and baryon
 number violating operator present.     However,  it has been pointed 
 out~\cite{Gomez:2005ii} that the hidden sector bundle of the work of Ref.~\cite{Braun:2005nv} is not 
 slope-stable which would require changing the hidden sector and will result in different
 phenomenological properties\cite{Braun:2005nv}. Further discussions of these models
 can be  found in \cite{Braun:2006ae,Bouchard:2006dn}.

 Among other attempts at realizing 4D string model building in heterotic strings is
  the work of Ref.~\cite{Kobayashi:2004ya}. The analysis is motivated by  orbifold
 GUTs discussed in the previous section. Specifically they consider the 5D SO(10)  models
 of Refs~\cite{Dermisek:2001hp,Kim:2002im} 
  with a bulk extension where the extra dimension is a 
 half circle $S^1/Z_2$.   The effective gauge group in 4 dimensions  is the 
 Pati-Salam group~\cite{Salam:1973uk}
 $SU(4)_C\times SU(2)_L\times SU(2)_R$. The model  has the interesting feature that three
 generations of matter  can be realized  with two generations  localized on the $Z_2$
 orbifold  fixed point while one generations propagates in the bulk. It predicts 
 a gauge-Yukawa unification at the 5D compactification scale. However, the model has a 
 problem  in that  there is no identifiable  symmetry for suppression of 
 dangerous proton decay operators.
\subsection{Proton decay in M theory compactifications}
As discussed already in the beginning of this  section M theory  is conjectured to be  the
source of  all string theories. The low energy limit of  this theory is the  
11 dimensional supergravity~\cite{Nahm:1977tg,Cremmer:1978km}
  formulated in the  late seventies. An interesting
phenomenon is that N=1 supersymmetry can be preserved if  one  
 compactifies the 11 dimensional supergravity to 4 dimensions  on a seven-compact  
 manifold X of $G_2$ holonomy. 
 But if X is a smooth  manifold then one obtains only an abelian gauge
 group and no chiral fermions~\cite{Shatashvili:1994zw,Papadopoulos:1995da}.
  How to get  a non-abelian gauge  symmetry in compactification of such a theory is  
non-trivial.  
One  way is  to compactify  M-theory on a manifold with boundary~\cite{hw}.
Another possibility is to get gauge  fields and chiral fermions from singularities 
in geometry~\cite{Atiyah:2001qf,Acharya:2001gy,Strominger:1996it}.
 Thus A-D-E orbifold  singularities can produce gauge fields~\cite{Witten:1995gx}
   and conifold
singularities  can produce chiral fields~\cite{Strominger:1995qi}. For  example,  consider  M-theory
on ${\cal R}^4\times X$, where X is the  manifold of $G_2$ holonomy. If X looks locally like
$Q\times {\cal R}^4/\Gamma$ where  $Q$ is  a  three-manifold, then one will  get gauge fields
on the singular set ${\cal R}^4\times Q$.  The  case $\Gamma =Z_5$  will  lead  to the $SU(5)$ 
gauge fields on the ${\cal R}^4\times Q$~\cite{Friedmann:2002ty,witten}\footnote{A detailed study
of these compactifications including the  $\Gamma =Z_5$ case has  been carried out in the
quantum moduli space of  M-theory compactifications in Refs.~\cite{Friedmann:2002ct,Friedmann:2003cd}.}.
\\

We discuss now proton decay in the above framework following closely the analysis of Friedmann and
Witten in  Ref.~\cite{Friedmann:2002ty}. 
 In the analysis of  models in
${\cal R}^4\times Q$, we  begin by assuming that in general quark-lepton multiplets
are  located at  different  points $q_i$, in the manifold Q.  Thus effective operator
for proton decay will arise from interactions of the type

\beqn
g_7^2  \int d^4x j_{\mu}(x;q_1) \tilde j^{\mu}(0;q_2) D(x;q_1;0;q_2)
\label{aug27m}
\eeqn
where
$j_{\mu}, \tilde j^{\mu}$ are  the currents and $D(x,q;y,q')$ is the gauge boson propagator function in the space 
 ${\cal R}^4\times Q$  and
satisfies the relation 

\beqn
(\Delta_{{\cal R}^4} +\Delta_Q) D(x,q;y,q') = \delta^4(x-y) \delta(q,q')
\label{aug27n}
\eeqn
For heavy gauge  bosons  one can use the conventional 'local' approximation where  we  put 
currents at the same spatial  point and  in that approximation  the effective operator  is 

\beqn
  j_{\mu}(0;q_1) \tilde j^{\mu}(0;q_2) g_7^2F(q_1,q_2)
\label{aug27o}
\eeqn
where  $F(q_1,q_2)=\int d^4x D(x,q_1;0,q_2)$.
Now $F(q_1,q_2)$ is bounded  for large  separation $|q_1-q_2|$ and for small separations as 
$q_1\to q_2$, one has $F(q_1,q_2)\to 1/4\pi |q_1-q_2|$. 
Thus in computing the dimension 
six operators for multiplets  residing at the same point in the compact space,  
the limit $q_2\to q_1$ is necessary which, however, is a singular limit. 
In a  realistic treatment
 a  cutoff should  emerge to render such an analysis a meaningful exercise. 
A rough fix  is to replace $1/|q_1-q_2|$ by $M_{11}$ and replace $g_7^2 F(q_1,q_2)$ as 
$q_2\to q_1$ by  $Cg_7^2M_{11}/4\pi$, where $C$ is a constant which in principle
can be  computed by the details of an M theory calculation. 
Using Eq.(\ref{aug27ka}) of Appendix I for $g_7^2M_{11}$ one finds 
an effective dimension six operator of the form~\cite{Friedmann:2002ty} 

\beqn
O_{eff}^{M-theory}=\sum_q 2\pi C j_{\mu}(q) \tilde j^{\mu}(q)  \alpha_G^{2/3}L_Q^{2/3} M_G^{-2}
\label{aug27p}
\eeqn

Eq.(\ref{aug27p}) contains an interaction of the type $10^2\overline{10}^2$ which   
gives rise to the decay $p\to e^+_L\pi^0$. Unlike the case of proton decay in 
intersecting D brane models~\cite{Klebanov:2003my} which will be discussed  next 
it is not possible to make a definitive statement here  whether this decay is enhanced or not 
relative  to what one expects in a grand unified theory due to the unknown constant C. 
One hopes that further progress in M theory calculations would allow one to  make a
more predictive statement. 

 We discuss now the decay $p\to e^+_R\pi^0$ which 
  arises from the interaction $10^2\bar 5^2$. If
$10$ and $\bar 5$ are located at different points in Q, one expects a  
suppression for this decay relative to $p\to e^+_L\pi^0$. 
It is important then to be able to detect the helicity of the 
outgoing charged  lepton to check on this model.
Finally, this class  of models have  a natural doublet-triplet splitting~\cite{witten-de} 
and also because  of  a discrete symmetry the 
 dimension five operators  from Higgs triplet exchange  do not arise~\cite{Friedmann:2002ty}.
 
\subsection{Proton decay in intersecting D brane models} 
An interesting class of models are those based on intersecting D  
 branes~\cite{berk,Blumenhagen:2000wh,Angelantonj:2000hi,Aldazabal,Blumenhagen:2000ea}
and attempts have been made  to build semi-realistic  models based on 
these~\cite{Blumenhagen:2001te,Koko1,Koko2,Cvet,Blumenhagen:2002gw,CvetPap}, 
and issues of gauge coupling unification, soft  breaking and possible  applications to
the real world have also been discussed~\cite{Blumenhagen:2003jy,Kors:2003wf,Grana:2003ek}
(For reviews see Ref.~\cite{Lust:2004ks,Kiritsis:2003mc,Blumenhagen:2005mu}).       
Here  we  follow closely the work of  Klebanov and Witten in Ref.~\cite{Klebanov:2003my} 
which investigates proton decay on SU(5) GUT  like models in Type IIA 
orientifolds with D6-branes (Also see in this context Ref.~\cite{Axenides:2003hs}).
 We will assume that proton decay from dimension four
and dimension five  operators which arise in supersymmetric GUT  theories are absent  
due to a symmetry in the model and thus we focus  on the dimension six operators. 
In the  analysis of  Ref.~\cite{Klebanov:2003my}
one assumes a  stack of D6  branes which intersect an 
orientifold  fixed  six-plane along the 3+1 directions. The above can be viewed as  a 
stack of D6 branes intersecting  an image set of D6$'$ branes on the covering space.
If the stack has five D6 branes, the covering space contains the $SU(5)\times SU(5)$
gauge group, and the open strings  are localized at the intersection and lie in
$(5,\bar 5)+ (\bar 5,5)$  representations.  An orientifold  projection gives an 
SU(5) theory with matter  in $10+\overline{10}$.   In 4 dimensional SU(5) grand unification
dimension six operators are  of type $5^2\bar  5^2$, $10\overline{10}5\bar 5$, and
$10^2\overline{10}^2$.   The $5^2\bar  5^2$ do not have  baryon and lepton number violation
and  $10\overline{10}5\bar 5$ operators  do not appear in the D brane analysis  being 
discussed here. However, $10^2\overline{10}^2$ operators do arise and we discuss their
contribution to proton decay.\\

The analysis is done in the covering space and for specificity it is  assumed  that  the
D6 branes are oriented in the 0123468 directions and the D6$'$-branes  intersect them along
the 0123 directions, resulting in a 3+1 dimensional intersecting brane world. 
The orientation in the six transverse directions are specified by the complex 
coordinates 
$z_1=x^4+ix^5,~~z_2= x^6+ix^7,~~z_3=x^8+ix^9$.
N=1  supersymmetry in (3+1) dimensions can be preserved if the rotations  act  on an
SU(3) matrix  on the three complex coordinates. A diagonal rotation that transforms
D6  branes to D6$'$ branes is  
\beqn
z_i\to e^{i\pi \theta_i} z_i,~~(i=1,2,3),
~~\sum_i  \theta_i = 2 ~~mod~2Z.
\label{1.12}
\eeqn
 An analysis of 4 fermion amplitude in Ref.~\cite{Klebanov:2003my} gives
\beqn
A_{st}= i\pi g_s \alpha' I(\theta_1, \theta_2,\theta_3)
\label{1.13}
\eeqn
 where

 \begin{eqnarray}
I(\theta_1, \theta_2,\theta_3)
&=& 2\int_0^{\infty} dt \prod_{i=1}^3 (\sin(\pi \theta_i))^{\frac{1}{2}} 
F(\theta_i,1-\theta_i;1;e^{-t}) \times \nonumber \\ 
&&[F(\theta_i,1-\theta_i;1,1-e^{-t})]^{-\frac{1}{2}}
 \label{1.14} \nonumber \\
  \end{eqnarray}
  and where F is a hypergeometric function. To fix the size of $g_s$ and $\alpha'$ one may 
 consider the gravitational action for a Type IIA superstring 
 
 \beqn
 ((2\pi)^{-7} \alpha^{'-4}\int d^{10}x \sqrt{-G}e^{-2\Phi} R
 \label{1.15}
 \eeqn
 where $\Phi$ is a dilaton field and the string coupling constant is
$g_s=e^{\Phi}$.
 Reduction to 4 dimensions  is necessary to make contact with the familiar 4 D quantities 
 such as the GUT coupling constant $\alpha_G$ and the GUT scale $M_G$. 
 The  details can be  found in Ref.~\cite{Klebanov:2003my}(see also Appendix I).  
Thus the relation connecting $\alpha'$ and $g_s$ with  $\alpha_G$ and $M_G$  is  given by
\beqn
\alpha'= \frac{\alpha_G^{{2}/{3}} L_Q^{{2}/{3}}} {4\pi^2 g_s^{{2}/{3}} M_G^2}
\label{aug24_12aa}
\eeqn
 where  $L_Q$ is the Ray-Singer~\cite{Friedmann:2002ty,ray,rs}
topological invariant of the compact three-manifold.
The Ray-Singer torsion is a  model dependent quantity and requires the specification of the 
compact three-manifold for its computation. 
Eliminating $\alpha'$ in Eq.(\ref{1.13}) using Eq.(\ref{aug24_12aa})  we can write $A_{st}$ 
in the form

\beqn
A_{st}= g_s^{{1}/{3}} \alpha_G^{{2}/{3}}
\frac{ L_Q^{{2}/{3}} I(\theta_1,\theta_2,\theta_3)}{4\pi M_G^2}
\label{aug24_12b}
\eeqn
To compare  the string calculation with the comparable result in a grand unification model
one can carry out a field theory analysis of the four-fermion scattering and here  one gets
\beqn
A_G= \frac{2\pi \alpha_G}{M_X^2}
\label{aug24_12c}
\eeqn
Eqs.(\ref{aug24_12b}) and (\ref{aug24_12c}) lead to the relation
\beqn
\frac{A_G}{A_{st}}= \frac{g_s^{1/3} L_Q^{2/3} I(\theta_1,\theta_2,\theta_3)}
{8\pi^2 \alpha_G^{1/3}} \frac{M_G^2}{M_X^2}
\label{aug24_12ba}
\eeqn
One can now  compare the life time for the decay mode $p\to e^+\pi^0$ in the string model
 compared to its life time in a  GUT model. One finds 
 \beqn
 \tau_{st}(p\to e^+\pi^0) = \tau_{GUT}(p\to e^+\pi^0) {C_{st}}  \frac{M_G^4}{M_X^4}
 \label{aug24_bb}
 \eeqn
 where $C_{st}$ is the  string enhancement factor of the proton lifetime and is given by
  
 \beqn
{ C_{st}}= \frac{1}{1-y}  (\frac {8\pi^2 \alpha_G^{1/3}}  {g_s^{1/3} L_Q^{2/3} I(\theta_1,\theta_2,\theta_3)})^2.
 \label{aug24_12bc}
 \eeqn
 Here y is the fraction of $p\to e^+_R\pi^0$ to $p\to e^+_L\pi^0$  which is
$ y=1/[1+(1+|V_{ud}|^2)^2]$ where $V_{ud}\sim 1$.
  The factor  $1/(1-y)$ is inserted in Eq.(\ref{aug24_12bc}) to take account of the missing 
  $p\rightarrow e^+_R\pi^0$ mode in the intersecting D brane model here.
     We note that the factor $M_X^{-4}$  cancels  out in the  product 
   $\tau_{GUT}(p\to e^+\pi^0) M_X^{-4}$,  and thus $\tau_{st}$ is determined  directly 
   in terms of $M_G$. In this sense $\tau_{st}$ is more  model independent since  it 
   depends  directly on $M_G$  rather than on the X gauge  boson mass. 
   \\

To numerically estimate the proton lifetime one may consider $Q=S^3/Z_k$ (lens space) where k in an integer.
 In this case~\cite{Friedmann:2002ty} 
 \beqn
 L_Q=  4ksin^2(5\pi m/k)
 \eeqn
where m is an integer such that
 5m is not divisible by k. For the case m=1, k=2, one finds  $L_Q=8$. The analysis
 of Ref.~\cite{Klebanov:2003my} finds I in the range $7-11$. Setting   
$g_s\sim 1$,  $\alpha_G=0.04$, $y=0.2$, and $M_G=M_X$ gives $C_{st}\simeq 0.5-1.2$.
 Since the current estimate of  the
 GUT  prediction is $\tau_{GUT}=1.6\times 10^{36}$yr for values of $\alpha_G=0.04$
 and $M_X=2\times 10^{16}$ GeV, one finds that the string prediction in
 this case  is  $ (0.8-1.9)\times 10^{36}$ years. 
The more recent analysis  of Ref.~\cite{Cvetic:2006iz} gives the range
 $(0.5-2.1)\times 10^{36}$ yr.
The current  experimental limit on this decay mode is
$\tau_{exp}(p\to e^+\pi^0)> 1.6\times 10^{33}$ yr (Table 1.
See, however, Ref.~\cite{Ganezer:2001qk} which gives  
 $\tau_{exp}(p\to e^+\pi^0)> 4.4\times 10^{33}$ yr). The next  generation of proton decay 
 experiment  using underground water Cherenkov detectors may improve the 
 experimental lower limit for this mode by a  factor of 10 close to $10^{35}$ yr~\cite{Jung:1999jq} which, however, 
 falls short  of the theory prediction above.
  However,  one must keep in mind the model dependence  of the theoretical prediction arising from the
 assumed values  of  $L_Q$, $g_s$, assumption on fermion mixings etc. Thus, for example, if $L_Q$ lies in the
 range 1-10~\cite{Burikham:2005wj}, then $C_{st}$  will lie in the  range 
 (0.4-19) which is a significant shift from the previous estimate.
\subsection{Nucleon stability in string landscape models}
The natural scale of vacuum energy density $\rho_V$  is the Planck scale 
while $\rho_{\rm obs}$ is much smaller.

 \beq
\rho_V \sim {\rm M_{\rm Pl}}^4,~~~~~\rho_{\rm obs}\leq (3\times 10^{-3} {\rm ev})^4 
\eeq
To fit observation this requires a fine tuning of order $O(10^{120})$
to get the observed scale. With softly broken SUSY of scale
$\rm M_S=O(1)$ TeV one gets 

\beq
\rho_V \sim \rm {M_{\rm S}}^4
\eeq
Here one needs a fine tuning of order $O(10^{60})$.
There are two ways out to resolve the problem. The first one is the possibility that some as yet
unknown  symmetry principle sets the vacuum energy effectively to zero. However, as 
exhibited  above one does not need an exactly vanishing value of vacuum energy but a small one,
and thus one would still need  to 
find a way to give the vacuum energy a tiny positive value consistent  with current
experiment. The second possibility is to invoke the anthropic principle. Thus  
Weinberg~\cite{Weinberg:1987dv}
has observed that the seeding of the galaxies 
requires that the  value of the cosmological constant lie in a rather restricted
range of the current value. The idea is that there are a large number of 
different vacua and the one we live in corresponds to a small value of
the cosmological constant. In this sense the current value of the vacuum energy becomes 
just an 'environmental'  parameter rather than something intrinsically fundamental.\\

Some support for the anthropic idea has come from studies of string landscapes~\cite{landscape,landscape1}.
We discuss now the idea of  string landscape briefly as such ideas  have  implications also for
string model building and for proton stability. 
As is well known a common feature of string models is the  presence  of many moduli. Often 
the moduli potential is flat leaving the moduli undetermined. Thus one needs  to 
lift the flat directions to fix or stabilize the moduli. This is the so called moduli stabilization 
problem. There  has been   
recent progress in this direction in that inclusion of fluxes in the  compactification of extra  
dimensions allow one to lift  the flat directions and with 
 fluxes turned  on it is possible to stabilize the moduli. An example of this phenomenon is the type IIB
string theory where one has  three  form RR and NS and fiveform RR fluxes which can
be  turned on in compactification. There are many choices for these fluxes and the possibilities
are very large. In the presence of the fluxes one has a non-vanishing tree level
 superpotential $W_0$ 
 which is moduli dependent~\cite{Gukov:1999ya}. In addition it  has been observed~\cite{kklt} 
 that  there will be in general a non-perturbative  contribution to the superpotential 
 $W_{NP}$  arising from strong coupling dynamics such as from gaugino condensation,
 instantons  etc which can be parametrized by $W_{NP}= Aexp(-c\rho)$ where 
 c depends on strong interaction dynamics and $\rho$  is a size
 modulus.  Together  the potential then will have the form~\cite{kklt}
 
 \beqn
 W=W_0 + A e^{-c\rho}
 \eeqn  
It is then possible to stabilize the moduli but one ends up with
  anti-de Sitter (AdS) vacua with a negative vacuum energy. However, with inclusion of 
 supersymmetry breaking it is possible to get de Sitter  vacua with positive energy. 
There are a huge number of allowed possible states. An order estimate can be  gotten 
as  follows~\cite{Dine:2004ct}: Consider an integer flux lattice of dimension K in Type IIB strings. 
The vectors in the 
lattice  $\vec n$ are constrained by $\vec n^2\leq L$  where  $L$ is an integer determined
by the tadpole cancellation condition. To compute the number of allowed states one 
computes the number of states in a K dimensional sphere with radius $\sqrt L$. This
results in  the number of  allowed states to be~\cite{Dine:2004ct}

\beqn
N_{\rm vac}\sim \frac{L^{K/2}}{\Gamma(K/2)}
\eeqn
With  $L\sim 10^3, ~~~K\sim 10^2$ one has $N_{\rm vac}\sim 10^{1000}$.
Thus there are a huge number of metastable de Sitter vacua.   
This huge number allows the possibility that the cosmological constant takes on fine grain 
values and there is a range in which the 
physically observed value of the cosmological constant could lie. 
Such calculations could be impacted by a further restriction  of proton stability 
by a study of the gauge group ranks~\cite{Kumar:2004pv,Conlon:2004ds}.
Further, 
the same principle may be used to fine tune the Higgs mass if the SUSY breaking scale was
high.
Specifically it is advocated that the scalars except for the
Higgs could all lie at the Planck scale and be superheavy while the light particles  would
consist of gauginos and Higgsinos~\cite{Arkani-Hamed:2004fb,Giudice:2004tc}. 
Unified models with landscape scenarios have been discussed in 
Refs.~\cite{Calmet:2004ck,bgns}.
\\

In the above scenario the  proton decay via dimension five
operators will be highly suppressed since the squarks and sleptons  fields  in 
split supersymmetry scenario are typically supermassive. 
 It  is interesting to ask how a large  mass  hierarchy in supersymmetry breaking 
 can arise  in string models. It turns  out that a natural  hierarchy in 
 supersymmetry breaking scales  can arise in D brane  models~\cite{antoniadis}
 and  more generally in string models with Fayet-Illiopoulos  D terms~\cite{Fayet:1974jb,lands2}. 
  One can illustrate this even in the framework of global
  supersymmmetry. Thus we consider extended gauge symmetry  $SU(3)_C\times SU(2)_L\times U(1)^n$, 
  where the extended $U(1)$ sector aside from the hypercharge  contains an anomalous $U(1)$, a situation
  which is quite common in string theory, where the anomaly in then canceled by 
  Green-Schwarz (GS) mechanism [The corresponding gauge boson develops a 
Stueckelberg mass and decouples (see e.g.~\cite{Klein:1999im})]. 
This provides  a motivation for inclusion of an FI term.  
  Including the FI term the D term potential in global supersymmetry takes the form

\beqn
V_D ~=~ \sum_a \frac{g_a^2}{2} D_a^2  = \sum_a \frac{g_a^2}{2} 
 \Big( \sum_i Q_a^i |\tilde f_i|^2 + \xi_a \Big)^2 
\eeqn
   A single extra $U(1)$ cannot lead to a SUSY hierarchy but with multiple extra $U(1)$'s 
   it is possible.  Thus consider  an extra $U(1)_X$  where
   we add two oppositely charged 
   scalars  $\pm 1$ under the extra $U(1)_X$ and assume an interaction of the type 
   $W_\pm = m \phi^+\phi^-$ in the superpotential~\cite{Dvali:1996rj,Binetruy:1996uv}. 
   Minimization yields
   $\langle \phi^+ \rangle =0$, and $\langle \phi^- \rangle^2 = \xi_X- {m^2}/{g_X^2}$.
   This leads to $\langle D_X \rangle ~=~ {m^2}/{g_X^2}$ and 
$\langle F_{\phi^+} \rangle ~=~ m \sqrt{\xi_X} +\, \cdots $  where  $F_{\phi^+}$ is the
supersymmetry breaking scale. The above analysis gives  for the 
scalar masses $m_i$ the result
$m_i^2 ~=~ \sum_a g_a^2 Q_a^i \langle D_a \rangle $ and  for the gaugino masses $m_{\lambda}$ 
the result $m_\lambda ~ \sim~ m \xi_X/M_{\rm Pl}^2$. 
In this case both the scalar and the gaugino masses are scaled by the same mass m and we 
find no hierarchy.  Thus for $m \sim \co({\rm TeV})$ and $\xi \sim \co(M_{\rm Pl}^2)$ 
and all masses are  at the electro-weak scale. 
 However, the situation  changes drastically if one considers multiple
anomalous $U(1)$'s. Here it is possible to split the masses of the 
scalars and the gauginos. A realistic scenario requires that one carry out the analysis within
the framework of supergravity unification. Then one finds that the condition that the vacuum
energy vanish requires that 
$\langle F_I \rangle ~\leq~ m_{{3}/{2}} M_{\rm Pl}$ and 
$\langle D_a \rangle ~\leq~ m_{{3}/{2}} M_{\rm Pl}$.  
Since the scalar masses are proportional to $\langle D_a \rangle^{1/2}$  one finds that
for  $m_{{3}/{2}}\sim\co({\rm TeV})$, the above relation implies~\cite{lands2}

\beqn 
m_{\tilde  f}\leq  \sqrt{m_{{3}/{2}} M_{\rm Pl}}\sim 10^{10-13} ~(\rm GeV)
\eeqn
which is the usual intermediate supersymmetry breaking scale 
that arises in  SUGRA models.  The F-term masses are   
$F_I/M_{\rm Pl}\sim\co({\rm TeV})$. Thus the scalar masses arising from the D terms are
much larger than the F term masses. In the context of the heterotic strings the 
FI parameter is of size $M_{Pl}^2$ and thus the gaugino mass is of size $M_{Pl}$ if its
mass arises from the above mechanism. However, the FI-parameter can in principle be of 
any size in  orientifold D-brane models, and thus the above problem is circumvented 
in orientifold D-brane models. In this case one has a hierarchical symmetry breaking
with scalars superheavy which put the proton decay rates much above  the current
experimental limits.
It is to be  noted  that large scalar masses naturally arise on the Hyperbolic branch (HB)
of radiative breaking  of the electro-weak 
symmetry~\cite{Chan:1997bi}. 
The quite interesting 
phenomenon is that it is possible to keep the parameter $\mu$ small while the scalar
masses get large. The parameter  $M_Z/\mu$ also provides at least one measure of fine
tuning and naturalness. Thus the larger $\mu$ gets, the more 
fine tuned is the radiative breaking. The fact that it is  possible to achieve
large $m_0$ while keeping $\mu$ small implies that  large scalar  masses can be
be construed as being natural when they arise on HB. Now numerical analysis indicates 
that scalar masses as large 10-20 TeV arise quite naturally 
on HB~\cite{Chan:1997bi}. It is interesting
then that the HB branch of radiative  breaking leads to a suppression of proton decay. 
\subsection{Proton decay from  black hole and wormhole effects}
Quantum gravity does not conserve  baryon number and thus can catalyze
proton decay. There is a  significant amount of literature
trying to analyze proton decay lifetime arising from such 
effects~\cite{Zeldovich:1976vq,Hawking:1979hw,Page:1980qm,Ellis:1983qm,Gilbert:1989nq,Adams:1996xe,Adams:1998gh,Adams:2000za}.	
		Thus in quantum gravity  one will  have not only the  exchange of  gravitons but also
 exchange of mini black holes and wormhole tunneling effects. 
  For example, the mass ($m_{\rm BH}$) of a mini black hole will be  typically the Planck mass,
	 and its Compton length typically the Planck  length 

	\beqn
	<m_{\rm BH}> \sim M_{\rm Pl}, ~~<L_{\rm BH}>\sim l_{\rm Pl}\sim 10^{-33} {\rm cm}
	\eeqn
	It is  possible  then that the 
	two quarks in the proton might end up falling into the mini black hole
	and since  one expects black holes  not to conserve baryon number, 
	such virtual black hole processes
	will lead to baryon number violating processes  such as
	\beqn
	 q+q\to \bar q +l, \cdot\cdot 
	\eeqn
	These  processes  can be simulated by effective four -fermi interaction, with an effective
	coupling scaled by the inverse of the quantum gravity scale $M_{QG}$. A typical proton lifetime  from
	such interactions will be  

	\beqn
	\tau_p \simeq  10^{36} {\rm yr} (\frac{M_{QG}}{10^{16} {\rm GeV}})^4.
	\label{qpdecay}
		\eeqn
	For $M_{QG}=M_{\rm Pl}$ the above leads  to a proton lifetime of $\sim 10^{45}$ yr. 
	A lifetime  of this  size is  certainly beyond the experimental  reach. However, 
	 it will have  significance in  determining the ultimate fate of the 
	universe~\cite{Dicus:1981ab,Adams:1996xe}. \\

	It is also instructive  to investigate proton decay from quantum gravity effects
	in the context of large  extra dimensions~\cite{Adams:2000za}. In theories of 
	large extra dimensions the fundamental scale is lowered. Now the geometry 
	of extra dimensions affects the physics of the virtual black holes and also 
	 the quark-lepton interactions.  Thus suppose the quarks can propagate in n extra
	dimensions rather than being confined to the four dimensional wall. Since the quarks
	can propagate in more dimensions they are less likely to encounter each other and
	this effectively  weakens their interactions. The above must be folded with the effect
	arising from the black holes now living in (4+n) dimensions. Together these 
	modify the proton lifetime so that~\cite{Adams:2000za}
	
	\beqn
	\tau_p\sim (\frac{M_{QG}}{m_p})^{4+n} m_p^{-1}.
	\eeqn  
	Using the current experimental data of $\tau_p>10^{33}$yr, one finds that $M_{QP}$ should
	satisfy the constraint~\cite{Adams:2000za}

	\beqn
	M_{QG}>10^{64/(4+n)} ~{\rm GeV}
	\eeqn
	The above implies that for  the case  when quarks are confined to the four dimensional wall, 
	so that n=0, one has  $M_{QG}>10^{16}$ GeV. Even for the case when $n=6$, one finds that $M_{QP}>2.5\times 10^3$ 
	TeV. These results appear to be disappointing from the point of view of observation of the
	fundamental scale $M_{QG}$ at accelerators. \\

\subsection{ $U(1)$ string symmetries and proton stability}
We are  already familiar  with the fact that in supersymmetric theories the \bl
dimension four operators  $QLD^c$, $U^cD^cD^c$, and $LLE$ are eliminated  by the
gauge $B-L$ symmetries~\cite{Weinberg:1981wj}. This is so because these operators  have   $B-L=-1$, and
an imposition of $B-L$ invariance does not allow these operators  to appear in the
superpotential. On the other hand dimension five operators $QQQL$ and $U^cU^cD^cE^c$ 
have  $B-L=0$. and thus imposition of $B-L$ invariance alone would not eliminate these
operators. While symmetries of SO(10) are not sufficient to suppress these operators, 
one may investigate if a larger group such as $E_6$ could provide the additional
$U(1)$ symmetry to suppress such operators. Indeed, $E_6\to SO(10)\times U(1)_{\psi}$
so there is an indeed an extra $U(1)$ that may help. However, on closer scrutiny one
finds that color triplets $H_3$ and $H_{\bar 3}$ arising from the 27 plet exist in
the spectrum and the exchange of these triplets will induce \bl dimension five
operators. Thus the symmetries arising from $E_6$ are not sufficient to suppress
the dangerous operators~\cite{Weinberg:1981wj,Pati:1996fn}. 
 It is possible, however, that string derived symmetries are  more powerful. 
This issue  has been explored at some length by Pati\cite{Pati:1996fn} 
with focus on the standard like
models by Faraggi~\cite{Faraggi:1991jr,Faraggi:1992rd,Faraggi:1992fa,Faraggi:1993su,Faraggi:1991be}
using free fermionic constructions~\cite{Antoniadis:1986rn,Kawai:1986ah,Kawai:1986va}.
In these models either the Higgs  triplets are  absent from the spectrum or the extra U(1) symmetries
suppress  their couplings with quarks and leptons.
Thus in the model of Ref.~\cite{Faraggi:1991jr,Faraggi:1992rd,Faraggi:1992fa,Faraggi:1993su}
one has  six  $U(1)$ factors, 
such that 

\beqn
\frac{1}{2}TrU_1=\frac{1}{2}TrU_2=\frac{1}{2} TrU_3=-TrU_4=-TrU_5=-TrU_6=12
\eeqn
so all the $U(1)$'s are anomalous. However, it is possible to choose five anomaly free 
and one anomalous combination [The anomaly free combinations can be chosen so that 
$ U_{\alpha}=U-1-U_2$ , $U_{\beta}=U_4-U_5$, $U_{\gamma}=U_4+U_5-2U_6$,
$U_{\delta}=U_1+U_2-2U_3$, and $U_{\epsilon}= U_1+U_2+U_3+2U_4+2U_5+2U_6$. 
The anomalous combination can be chosen so that 
$U_A$=$U_1+U_2-2U_3$.].

\begin{table}[h]
\begin{center}
\begin{tabular}{|r|r|r|r|}
\hline\hline
  fields$_{\rm }$(generations)  & $Q_{\delta}$  &  $Q_{\epsilon}$ & $Q_{\delta}+ Q_{\epsilon}$ \\
\hline
 Q(1,2)  & 1/2 & -1/2 & 0\\ 
L(1,2) & 1/2 & 3/2 & 2\\
$U^C, E^C$(1,2) & 1/2 & 3/2 & 2\\
$D^C, N^C_R(1,2)$ & 1/2 & -1/2 & 0 \\ 
\hline
 Q(3)  & -1 & -1/2 & 0\\ 
L(3) & -1& 3/2 & 2\\
$U^C, E^C$(3) &-1 & 3/2 & 2\\
$D^C, N^C_R(3)$ & -1 & -1/2 & 0 \\ 
\hline
\end{tabular}
\end{center}
\caption{ The U(1) quantum numbers of the fields in a class of string 
derived models~\cite{Pati:1996fn}}
\label{fara-pati}
\end{table}
From Table (\ref{fara-pati}) one finds  that \bl dimension four  operators $U^CD^CD^C$, 
$QLD^C$, and $LLE^C$ are not allowed  if one requires  invariance  under $Q_{\delta} +Q_{\epsilon}$.
Further, \bl dimension five operators $QQQL$ are also eliminated if one requires invariance 
under  $Q_{\delta} +Q_{\epsilon}$.  For the case of $U^CU^CD^CE^C$, this operators 
is eliminated  for all cases under the $Q_{\delta} +Q_{\epsilon}$  invariance except when 
all four fields are from generation 3. However, here if one requires that in addition one 
also has invariance  under  either $Q_{\delta}$ or $Q_{\epsilon}$ then these dimension five
operators are also eliminated and thus there in no proton decay from this set of operators.
At the same time some combination of safe operators  such as $LLH_iH_j$ where $H_i$ are 
the Higgs doublets  of the model are allowed. This operator violates lepton number  
but is desirable as it enters in the neutrino mass matrix.  Thus here one  finds that 
a combination of the symmetries  generated by $Q_{\delta}$ and $Q_{\epsilon}$ eliminate
dangerous \bl operators but allow for desirable lepton number violating  operators. 
So in this sense the string derived symmetries  are  more powerful than the 
symmetries  that  can be gotten from the grand unified models. 
While the additional exact $U(1)$ gauge symmetries suppress proton decay they also
bring in additional massless modes which are  not acceptable on phenomenological
grounds. Thus one  must break these  symmetries  spontaneously  and such breakings
lead to additional $Z'$ gauge  bosons whose masses depend on the scale of spontaneous 
breaking.

\subsection{Discrete Symmetries and proton stability}
Dimension 4 and dimension 5 proton decay operators can be eliminated by specific choice  of 
discrete  symmetries. However, if the symmetries are global they would  not be 
respected by quantum gravity~\cite{Zeldovich:1976vq, Hawking:1979hw,Page:1980qm}
specifically, for example in virtual blackhole exchange and in wormhole tunneling,
and thus such phenomena can lead to new sources for proton decay~\cite{Gilbert:1989nq}. 
The way out of this problem suggested by Krauss and Wilczek~\cite{Krauss:1988zc}
is to use discrete gauge symmetries.   An example of this phenomenon  is a $U(1)$ gauge theory
where the gauge invariance is broken by condensation of a scalar Higgs field $\xi$ with charge $NQ$
while the charges of the remaining fields $\psi_i$  in the theory are all $Q$. In this case  
one will have after condensation of the Higgs field a residual $Z_N$ symmetry

\beqn
U(1) \to Z_N,~~~ \psi_i\to e^{\frac{2\pi i}{N}} \psi_i
 \label{kw}
\eeqn   
So $Z_N$ is just the residual symmetry that is a remnant of the broken abelian gauge  
symmetry. As pointed out by Krauss  and Wilczek, although Eq.(\ref{kw}) looks very 
much like a global  symmetry,  the  fact that it is remnant of  a  local symmetry
means that it is protected against even worm hole tunneling and black hole 
interactions. Consequently such symmetries are then an ideal instrument to protect
the proton against dangerous decays via virtual black hole exchanges.  
Ibanez and Ross (IR)~\cite{Ibanez:1991pr,Ibanez:1991hv}  
have  analyzed the  generalized $Z_N$ parities for the standard 
model  superfields  such that $\psi_i\to exp(2\pi i  \alpha_i/N) \psi_i$ where

\beqn
\{\psi_i\} = (Q,u^c, d^c, L, e^c),~~
\{\alpha_i\}= (\alpha_Q, \alpha_{u^c}, \alpha_{d^c}, \alpha_L, \alpha_{e^c})
\eeqn
Not listed  above  are the  Higgs superfields $H_1, H_2$ whose charges  are determined via
their couplings  with  the SM fields. Since  each of the charges  assume N values
there are $N^5$ possibilities. However, not all are independent. IR reduce this set by
imposing the  restriction that all elements related by hypercharge rotation 
$exp(2\pi i Y/N)$ are equivalent, which corresponds to an invariance  under the
shift $\vec \alpha\to \vec\alpha +(1,-4,2,3,-6)$ mod N. Further, the constraint
that the Higgs field H couple to $Qd^c$ and $Le^c$  imposes the constraint
$\alpha_Q+\alpha_d=\alpha_L+\alpha_e$. With the above constraints there is a reduction
in the allowed number of possibilities.\\

 The symmetries can be classified according to the constraints  they impose on dimension four
operators. These are~\cite{Ibanez:1991pr}:(i) symmetries which forbid both lepton and  baryon number
violation.  
 These  require the constraint $\alpha_{u^c}+2\alpha_{d^c}\neq 0$ (mod N)
and $2\alpha_L+\alpha_e\neq 0$ (mod N). Specifically they forbid cubic 
interactions $u^cd^cd^c$ and $LLe^c$ in the superpotential. One might call these
constraints generalized matter  parity constraints (GMP);
(ii) symmetries  which forbid lepton number violation but allow for
baryon number violation violation, i.e., $u^cd^cd^c$  is allowed but $LLe^c$ is forbidden
in the superpotential. 
 These require the constraint
$\alpha_u+2\alpha_d=0$ (mod N), $2\alpha_L+\alpha_e\neq  0$ (mod N).
One might call this the generalized lepton parity (GLP);  
(iii) symmetries which allow for the lepton number but  not  the baryon number violation, i.e.,
$u^cd^cd^c$  is forbidden  but $LLe^c$ is allowed in the superpotential. 
These require $\alpha_u+2\alpha_d\neq  0$ (mod N), $2\alpha_L+\alpha_e=0$  (mod N).
One might call this the generalized baryon parity (GBP);
and finally (iv) symmetries which allow both lepton number and baryon number violation.
Here both $u^cd^cd^c$  and  $LLe^c$ are forbidden and 
 the constraints are $\alpha_u+2\alpha_d=0 $ (mod N),  
$2\alpha_L+\alpha_e=0$ (mod N). Possibility (iv) is  excluded  as  it allows
for dangerous proton decay. \\

Further constraints arise from anomaly cancellation conditions. Analogous to the anomaly
cancellation condition for gauge symmetries, there are also anomaly cancellation conditions 
for discrete symmetries arising as remnants of gauge symmetries [The discrete 
gauge anomalies can be understood in the low energy theory in terms of instantons and 
are required for the consistency of the low energy discrete gauge 
theory~\cite{Preskill:1991kd,Banks:1991xj}].  
 One might call these 
discrete gauge anomaly cancellation conditions~\cite{Ibanez:1991hv}. 
The $Z_N$  arising from the  extra
U(1) must be considered in conjunction with $SU(3)\times SU(2)\times U(1)$ of the
standard model. Consequently all non-trivial anomalies involving $Z_N$ and factors 
of $SU(3)$, $SU(2)$ and $U(1)_Y$ must be considered. Thus typically we will have anomalies
from the combinations $Z_N^3$, $Z_N^2\times U(1)_Y$, $Z_N\times U(1)_Y^2$, 
$Z_N\times SU(M)\times SU(M)$
(M=2,3) as well as mixed $Z_N$-gravitational anomalies.
An analysis with inclusion of anomaly cancellation constraints shows that with the 
particle content of the minimal supersymmetric standard model there are two discrete anomaly free 
generalized parities. One of these is the familiar $Z_2$ R parity symmetry, while the second is
a $Z_3$ symmetry $B_3$ which allows for lepton number  violation. The  phase assignment
for this symmetry are $(1,\alpha^2, \alpha, \alpha^2,\alpha^2)$  for elements $(Q,u^c,d^c,L,e^c)$.
The analysis of  Ref.~\cite{Ibanez:1991pr}
 assumed that the hypercharge  is unbroken. However, hypercharge is a broken 
symmetry below the electroweak scale after the Higgs field gets a VEV. The analysis of 
Ref.~\cite{Pawl:2005mj}
re-examined the IR analysis  without the assumption of an unbroken hypercharge
symmetry. Here  the constraints that arise from the fermion mass terms give 
\beqn
\alpha_{Q}+ \alpha_{u^c}=\alpha_{Q}+\alpha_{d^c}=\alpha_L+\alpha_{e^c}=\alpha_L+\alpha_{\nu^c}=0
~~(mod ~N)
\eeqn
where we have assumed  generational independence. 

The above gives 
$\vec\alpha $=$(\alpha_Q, -\alpha_Q,$  $ -\alpha_Q, $ $\alpha_L,$ $ -\alpha_L, -\alpha_L)$ where
the elements corresponds to the set $(Q,u^c,d^c, L, e^c,\nu^c)$. 
 The analysis of Ref.~\cite{Pawl:2005mj} requires  that dimension five operators $QQQL$ and $u^cd^cd^ce^c$ be
 absent which leads to the constraint $3\alpha_Q +\alpha_L=0$ (mod N) and in addition requires
 that the neutron-antineutron oscillation mediated by operators $uddudd$ be absent which implies
 $6\alpha_Q\neq 0$ (mod N). Along with the above there is one anomaly cancellation 
 condition from the $Z_N\times SU(2)\times SU(2)$ sector which gives $9\alpha_Q+3\alpha_L=0$
 (mod N). The lowest N consistent with the above is N=9 and the allowed 
 $(\alpha_Q, \alpha_L)$ sets are $(1,0)$, $(1,3)$, $(2,0)$, $(2,6)$, $(4,0)$ and
 $(4,3)$.   These choices suffice to suppress  proton decay from 
 dimension 4 and dimension 5  operators  as well as  from gravitationally induced 
  wormhole tunneling and blackholes induced processes.\\
  
 A more recent analysis shows that the conclusion of IR that only $B_3$ symmetry 
 (also called $B_3$-triality) 
  forbids the  problematic dimension five proton decay operators is a consequence
  of the restriction to $Z_N$ for N=2,3 discrete symmetries. The more general analysis
  of Ref.~\cite{Dreiner:2005rd} has extended  the  work of IR to arbitrary values of N. 
   In doing so the authors of Ref.~\cite{Dreiner:2005rd} find 22 new anomaly-free  
  discrete gauge symmetries. After imposition of the phenomenological requirements 
 (i) the presence of the mu-term in the superpotential,
 (ii) baryon-number conservation up to dimension-five operators, and
(iii) the presence of the See-Saw neutrino mass term LHLH, 
  they are left with only two anomaly-free discrete gauge symmetries. These  
  are the $B_3$ symmetry discussed above  and in addition  a new symmetry which the authors
  call proton-hexality, $P_6$. This symmetry is a  $Z_6$ symmetry and reproduces 
  the low-energy R-parity conserving superpotential without the undesirable 
  dimension-five proton decay operators. Thus the main problem of the MSSM with R-parity 
  with respect to proton decay is solved with proton hexality symmetry..

  In the context of string theory an interesting issue in model building concerns the 
  question if the imposition of the anomaly cancellation condition is always essential. 
  It may be that in string models all anomalies in discrete symmetries 
  are cancelled~\cite{Dine:2004dk} 
  by the Green-Schwarz mechanism~\cite{Green:1984sg}. 
  In that case one may obviate  the necessity of imposing the anomaly cancellation
  condition. 

\section{Other Aspects}
In this section we discuss a number of topics related to proton stability. 
One important issue concerns the relationship of  proton stability to  neutrino 
masses and mixings in grand unified models.  This topic is discussed in Sec.(8.1) in the context
of SO(10) grand unified models. Another phenomena which is closely associated
with proton stability in supersymmetric grand unified models is dark matter. 
Thus grand unified models with R parity automatically have the LSP which is 
absolutely stable and if the LSP is neutral, it becomes a candidate for dark matter.
It turns out that in such a circumstance severe constraints exist in obtaining 
amounts of dark matter consistent with experiment and at the same time 
achieving proton lifetime consistent with data. This topic is discussed in Sec.(8.2).
A discussion of exotic \bl~violation is given Sec.(8.3) where  
$\Delta B=3$ such as $^3H\to e^+\pi^0$, and \bl
 ~violation involving higher generations, e.g., $p\to  \tau^*\to \bar \nu_{\tau} \pi^+$,
are discussed. Also discussed in this section is proton decay via
monopole catalysis where $M+p\to M+e^+ + {\rm mesons}$.
Speculations on proton decay and the ultimate
fate of the universe are discussed in Sec.(8.4). 

\subsection{Neutrino masses and proton decay}
As pointed out above an important issue concerns the implications of neutrino masses and mixings
 for proton decay lifetime.  A grand unified model such as $SO(10)$  has a 
right handed neutrino which is an SU(5) singlet along with a left handed  neutrino, 
which resides in the $\bar 5$ in the decomposition $16=1+\bar 5+10$. This
allows  for the possibility of both Dirac and Majorana type mass terms 
for the neutrino states. Together, they combine to produce neutrino masses
by the See-Saw type mechanism. The See-Saw contains information on the 
nature of unification and thus a study of neutrino masses may also have
implication for proton decay in unified models. It is thus desirable to discuss
the issue of neutrino masses. We first summarize the  current status of neutrino
oscillation experiments which yield results on masses and mixings 
of neutrinos~\cite{nureviews}. 
We will then discuss the theoretical aspects relevant for grand  unification 
and proton stability. The flavor states of the neutrino can be related to the
mass diagonal states by

\begin{equation}
\left(\nu_e,  ~\nu_{\mu}, ~\nu_{\tau}\right)= U 
\left(\begin{array}{c}
\nu_1\\ 
\nu_2\\
\nu_3\end{array}\right)
\end{equation}
where $U$ is a unitary matrix which can be parameterized in terms of three mixing angles 
$\theta_{12}, \theta_{23}, \theta_{13}$  and one phase. The natural range for the angles
are $ 0\leq \theta_{ij}\leq \pi/2$ and $0\leq  \delta \leq 2\pi$.  An explicit parameterization 
is 
\beqn
U=U_{23} U_{13}U_{12}
\eeqn
where $U_{ij}$ are defined by 
\beqn
U_{12}= \left(\begin{array}{ccc}
c_{12} & s_{12}  & 0 \\
-s_{12} & c_{12} & 0\\ 
0 & 0 & 1\end{array}\right),~~U_{13}=
\left(\begin{array}{ccc}
c_{13} & 0 & s_{13} e^{i\delta}\\ 
0 &  1 & 0\\
 -s_{13} e^{- i \delta} &  0 & c_{13}\end{array}\right)\nonumber
\eeqn
\beqn 
U_{23}= \left(
\begin{array}{ccc}
  1 & 0  & 0 \\
0 & c_{23} & s_{23}\\ 
0 & -s_{23} & c_{23}\end{array}\right)
\eeqn
where $c_{ij}=\cos(\theta_i-\theta_j)$ and $s_{ij}=\sin(\theta_i-\theta_j)$.
The neutrino mass  matrix $m_{\nu}$ in the flavor diagonal basis is related to the
mass matrix  $m_{\nu}^D$ in the mass diagonal basis by 
\beqn
m_{\nu}= U^* m_{\nu}^D U^{\dagger}
\eeqn
The solar neutrino and the atmospheric  
neutrino data give~\cite{solar,SNO1,atm} 
\beqn
\Delta m^2_{sol} =(5.4-9.5)\times 10^{-5}~ eV^2,~~ 
\Delta m^2_{atm} =(1.4-3.7)\times 10^{-3}~ eV^2
\eeqn
Within the three neutrino-generations fits to the data give 
\beqn
\Delta m^2_{sol}=||m_2|^2-|m_1|^2|,~~\Delta m^2_{atm}=||m_3|^2-|m_2|^2|\nonumber\\
\sin^2\theta_{12}=(0.23-0.39), ~~\sin^2\theta_{23}=(0.31-0.72), ~~\sin^2\theta_{13}<0.054
\label{nu-mixing}
\eeqn
The neutrino oscillation experiments measure only the mass 
squared differences and cannot tell us about 
the absolute  value of the neutrino masses. 

Information on the absolute values comes from other sources. Thus 
 neutrinoless double beta decay gives an upper limit of~\cite{doublebeta1,doublebeta2}  
\beqn
|m_{ee}|<(0.2-0.5) eV
\eeqn
where  
\beqn
 m_{ee}=(1-s_{13}^2)(m_{\nu_1}c_{12}^2  + m_{\nu_2}s_{12}^2) + m_{\nu_3} e^{2i\delta} s_{13}^2
\eeqn
while the WMAP collaboration gives~\cite{wmap1,wmap2}  
\beqn
\sum_i|m_{\nu_i}|< (0.7-1) eV
\eeqn
A variety of neutrino mass patterns are possible. Some possibilities that present themselves
are\\

\noindent
 (a) $|m_{\nu_3}|>> |m_{\nu_1,\nu_2}|$\\
(b) $|m_{\nu_1}|\sim |m_{\nu_2}|, |m_{\nu_1,\nu_2}|>>|m_{\nu_3}|$\\
(c) $|m_{\nu_1}|\sim |m_{\nu_2}|\sim  |m_{\nu_3}|$, 
$|m_{\nu_1,\nu_2,\nu_3}|>> ||m_{\nu_i}| -|m_{\nu_j}| |$\\

\noindent
The remarkable aspect of Eq.(\ref{nu-mixing}) is that  the mixing angles $\theta_{12}$ and
$\theta_{23}$ are  large with $\theta_{23}$ being close to maximal while $\theta_{13}$ is 
small. This is quite in contrast  to the case of mixings for the case  of the quarks
and it appears difficult  a priori to see how the neutrino mass textures and the quark mass textures 
could arise  from the same unified structure. However, such a conclusion may be hasty as the 
neutrino masses have a more intricate structure. Thus unified models typically produce
 Dirac neutrino masses $M_D$, LL type neutrino masses  $M_{LL}$, and RR type neutrino masses $M_{RR}$
 which combine to produce the neutrino mass matrix 
 \beqn
 m_{\nu}= M_{LL} - M_D M_{RR}^{-1} M_D^T
 \eeqn 
 The  second term involving $M_{RR}$ is the so called Type I See-Saw contribution
 while the first is the Type II See-Saw contribution. We see then that the neutrino
 mass matrix is more complex than the corresponding ones in the 
 quark-sector. While $M_D$ has a direct correspondence with
 the quark-lepton textures, their  connection with  $M_{LL}$ and $M_{RR}$ 
 is more dependent. Further, the matrices $M_{LL}$ and $M_{RR}$  
  can be helpful in connecting the
 two  very different type of hierarchies, i.e,, the  hierarchies in the quark 
 sector vs those in the neutrino sector. For example, it is proposed that 
 $M_{RR}$ textures may have  a hierarchy similar to the hierarchy in the
 Yukawa sector. The simplest such possibility is~\cite{Dermisek:2004tx}
   \beqn
 M_{RR}= M_R~ diag \left(\epsilon_{1R}, \epsilon_{2R}, 1 \right)
 \label{mrr}
 \eeqn 
  which leads to  
\begin{equation}
M_{RR}^{-1}= \left(M_R\right)^{-1}
\left(\begin{array}{ccc}
\epsilon_{1R}^{-1} & 0 & 0\\
0 & \epsilon_{2R}^{-1} & 0\\ 
0 & 0 & 1\end{array}\right)
\label{mrrinverse}
\end{equation}
With  $M_{RR}^{-1}$ of the form given by Eq.(\ref{mrrinverse}) and $\epsilon_{1R}<<\epsilon_{2R}<<1$   
  it is possible to generate the neutrino textures compatible with data. Such  possibilities
   along with a variety of others have been investigated within the $SO(10)$ grand 
   unification~\cite{nuso10}.
   Over the years attempts have also been made to understand neutrino masses within
   string models~\cite{nustring} and this effort is likely to grow with 
    discovery of additional realistic or semi-realistic  string based models.\\

We turn now to the connection of the neutrino masses to proton decay. It turns out that
the connection between the two is very  much model dependent.
This connection can vary from one extreme of little or no connection to a strong correlation. 
 To begin with   in the general analysis of dimension
   five operators  in MSSM  it is possible to suppress proton decay from dimension
   five operators by the elimination of certain operators while allowing for 
   lepton number violating operators  such as 
  \beqn
 f_{ij} \frac{1}{M} L_iL_j\phi\phi'
\label{llmass}
\eeqn
where $L_i$ are lepton doublets and $\phi,\phi'$ are Higgs doublets. A VEV formation
for the Higgs doublets then produces $M_{LLij}\nu_{L_i}\nu_{L_j} $ where 
$M_{LLij}= <\phi><\phi'>/M$. With $<\phi>\sim <\phi'>\sim M_{EW}$ and $M\sim M_G$ 
one can generate neutrino masses in the O($10^{-1}-10^{-5}$) eV range
which is a large enough range to accommodate a variety of possible scenarios. 
A mass term of type Eq.(\ref{llmass}) can arise naturally in a variety of $SU(5)$ and
$SO(10)$ models. 

  In SU(5) models the right handed neutrino is absent in the $\bar 5$ and the $10$ plet 
  representations but can be added to the spectrum by hand in an ad hoc fashion. 
  Because of this there is  no correlation between the neutrino masses and proton 
  decay in minimal SU(5) models in the case of Type I see-saw. However, 
in the case of Type II see-saw one can use $15_H$ in $SU(5)$, 
and there is a correlation between $B-L$ non-conserving channels for proton decay 
and neutrino masses (see, for example, Ref.~\cite{Dorsner:2005fq}).    
In SO(10) the right handed neutrino appears as a
  basic element of the 16 plet representation and because of this typically there
  is some correlation between neutrino masses and proton decay. For example, 
in SO(10) the textures for the Dirac neutrino masses are directly correlated with the  
    up quark mass  textures and arise from the same common couplings of the Higgs fields
    with matter. Further, in SO(10) the Majorana masses for the right handed neutrinos can arise
    from the $\overline{126}_H$ interaction with matter, i.e., from the  couplings
    \beqn
     \lambda_{\overline{126}} 16_i16_j\overline{126}_H
    \eeqn
   since  $\overline{126}$ contains an $SU(5)$ singlet as  can be seen from the $SU(5)$ 
   decomposition of $\overline{126}$
   \beqn
   \overline{126}=  1+ \bar 5+ 10+ \overline{15} +45+ \overline{50}
   \eeqn
    Alternately, one can generate Majorana  masses from the couplings  of matter with
     $\overline{16}_H$ of Higgs, ie.,
     \beqn
      \lambda_{\overline{16}}'  \frac{1}{M} 16_i16_j\overline{16}_H\overline{16}_H     
     \eeqn
    For  example, with $<\overline{16}_H>\sim M_G$ and $M\sim M_{\rm Pl}$, one has
    $M_{RR}\sim 10^{12-14}$ GeV which is  typically the intermediate  scale  mass. 
     Thus with restricted number of couplings a strong correlation between the 
     neutrino masses and proton decay can arise. However, the degree of correlation
     between the two phenomena depends on the number of assumed interactions. Thus, for
     example, in  the SO(10) model of Ref.~\cite{Dutta:2004zh}
      the  Higgs fields  that 
   couple with matter  consist  of $10_H$, $120_H$, and $\overline{126}$ while  the
   breaking of $SO(10)$ includes  the $210_H$ representation. In this case 
   it  is possible to suppress proton decay from dimension five operators
   to the current experimental level and at the same time get 
   consistency with the solar and the atmospheric neutrino oscillation data. 
\\  
A model where predictions of proton decay are connected with the prediction of 
neutrino masses  is discussed in  Ref.~\cite{Babu:1997js}.  
Here a new source of proton decay from dimensions five operators is suggested  
which arises from $\overline{126}_H$ couplings. Normally the mediation of $\overline{126}_H$ does
not produce dimension five operators, the reason being that the  $\overline{126}_H$ 
mass  term involves  $\overline{126}_H$  and $126_H$ and $126_H$ has no couplings with 16-plet 
of  matter.  However, consider the couplings where $126_H$ and $\overline{126}_H$ couple  with
a $54_H$ with $SO(10)$ invariant couplings
\beqn
W_{H}'= \lambda(126_H.126_H.54_H) +\bar  \lambda (\overline{126}_H \overline{126}_H 54_H)
\label{126_54}
\eeqn
Now  in the $SU(2)_L\times SU(2)_R\times SU(4)_C$ ($G_{224}$) decomposition, the $54$ plet has the 
decomposition: $54=(1,1,1)+ (3,3,1)+(1,1,20)+(2,2,6)$. Similarly, $\overline{126}$ has the decomposition
$\overline{126}=(1,3,\overline{10})+(3,1,10)+(2,2,15)+(1,1,6)$. Here  $(1,1,6)$ contains the color 
triplet and the color anti-triplet. If the $54_H$ acquires a  VEV in the $(1,1,1)$ direction then
the  superpotential of Eq.(\ref{126_54}) generates a (1,1,6).(1,1,6) mass term for the Higgs color triplets
and color anti-triplets. This mass term will mix with the mass term from $126_H\overline{126}_H$
and produce an effective color triplet mass  $M_{H_3}$ to suppress the dimension five operators. 
If we assume 
\beqn
W_{126}=f_{ij} (16_i16_j)\overline{126}_H 
\eeqn
 then the size of these  couplings $f_{ij}$ can be estimated. One can assume that the size of all 
 the VEVs including the VEV of the $45$, $54$ and  $\overline{126}_H$ which break the GUT symmetry 
 are order the GUT scale $M_G=2\times 10^{16}$ GeV. With the
 assumption of a universal Majorana mass of 
 $(1-3)\times 10^{12}$ GeV, one finds $f_{ij}\sim 10^{-4}$. With these parameters one finds dimension
 five operators with  strengths which can generate proton decay at observable rates.
  With the above assumptions one of the predictions of the model is a non-hierarchical  nature of
 the  couplings which lead to predictions such as~\cite{Babu:1997js} $\Gamma(l^+ K^0):\Gamma(l^+ \pi^0)\simeq 2:1$. 
A later analysis involving $16_H+\overline{16}_H$ rather than $126_H+\overline{126}_H$ is given
 in Ref.~\cite{Babu:1998wi} which appears  to give more realistic pattern of fermion masses and 
 mixings.  
 However, it is clear that the predictions from this sector depend strongly on the nature
 of the neutrino sector and thus different assumptions on couplings in this sector, specifically, 
 for example, on the nature of the Majorana mass matrix will lead to different predictions 
 on proton decay modes.
The inclusion of Planck slop can modify the correlation between proton
    decay and neutrino masses. Thus the
    addition of higher dimensional operators whose number increases 
   sharply with dimensionality typically  weaken the correlation between
    proton decay and neutrino masses  because of the greater arbitrariness
    that such operators bring in.  Finally, in the string framework there
    is no logical necessity for proton decay operators  to be correlated  
    with neutrino masses. 

\subsection{Proton stability and dark matter}
 There is a strong correlation between dark matter and proton stability in 
  supersymmetric theories. One may recall that in MSSM one introduces the 
  R parity symmetry to suppress dangerous proton decay from dimensions four
  operators and this R parity then serves to make the lowest mass supersymmetric 
  particle (LSP) absolutely stable. Further, if the LSP is neutral it becomes a candidate
  for dark matter~\cite{Goldberg:1983nd,Ellis:1983ew}. Quite remarkably one finds that in a 
  large class of supergravity 
  based models the LSP is a neutralino which then becomes  a candidate for 
  non-baryonic cold dark matter (CDM)~\cite{Arnowitt:1992aq}. 
    On the experimental side the Wilkinson Microwave Anisotropy Probe (WMAP) has placed stringent 
bounds on the amount of cold dark matter  present in the universe.
The analysis of WMAP data gives~\cite{Bennett:2003bz,Spergel:2003cb,Lahanas:2003bh}
\beqn
\Omega_{CDM} h^2 =0.1126^{+0.008}_{-0.009} \;,
\label{wmap}
\eeqn
where $\Omega_{CDM} =\rho_{CDM}/\rho_c$, and where $\rho_{CDM}$ is
the matter density of cold dark matter and $\rho_c$ is the 
critical matter density needed to close the universe, and 
$h$ is the Hubble parameter measured in units of 100km/s/Mpc.
A reasonable assumption is that our Milky way contains cold dark matter in 
similar amounts and this has led to much experimental  activity for the detection
of  cold dark matter in terrestrial experiments and larger experiments for future
are also being proposed.  On the theoretical side
the result of Eq.(\ref{wmap}) puts a stringent constraint on the  unified models.
Not only do the unified models  need to provide a candidate for the CDM, but also
predict a CDM in the amounts consistent  with Eq.~(\ref{wmap}).
It  is then of interest to investigate what correlations exist between dark matter
and proton stability in some of the current unified models of particle interactions.
In spite of the very strong connection between dark matter and proton stability, there are
only a few detailed studies exploring these constraints~\cite{Arnowitt:1998uz}. 
In supergravity unified models,
where proton decay via dimension five operators is  allowed, the connection between 
proton  stability and dark matter arises since both depend strongly on the soft breaking
sector. Typically, proton stability requires sparticle  spectrum to be heavy to suppress
proton decay while dark matter prefers  a  lighter  sparticle spectrum to facilitate
efficiently  excess CDM produced in the early universe. Thus the requirement that the 
constraints be  satisfied simultaneously limits severely the parameter space of the model.
In models with universal soft breaking proton decay is governed by 
$m_{1/2}/m_0^2$ which very roughly requires larger values of $m_0$ and relatively smaller 
values of gaugino masses. But large values of $m_0$ and squark masses tend to suppress the
annihilation of neutralinos.  
Thus typically satisfaction of the proton decay constraints renders the detection of 
dark matter more difficult~\cite{Arnowitt:1998uz}. 
Recently such connections have  also been explored in other 
scenarios with large extra dimensions~\cite{Mohapatra:2002ug,Agashe:2004ci,Servant:2004ke,Goldman:2004hk}.


\subsection{Exotic $B \ \& \ L$ violation }
\subsubsection{$|\Delta B| \ > \ 1$ violation and other non-standard $B\& L$ violation}
In Sec.(\ref{ued}) it was  found that in models with two universal 
extra dimensions, the surviving
discrete $Z_8$ symmetry which is a remnant of the $U(1)_{45}$ symmetry in the extra dimensions
$x_4$ and $x_5$ suppresses the  dimension six baryon and lepton number violation  
 but does allow such operators at high order. 
 One possibility is to dispense with the 
extra dimensional constructions  and simply focus on  discrete symmetries  to 
  generate the appropriate constraints. An analysis along this 
line is given in Ref.~\cite{Babu:2003qh} where an anomaly free 
$Z_6$ symmetry is imposed and it is shown that such a symmetry can emerge from
  $(I_R^3+L_i+L_j-2L_k)$ where $L_i$ is the lepton number for the ith generation. With the 
  $Z_6$ symmetry all $\Delta B=1$ and $\Delta B=2$ effective operators are forbidden but 
  $\Delta B=3$ operators are allowed and these give rise to some very exotic processes. 
  To illustrate this symmetry one may consider the  interaction 
  \beqn
  {\cal L}_Y = Qu^cH + Qd^cH^*+ l\nu^cH^* +l\nu^cH+ M_R\nu^c\nu^c
  \label{z6lagrangian}
  \eeqn
  where a Majorana mass term has  been included for generating the See-Saw type neutrino masses. 
  With  the  $Z_6$ charge assignments
  \beqn
  Q(6), u^c(5), d^c(1), l(2), e^c(5), \nu^c(3), H(1)
  \eeqn
 the Lagrangian of Eq.(\ref{z6lagrangian}) is invariant under the $Z_6$ symmetry. 
 With the above charge assignments the $Z_6$ discrete group is anomaly free.
 Now it can be easily seen that the $Z_6$ is a subgroup of $U(1)_{2Y-B+L}$ since
 under $U(1)_{2Y-B+L}$, the fields have the quantum numbers: 
$Q(0)$, $u^c(-1)$, $d^c(1)$, $l(2)$, $e^c(-1)$, $\nu^c(-3)$, $H(1)$ as can be easily checked
by recalling the B, L and Y quantum numbers for these fields. The above implies that
any effective operator allowed by the $Z_6$ symmetry must satisfy the constraint

\beqn
  \Delta (2  Y -B +2 L)=0 ~mod~ 6.
\label{z6selection}
\eeqn
Using the invariance  under $U(1)_Y$ ($\Delta Y=0$) it is then easily seen that
$\Delta B=1$ and $\Delta B=2$ effective operators are forbidden but 
$\Delta B=3$ effective operators  are allowed. Examples of such operators include
\beqn
 \frac{1}{\Lambda^{11}} \{Q  \bar u^{c3} \bar d^{c5} l, 
 Q^5  \bar d^{c4} l, Q^8\bar d^c\bar e^c, ..\}
\eeqn
These lead to processes such as~\cite{Babu:2003qh} 
\beqn
^3H\to e^+\pi^0, ~~^3He\to e^+\pi^+
\eeqn
It is now seen that an estimate of the triple nucleon decay lifetime is proportional to
$\Lambda^{22}$ and a quantitative analysis shows that even a $\Lambda\sim 10^{2}$ GeV
is sufficient to suppress the process to current experimental limits~\cite{Babu:2003qh}.

Among other models where non-standard $B\& L$ violation occurs is the analysis of 
Ref.~\cite{Kovalenko:2002eh}
 where a  generic lepto-quark extension of the Standard Model is 
considered. Here one finds dimension nine operators of the type 
\beqn
(\bar \nu P_Rd)(\bar e P_R d)(\overline{u^c}P_R e)
\eeqn 
which induce $\Delta L=-\Delta B=1$ proton decay producing decay channels of the
type
\beqn
p\to e^-e^+\nu \pi^0\pi^+,  e^-\bar \nu \nu \pi^+ \pi^+, e^-e^+\nu \pi^+, \nu\nu\bar \nu \pi^+
\eeqn

\subsubsection{$B \ \& \ L$ violation involving higher generations}
Another phenomenon concerns baryon number violation (BNV) 
involving decays of higher generations~\cite{Marciano:1994bg,Hou:2005iu}.
It was noted in Ref.~\cite{Marciano:1994bg} that an estimate of baryon number  
violating $\tau$ decays can be given by using limits on proton decay lifetimes.
 This can be  done using dimension
six operators which have all been classified. 
Let us label these  operators
by $O^{n}_{ijkl}$ where i,j,k,l are generation indices. The effective interaction that
governs baryon and lepton number violating processes is then 
\beqn
\sum_{n=1}^{6}  C_{ijkl}^{(n)} {\cal O}_{ijkl}^{(n)}
\eeqn
where n=1,..,6 indicates the different types of dimension 
six operators.  Now a possible decay mode of the proton is through an off-shell $\tau^*$ 
such that
\beqn
p\to \tau^*\to \bar \nu_{\tau} \pi^+
\eeqn
The effective $C_{uud\tau}$ coupling that enters this process can be constrained by the
current limit on 
$\tau(p\to \overline{\nu} \pi^+) \ > \ 2.5\times  10^{31}$ year 
to yield~\cite{Hou:2005iu} $C_{uud\tau}\leq 6\times 10^{-24}$ GeV$^{-2}$. The same coefficient then can be 
utilized to compute the decay branching ratio of $\tau\to p\pi^0$ and one 
finds~\cite{Hou:2005iu} 
\beqn
B(\tau\to p\pi^0)\leq 5.9\times 10^{-39}
\label{bnvtaudecay}
\eeqn
Similar considerations also  apply to a variety of other decay modes such as
$\tau \to \bar p K^0$, $\tau \to \bar  p \gamma$.  However, as  one can see from 
Eq.(\ref{bnvtaudecay}) the branching ratio for such decays is extremely small 
and one has no hope of observing such decays at colliders. 
Decay  modes of the above 
type have also been calculated in D and B decays such as  
$D^0\to \bar p l^+$ and $B^0\to \Lambda_c^+l^-$ and the branching ratios for these
are also highly suppressed as expected~\cite{Hou:2005iu}.
Still there  have
been searches for such decays to put  experimental limits of
BNV processes in $\tau$ decays. Thus the CLEO collaboration~\cite{Godang:1999ge} has looked for 
five modes of the  $\tau$ lepton that violate baryon and lepton number while preserving 
$B-L$. These searches which yield negative results include the decay modes 
$\tau^-\to \bar p \gamma$, $\bar  p\pi^0$, $\bar p\pi^0\pi^0$,  $\bar p \eta$ 
and $\bar p\pi^0\eta$.
\subsubsection{Monopole catalyzed proton decay}
The existence of magnetic monopoles~\cite{Dirac:1931kp} is a generic prediction of  
grand unified theories (GUTs). The magnetic monopoles appear in 
the early universe at the phase transition corresponding 
to the breaking of the unified gauge group ($G\to H\times U(1))$~\cite{'tHooft:1974qc}. 
The mass of the magnetic monopoles $M_m$ is related to the mass of the 
superheavy gauge bosons which mediate nucleon 
decay, $M_m \ \geq \ M_V / \alpha_{GUT}$.  
The GUT magnetic monopoles have a complex structure: 
a very small core ($r \sim 10^{-29}$ cm), 
an electroweak region, a confinement region, 
a fermion-antifermion condensate region, and for $r \geq 3$ fm it behaves 
as a point particle generating a magnetic field $B=g/r^2$.  

A remarkable property of monopoles discovered by Rubakov and  Callan is that
monopoles can catalyze proton 
decay~\cite{Rubakov:1981rg,Callan:1982au,Rubakov:1988aq,Goldhaber:1985cc}.
The catalysis of the proton decay is due to the interaction 
of the GUT monopole core which at the quark  level leads to the reaction
$d_L+M\to  e^+_L+\bar u_R+\bar u_R$ and at the nucleon level leads to the process

\begin{displaymath}
 M \ + \ p \ \to \ M \ + \ e^+ \ + \ mesons 
\end{displaymath}
The above phenomenon is caused 
by boundary conditions which must be imposed 
on fermion fields at the monopole core. 
These boundary conditions  mix quarks and leptons and cause 
the monopole to have an indefinite baryon number.  
An equally remarkable property of monopole catalysis is that the scattering 
amplitude is not suppressed by a factor $1/M_X$, i.e., by inverse power of the
unification mass.  However, it is  difficult to predict with 
precision the rate of the proton decay induced 
by the monopole~\cite{Brihaye:2002nz}.  Current estimates for the 
catalysis cross-sections lie in the range $10^{-27}cm^2-10^{-21}cm^2$. 
Typically Big Bang cosmology leads to an abundance of monopoles,
while realistic estimates with  $M_m\sim 10^{16}$ GeV lead to a number density
$n_m<10^{-14}n_p$ where $n_p$ is the number density of the proton~\cite{Preskill:1979zi}. This
is the familiar monopole problem of grand unification to which inflationary 
cosmology provides a solution~\cite{Guth:1980zm}. It is still important from the experimental 
view point to put limits  on the magnetic monopole flux. Since monopoles are
heavy one expects the monopoles to be non-relativistic with $\beta=\frac{v}{c}<<1$.
The most recent bounds on the monopole flux come from the MACRO collaboration 
which has put upper limits on the magnetic monopole flux at 
the level of $\sim 3 \times 10^{-16} \ cm^{-2} \ s^{-1} \ sr^{-1}$ for
$\beta$ lying in the range 
 $1.1 \times 10^{-4} \ \leq \beta \leq 5 \times 10^{-3}$, based 
on the search for catalysis events in the MACRO 
data~\cite{Ambrosio:2002qu}. 
   
\subsection{Proton decay and the ultimate fate of the universe}
Since quantum gravity effects could destabilize the proton, the eventual fate of 
	 the universe would be governed by  the proton lifetime~\cite{Dicus:1981ab,Adams:1996xe}. 
	 Thus, for example, over long
	 time span white  dwarfs and neutron stars will be powered by proton decay. The proton
	 decay mode $p\to  e^+\pi^0$  within a white dwarf will result in the process   

	\beqn
	p+e^-\to \gamma +\gamma+\gamma +\gamma 
	\eeqn
     where two of the $\gamma$'s  come from the decay of the $\pi^0$ and the other two arise
     from the annihilation of $e^++e^-$, and where the 
     energy of each of the photons will be 
     $\sim m_p/4$.  The photons have a short mean free path and will quickly thermalize.
      Other decay modes would involve neutrinos which would escape. An estimate of the 
      luminosity of the white  dwarf powered by proton decay gives~\cite{Adams:1996xe}
      
      \beqn
      L_{*}\simeq 10^{-22} L_{\odot} (\frac{10^{35}}{\tau_P ~{\rm  yrs}})       
      \eeqn
      where $L_{\odot}$ stands for stellar luminosity.
      The white dwarf luminosity arising from  proton decay is indeed extremely small
      relative to the solar luminosity. If we assume that the white dwarf consists of
      N  nucleons  initially, then the time for it  to deplete to $N_0$ 
      because of nucleon disintegration is given by~\cite{Adams:1996xe}

      \beqn
      \tau =\tau_P ln(\frac{N}{N_0})      
      \eeqn 
      For $N\sim 10^{57}$ and
     $N_0 =1$  one finds $\tau \sim 131 \tau_P$. A similar analysis holds for neutron stars and
     for planets although the evolution of the neutron star under nucleon decay processes
     is more involved. 

\section{Summary and Outlook}
We summarize now the main conclusions of this report. \\

\noindent
\underline{Non-supersymmetric grand unification }\\
In non-supersymmetric models proton decay proceeds via dimension six  operators
which are induced by gauge interactions and via exchange of scalar lepto-quarks.
In these models one needs an extreme fine tuning to get light Higgs doublets, which
however, may be justified in the context of string landscape models. An
analysis of proton lifetime requires that one first address properly the fermion mass
and mixing issues to predict in a realistic fashion proton lifetime. These issues 
are discussed in detail in Sec.(3) where it is shown that some of the non-supersymmetric
unified models may still pass the  stringent experimental proton lifetime constraints.
As an example one may consider a simple extension of the  
Georgi-Glashow~\cite{Georgi:1974sy} model with a Higgs sector  
composed of $5_H$, $24_H$, and 
$15_H$~\cite{Dorsner:2005fq}. In this case one finds an 
upper-bound on the total proton decay lifetime in
this scenario of $\tau_p \ \leq \ 1.4 \times 10^{36}$ 
years~\cite{Dorsner-Ricardo}. More discussion 
of this topic is given in Sec.(5.6).
\\

\noindent
\underline {SUSY and SUGRA grand unified models}
\\
Supersymmetric unified  models have several advantages over the 
non - supersymmetric models. The Higgs sector of the theory is 
free of quadratic divergences and no extreme fine tuning as
in non-supersymmetric models is needed. Globally supersymmetric
unified models are not viable because of the difficulty of 
breaking  supersymmetry which is overcome in supergravity 
unified (SUGRA) models. Interestingly supergravity models also 
allow for radiative breaking of the electro-weak symmetry
which is accomplished without the addition of ad hoc tachyonic
mass terms as is done in non-supersymmetri models. SUGRA models
predict a sparticle spectrum in the TeV mass range accessible 
at accelerators and such spectrum is consistent with the gauge
coupling unification. An apparent drawback of supersymmetric
models is the possibility of proton decay via dimension 4 
operators which, however, can be eliminated by an R parity 
invariance. Proton decay  dimension five operators
still remains and typically dominates over proton decay from
 dimension six
operators. This puts stringent limits on the allowed parameter
space of the theory to be consistent with experiment.
In Sec.(4) a number of topics were considered. They include
the constraints on R parity violating interactions using
experimental bounds, doublet-triplet splitting, and an analysis of
proton decay in $SU(5)$ and $SO(10)$  models.
\\

\noindent
\underline {Tests of grand unification}\\
In grand unified models predictions of the proton lifetime are
intimately tied with the fermion masses and mixings  since they 
arise from the same common Yukawas interactions.
In a more technical language one needs to have realistic Yukawa
 textures. In an analogous fashion, the Higgs triplet sector also
has  textures which are in general different from textures in 
the Higgs doublet sector and these enter into proton lifetime
predictions (Sec.5.1). A phenomenon which can affect proton lifetime 
in supergravity models is that of gravitational smearing. It arises
from the possibility of a non-trivial gauge kinetic energy functions
which can split the gauge coupling constants at the unification
scale. These splittings effectively modify the heavy thresholds
and specifically the Higgs triplet mass and consequently affect
proton lifetime (Sec.5.2). The masses of the Higgs triplet and
other heavy thresholds are also constrained by the gauge coupling
unification constraints but the analysis depends sensitively on the 
inputs (Sec.5.3). The important topic of testing grand unification
through proton decay modes was discussed in Sec.5.4 with special
attention to the gauge groups $SU(5)$, $SO(10)$ and flipped 
$SU(5)$. An investigation of the conditions under which gauge
dimension six  proton decay can be eliminated in flipped $SU(5)$ 
is given in Sec.5.5. An  analysis of the upper bounds on
proton decay lifetimes in GUT models is given in Sec.5.6 where
it is shown that it is possible to find a model independent 
upper bound on the total proton decay lifetime. Such bounds
are useful in testing unified models.\\

\noindent
\underline {Grand unified models in extra dimensions}\\
The most attractive feature of extra dimensional models is that they 
provide a 
mechanism for a natural doublet-triplet splitting where
one achieves a light Higgs doublet necessary for electroweak
symmetry breaking while the Higgs triplet becomes 
superheavy. A large number of models in 5D and 6D with gauge
groups $SU(5)$, $SO(10)$, $SU(6)$ and $SU(3)^3$  have been 
investigated which, however, differ vastly 
in their predictions for proton decay. For example,
proton decay from dimension 4 and dimension 5 operators can be 
killed in some models by a residual $U(1)_R$ symmetry which 
leaves the exchange of $X$ and $Y$ gauge bosons as the main 
source of proton decay. However, proton decay from these
is typically dependent on the way matter is located in the 
extra dimensions. As discussed in Sec.(6)  if, for example, the matter fields 
 propagate in the bulk, then a full generation of 
quarks  and leptons must arise from split multiplets which have 
no normal $X$ and $Y$ gauge interactions among them. In such 
models  proton decay can arise only via higher than six dimensional 
operators and is suppressed. 
The usual dimension six operators can also 
be forbidden by location of matter on certain brains. For example, for 
the $SO(10)$ case placing all three generations on the \ps brane will 
give vanishing dimension six operators from the normal $X$ and $Y$ exchanges
since the wave functions for  the $X$  and $Y$ gauge  bosons vanish on 
the  \ps brane. However, with other choices  of locating matter on 
branes,  one will  have in general proton decay from dimension six operators.
Additionally proton decay can arise from derivative couplings. Consequently, 
predictions of proton decay in higher dimensional models vary over a 
wide range, from highly suppressed to the possibility of observation 
in the next generation of experiment. 

We emphasize,however, that the branching ratios into various modes can be used 
as probes of models including extra dimensional models. As an example in Sec.(6),
we discussed the work of Ref.~\cite{Buchmuller:2004eg}  which investigates a specific 
model in 6D where the three generations of 16 plets of matter are located at different branes: 
 generation 1 is placed
on the $SU(5)\times U(1)$ brane, generation 2 is placed  on the flipped $SU(5)\times U(1)$ brane,
and generation 3 is placed on the \ps brane.  With additional assumptions  regarding the
Higgs structure and  flavor sector  of the theory,  the model predicts the dominant proton decay
branching ratios so that~\cite{Buchmuller:2004eg} 
$BR(\pi^0e^+)=(71-75)\%$, $BR(\bar \nu \pi^+)=(19-33)\%$, and  
$BR(\mu^+\pi^0) = (4-5)\%$ (Sec.6.4).
Clearly the branching ratios provide important signatures for testing the models.
Another example, is proton decay in universal extra dimension models  where
in a class of such models one finds~\cite{Appelquist:2000nn}                                    
$p\to \pi^+\pi^+e^-\nu\nu,~~\pi^+\pi^+\mu^-\nu\nu$ (Sec.6.6).
Again such signatures provide a possible avenue to differentiate among various 
classes of models if proton decay is observed and branching ratios measured.
\\

\noindent
\underline {String unified models }
\\ 
 There  are  five types  of  known string theories:
 Type I, Type IIA, Type IIB, $SO(32)$ heterotic and $E_8\times E_8$ heterotic 
 which are connected by a web  of dualities and may have a common origin in a more 
 fundamental theory - the M theory. Of these the $E_8\times E_8$ heterotic 
 case has been investigated the most from the point of view of model building but 
 considerable progress has also occurred  recently in model building based
 on Type IIA and Type IIB. In Sec.(7) we discussed the status of proton 
 decay in a class of Calabi-Yau compactifications of the heterotic string.
 The dominant decay mode of the proton in these models as in supersymmetric
 $SU(5)$ is  $p\to  \bar \nu K^+$ but further progress in needed in computations
 of the Kahler potential to make precise lifetime predictions. There has been 
 a revival of interest recently in the heterotic string models.  
 One class of models is successful in achieving the MSSM particle spectrum without 
 exotics, and  here are no dimension four, five, or six  proton decay operators 
 in these models. For $k>1$ Kac-Moody  string models, generally dimension four proton decay 
 operators are absent due to the underlying gauge and discrete symmetries of the model 
 but dimension
 five proton decay operators are present~\cite{Kakushadze:1997mc}. 
 However, it is difficult to get realistic 
 quark-lepton textures in these models and hence difficult to make reliable estimates
 of proton decay lifetime in these models~\cite{Kakushadze:1997mc}. 
 
 An interesting recent result on proton decay comes from the M theory analysis discussed in 
 Sec.(7.5) where one considers M theory on ${\cal R}^4\times X$, where X is the  manifold 
 of $G_2$ holonomy~\cite{Friedmann:2002ty}.  If X looks locally like
$Q\times {\cal R}^4/\Gamma$ where  $Q$ is  a  three-manifold, then one will  get gauge fields
on the singular set ${\cal R}^4\times Q$.  The  case $\Gamma =Z_5$   leads  to the $SU(5)$ 
gauge fields on the ${\cal R}^4\times Q$~\cite{witten,Friedmann:2002ct}. 
Here  the assumption that the 
quark-lepton multiplet are in general located at different points in the  manifold
$Q$ leads to the prediction that the decay $p\to e_L^+\pi^0$ which arises from the
interaction $10^2 \overline{10}^2$ is enhanced relative  to $p\to e^+_R+\pi^0$ which arises 
from $10^2\bar 5^2$.  Since  $10$ and $\bar 5$ are located at different points  in $Q$
the $e^+_R$ mode is in general suppressed~\cite{Friedmann:2002ty}. 
Unfortunately, the decay lifetime is not predicted 
due  to unknown normalization factors in the effective proton decay dimension six
operator that arises from M theory. Further, it remains to be seen if  experiment can be
geared to measure the polarization of the exiting charged lepton.
Another interesting analysis of proton decay is the one based on intersecting D branes~\cite{Klebanov:2003my} 
which investigates  proton decay on SU(5) GUT  like models in Type IIA 
orientifolds with D6-branes. Here the analysis of the proton decay mode $p\to e^+\pi^0$ gives
a lifetime which may lie within reach of the next generation experiment (see Sec.(7.5)). \\

\noindent
\underline{Proton decay from  black hole and wormhole effects}
Quantum gravity does not conserve  baryon number and thus can catalyze
proton decay. Such an effect can arise  from virtual black hole exchange and
wormhole tunneling (Sec.(7.7)). It is then possible that the 
	two quarks in the proton might end up falling into the mini black hole
	and since  one expects black holes  not to conserve baryon number, a process such 
	 as this can lead to baryon number violation through 
	 $q +q\to l+\nu$ and  $q+q\to \bar q +l$ and consequently to proton decay. 
	 If the scale of quantum gravity  $M_{QG}=M_{\rm Pl}$, the proton lifetime will be
	 very high, i.e., $\sim 10^{45}$ years and outside the realm of 
experimental observation. However, such lifetimes still have significance 
in  determining the ultimate fate of the universe. 
\\

\noindent
\underline{Outlook}
 Search for proton decay should continue as one of the prime 
 experimental  efforts as it probes the nature of particle  interactions
 at extremely short distances which the accelerators can never hope to
 reach.  Fortunately there are proposals already  being pursued 
 which will improve the sensitivity of the proton decay searches by an order of
 magnitude or more. Chief among these are the HYPERK, UNO, MEMPHYS, ICARUS,
 LANNDD at the Deep Underground Science and Engineering Laboratory (DUSEL), and LENA.
 On the theoretical side one finds that in general predictions of absolute proton
 lifetime in unified models contain  significant uncertainties. These 
 arise from uncertainties in extrapolations from the GUT/string scale to the proton decay 
 scale of $m_p\sim 1$  GeV,  uncertainties  in the 3-quark matrix element between the proton and the
 vacuum state, uncertainties due to the quark-lepton textures and  uncertainties
 due to the approximation of using the effective Lagrangian to compute prediction of
 dimension six operators. 
  However, models do better in predicting the relative branching ratios since these
  are subject to a smaller subset of uncertainties. Thus even with a fuzzy knowledge
  of the absolute decay rates, one can use branching ratios as an
 instrument for differentiating models.
 Examples of this  possibility are provided  by  the $e^+\pi^0$ for the non-supersymmetric minimal $SU(5)$ model, 
 by $\bar\nu K^+$ mode for the minimal SUSY $SU(5)$ model, by the   branching ratios 
for the specific six dimensional model of Ref.~\cite{Buchmuller:2004eg}
 and by the modes $\pi^+\pi^+e^-\nu\nu,~~\pi^+\pi^+\mu^-\nu\nu$ for UED models,
 and by the dominance of $e^+_L\pi^0$ over $e^+_R\pi^0$ for the M-theory model
 of Ref.~\cite{Friedmann:2002ty}. 
 
 The preceding discussion points up that given sufficient
 data one can distinguish among a variety of unified models arising from 4D, 5D. 6D and
 from strings and branes. However, as one of the main observations of this report
 it is imperative that more theoretical effort is needed in the prediction
 of absolute rates to coincide  with the larger experimental effort in 
 improving proton lifetime sensitivities by an order  of magnitude or more. It is  only then
 that the maximum benefit from the new generation of proton decay experiment will
 accrue. 
 In summary, if proton decay is found it will winnow down the allowed 
 set of unified models. Further, as exhibited in this report  a detailed  knowledge
 of its  decay  modes will help to test specific  grand unified, string and 
 M theory scenarios.  
  Even if no proton decay is found in the next generation
 experiments, the improved theoretical predictions and the improved 
 experimental lower limits will eliminate or more stringently  constrain unification 
 models of particle interactions and gravity.
\section{Summary Table of p decay predictions}
It is useful to summarize in a tabular form the range of  theory predictions  
for the proton lifetime vs the current experimental limits.  As the report 
exhibits the literature on this topic is enormous, and in the Table below 
we show only a sample of the models discussed in the report.  Thus the Table
should be  used only as a guide to the more detailed discussion  in the body of
the  report.    Further,  the numerical estimates  on the lifetimes exhibited    are gotten
 under  specific assumptions which should be kept in mind while using these estimates.  
 For example, in entry (3) a different choice  of the Ray-Singer torsion can change the estimate
 by more than a factor of 10.
 In entry (7) we have used in Eq.(\ref{21.8b})  the value of the compactification scale
 $M_C=2\times 10^{16}$ GeV.  Even a factor of  2 shift on $M_C$ would  modify the estimate by
 more than an order of magnitude. 
 In entry (10)  we have  used  in Eq.(\ref{qpdecay}) the 
 value $M_{QP}=M_{\rm Pl}$ for the scale of quantum gravity.  
   In  entry  (12) we have assumed that the coupling f that appears
 in the Calabi-Yau manifolds is $\sim .05$ in the relation Eq.(\ref{Dexchange}). 
 In entry (16) the proton lifetime in proportional to $\Lambda^{22}$ where $\Lambda$ is the
 scale to which the 6D effective theory is valid. Thus the predictions are highly unstable, 
and the estimate can change by six orders of magnitude by a shift in the value of
$\Lambda$ by a factor of  2. 
 In entries (15) and (17), the  terminology 'suppressed' implies  that the parameters of the
 respective  models  can be adjusted to suppress proton decay to the current experimental limits. 
Thus, in these models the possibility exists of detecting proton decay just beyond the current limits. 
     Conservatively,  overall  the estimates given should be  considered accurate to no better than 
a factor of  10.  Finally, we note that in the Table we have presented only some sample set of 
decay modes.  Models typically have many more  such as  $\bar \nu \pi^+$, $\mu^+ K^0$ etc. 
The reader is directed to the original references listed in the Table and in the various sections,
for estimates on these decay modes.  

$$  $$
\begin{center} \begin{tabular}{|c|c|c|c|}
\multicolumn{4}{c}{Summary of proton decay  lifetime estimates  in years  for various models } \\
\hline
Mode & Reference  & Lifetime estimate & Exp. limit  \\
\hline
1. $p\to e^+\pi^0$  &\cite{Hisano:2000dg}   & $1.6\times 10^{34}$  &  $1.6\times 10^{33}$  \\
\hline 
2.     $p\to e^+\pi^0$                           &       \cite{Lee:1994vp}              &    $10^{33-38}$&                                                        \\
\hline                
3. $p\to e^+\pi^0$ &\cite{Klebanov:2003my} &    $(0.8-1.9)\times 10^{36}$   &                                                               
                                 \\
\hline 
4.   $p\to e^+\pi^0$                            &  \cite{Kim:2002im}       &     $\sim 7\times 10^{33\pm 2}$              &                                                                             \\
\hline 
5.  $p\to e^+\pi^0$                      &  \cite{jcpati}         &    $\sim 5\times 10^{35\pm 1}$        &                                                                              \\
\hline 
6.   $p\to e^+\pi^0$                      &  \cite{Buchmuller:2004eg}         &    $\sim 5\times 10^{34\pm 1}$        &                                                                              \\
\hline
7.   $p\to e^+\pi^0$                      &  \cite{Hebecker:2002rc}         &    $\sim 4\times 10^{36}$        &                                                                              \\
\hline 
8. $p\to e^+\pi^0,\bar{\nu} K^+ , ...$                               & \cite{Dorsner:2005fq,Dorsner-Ricardo}&    $\leq 1.4\times 10^{36}$                       &                                                                              \\
\hline 
9. $p\to e^+\pi^0$ &\cite{Dorsner:2006dj}  &    $\sim 10^{37}$         &                                                                              \\

\hline 
10. $p\to e^+\pi^0$ etc &  Black holes (Sec.7.7)  &    $\sim 10^{45}$         &                                                                              \\
\hline 
11.  $p\to \bar \nu K^+$    &    \cite{Nath:1985ub,Murayama:2001ur,Bajc:2002bv,Bajc:2002pg,Emmanuel-Costa:2003pu}         
    &    $\sim 10^{34}$        &     $  6.7\times 10^{32}$                                                                                \\
\hline 
12.   $p\to \bar \nu K^+$         &   \cite{Arnowitt:1989ud}         &    $\sim 10^{34}$        &                                                                              \\
\hline 
  13.  $p\to \bar \nu K^+$                             &\cite{Lucas:1996bc}&    $(6.6-3\times 10^2)10^{33}$          &                                                                              \\
\hline 
 14.   $p\to \bar \nu K^+$            &\cite{jcpati,Babu:1998wi} &    $(1/3-2)\times 10^{34}$          &                                                                              \\
\hline 
15.    $p\to \bar \nu K^+$             & \cite{ Dutta:2004zh,bgns}    &  suppressed  &                                                                             \\
\hline 
16.   $p\to \pi^+\pi^+l^-\nu\nu$       &\cite{Appelquist:2001mj}& $\geq 10^{35}$           &         $  3\times 10^{31}$                               \\
\hline
17. $p\to e^-e^+\nu \pi^+$ etc & \cite{Kovalenko:2002eh}  & suppressed  & $  1.5\times 10^{25}$ \\
\hline 
18. $p (n)\to \gamma + e^+ (\bar \nu)$  &\cite{ngamma}   &  $>10^{38\pm 1}$    &      $  6.7\times 10^{32}$                                                                        \\
\hline
\hline
\end{tabular}
\label{leptonicmodes}
\end{center}
$$ $$

\section*{Acknowledgments}
Communications and discussions with many colleagues on topics discussed  
in this report are acknowledged. They  include  Kaustubh Agashe, Richard Arnowitt, 
Kaladi Babu, Jonathan Bagger, Wilfried Buchmuller, Jean -Eric Campagne, 
 David Cline, Ali Chamseddine, Mirjam Cvetic, Ron Donagi, 
 Herbert Dreiner, Alon Faraggi, Paul Frampton, Tamar Friedmann, Takeshi Fukuyama,
Ilia Gogoladze, Haim Goldberg, Tarek Ibrahim, Christos Kokorelis, 
Boris Kors,  P. Minkowski, Peter Nilles, Burt Ovrut,  Tony Pantev, Jogesh Pati, 
Tomasz Taylor, Stuart Raby  and Raza Syed.
Especially fruitful were  many interactions with colleagues 
during the 'Planck 05' conference May 23-28, 2005 at ICTP, 
Trieste, Italy and the  'String Phenomenology 2005' Workshop 
June 13-18, 2005 in  Munich. Part of the work of the report 
was done while one of the authors (PN) was at the Max Planck 
Institute for Physics in Munich and acknowledges 
support from the Alexander von Humboldt Foundation 
during this period.  The research of PN is supported in 
part by the NSF grant PHY-0546568. 
P.~F.~P. would like to thank Borut Bajc, Marco Aurelio Diaz, 
Ilja Dorsner, Manuel Drees,  D. Emmanuel Costa,  
R. Gonz\'alez Felipe, German Rodrigo and Goran Senjanovi\'c 
for collaboration and discussions. 
The work of P.~F.~P. has been supported by FCT, through the 
project CFTP, POCTI.SFA.2.777 and by a fellowship under the project 
POCTI/FNU/44409/2002. One of the authors 
(P.~F.~P) thanks G. Walsch and A. Gay Cabrera 
for strong support during this period. 
\begin{appendix}

\section*{APPENDICES}
\section{Mathematical aspects of $SU(5)$ and $SO(10)$ unification}
In this appendix we will give some technical details of group theory 
that will facilitate reading the main body of the report. 
We begin by discussing $SU(5)$ where a single generation of 
quarks and  leptons is placed in the $\bar 5$ and the 
$10$ plet of $SU(5)$. The particle decomposition of $\bar 5$ is

\begin{eqnarray}
{\overline {5}}= \left(\begin{array}{c}d_{ L a}^c\\
e^-_L\\  -\nu_{eL}
  \end{array}\right)
\end{eqnarray}
where  sub a is the color  index. For the 10 plet of $SU(5)$ we have

\begin{eqnarray}
10 =\left(\begin{array}{ccccc}0 & u_3^c &-u_2^c  &-u^1 &-d^1\\ -u_3^c &0
&u_1^c  &-u^2 &-d^2\\ u_2^c &-u_1^c &0  &-u^3 &-d^3 \\
 u^1 &u^2 &u^3  &0 &e^+ \\ d^1 &d^2 &d^3 &-e^+
&0
  \end{array}
 \right)_L
\end{eqnarray}
To recover the interaction of the Standard Model particles from their $SU(5)$ invariant couplings  
one needs to carry out their $SU(3)_C\times SU(2)_L\times U(1)_Y$ invariant reduction. 
 Here we will illustrate the basic technique for  the reduction of $SU(5)$
 tensors  into tensors which are  irreducible under 
  $SU(3)_C\times SU(2)_L\times U(1)_Y$. First  it  is useful to record  the tensorial  representations  of  
    irreducible representations of $SU(5)$ which commonly surface in model building based  on the 
    group $SU(5)$. As we  have seen the matter falls in the $SU(5)$ representations $\bar 5_M +10_M$
    while, the Higgs could be  in any of the fields $5_H, \bar 5_H, 10_H, \overline{10}_H, 24_H$,
    $45_H$, $\overline{45}_H$, 
    $50_H, \overline{50}_H$, $75_H$ etc.  The tensors representing these are  
    \beqn
    5^i, \bar 5_i, 10^{ij}, \overline{10}_{ij}, 24^i_j, 45^{ij}_k,
    \overline{45}^i_{jk},
      50^{ijk}_{lm}, \overline{50}^{ij}_{klm}, 
    75^{ij}_{kl}
    \eeqn
    where one has  anti-symmetry in all the sub indices and  in all the super indices.
    Further, the  $24^i_j$ is traceless, while the 50-plet, $\overline{50}$ plet,
    and the 75-plet satisfy the following constraints 
    \beqn
    \sum_{n=1}^{5} 45^{in}_{n}=0 = \sum_{n=1}^{5} \overline{45}^{n}_{in};~~    
    \sum_{n=1}^{5} 50^{in}_{jkn}=0 = \sum_{n=1}^{5} \overline{50}^{ijn}_{kn};~~
    \sum_{n=1}^{5} 75^{in}_{jn} =0
    \eeqn 
      The following decomposition is useful in the reduction of the $SU(5)$ irreducible tensors
      into irreducible  components
      under  $SU(3)_C\times SU(2)_L\times U(1)_Y$          
   \beqn
   \delta^i_j= \sum_{a=1}^{3} \delta^i_a\delta^a_j + \sum_{\alpha=4}^{5} \delta^i_{\alpha}
   \delta^{\alpha}_j
   \eeqn
     where $a=1,2,3$ is the color index and $\alpha =4,5$ is the $SU(2)$ index.  Thus consider
     the 24-plet which has  the  following 
$SU(3)_C\times SU(2)_L\times U(1)_Y$ decomposition
    \beqn
    24=(1,1,0)+ (1,3,0) +(8,1,0) + (3,2,-5/3) + (\bar 3,2,5/3) 
   \label{24rep}
    \eeqn  
     Using the above  technique $(1,1,0)$ takes  the form
     \beqn
     24^i_j(1,1,0)= \sqrt{\frac{2}{15}}
      (\sum_{a=1}^{3}\delta^i_a\delta^a_j -\frac{3}{2} 
      \sum_{\alpha =4}^{5}\delta^i_{\alpha}\delta^{\alpha}_j) \sigma_{(110)}
     \eeqn 
 where $\sigma_{(110)}$ is the $SU(3)_C\times SU(2)_L\times U(1)$ singlet field 
   and the other components in Eq.(\ref{24rep}) can be similarly gotten.    
     \\ 
  
 We now give some mathematical background relevant for the group $SO(10)$.
 For reasons of computation of $SO(10)$ couplings it is found useful to 
 decompose them in the more familiar $SU(5)$ representations. We  begin
 by defining the 45 generators of $SO(10)$ in the spinor representation 
 so that 
  \begin{equation}
 \Sigma_{\mu\nu}=\frac{1}{2i}[\Gamma_\mu,\Gamma_\nu]
 \end{equation}
 where  elements  $\Gamma_{\mu}$  ($\mu=1,2,...,10$) which satisfy Clifford algebra 
\begin{equation}
\{\Gamma_{\mu},\Gamma_{\nu}\}=2\delta_{\mu\nu}.
\end{equation}
It is convenient to define $\Gamma_{\mu}$ in terms of creation and destruction operators,
$b_i$ and $b_i^{\dagger}$ ($i=1,2,...,5$)~\cite{Mohapatra:1979nn,Wilczek:1981iz} so that 
\begin{equation}
\Gamma_{2i}= (b_i+ b_i^{\dagger});~~~ \Gamma_{2i-1}= -i(b_i-
b_i^{\dagger})
\end{equation}
where
\begin{equation}
\{b_i,b_j\}=0;~~~\{b_i,b_j^{\dagger}\}=\delta_{i}^j; ~~~\{b_i^{\dagger},b_j^{\dagger}\}=0
\end{equation}
and where the $SU(5)$ singlet state  $|0>$  satisfies $b_i|0>=0$.
One can define an $SO(10)$ chirality operator  $(1\pm \Gamma_0)/2$ where 
$\Gamma_0=i^5 \Gamma_1\Gamma_2...\Gamma_{10}$  so that the 
 32 plet spinor of $SO(10)$ can be split into semi-spinors 
 $\Psi_{(\pm)\acute{a}}$ ($\acute{a}=1,2,3$ is  the generation index) which are eigen -states of
 of $SO(10)$ chirality. 
 \beqn
 \Psi_{(\pm)\acute{a}}=  \frac{1}{2}[1\pm \Gamma_0] \Psi_{\acute{a}}
 \eeqn 
 Now $\Psi_{(\pm)\acute{a}}$
transforms as a 16($\overline{16}$)-dimensional irreducible
representation of $SO(10)$. They can be expanded in $SU(5)$ decomposition 
so that  $16=1+\overline{5}+10$($\overline{16}=1+5+\overline{10}$) and are given by
\begin{equation}
|16_{\acute{a}} >=|0>{1}_{\acute{a}}+\frac{1}{2}b_i^{\dagger}b_j^{\dagger}|0>{
10}_{\acute{a}}^{ij} +\frac{1}{24}\epsilon^{ijklm}b_j^{\dagger}
b_k^{\dagger}b_l^{\dagger}b_m^{\dagger}|0>{\bar 5}_{\acute{a}i}
\end{equation}

\begin{equation}
|\overline{16}_{\acute{a}}>=b_1^{\dagger}b_2^{\dagger}b_3^{\dagger}b_4^{\dagger}b_5^{\dagger}|0>{
1'}_{\acute{a}}+\frac{1}{12}\epsilon^{ijklm}b_k^{\dagger}b_l^{\dagger}b_m^{\dagger}|0>{
10'}_{\acute{a}ij}+b_i^{\dagger}|0>{5'}_{\acute{a}}^i
\end{equation}
One generation of quarks and leptons can be identified as residing in a single $16$ plet 
representation of $SO(10)$, i.e., in the $\bar 5$ and $10$ SU(5) multiplets, while
the $SU(5)$ singlet field is a right handed neutrino which is needed in generation of
neutrino masses in a See-Saw mechanism. 
One may also define   a charge conjugation operator in $SO(10)$ by 

\begin{equation}
B=\prod_{\mu =odd}\Gamma_{\mu}= -i\prod_{k=1}^5
(b_k-b_k^{\dagger})
\end{equation}
This operator is needed in forming the $SO(10)$ invariant interactions.\\

In building models using  $SO(10)$  grand unification, one needs
the explicit decomposition of the $SO(10)$ invariant couplings in terms of
the Standard  Model fields. This task in facilitated  by decomposition of the
$SO(10)$ invariant couplings in terms of the $SU(5)$ invariant couplings, since
$SU(5)$  invariant couplings can be  easily decomposed in terms of the Standard 
Model states. The decomposition of the $SO(10)$ invariant couplings in terms of
$SU(5)$ invariant couplings can be  easily achieved  by use of the so called
Basic Theorem~\cite{Nath:2001uw,Nath:2001yj,Syed:2005gd,Syed:2004if,Nath:2005bx} which we  explain  briefly 
below.
We note that an $SO(10)$ invariant vertex can be expanded in a specific set of $SU(5)$
reducible tensors ${\Phi}_{c_k}$ and ${\Phi}_{\overline c_k}$ defined as  follows:
${\Phi}_{c_k}\equiv{\Phi}_{2k}+i{\Phi}_{2k-1},~
{\Phi}_{\overline c_k}\equiv{\Phi}_{2k}-i{\Phi}_{2k-1}$.
We can extend the above easily to define the quantity $\Phi_{c_ic_j\bar c_k..}$
which has an arbitrary number of barred and unbarred indices where each
 $c$ index is defined  so that
${\Phi}_{c_ic_j\overline
c_k...}={\Phi}_{2ic_j\overline c_k...}+i{\Phi}_{2i-1c_j\overline
 c_k...}~$etc.
 The above implies  that  the quantity  $\Phi_{c_ic_j\overline
c_k...c_N}$ is a sum of
 $2^N$ terms gotten by expanding all the c indices.
$\Phi_{c_ic_j\overline c_k...c_n}$ is completely anti-symmetric in
the interchange of its c indices whether unbarred or barred:
 ${\Phi}_{c_i\overline c_jc_k...\overline c_n}=-{\Phi}_{c_k\overline c_jc_i...\overline c_n}$.
Further, $ {\Phi}^*_{c_i\overline c_jc_k...\overline
c_n}={\Phi}_{\overline c_ic_j\overline
  c_k...c_n}$ etc.
    It is now clear  that the quantity
$\Phi_{c_ic_j\overline c_k...c_n}$ transforms like a reducible
representation of $SU(5)$.  This reducible representation can be further
decomposed into a sum of irreducible tensors. Thus the procedure is that
one first computes the $SO(10)$ invariant couplings in terms of the $SU(5)$
reducible tensors  and then decomposes them further in terms of the
the irreducible tensors. The above procedure can be summarized in terms of 
the following result in a compact form: The $SO(10)$ invariant vertex 
$\Gamma_{\mu}\Gamma_{\nu}\Gamma_
{\lambda}..\Gamma_{\sigma}$ $\Phi_{\mu\nu\lambda ..\sigma}$, where
$\Phi_{\mu\nu\lambda ..\sigma}$ is a tensor field, 
can  be expanded as follows~\cite{Nath:2001uw} 
 
\begin{eqnarray}
\Gamma_{\mu}\Gamma_{\nu}\Gamma_{\lambda}..\Gamma_{\sigma}
\Phi_{\mu\nu\lambda ...\sigma}&=& b_i^{\dagger}
b_j^{\dagger}b_k^{\dagger}...b_n^{\dagger} \Phi_{c_ic_jc_k...c_n}
+ ( b_i b_j^{\dagger}b_k^{\dagger}...b_n^{\dagger}
\Phi_{\overline c_ic_jc_k...c_n} \nonumber\\  &+& ~perms )
+\left(b_i b_jb_k^{\dagger}...b_n^{\dagger} \Phi_{\overline
c_i\overline c_jc_{k}...c_n}+~perms\right)+ ...
+  \nonumber\\ & & \left(b_ib_jb_k...b_{n-1}b_n^{\dagger}\Phi_{\overline
c_i\overline c_j
\overline c_k...\overline c_{n-1}c_n}+~perms\right)\nonumber\\
&+& b_ib_jb_k...b_n \Phi_{\overline c_i\overline c_j\overline
c_k...\overline c_n}
\end{eqnarray}
The quantity $\Phi_{c_ic_j\overline c_k...c_n}$ transforms like a
reducible representation of $SU(5)$ and can  be further decomposed
into  irreducible $SU(5)$ parts. The above technique is easily extended
to the expansion of an $SO(2N)$ vertex in terms of $SU(N)$ vertices.

With the above technique the cubic couplings in the 
superpotential involving 16-plet of 
matter and the $10$, $120$ and $\overline{126}$ of Higgs fields, 
and cubic couplings in the Lagrangian involving the 16-plet of matter fields  and the
$45$ plet of gauge fields can be computed.  We give now the explicit computations.
 For the $16-16-10$ couplings one finds the following expansion in their $SU(5)$  decomposed
 form
 \begin{eqnarray}
  W^{(10)}=(2\sqrt 2 i)f^{(+)}_{ab}(10^{ij}_a\bar 5_{ib}\bar 5_{Hj}
  -1_{a}\bar 5_{ib}5_H^i + \frac{1}{8}\epsilon_{ijklm}10^{ij}_a10^{kl}_b 5_H^m)
\label{161610}
\end{eqnarray} 
where the 10-plet of $SO(10)$  Higgs fields is decomposed in $SU(5)$ representations so that
$10_H=5_H+\bar 5_H$. In Eq.(\ref{161610}) the Higgs fields are identified  with
the subscript H while the remaining fields are the matter fields. 
 In analyzing the 16-16-120 coupling in the superpotential in terms of 
 $SU(5)$ representations we note that the 120-plet representation can be  decomposed
 in $SU(5)$ representations as follows: $120=5+\bar 5$ + $10+\overline{10}$+$45 +\overline{45}$.
 Thus  one has 
 
 \begin{eqnarray}
W^{(120)}=i\frac{2}{\sqrt 3}f_{\acute{a}\acute{b}}^{(-)} [
2 (1_{\acute{a}} 5_{i\acute{b}}5_H^i) +
10^{ij}_{\acute {a}}1_{\acute{b}}10_{Hij}
+5_{i\acute a}5_{j\acute b}10_H^{ij}\nonumber\\
-10^{ij}_{\acute a}\bar 5_{i\acute b}\bar 5_{Hj}+  \bar 5_{i\acute a}10_{\acute b}^{jk}\bar 45^i_{Hjk}
-\frac{1}{4}\epsilon_{ijklm}10_{\acute a}^{ij}10_{\acute b}^{mn}45^{kl}_{Hn} ]  
\end{eqnarray}
 where the fields with subscripts H are the Higgs fields in $SU(5)$ representations. 
 In decomposing the vertex involving the $\overline{126}$ coupling  we  note that 
 the $\overline{126}$ and ${126}$ have the    $SU(5)$   decompositions: 
 $\overline{126}= 1+5  +\overline{10} + 15 +\overline{45} + 50$  while 
${126}= 1+\bar 5 +{10} +\overline{15} + {45} + \overline{50}$. 
 The $16-16-\overline{126}$ vertex can be expanded as follows
 \begin{eqnarray}
W^{(\bar{126})}&=& if_{\acute{a}\acute{b}}^{(+)} {\frac{\sqrt 2}{\sqrt {15}}[-\sqrt 2}
(1_{\acute a} 1_{\acute b}1_H) -\sqrt 3  (1_{\acute a} \bar 5_{i\acute b}5^i_H) 
+1_{\acute a}10^{ij}_{\acute b}10_{Hij}\nonumber\\
&-& \frac{1}{8\sqrt 3}10_{\acute a}^{ij}10_{\acute b}^{kl}5^m_H\epsilon_{ijklm} 
-\bar 5_{i\acute a}\bar 5_{j\acute b} 15^{ij}_{HS} 
+10^{ij}_{\acute a}\bar 5_{\acute{b}k}\overline{45}^k_{Hij} \nonumber\\
&-& \frac{1}{12\sqrt 2}\epsilon_{ijklm}10_{\acute a}^{lm}10_{\acute b}^{rs}  50^{ijk}_{Hrs}]
\end{eqnarray}
where again the Higgs fields have been displayed with a subscript H while the other fields 
 are matter fields. 
 
The SO(10) gauge invariant couplings involve  the couplings of the 45 plet of gauge  vector
 bosons with 16-plet of matter. The supersymmetric Yang-Mills part of the Lagrangian 
 in superfield notation is 
\begin{equation}
 \int d^2\theta~tr(W^{\alpha}W_{\alpha}))
+ \int d^2\bar{\theta}~
tr(\overline{W}_{\dot{\alpha}}\overline{W}^{\dot{\alpha}})  
\end{equation}
 where  $W_{\alpha}$ is the field strength
 chiral spinor superfield. Since we are interested in dimension six fermion operators
 arising from these interactions, such interactions  arise only from the elimination
 of gauge vector bosons. Thus we exhibit only the gauge vector boson interactions 
 of the 45 gauge vectors  $V_{A\mu\nu}$  where A is a Lorentz index (A=0,1-3)
\begin{equation}
{\mathsf L}^{(45)}=\frac{1}{i}\frac{1}{2!}g^{^{(45)}}_{\acute{a}\acute{b}}
<\Psi_{(+)\acute a}|\gamma^0\gamma^A
\Sigma_{\mu\nu}|\Psi_{(+)\acute b}>V_{A\mu\nu}
\end{equation}
Here $\gamma^A $ spans the Clifford algebra
associated with the Lorentz group,  and $g$ is the gauge coupling constant.
Now in $SU(5)$ decomposition the 45 plet of SO(10) can be decomposed as follows
\begin{equation}
45= 1+ 10 + \overline{10} + 24
\end{equation}
We exhibit the $16-\overline{16}-45$ couplings in the decomposed  form 

\begin{eqnarray}
{\mathsf L}^{(45)}&=&g^{^{(45)}}_{\acute{a}\acute{b}}[\sqrt 5\left(-\frac{3}{5}
\overline{\bar 5_{\acute a}}^i\gamma^A \bar 5_{\acute{b}i}+
\frac{1}{10}\overline{10}_{\acute{a}ij}
\gamma^A 10_{\acute b}^{ij}+
\overline{1_{\acute a}}\gamma^A 1_{\acute b}\right){\mathsf V}_{A} \nonumber\\
&+&{\frac{1} {\sqrt 2}}\left(\overline {1}_{\acute a} \gamma^{A} {10}_{\acute b}^{lm}
+{\frac{1}{2}}\epsilon^{ijklm}\overline {10}_{\acute{a}ij}\gamma^A{\bar 5}_{\acute{b}k}\right) 
{\mathsf V}_{Alm}\nonumber\\
&-&\frac{1}{\sqrt 2}\left(
{\overline{10}}_{\acute{a}lm}\gamma^A{1}_{\acute b} 
+ \frac{1}{2}\epsilon_{ijklm}
 {\overline{\bar 5}}_{\acute a}^i\gamma^A{10}_{\acute b}^{jk}\right){\mathsf
V}_A^{lm} \nonumber\\
&+&\sqrt{2}\left(\overline {10}_{\acute{a}ik}\gamma^A {10}_{\acute b}^{kj}+
{\overline{\bar 5}}_{\acute a}^j\gamma^A{\bar 5}_{\acute{b}i}\right){\mathsf V}_{Aj}^i]. 
\end{eqnarray}
where $V_A$, $V_A^{ij}$, $V_{Aij}$, $V^i_{Aj}$ are the 1, 10, 
$\overline{10}$, and 24 plets of SU(5). The same technique can be used to compute
the interactions involving Higgs fields lying in  representations  $10$, $45$, $54$, $120$, 
$\overline{126}$,  $210$ (For later works using different techniques, 
see~\cite{Aulakh:2002zr,Fukuyama:2004ps,Aulakh:2004hm,Bajc:2004xe}).

We discuss now briefly the vector-spinor $\overline{144} (144)$ which requires  special
care~\cite{Nath:2005bx}. The reason for this is that the $\overline{144} (144)$ arise  via a constraint
on the reducible vector spinor $\overline{160} (160)$. Thus the 
 ${\overline {160}}$ vector-spinors has an  expansion  in $SU(5)$ oscillator modes so that:
\begin{equation}\label{bar160spinor}
|\Psi_{(+)\acute{a}\mu}>=|0>{\bf
P}_{\acute{a}\mu}+\frac{1}{2}b_i^{\dagger}b_j^{\dagger}|0>{\bf
P}_{\acute{a}\mu}^{ij} +\frac{1}{24}\epsilon^{ijklm}b_j^{\dagger}
b_k^{\dagger}b_l^{\dagger}b_m^{\dagger}|0>{\bf P}_{\acute{a}i\mu}
\end{equation}
while the 
${\overline {160}}$ vector-spinor has an expansion in $SU(5)$ oscillator modes so that:

\begin{equation}\label{160spinor}
|\Psi_{(-)\acute{b}\mu}>=b_1^{\dagger}b_2^{\dagger}
b_3^{\dagger}b_4^{\dagger}b_5^{\dagger}|0>{\bf Q}_{\acute{b}\mu}
+\frac{1}{12}\epsilon^{ijklm}b_k^{\dagger}b_l^{\dagger}
b_m^{\dagger}|0>{\bf Q}_{\acute{b}ij\mu}+b_i^{\dagger}|0>{\bf
Q}_{\acute{b}\mu}^i
\end{equation}
where  $i,j,k,l,m,...=1,2,...,5$ are
$SU(5)$ indices, 
$\mu,\nu,\rho,...=1,2,...,10$ are $SO(10)$ indices, while $\acute{a}, \acute{b},
\acute{c}, \acute{d}=1,2,3$ are generation indices. 
The $SU(5)$
field content of $\overline{160}$ multiplet is
\begin{eqnarray}
\label{su5decompositionofbar160,160}
\overline{160} (\Psi_{(+)\mu})&=& 1({\bf {\widehat P}})+\bar 5({\bf
P}_{i})+5 ({\bf P}^i)+5({\bf {\widehat P}}^i)+{\overline
{10}}({\bf P}_{ij}) +{\overline {10}}({\bf {\widehat
P}}_{ij}) \nonumber\\
&+& {\overline {15}}({\bf P}_{ij}^{(S)})
+24 ({\bf P}^i_j)+{\overline {40}}({\bf { P}}_{jkl}^i)+45 ({\bf
P}^{ij}_k)
\end{eqnarray}
while the $SU(5)$
field content of $\overline{160}$ multiplet is

\begin{eqnarray}
160 (\Psi_{(-)\mu})&=& 1({\bf {\widehat Q}})+5({\bf Q}^{i})+\bar 5
({\bf Q}_i)+\bar 5({\bf {\widehat Q}}_i)+10({\bf Q}^{ij}) +10({\bf
{\widehat
Q}}^{ij}) \nonumber\\
&+& 15({\bf Q}^{ij}_{(S)})
+24 ({\bf Q}^i_j)+40({\bf { Q}}^{ijk}_l)+{\overline {45}} ({\bf
Q}_{jk}^i)
\end{eqnarray}
To get the $\overline{144}144$ multiplets these must be  subject to the
constraint
 \beqn \Gamma_{\mu} |{\Upsilon}_{(\pm)\mu}>=0
\label{144constraint} \eeqn
Imposing these  constraints on the $\overline {160}$ multiplet gives

\begin{eqnarray}
\Gamma_{\mu}|\Psi_{(+)\mu}>&=& b_1^{\dagger}b_2^{\dagger}
b_3^{\dagger}b_4^{\dagger}b_5^{\dagger}|0>{\widehat {\bf P}}
+\frac{1}{12}\epsilon^{ijklm}b_k^{\dagger}b_l^{\dagger}
b_m^{\dagger}|0>\left({\bf P}_{ij}+6{\widehat {\bf
P}}_{ij}\right) \nonumber\\
&+& b_i^{\dagger}|0>\left({\bf P}^{i}+{\widehat {\bf P}}^{i}\right)
\end{eqnarray}
Thus to get the $\overline {144}$ spinor,
$|\Upsilon_{(+)\mu}>$ the following constraints  must  be imposed
on the components  in $|\Psi_{(+)\mu}>$
\begin{eqnarray}\label{constraints}
{\widehat {\bf P}}=0,~~~{\widehat {\bf P}}^i=-{\bf
P}^i,~~~{\widehat
{\bf P}}_{ij}=-\frac{1}{6}{\bf P}_{ij}
\end{eqnarray}
Similarly to reduce the 160-plet $|\Upsilon_{(-)\mu}>$
to $144$-plet we need to impose the constraints  

\begin{eqnarray}
{\widehat {\bf Q}}=0,~~~{\widehat {\bf Q}}_i=-{\bf
Q}_i,~~~{\widehat {\bf Q}}^{ij}=-\frac{1}{6}{\bf Q}^{ij}
\end{eqnarray}

The $SO(10)$ invariant cubic couplings in the superpotential
 involving two  vector-spinors
and  the tensors 1, 10, 45, 120, 210, and ${\overline
{126}}$ plet of Higgs etc can be written analogous to the 
couplings of the 16-plet spinor. We display the couplings for the
1,45, 10 tensors. The couplings for these are

\begin{eqnarray}
{\mathsf
W}^{(1)}=h^{^{(1)}}_{\acute{a}\acute{b}}<\Upsilon^{*}_{(-)\acute{a}\mu}|B|\Upsilon_{(+)\acute{b}\mu}>\Phi,
\nonumber\\
{\mathsf
W}^{(10)}=h^{^{(10)}}_{\acute{a}\acute{b}}<\Upsilon^{*}_{(+)\acute{a}\mu}|B\Gamma_{\nu}|\Upsilon_{(+)\acute{b}\mu}>
\Phi_{\nu},
\nonumber\\
{\mathsf
W}^{(45)}=\frac{1}{2!}h^{^{(45)}}_{\acute{a}\acute{b}}<\Upsilon^{*}_{(-)\acute{a}\mu}|B\Sigma_{\rho\sigma}|\Upsilon_{(+)\acute{b}
\mu}>\Phi_{\rho\sigma}.
\end{eqnarray}
Here $\Phi$ is the 1-plet,  $\Phi_{\nu}$  is the 10 plet, and $\Phi_{\rho\sigma}$ is the 45-plet
Higgs field. A detailed computation of these
and other couplings is  given in Ref.~\cite{Nath:2005bx}.

\section{$d=5$ contributions to the decay of the proton}
In this Appendix we present the complete set of diagrams
responsible for $d=5$ nucleon decay in supersymmetric scenarios. 
In this case proton decay is mediated by scalar leptoquarks 
and their superpartners. The relevant interactions for proton 
decay are the following:
\begin{eqnarray}
W &=& \hat{Q} \ \underline{A} \ \hat{Q} \ \hat{T} \ + 
\ \hat{U}^C \ \underline{B} \ \hat{E}^C \ \hat{T} \ \\
& + & \ \hat{Q} \ \underline{C} \ L \ \hat{\overline{T}} \ + \ 
\hat{U}^C \ \underline{D} \ \hat{D}^C \  \hat{\overline{T}} \ + \ M_T \hat{T} \hat{\overline{T}}
\end{eqnarray}
where we use the conventional notation for all MSSM superfields. 
The superfields $\hat{T}$, and $\hat{\overline{T}}$ transform 
as $(\bf{3},1,-2/3)$, and $(\overline{\bf{3}},1,2/3)$, 
respectively~\cite{Bajc:2002bv}.

\vskip 0.5cm
\noindent
\begin{center}
{\bf \underline{ Decay Channels:}}
\end{center}
\vspace{0.2cm}
\begin{center}
$p \ \to \ (K^+,\pi^+,\rho^+) \ \bar\nu_i$, and 
$n \ \to \ (\pi^0,\rho^0,\eta,\omega,K^0) \ \bar\nu_i$, 
with $i=1,2,3$. 
\end{center}

\begin{equation}
\label{nw1}
\def\trgor{$\tilde T$}
\def\trdol{$\tilde{\bar T}$}
\def\sfgor{$\tilde t$}
\def\sfdol{$\tilde\tau$}
\def\spgor{$\tilde w^+$}
\def\spdol{$\tilde w^-$}
\def\iena{$d_{1,2}$}
\def\idva{$u$}
\def\itri{$\nu_i$}
\def\isti{$d_{2,1}$}
\input{box}
\propto\;\;\;\;\;
(D^T\underline A^S\tilde U)_{13,23}(\tilde U^\dagger D)_{32,31}
(N^T\tilde E^*)_{i3}(\tilde E^T\underline C^TU)_{31}
\end{equation}

\begin{equation}
\def\tr{$T$}
\def\sfgor{$\tilde t$}
\def\sfdol{$\tilde\tau$}
\def\spgor{$\tilde w^+$}
\def\spdol{$\tilde w^-$}
\def\iena{$d_{1,2}$}
\def\idva{$u$}
\def\itri{$\nu_i$}
\def\isti{$d_{2,1}$}
\input{triangle}
\propto\;\;\;\;\;
(D^T\underline A^S U)_{11,21}(N^T\tilde E^*)_{i3}
(\tilde E^T\underline C^T\tilde U)_{33}(\tilde U^\dagger D)_{32,31}
\end{equation}

\begin{equation}
\def\trgor{$\tilde T$}
\def\trdol{$\tilde{\bar T}$}
\def\sfgor{$\tilde t$}
\def\sfdol{$\tilde b$}
\def\spgor{$\tilde w^+$}
\def\spdol{$\tilde w^-$}
\def\iena{$d_{1,2}$}
\def\idva{$\nu_i$}
\def\itri{$u$}
\def\isti{$d_{2,1}$}
\input{box}
\propto\;\;\;\;\;
(D^T\underline A^S \tilde U)_{13,23}(\tilde U^\dagger D)_{32,31}
(U^T\tilde D^*)_{13}(\tilde D^T\underline CN)_{3i}
\end{equation}

\begin{equation}
\def\tr{$\bar T$}
\def\sfgor{$\tilde t$}
\def\sfdol{$\tilde b$}
\def\spgor{$\tilde w^+$}
\def\spdol{$\tilde w^-$}
\def\iena{$d_{1,2}$}
\def\idva{$\nu_i$}
\def\itri{$u$}
\def\isti{$d_{2,1}$}
\input{triangle}
\propto\;\;\;\;\;
(D^T\underline CN)_{1i,2i}(U^T\tilde D^*)_{13}
(\tilde D^T\underline A^S \tilde U)_{33}(\tilde U^\dagger D)_{32,31}
\end{equation}

\begin{equation}
\def\trgor{$\tilde T$}
\def\trdol{$\tilde{\bar T}$}
\def\sfgor{$\tilde t$}
\def\sfdol{$\tilde b$}
\def\spgor{$\tilde{\bar h}_-^\dagger$}
\def\spdol{$\tilde h_+^\dagger$}
\def\iena{$d_{1,2}$}
\def\idva{$\nu_i$}
\def\itri{${\bar u}^c$}
\def\isti{${\bar d}_{2,1}^c$}
\input{box}
\propto\;\;\;\;\;
(D^T\underline A^S \tilde U)_{13,23}(\tilde U^\dagger Y_D^*D_c^*)_{32,31}
(U_c^\dagger Y_U^\dagger\tilde D^*)_{13}(\tilde D^T\underline CN)_{3i}
\end{equation}

\begin{equation}
\label{nh2}
\def\tr{$\bar T$}
\def\sfgor{$\tilde t$}
\def\sfdol{$\tilde b$}
\def\spgor{$\tilde{\bar h}_-^\dagger$}
\def\spdol{$\tilde h_+^\dagger$}
\def\iena{$d_{1,2}$}
\def\idva{$\nu_i$}
\def\itri{${\bar u}^c$}
\def\isti{${\bar d}_{2,1}^c$}
\input{triangle}
\propto\;\;\;\;\;
(D^T\underline CN)_{1i,2i}(U_c^\dagger Y_U^\dagger\tilde D^*)_{13}
(\tilde D^T\underline A^S \tilde U)_{33}(\tilde U^\dagger Y_D^*D_c^*)_{32,31}
\end{equation}

\begin{equation}
\label{nh3}
\def\trgor{$\tilde T^\dagger$}
\def\trdol{$\tilde{\bar T}^\dagger$}
\def\sfgor{$\tilde\tau^c$}
\def\sfdol{$\tilde t^c$}
\def\spgor{$\tilde{\bar h}_-$}
\def\spdol{$\tilde h_+$}
\def\iena{${\bar u}^c$}
\def\idva{${\bar d}_{2,1}^c$}
\def\itri{$d_{1,2}$}
\def\isti{$\nu_i$}
\input{box}
\propto\;\;\;\;\;
(U_c^\dagger\underline B^*\tilde E_c^*)_{13}(\tilde E_c^TY_EN)_{3i}
(D^TY_U\tilde U_c)_{13,23}(\tilde U_c^\dagger\underline D^*D_c^*)_{32,31}
\end{equation}

\begin{equation}
\label{nh4}
\def\tr{$\bar T$}
\def\sfgor{$\tilde\tau^c$}
\def\sfdol{$\tilde t^c$}
\def\spgor{$\tilde{\bar h}_-$}
\def\spdol{$\tilde h_+$}
\def\iena{${\bar u}^c$}
\def\idva{${\bar d}_{1,2}^c$}
\def\itri{$d_{2,1}$}
\def\isti{$\nu_i$}
\input{triangle}
\propto\;\;\;\;\;
(U_c^\dagger\underline D^* D_c^*)_{11,12}(D^TY_U\tilde U_c)_{23,13}
(\tilde U_c^\dagger\underline B^*\tilde E_c^*)_{33}(\tilde E_c^TY_EN)_{3i}
\end{equation}

\begin{equation}
\label{nh01}
\def\trgor{$\tilde T$}
\def\trdol{$\tilde{\bar T}$}
\def\sfgor{$\tilde t$}
\def\sfdol{$\tilde b$}
\def\spgor{$\tilde h_0^\dagger$}
\def\spdol{$\tilde{\bar h}_0^\dagger$}
\def\iena{$d_{1,2}$}
\def\idva{$\nu_i$}
\def\itri{${\bar d}_{2,1}^c$}
\def\isti{${\bar u}^c$}
\input{box}
\propto\;\;\;\;\;
(D^T\underline A^S \tilde U)_{13,23}(\tilde U^\dagger Y_U^*U_c^*)_{31}
(D_c^\dagger Y_D^\dagger\tilde D^*)_{23,13}(\tilde D^T\underline CN)_{3i}
\end{equation}

\begin{equation}
\def\tr{$\bar T$}
\def\sfgor{$\tilde b$}
\def\sfdol{$\tilde t$}
\def\spgor{$\tilde{\bar h}_0^\dagger$}
\def\spdol{$\tilde h_0^\dagger$}
\def\iena{$d_{1,2}$}
\def\idva{$\nu_i$}
\def\itri{${\bar u}^c$}
\def\isti{${\bar d}_{2,1}^c$}
\input{triangle}
\propto\;\;\;\;\;
(D^T\underline CN)_{1i,2i}(U_c^\dagger Y_U^\dagger\tilde U^*)_{13}
(\tilde U^T\underline A^S \tilde D)_{33}(\tilde D^\dagger Y_D^*D_c^*)_{32,31}
\end{equation}

\begin{equation}
\def\trgor{$\tilde T$}
\def\trdol{$\tilde{\bar T}$}
\def\sfgor{$\tilde t$}
\def\sfdol{$\tilde b$}
\def\spgor{$\tilde V_0$}
\def\spdol{$\tilde V_0$}
\def\iena{$d_{1,2}$}
\def\idva{$\nu_i$}
\def\itri{$d_{2,1}$}
\def\isti{$u$}
\input{box}
\propto\;\;\;\;\;
(D^T\underline A^S \tilde U)_{13,23}(\tilde U^\dagger U)_{31}
(D^T\tilde D^*)_{23,13}(\tilde D^T\underline CN)_{3i}
\end{equation}

\begin{equation}
\def\tr{$\bar T$}
\def\sfgor{$\tilde t$}
\def\sfdol{$\tilde b$}
\def\spgor{$\tilde V_0$}
\def\spdol{$\tilde V_0$}
\def\iena{$d_{1,2}$}
\def\idva{$\nu_i$}
\def\itri{$d_{2,1}$}
\def\isti{$u$}
\input{triangle}
\propto\;\;\;\;\;
(D^T\underline CN)_{1i,2i}(D^T\tilde D^*)_{23,13}
(\tilde D^T\underline A^S \tilde U)_{33}(\tilde U^\dagger U)_{31}
\end{equation}

\begin{equation}
\def\trgor{$\tilde{\bar T}$}
\def\trdol{$\tilde T$}
\def\sfgor{$\tilde\nu$}
\def\sfdol{$\tilde b$}
\def\spgor{$\tilde V_0$}
\def\spdol{$\tilde V_0$}
\def\iena{$d_{1,2}$}
\def\idva{$u$}
\def\itri{$d_{2,1}$}
\def\isti{$\nu_i$}
\input{box}
\propto\;\;\;\;\;
(D^T\underline C\tilde N)_{13,23}(\tilde N^\dagger N)_{3i}
(D^T\tilde D^*)_{23,13}(\tilde D^T\underline A^S U)_{31}
\end{equation}

\begin{equation}
\label{nv4}
\def\tr{$T$}
\def\sfgor{$\tilde\nu$}
\def\sfdol{$\tilde b$}
\def\spgor{$\tilde V_0$}
\def\spdol{$\tilde V_0$}
\def\iena{$u$}
\def\idva{$d_{1,2}$}
\def\itri{$d_{2,1}$}
\def\isti{$\nu_i$}
\input{triangle}
\propto\;\;\;\;\;
(U^T\underline A^S D)_{11,12}(D^T\tilde D^*)_{23,13}
(\tilde D^T\underline C\tilde N)_{33}(\tilde N^\dagger N)_{3i}
\end{equation}

\begin{equation}
\label{nv5}
\def\trgor{$\tilde{\bar T}$}
\def\trdol{$\tilde T$}
\def\sfgor{$\tilde\nu$}
\def\sfdol{$\tilde t$}
\def\spgor{$\tilde V_0$}
\def\spdol{$\tilde V_0$}
\def\iena{$d_{1,2}$}
\def\idva{$d_{2,1}$}
\def\itri{$u$}
\def\isti{$\nu_i$}
\input{box}
\propto\;\;\;\;\;
(D^T\underline C\tilde N)_{13,23}(\tilde N^\dagger N)_{3i}
(U^T\tilde U^*)_{13}(\tilde U^T\underline A^S D)_{32,31}
\end{equation}

\vskip 1cm

\noindent

\begin{center}
{\bf \underline{Decay Channels:}}
\end{center}

\vspace{0.5cm}

\begin{center}
$p \ \to \ (K^0,\pi^0,\eta,\rho^0,\omega) \ e_i^+$, and 
$n \ \to \ (K^-,\pi^-,\rho^-) \ e_i^+$ 
with $i=1,2$.
\end{center}

\begin{equation}
\def\trgor{$\tilde{\bar T}$}
\def\trdol{$\tilde T$}
\def\sfgor{$\tilde\nu$}
\def\sfdol{$\tilde b$}
\def\spgor{$\tilde w^+$}
\def\spdol{$\tilde w^-$}
\def\iena{$d_{1,2}$}
\def\idva{$u$}
\def\itri{$u$}
\def\isti{$e_i$}
\input{box}
\propto\;\;\;\;\;
(D^T\underline C\tilde N)_{13,23}(\tilde N^\dagger E)_{3i}
(U^T\tilde D^*)_{13}(\tilde D^T\underline A^S U)_{31}
\end{equation}

\begin{equation}
\def\tr{$T$}
\def\sfgor{$\tilde\nu$}
\def\sfdol{$\tilde b$}
\def\spgor{$\tilde w^+$}
\def\spdol{$\tilde w^-$}
\def\iena{$d_{1,2}$}
\def\idva{$u$}
\def\itri{$u$}
\def\isti{$e_i$}
\input{triangle}
\propto\;\;\;\;\;
(D^T\underline A^S U)_{11,21}(U^T\tilde D^*)_{13}
(\tilde D^T\underline C\tilde N)_{33}(\tilde N^\dagger E)_{3i}
\end{equation}

\begin{equation}
\def\trgor{$\tilde T$}
\def\trdol{$\tilde{\bar T}$}
\def\sfgor{$\tilde b$}
\def\sfdol{$\tilde t$}
\def\spgor{$\tilde w^-$}
\def\spdol{$\tilde w^+$}
\def\iena{$u$}
\def\idva{$e_i$}
\def\itri{$d_{1,2}$}
\def\isti{$u$}
\input{box}
\propto\;\;\;\;\;
(U^T\underline A^S \tilde D)_{13}(\tilde D^\dagger U)_{31}
(D^T\tilde U^*)_{13,23}(\tilde U^T\underline CE)_{3i}
\end{equation}

\begin{equation}
\def\tr{$\bar T$}
\def\sfgor{$\tilde b$}
\def\sfdol{$\tilde t$}
\def\spgor{$\tilde w^-$}
\def\spdol{$\tilde w^+$}
\def\iena{$u$}
\def\idva{$e_i$}
\def\itri{$d_{1,2}$}
\def\isti{$u$}
\input{triangle}
\propto\;\;\;\;\;
(U^T\underline CE)_{1i}(D^T\tilde U^*)_{13,23}
(\tilde U^T\underline A^S \tilde D)_{33}(\tilde D^\dagger U)_{31}
\end{equation}

\begin{equation}
\def\trgor{$\tilde{\bar T}^\dagger$}
\def\trdol{$\tilde T^\dagger$}
\def\sfgor{$\tilde t^c$}
\def\sfdol{$\tilde b^c$}
\def\spgor{$\tilde h_+$}
\def\spdol{$\tilde{\bar h}_-$}
\def\iena{${\bar e}_i^c$}
\def\idva{${\bar u}^c$}
\def\itri{$u$}
\def\isti{$d_{1,2}$}
\input{box}
\propto\;\;\;\;\;
(E_c^\dagger\underline B^\dagger\tilde U_c^*)_{i3}
(\tilde U_c^TY_U^TD)_{31,32}(U^TY_D\tilde D_c)_{13}
(\tilde D_c^\dagger\underline D^\dagger U_c^*)_{31}
\end{equation}

\begin{equation}
\def\tr{$T$}
\def\sfgor{$\tilde t^c$}
\def\sfdol{$\tilde b^c$}
\def\spgor{$\tilde h_+$}
\def\spdol{$\tilde{\bar h}_-$}
\def\iena{${\bar e}_i^c$}
\def\idva{${\bar u}^c$}
\def\itri{$u$}
\def\isti{$d_{1,2}$}
\input{triangle}
\propto\;\;\;\;\;
(E_c^\dagger\underline B^\dagger U_c^*)_{i1}(U^TY_D\tilde D_c)_{13}
(\tilde D_c^\dagger\underline D^\dagger\tilde U_c^*)_{33}
(\tilde U_c^TY_U^TD)_{31,32}
\end{equation}

\begin{equation}
\def\trgor{$\tilde T$}
\def\trdol{$\tilde{\bar T}$}
\def\sfgor{$\tilde b$}
\def\sfdol{$\tilde t$}
\def\spgor{$\tilde h_+^\dagger$}
\def\spdol{$\tilde{\bar h}_-^\dagger$}
\def\iena{$u$}
\def\idva{$e_i$}
\def\itri{${\bar d}_{1,2}^c$}
\def\isti{${\bar u}^c$}
\input{box}
\propto\;\;\;\;\;
(U^T\underline A^S \tilde D)_{13}(\tilde D^\dagger Y_U^*U_c^*)_{31}
(D_c^\dagger Y_D^\dagger\tilde U^*)_{13,23}(\tilde U^T\underline CE)_{3i}
\end{equation}

\begin{equation}
\def\tr{$\bar T$}
\def\sfgor{$\tilde b$}
\def\sfdol{$\tilde t$}
\def\spgor{$\tilde h_+^\dagger$}
\def\spdol{$\tilde{\bar h}_-^\dagger$}
\def\iena{$u$}
\def\idva{$e_i$}
\def\itri{${\bar d}_{1,2}^c$}
\def\isti{${\bar u}^c$}
\input{triangle}
\propto\;\;\;\;\;
(U^T\underline CE)_{1i}(D_c^\dagger Y_D^\dagger\tilde U^*)_{13,23}
(\tilde U^T\underline A^S \tilde D)_{33}(\tilde D^\dagger Y_U^*U_c^*)_{31}
\end{equation}

\begin{equation}
\def\trgor{$\tilde{\bar T}$}
\def\trdol{$\tilde T$}
\def\sfgor{$\tilde\nu$}
\def\sfdol{$\tilde b$}
\def\spgor{$\tilde{\bar h}_-^\dagger$}
\def\spdol{$\tilde h_+^\dagger$}
\def\iena{$d_{1,2}$}
\def\idva{$u$}
\def\itri{${\bar u}^c$}
\def\isti{${\bar e}_i^c$}
\input{box}
\propto\;\;\;\;\;
(D^T\underline C\tilde N)_{13,23}(\tilde N^\dagger Y_E^\dagger E_c^*)_{3i}
(U_c^\dagger Y_U^\dagger\tilde D^*)_{13}(\tilde D^T\underline A^S U)_{31}
\end{equation}

\begin{equation}
\def\tr{$T$}
\def\sfgor{$\tilde\nu$}
\def\sfdol{$\tilde b$}
\def\spgor{$\tilde{\bar h}_-^\dagger$}
\def\spdol{$\tilde h_+^\dagger$}
\def\iena{$d_{1,2}$}
\def\idva{$u$}
\def\itri{${\bar u}^c$}
\def\isti{${\bar e}_i^c$}
\input{triangle}
\propto\;\;\;\;\;
(D^T\underline A^S U)_{11,21}(U_c^\dagger Y_U^\dagger\tilde
D^*)_{13}(\tilde D^T\underline C\tilde N)_{33}
(\tilde N^\dagger Y_E^\dagger E_c^*)_{3i}
\end{equation}

\begin{equation}
\def\trgor{$\tilde T$}
\def\trdol{$\tilde{\bar T}$}
\def\sfgor{$\tilde b$}
\def\sfdol{$\tilde t$}
\def\spgor{$\tilde{\bar h}_0^\dagger$}
\def\spdol{$\tilde h_0^\dagger$}
\def\iena{$u$}
\def\idva{$e_i$}
\def\itri{${\bar u}^c$}
\def\isti{${\bar d}_{1,2}^c$}
\input{box}
\propto\;\;\;\;\;
(U^T\underline A^S \tilde D)_{13}(\tilde D^\dagger Y_D^* D_c^*)_{31,32}
(U_c^\dagger Y_U^\dagger\tilde U^*)_{13}(\tilde U^T\underline CE)_{3i}
\end{equation}

\begin{equation}
\def\tr{$\bar T$}
\def\sfgor{$\tilde t$}
\def\sfdol{$\tilde b$}
\def\spgor{$\tilde h_0^\dagger$}
\def\spdol{$\tilde{\bar h}_0^\dagger$}
\def\iena{$u$}
\def\idva{$e_i$}
\def\itri{${\bar d}_{1,2}^c$}
\def\isti{${\bar u}^c$}
\input{triangle}
\propto\;\;\;\;\;
(U^T\underline CE)_{1i}(D_c^\dagger Y_D^\dagger\tilde D^*)_{13,23}
(\tilde D^T\underline A^S \tilde U)_{33}(\tilde U^\dagger Y_U^*U_c^*)_{31}
\end{equation}

\begin{equation}
\def\trgor{$\tilde{\bar T}$}
\def\trdol{$\tilde T$}
\def\sfgor{$\tilde t^c$}
\def\sfdol{$\tilde\tau^c$}
\def\spgor{$\tilde h_0$}
\def\spdol{$\tilde{\bar h}_0$}
\def\iena{${\bar d}_{1,2}^c$}
\def\idva{${\bar u}^c$}
\def\itri{${\bar e}_i$}
\def\isti{$u$}
\input{box}
\propto\;\;\;\;\;
(D_c^\dagger\underline D^\dagger\tilde U_c^*)_{13,23}
(\tilde U_c^TY_U^TU)_{31}(E^TY_E^T\tilde E_c)_{i3}
(\tilde E_c^\dagger\underline B^\dagger U_c^*)_{31}
\end{equation}

\begin{equation}
\def\tr{$\bar T$}
\def\sfgor{$\tilde t^c$}
\def\sfdol{$\tilde\tau^c$}
\def\spgor{$\tilde h_0$}
\def\spdol{$\tilde{\bar h}_0$}
\def\iena{${\bar d}_{1,2}^c$}
\def\idva{${\bar u}^c$}
\def\itri{${\bar e}_i$}
\def\isti{$u$}
\input{triangle}
\propto\;\;\;\;\;
(D_c^\dagger\underline D^\dagger U_c^*)_{11,21}
(E^TY_E^T\tilde E_c)_{i3}(\tilde E_c^\dagger\underline B^\dagger
\tilde U_c^*)_{33}(\tilde U_c^TY_U^TU)_{31}
\end{equation}

\begin{equation}
\def\trgor{$\tilde{\bar T}$}
\def\trdol{$\tilde T$}
\def\sfgor{$\tilde b^c$}
\def\sfdol{$\tilde t^c$}
\def\spgor{$\tilde{\bar h}_0$}
\def\spdol{$\tilde h_0$}
\def\iena{${\bar u}^c$}
\def\idva{${\bar e}_i^c$}
\def\itri{$u$}
\def\isti{$d_{1,2}$}
\input{box}
\propto\;\;\;\;\;
(U_c^\dagger\underline D^*\tilde D_c^*)_{13}
(\tilde D_c^TY_D^TD)_{31,32}(U^TY_U\tilde U_c)_{13}
(\tilde U_c^\dagger\underline B^* E_c^*)_{3i}
\end{equation}

\begin{equation}
\def\tr{$T$}
\def\sfgor{$\tilde t^c$}
\def\sfdol{$\tilde b^c$}
\def\spgor{$\tilde h_0$}
\def\spdol{$\tilde{\bar h}_0$}
\def\iena{${\bar e}_i^c$}
\def\idva{${\bar u}^c$}
\def\itri{$d_{1,2}$}
\def\isti{$u$}
\input{triangle}
\propto\;\;\;\;\;
(E_c^\dagger\underline B^\dagger U_c^*)_{i1}(D^TY_D\tilde D_c)_{13,23}
(\tilde D_c^\dagger\underline D^\dagger\tilde U_c^*)_{33}
(\tilde U_c^TY_U^TU)_{31}
\end{equation}

\begin{equation}
\def\trgor{$\tilde{\bar T}$}
\def\trdol{$\tilde T$}
\def\sfgor{$\tilde t$}
\def\sfdol{$\tilde\tau$}
\def\spgor{$\tilde h_0^\dagger$}
\def\spdol{$\tilde{\bar h}_0^\dagger$}
\def\iena{$d_{1,2}$}
\def\idva{$u$}
\def\itri{${\bar e}_i^c$}
\def\isti{${\bar u}^c$}
\input{box}
\propto\;\;\;\;\;
(D^T\underline A^S \tilde U)_{13,23}
(\tilde U^\dagger Y_U^*U_c^*)_{31}(E_c^\dagger Y_E^*\tilde E^*)_{i3}
(\tilde E^T\underline C^T U)_{31}
\end{equation}

\begin{equation}
\def\tr{$T$}
\def\sfgor{$\tilde\tau$}
\def\sfdol{$\tilde t$}
\def\spgor{$\tilde{\bar h}_0^\dagger$}
\def\spdol{$\tilde h_0^\dagger$}
\def\iena{$u$}
\def\idva{$d_{1,2}$}
\def\itri{${\bar u}^c$}
\def\isti{${\bar e}_i^c$}
\input{triangle}
\propto\;\;\;\;\;
(U^T\underline A^S D)_{11,12}
(U_c^\dagger Y_U^\dagger\tilde U^*)_{13}(\tilde U^T\underline
C\tilde E)_{33}(\tilde E^\dagger Y_E^\dagger E_c^*)_{3i}
\end{equation}

\begin{equation}
\def\trgor{$\tilde T$}
\def\trdol{$\tilde{\bar T}$}
\def\sfgor{$\tilde b$}
\def\sfdol{$\tilde t$}
\def\spgor{$\tilde V_0$}
\def\spdol{$\tilde V_0$}
\def\iena{$u$}
\def\idva{$e_i$}
\def\itri{$u$}
\def\isti{$d_{1,2}$}
\input{box}
\propto\;\;\;\;\;
(U^T\underline A^S \tilde D)_{13}(\tilde D^\dagger D)_{31,32}
(U^T\tilde U^*)_{13}(\tilde U^T\underline CE)_{3i}
\end{equation}

\begin{equation}
\def\tr{$\bar T$}
\def\sfgor{$\tilde b$}
\def\sfdol{$\tilde t$}
\def\spgor{$\tilde V_0$}
\def\spdol{$\tilde V_0$}
\def\iena{$u$}
\def\idva{$e_i$}
\def\itri{$u$}
\def\isti{$d_{1,2}$}
\input{triangle}
\propto\;\;\;\;\;
(U^T\underline CE)_{1i}(U^T\tilde U^*)_{13}
(\tilde U^T\underline A^S \tilde D)_{33}(\tilde D^\dagger D)_{31,32}
\end{equation}

\begin{equation}
\def\trgor{$\tilde{\bar T}$}
\def\trdol{$\tilde T$}
\def\sfgor{$\tilde\tau$}
\def\sfdol{$\tilde t$}
\def\spgor{$\tilde V_0$}
\def\spdol{$\tilde V_0$}
\def\iena{$u$}
\def\idva{$d_{1,2}$}
\def\itri{$u$}
\def\isti{$e_i$}
\input{box}
\propto\;\;\;\;\;
(U^T\underline C\tilde E)_{13}(\tilde E^\dagger E)_{3i}
(U^T\tilde U^*)_{13}(\tilde U^T\underline A^S D)_{31,32}
\end{equation}

\begin{equation}
\def\tr{$T$}
\def\sfgor{$\tilde\tau$}
\def\sfdol{$\tilde t$}
\def\spgor{$\tilde V_0$}
\def\spdol{$\tilde V_0$}
\def\iena{$u$}
\def\idva{$d_{1,2}$}
\def\itri{$u$}
\def\isti{$e_i$}
\input{triangle}
\propto\;\;\;\;\;
(U^T\underline A^S D)_{11,12}(U^T\tilde U^*)_{13}
(\tilde U^T\underline C\tilde E)_{33}(\tilde E^\dagger E)_{3i}
\end{equation}

\begin{equation}
\def\trgor{$\tilde{\bar T}^\dagger$}
\def\trdol{$\tilde T^\dagger$}
\def\sfgor{$\tilde b^c$}
\def\sfdol{$\tilde t^c$}
\def\spgor{$\tilde V_0^\dagger$}
\def\spdol{$\tilde V_0^\dagger$}
\def\iena{${\bar u}^c$}
\def\idva{${\bar e}_i^c$}
\def\itri{${\bar u}^c$}
\def\isti{${\bar d}_{1,2}^c$}
\input{box}
\propto\;\;\;\;\;
(U_c^\dagger\underline D^*\tilde D_c^*)_{13}(\tilde D_c^T D_c^*)_{31,32}
(U_c^\dagger\tilde U_c)_{13}(\tilde U_c^\dagger\underline B^*E_c^*)_{3i}
\end{equation}

\begin{equation}
\def\tr{$T$}
\def\sfgor{$\tilde b^c$}
\def\sfdol{$\tilde t^c$}
\def\spgor{$\tilde V_0^\dagger$}
\def\spdol{$\tilde V_0^\dagger$}
\def\iena{${\bar u}^c$}
\def\idva{${\bar e}_i^c$}
\def\itri{${\bar u}^c$}
\def\isti{${\bar d}_{1,2}^c$}
\input{triangle}
\propto\;\;\;\;\;
(U_c^\dagger\underline B^*E_c^*)_{1i}(U_c^\dagger\tilde U_c)_{13}
(\tilde U_c^\dagger\underline D^*\tilde D_c^*)_{33}
(\tilde D_c^TD_c^*)_{31,32}
\end{equation}

\begin{equation}
\def\trgor{$\tilde T^\dagger$}
\def\trdol{$\tilde{\bar T}^\dagger$}
\def\sfgor{$\tilde\tau^c$}
\def\sfdol{$\tilde t^c$}
\def\spgor{$\tilde V_0^\dagger$}
\def\spdol{$\tilde V_0^\dagger$}
\def\iena{${\bar u}^c$}
\def\idva{${\bar d}_{1,2}^c$}
\def\itri{${\bar u}^c$}
\def\isti{${\bar e}_i^c$}
\input{box}
\propto\;\;\;\;\;
(U_c^\dagger\underline B^*\tilde E_c^*)_{13}(\tilde E_c^TE_c^*)_{3i}
(U_c^\dagger\tilde U_c)_{13}
(\tilde U_c^\dagger\underline D^*D_c^*)_{31,32}
\end{equation}

\begin{equation}
\def\tr{$\bar T$}
\def\sfgor{$\tilde\tau^c$}
\def\sfdol{$\tilde t^c$}
\def\spgor{$\tilde V_0^\dagger$}
\def\spdol{$\tilde V_0^\dagger$}
\def\iena{${\bar u}^c$}
\def\idva{${\bar d}_{1,2}^c$}
\def\itri{${\bar u}^c$}
\def\isti{${\bar e}_i^c$}
\input{triangle}
\propto\;\;\;\;\;
(U_c^\dagger\underline D^* D_c^*)_{11,12}
(U_c^\dagger\tilde U_c)_{13}
(\tilde U_c^\dagger\underline B^*\tilde E_c^*)_{33}
(\tilde E_c^T E_c^*)_{3i}
\end{equation}
\\
\\
\\
\\

where:

\beq
\underline{A}^S=\underline{A} + \underline{A}^T
\eeq
\\
\\
\\
\\
For the case when there are more than one pair of Higgs triplet and
anti-triplet, as is often the case in $SO(10)$ models, one must
go to the mass diagonal basis for these fields to compute the
dimension five operators generated by their elimination. 
 The preceding diagrammatic analysis exhibits 
that the baryon and lepton number violating $d=5$  operators 
 are quite model dependent and generic predictions are not possible. 
 Specifically the proton lifetime predictions depend on the structure of the 
Higgs sector, and on the mixing between the fermion and the sfermions. 
However,  calculations with greater predictivity are possible if the model is fully 
well defined, including the Higgs sector, and the supersymmetry
breaking sector is well defined as, for example, is the case for the minimal supergravity
  model (mSUGRA). 
 
\section{Dressing of the d=5 operators}
In this appendix we exhibit the expressions for the dimension 
six operators after the dressing of the $d=5$ operators.  The full analysis
of the dressings of LLLL and RRRR dimension five operators including the
chargino, gluino, and neutralino exchanges and including the  sfermion 
mixings is given in the context of supergravity grand unification
in Ref.\cite{Nath:1985ub}, and later in Refs.\cite{Hisano:1992jj,Goto:1998qg,Fukuyama:2004xs}. 
The dressings are  carried
out at the electroweak scale which one may take as the average scale of 
sparticle masses ($M_{\mathrm{SUSY}}$). Here we exhibit  the results  in a compact
form (See references~ \cite{Nath:1985ub,Hisano:1992jj,Goto:1998qg,Fukuyama:2004xs}):
\bea
{\mathcal L}_5 &=& 
C_{L}^{(\tilde{u} \tilde{d} u e) abij}
\widetilde{u}_{a} \widetilde{d}_{b} u_{Li} e_{Lj}
+C_{L}^{(\tilde{u} \tilde{u} d e) abij}
\frac{1}{2}
\widetilde{u}_{a} \widetilde{u}_{b} d_{Li} e_{Lj}
\nonumber\\
&+&
C_{R}^{(\tilde{u} \tilde{d} u e) abij}
\widetilde{u}_{a} \widetilde{d}_{b} u_{Ri} e_{Rj}
+C_{R}^{(\tilde{u} \tilde{u} d e) abij}
\frac{1}{2}
\widetilde{u}_{a} \widetilde{u}_{b} d_{Ri} e_{Rj}
\nonumber\\
&+&
C_{L}^{(\tilde{u} \tilde{d} d \nu) abij}
\widetilde{u}_{a} \widetilde{d}_{b} d_{Li} \nu_{Lj}
+C_{L}^{(\tilde{d} \tilde{d} u \nu) abij}
\frac{1}{2}
\widetilde{d}_{a} \widetilde{d}_{b} u_{Li} \nu_{Lj}
\nonumber\\
&+&
C_{L}^{(\tilde{u} \tilde{e} u d) abij}
\widetilde{u}_{a} \widetilde{e}_{b} u_{Li} d_{Lj}
+C_{L}^{(\tilde{d} \tilde{e} u u) abij}
\frac{1}{2}
\widetilde{d}_{a} \widetilde{e}_{b} u_{Li} u_{Lj}
\nonumber\\
&+&
C_{R}^{(\tilde{u} \tilde{e} u d) abij}
\widetilde{u}_{a} \widetilde{e}_{b} u_{Ri} d_{Rj}
+C_{R}^{(\tilde{d} \tilde{e} u u) abij}
\frac{1}{2}
\widetilde{d}_{a} \widetilde{e}_{b} u_{Ri} u_{Rj}
\nonumber\\
&+&
C_{L}^{(\tilde{d} \tilde{\nu} u d) abij}
\widetilde{d}_{a} \widetilde{\nu}_{b} u_{Li} d_{Lj}
+C_{L}^{(\tilde{u} \tilde{\nu} d d) abij}
\frac{1}{2}
\widetilde{u}_{a} \widetilde{\nu}_{b} d_{Li} d_{Lj}
\eea
These coefficients are obtained from the coefficients 
of the original dimension-five operators including 
their renormalization from $M_{GUT}$ to $M_{\mathrm{SUSY}}$. 
For the renormalization of the $d=5$ operators see the next 
appendix.    
After the sparticles dressing, we obtain the following 
dimension-six operators for nucleon decays:
\bea
{\mathcal L}_6&=&
C_{LL}^{(udue)ij} (u_{L}d_{Li})(u_{L}e_{Lj})
+C_{RL}^{(udue)ij} (u_{R}d_{Ri})(u_{L}e_{Lj})
\nonumber\\
&+&
C_{LR}^{(udue)ij} (u_{L}d_{Li})(u_{R}e_{Rj})
+C_{RR}^{(udue)ij} (u_{R}d_{Ri})(u_{R}e_{Rj})
\nonumber\\
&+&
C_{LL}^{(udd\nu)ijk} (u_{L}d_{Li})(d_{Lj}\nu_{Lk})
+C_{RL}^{(udd\nu)ijk} (u_{R}d_{Ri})(d_{Lj}\nu_{Lk})
\nonumber\\
&+&
C_{RL}^{(ddu\nu)ijk} \frac{1}{2} (d_{Ri}d_{Rj})(u_{L}\nu_{Lk})
\eea
For the dimension-six operator, we have three contributions 
according to the dressed sparticles. Thus, for example,  
\beq
C_{LL}^{(udue)ij}
=C_{LL}^{(udue)ij}(\widetilde{g})
+C_{LL}^{(udue)ij}(\widetilde{\chi}^{0})
+C_{LL}^{(udue)ij}(\widetilde{\chi}^{\pm})
\eeq
and the same for the rest of the coefficients.
After the dressing we have the following expressions: 
\beq
C_{LL}^{(udue)ij}(\widetilde{g})
=\frac{4}{3}\frac{1}{m_{\widetilde{g}}}
C_L^{(\tilde{u}\tilde{d}ue)ab1j}\Gamma(G)_{1a}^{L(u)}\Gamma(G)_{ib}^{L(d)}
I\left(\frac{m_{\widetilde{g}}^2}{m_{\widetilde{u}_{a}}^2}, 
\frac{m_{\widetilde{g}}^2}{m_{\widetilde{d}_{b}}^2} \right),
\eeq
\bea 
C_{LL}^{(udue)ij}(\widetilde{\chi}^{\pm})
&=&\frac{1}{m_{\widetilde{\chi}_m^{+}}}
\left[-C_L^{(\tilde{u}\tilde{d}ue)ab1j}\Gamma(C)_{1mb}^{L(u)}
\Gamma(C)_{ima}^{L(d)}
I\left(\frac{m_{\widetilde{\chi}_m^{+}}^2}{m_{\widetilde{u}_{a}}^2}, 
\frac{m_{\widetilde{\chi}_m^{+}}^2}{m_{\widetilde{d}_{b}}^2} \right)
\right.
\nonumber
\eea
\beq
+\left. C_L^{(\tilde{d}\tilde{\nu}ud)ab1i}\Gamma(C)_{1ma}^{L(u)}
\Gamma(C)_{jmb}^{L(e)}
I\left(\frac{m_{\widetilde{\chi}_m^{+}}^2}{m_{\widetilde{d}_{a}}^2}, 
\frac{m_{\widetilde{\chi}_m^{+}}^2}{m_{\widetilde{\nu}_{b}}^2} \right)
\right],
\eeq 
\bea
C_{LL}^{(udue)ij}(\widetilde{\chi}^{0})
&=&\frac{1}{m_{\widetilde{\chi}_m^{0}}}
\left[C_L^{(\tilde{u}\tilde{d}ue)ab1j}\Gamma(N)_{1ma}^{L(u)}
\Gamma(N)_{imb}^{L(d)}
I\left(\frac{m_{\widetilde{\chi}_m^{0}}^2}{m_{\widetilde{u}_{a}}^2}, 
\frac{m_{\widetilde{\chi}_m^{0}}^2}{m_{\widetilde{d}_{b}}^2} \right)
\right.
\nonumber
\eea
\beq
+
\left.
C_L^{(\tilde{u}\tilde{e}ud)ab1i}\Gamma(N)_{1ma}^{L(u)}\Gamma(N)_{jmb}^{L(e)}
I\left(\frac{m_{\widetilde{\chi}_m^{0}}^2}{m_{\widetilde{u}_{a}}^2}, 
\frac{m_{\widetilde{\chi}_m^{0}}^2}{m_{\widetilde{e}_{b}}^2}\right)
\right],
\eeq
\beq
C_{RL}^{(udue)ij}(\widetilde{g})
=\frac{4}{3}\frac{1}{m_{\widetilde{g}}}
C_L^{(\tilde{u}\tilde{d}ue)ab1j}\Gamma(G)_{1a}^{R(u)}\Gamma(G)_{ib}^{R(d)}
I\left(\frac{m_{\widetilde{g}}^2}{m_{\widetilde{u_{a}}}^2}, 
\frac{m_{\widetilde{g}}^2}{m_{\widetilde{d_{b}}}^2} \right),
\eeq
\beq
C_{RL}^{(udue)ij}(\widetilde{\chi}^{\pm})
=-\frac{1}{m_{\widetilde{\chi}_m^{+}}}
C_L^{(\tilde{u}\tilde{d}ue)ab1j}\Gamma(C)_{1mb}^{R(u)}\Gamma(C)_{ima}^{R(d)}
I\left(\frac{m_{\widetilde{\chi}_m^{+}}^2}{m_{\widetilde{u}_{a}}^2}, 
\frac{m_{\widetilde{\chi}_m^{+}}^2}{m_{\widetilde{d}_{b}}^2} \right),
\eeq
\bea
C_{RL}^{(udue)ij}(\widetilde{\chi}^{0})
&=&\frac{1}{m_{\widetilde{\chi}_m^{0}}}
\left[C_L^{(\tilde{u}\tilde{d}ue)ab1j}\Gamma(N)_{1ma}^{R(u)}
\Gamma(N)_{imb}^{R(d)}
I\left(\frac{m_{\widetilde{\chi}_m^{0}}^2}{m_{\widetilde{u}_{a}}^2}, 
\frac{m_{\widetilde{\chi}_m^{0}}^2}{m_{\widetilde{d}_{b}}^2} \right)
\right.
\nonumber
\eea
\beq
+
\left.
C_R^{(\tilde{u}\tilde{e}ud)ab1i} \Gamma(N)_{1ma}^{L(u)} 
\Gamma(N)_{jmb}^{L(e)}
I\left(\frac{m_{\widetilde{\chi}_m^{0}}^2}{m_{\widetilde{u}_{a}}^2}, 
\frac{m_{\widetilde{\chi}_m^{0}}^2}{m_{\widetilde{e}_{b}}^2}\right)
\right],
\eeq
\beq
C_{LR}^{(udue)ij}(\widetilde{g})
=\frac{4}{3}\frac{1}{m_{\widetilde{g}}}
C_R^{(\tilde{u}\tilde{d}ue)ab1j}\Gamma(G)_{1a}^{L(u)}\Gamma(G)_{ib}^{L(d)}
I\left(\frac{m_{\widetilde{g}}^2}{m_{\widetilde{u}_{a}}^2}, 
\frac{m_{\widetilde{g}}^2}{m_{\widetilde{d}_{b}}^2} \right),
\eeq 
\bea
C_{LR}^{(udue)ij}(\widetilde{\chi}^{\pm})
&=&\frac{1}{m_{\widetilde{\chi}_m^{+}}}
\left[-C_R^{(\tilde{u}\tilde{d}ue)ab1j}\Gamma(C)_{1mb}^{L(u)}
\Gamma(C)_{ima}^{L(d)}
I\left(\frac{m_{\widetilde{\chi}_m^{+}}^2}{m_{\widetilde{u}_{a}}^2}, 
\frac{m_{\widetilde{\chi}_m^{+}}^2}{m_{\widetilde{d}_{b}}^2} \right)
\right.
\nonumber
\eea
\beq
+
\left.
C_L^{(\tilde{d}\tilde{\nu} ud)ab1i}\Gamma(C)_{1ma}^{R(u)}
\Gamma(C)_{jmb}^{R(e)}
I\left(\frac{m_{\widetilde{\chi}_m^{+}}^2}{m_{\widetilde{d}_{a}}^2}, 
\frac{m_{\widetilde{\chi}_m^{+}}^2}{m_{\widetilde{\nu}_{b}}^2}\right)
\right],
\eeq
\bea
C_{LR}^{(udue)ij}(\widetilde{\chi}^{0})
&=&\frac{1}{m_{\widetilde{\chi}_m^{0}}}
\left[C_R^{(\tilde{u}\tilde{d}ue)ab1j} \Gamma(N)_{1ma}^{L(u)}
\Gamma(N)_{imb}^{L(d)}
I\left(\frac{m_{\widetilde{\chi}_m^{0}}^2}{m_{\widetilde{u}_{a}}^2}, 
\frac{m_{\widetilde{\chi}_m^{0}}^2}{m_{\widetilde{d}_{b}}^2} \right)
\right.
\nonumber 
\eea
\beq
+
\left.
C_L^{(\tilde{u}\tilde{e}ud)ab1i}\Gamma(N)_{1ma}^{R(u)} \Gamma(N)_{jmb}^{R(e)}
I\left(\frac{m_{\widetilde{\chi}_m^{0}}^2}{m_{\widetilde{u}_{a}}^2}, 
\frac{m_{\widetilde{\chi}_m^{0}}^2}{m_{\widetilde{e}_{b}}^2}\right)
\right],
\eeq
\beq
C_{RR}^{(udue)ij}(\widetilde{g})
=\frac{4}{3}\frac{1}{m_{\widetilde{g}}}
C_R^{(\tilde{u}\tilde{d}ue)ab1j} \Gamma(G)_{1a}^{R(u)} \Gamma(G)_{ib}^{R(d)}
I\left(\frac{m_{\widetilde{g}}^2}{m_{\widetilde{u_{a}}}^2}, 
\frac{m_{\widetilde{g}}^2}{m_{\widetilde{d}_{b}}^2} \right),
\eeq
\beq
C_{RR}^{(udue)ij}(\widetilde{\chi}^{\pm})
=-\frac{1}{m_{\widetilde{\chi}_m^{+}}}
C_R^{(\tilde{u}\tilde{d}ue)ab1j} \Gamma(C)_{1mb}^{R(u)}\Gamma(C)_{ima}^{R(d)}
I\left(\frac{m_{\widetilde{\chi}_m^{+}}^2}{m_{\widetilde{u}_{a}}^2}, 
\frac{m_{\widetilde{\chi}_m^{+}}^2}{m_{\widetilde{d}_{b}}^2} \right),
\eeq 
\bea
C_{RR}^{(udue)ij}(\widetilde{\chi}^{0})
&=&\frac{1}{m_{\widetilde{\chi}_m^{0}}}
\left[C_R^{(\tilde{u}\tilde{d}ue)ab1j} \Gamma(N)_{1ma}^{R(u)}
\Gamma(N)_{imb}^{R(d)}
I\left(\frac{m_{\widetilde{\chi}_m^{0}}^2}{m_{\widetilde{u}_{a}}^2}, 
\frac{m_{\widetilde{\chi}_m^{0}}^2}{m_{\widetilde{d}_{b}}^2} \right)
\right.
\nonumber 
\eea
\beq
+
\left.
C_R^{(\tilde{u}\tilde{e}ud)ab1i} \Gamma(N)_{1ma}^{R(u)} 
\Gamma(N)_{jmb}^{R(e)}
I\left(\frac{m_{\widetilde{\chi}_m^{0}}^2}{m_{\widetilde{u}_{a}}^2}, 
\frac{m_{\widetilde{\chi}_m^{0}}^2}{m_{\widetilde{e}_{b}}^2}\right)
\right],
\eeq
\bea
C_{LL}^{(udd\nu)ijk}(\widetilde{g})
&=&\frac{4}{3}\frac{1}{m_{\widetilde{g}}}
\left[C_L^{(\tilde{u}\tilde{d}d\nu)abjk} \Gamma(G)_{1a}^{L(u)} 
\Gamma(G)_{ib}^{L(d)}
I\left(\frac{m_{\widetilde{g}}^2}{m_{\widetilde{u}_{a}}^2}, 
\frac{m_{\widetilde{g}}^2}{m_{\widetilde{d}_{b}}^2} \right)
\right.
\nonumber 
\eea
\beq
+
\left.
C_L^{(\tilde{d}\tilde{d}u\nu)ab1k}\Gamma(G)_{ja}^{L(d)}\Gamma(G)_{ib}^{L(d)}
I\left(\frac{m_{\widetilde{g}}^2}{m_{\widetilde{d}_{a}}^2}, 
\frac{m_{\widetilde{g}}^2}{m_{\widetilde{d}_{b}}^2} \right)
\right],
\eeq  
\bea
C_{LL}^{(udd\nu)ijk}(\widetilde{\chi}^{\pm})
&=&\frac{1}{m_{\widetilde{\chi}_m^{+}}}
\left[-C_L^{(\tilde{u}\tilde{d}d\nu)abjk} \Gamma(C)_{1mb}^{L(u)} 
\Gamma(C)_{ima}^{L(d)}
I\left(\frac{m_{\widetilde{\chi}_m^{+}}^2}{m_{\widetilde{u}_{a}}^2}, 
\frac{m_{\widetilde{\chi}_m^{+}}^2}{m_{\widetilde{d}_{b}}^2} \right)
\right.
\nonumber 
\eea
\beq
+
\left.
C_L^{(\tilde{u}\tilde{e}ud)ab1i}\Gamma(C)_{jma}^{L(d)} 
\Gamma(C)_{kmb}^{L(\nu)}
I\left(\frac{m_{\widetilde{\chi}_m^{+}}^2}{m_{\widetilde{u}_{a}}^2}, 
\frac{m_{\widetilde{\chi}_m^{+}}^2}{m_{\widetilde{e}_{b}}^2}\right)
\right],
\eeq
\bea
C_{LL}^{(udd\nu)ijk}(\widetilde{\chi}^{0})
&=&\frac{1}{m_{\widetilde{\chi}_m^{0}}}
\left[C_L^{(\tilde{u}\tilde{d}d\nu)abjk} \Gamma(N)_{1ma}^{L(u)} 
\Gamma(N)_{imb}^{L(d)}
I\left(\frac{m_{\widetilde{\chi}_m^{0}}^2}{m_{\widetilde{u}_{a}}^2}, 
\frac{m_{\widetilde{\chi}_m^{0}}^2}{m_{\widetilde{d}_{b}}^2} \right)
\right.
\nonumber\\
&+&
C_L^{(\tilde{d}\tilde{d}u\nu)ab1k} \Gamma(N)_{jma}^{L(d)}\Gamma(N)_{imb}^{L(d)}
I\left(\frac{m_{\widetilde{\chi}_m^{0}}^2}{m_{\widetilde{d}_{a}}^2}, 
\frac{m_{\widetilde{\chi}_m^{0}}^2}{m_{\widetilde{d}_{b}}^2}\right)
\nonumber\\
&+&
C_L^{(\tilde{d}\tilde{\nu} ud)ab1i} \Gamma(N)_{jma}^{L(d)} 
\Gamma(N)_{kmb}^{L(\nu)}
I\left(\frac{m_{\widetilde{\chi}_m^{0}}^2}{m_{\widetilde{d}_{a}}^2}, 
\frac{m_{\widetilde{\chi}_m^{0}}^2}{m_{\widetilde{\nu}_{b}}^2}\right)
\nonumber 
\eea
\beq
+
\left.
C_L^{(\tilde{u}\tilde{\nu} dd)abji} \Gamma(N)_{1ma}^{L(u)} 
\Gamma(N)_{kmb}^{L(\nu)}
I\left(\frac{m_{\widetilde{\chi}_m^{0}}^2}{m_{\widetilde{u}_{a}}^2}, 
\frac{m_{\widetilde{\chi}_m^{0}}^2}{m_{\widetilde{\nu}_{b}}^2}\right)
\right],
\eeq
\beq
C_{RL}^{(udd\nu)ijk}(\widetilde{g})
=\frac{4}{3}\frac{1}{m_{\widetilde{g}}}
C_L^{(\tilde{u}\tilde{d}d\nu)abjk} \Gamma(G)_{1a}^{R(u)} 
\Gamma(G)_{ib}^{R(d)}
I\left(\frac{m_{\widetilde{g}}^2}{m_{\widetilde{u}_{a}}^2}, 
\frac{m_{\widetilde{g}}^2}{m_{\widetilde{d}_{b}}^2} \right),
\eeq
\bea
C_{RL}^{(udd\nu)ijk}(\widetilde{\chi}^{\pm})
&=&\frac{1}{m_{\widetilde{\chi}_m^{+}}}
\left[-C_L^{(\tilde{u}\tilde{d}d\nu)abjk}\Gamma(C)_{1mb}^{R(u)} 
\Gamma(C)_{ima}^{R(d)}
I\left(\frac{m_{\widetilde{\chi}_m^{+}}^2}{m_{\widetilde{u}_{a}}^2}, 
\frac{m_{\widetilde{\chi}_m^{+}}^2}{m_{\widetilde{d}_{b}}^2} \right)
\right.
\nonumber 
\eea
\beq
+
\left.
C_R^{(\tilde{u}\tilde{e}ud)ab1i} \Gamma(C)_{jma}^{L(d)}
\Gamma(C)_{kmb}^{L(\nu)}
I\left(\frac{m_{\widetilde{\chi}_m^{+}}^2}{m_{\widetilde{u}_{a}}^2}, 
\frac{m_{\widetilde{\chi}_m^{+}}^2}{m_{\widetilde{e}_{b}}^2}\right)
\right],
\eeq
\beq
C_{RL}^{(udd\nu)ijk}(\widetilde{\chi}^{0})
=\frac{1}{m_{\widetilde{\chi}_m^{0}}}
C_L^{(\tilde{u}\tilde{d}d\nu)abjk} \Gamma(N)_{1ma}^{R(u)} 
\Gamma(N)_{imb}^{R(d)}
I\left(\frac{m_{\widetilde{\chi}_m^{0}}^2}{m_{\widetilde{u}_{a}}^2}, 
\frac{m_{\widetilde{\chi}_m^{0}}^2}{m_{\widetilde{d}_{b}}^2} \right),
\eeq
\beq
C_{RL}^{(ddu\nu)ijk}(\widetilde{g})
=\frac{4}{3}\frac{1}{m_{\widetilde{g}}}
C_L^{(\tilde{d}\tilde{d}u\nu)ab1k} \Gamma(G)_{ia}^{R(d)} 
\Gamma(G)_{jb}^{R(d)}
I\left(\frac{m_{\widetilde{g}}^2}{m_{\widetilde{d}_{a}}^2}, 
\frac{m_{\widetilde{g}}^2}{m_{\widetilde{d}_{b}}^2} \right),
\eeq
\beq
C_{RL}^{(ddu\nu)ijk}(\widetilde{\chi}^{\pm})
=0,
\eeq
\beq
C_{RL}^{(ddu\nu)ijk}(\widetilde{\chi}^{0})
=\frac{1}{m_{\widetilde{\chi}_m^{0}}}
C_L^{(\tilde{d}\tilde{d}u\nu)ab1k} \Gamma(N)_{ima}^{R(d)} 
\Gamma(N)_{jmb}^{R(d)}
I\left(\frac{m_{\widetilde{\chi}_m^{0}}^2}{m_{\widetilde{d}_{a}}^2}, 
\frac{m_{\widetilde{\chi}_m^{0}}^2}{m_{\widetilde{d}_{b}}^2} \right).
\eeq
where the loop function is defined by:
\beq
I(a,b) \equiv \frac{1}{16\pi^2 }\frac{a \, b}{a-b} 
\left(\frac{1}{1-a}\log a-\frac{1}{1-b}\log b \right).
\eeq
For the dimension-five operators, we have the following expressions 
(using the following notation for the anti-symmetric tensor, 
$C^{[ijk]l} \equiv C^{ijkl}-C^{kjil}.$)  
\\
\beq
C_{L}^{(\tilde{u} \tilde{d} u e) abij}
=
C_L^{[ijk]l}
(O_{\widetilde{u}}^{*})_{ak} (O_{\widetilde{d}}^{*})_{bl}
\eeq
\beq
C_{L}^{(\tilde{u} \tilde{u} d e) abij}
=
C_L^{[kjl]m}
(O_{\widetilde{u}}^{*})_{ak} (O_{\widetilde{u}}^{*})_{bl}
(V_{\mathrm{CKM}})_{im}
\eeq
\beq
C_{R}^{(\tilde{u} \tilde{d} u e) abij}
=
(C_R^{*klji}-C_R^{*iljk})
(O_{\widetilde{u}}^{*})_{a,k+3} (O_{\widetilde{d}}^{*})_{b,l+3}
\eeq
\beq
C_{R}^{(\tilde{u} \tilde{u} d e) abij}
=
(C_R^{*klji}-C_R^{*iljk})
(O_{\widetilde{u}}^{*})_{a,k+3} (O_{\widetilde{u}}^{*})_{b,l+3}
\eeq
\beq 
C_{L}^{(\tilde{u} \tilde{d} d \nu) abij}
=
(C_L^{mnkl}-C_L^{lknm})
(O_{\widetilde{u}}^{*})_{ak} (O_{\widetilde{d}}^{*})_{bl}
(V_{\mathrm{CKM}})_{im} (V_{\mathrm{PMNS}})_{jn}
\eeq
\beq 
C_{L}^{(\tilde{d} \tilde{d} u \nu) abij}
=
(C_L^{lnik}-C_L^{knil})
(O_{\widetilde{d}}^{*})_{ak} (O_{\widetilde{d}}^{*})_{bl}
(V_{\mathrm{PMNS}})_{jn}
\eeq
\beq
C_{L}^{(\tilde{u} \tilde{e} u d) abij}
=
C_L^{[kli]m}
(O_{\widetilde{u}}^{*})_{ak} (O_{\widetilde{e}}^{*})_{bl}
(V_{\mathrm{CKM}})_{jm}
\eeq
\beq
C_{L}^{(\tilde{d} \tilde{e} u u) abij}
=
C_L^{[ilj]k}
(O_{\widetilde{d}}^{*})_{ak} (O_{\widetilde{e}}^{*})_{bl}
\eeq 
\beq
C_{R}^{(\tilde{u} \tilde{e} u d) abij}
=
(C_R^{*jkli}-C_R^{*kjli})
(O_{\widetilde{u}}^{*})_{a,k+3} (O_{\widetilde{e}}^{*})_{b,l+3}
\eeq 
\beq
C_{R}^{(\tilde{d} \tilde{e} u u) abij}
=
(C_R^{*jkli}-C_R^{*iklj})
(O_{\widetilde{d}}^{*})_{a,k+3} (O_{\widetilde{e}}^{*})_{b,l+3}
\eeq
\beq
C_{L}^{(\tilde{d} \tilde{\nu} u d) abij}
=
(C_L^{klim}-C_L^{mlik})
(O_{\widetilde{d}}^{*})_{ak} (O_{\widetilde{\nu}}^{*})_{bl}
(V_{\mathrm{CKM}})_{jm}
\eeq 
\beq
C_{L}^{(\tilde{u} \tilde{\nu} d d) abij}
=
(C_L^{nlkm}-C_L^{mlkn})
(O_{\widetilde{u}}^{*})_{ak} (O_{\widetilde{\nu}}^{*})_{bl}
(V_{\mathrm{CKM}})_{im} (V_{\mathrm{CKM}})_{jn}
\eeq
where:
\beq
\Gamma(G)_{ia}^{R(u)} =
g_3 (O_{\tilde{u}})_{a,i+3}
\eeq 
\beq
\Gamma(G)_{ia}^{L(u)} = 
g_3 (O_{\tilde{u}})_{ai}
\eeq 
\beq
\Gamma(G)_{ia}^{R(d)} =
g_3 (O_{\tilde{d}})_{a,i+3}
\eeq 
\beq
\Gamma(G)_{ia}^{L(d)} =
g_3 (O_{\tilde{d}})_{ak} (V_{\mathrm{CKM}})_{ki} 
\eeq
\beq
\Gamma(C)_{ima}^{R(u)} =
g \frac{m_{u_i}}{\sqrt{2} M_{\mathrm{W}} \sin \beta} 
(O_{+}^{\dagger})_{m 2} (O_{\tilde d})_{ai} 
\eeq 
\beq
\Gamma(C)_{ima}^{L(u)} = 
g \left\{(O_{-}^{\dagger})_{m1} (O_{\tilde d})_{ai}
-
\frac{m_{d_k}}{\sqrt{2} M_{\mathrm{W}} \cos \beta} 
(O_{-}^{\dagger})_{m 2} (O_{\tilde d})_{a,k+3} 
(V_{CKM}^{\dagger})_{ki}\right\}  
\eeq 
\beq
\Gamma(C)_{ima}^{R(d)} =
- g \frac{m_{d_i}}{\sqrt{2} M_{\mathrm{W}} \cos \beta} 
(O_{-})_{2 m} (O_{\tilde u})_{ak} (V_{CKM})_{ki}  
\eeq 
\beq
\Gamma(C)_{ima}^{L(d)} = 
g \left\{(O_{+})_{1m} (O_{\tilde u})_{ak}
+
\frac{m_{u_k}}{\sqrt{2} M_{\mathrm{W}} \sin \beta} 
(O_{+})_{2 m} (O_{\tilde u})_{a,k+3} \right\} 
(V_{\mathrm{CKM}})_{ki}
\eeq
\beq
\Gamma(C)_{ima}^{L(\nu)} = 
g \left\{-(O_{-}^{\dagger})_{m1} (O_{\tilde e})_{ak}
+ \frac{m_{e_k}}{\sqrt{2} M_{\mathrm{W}} \cos \beta} 
(O_{-}^{\dagger})_{m 2} (O_{\tilde e})_{a,k+3} \right\} 
(V_{\mathrm{PMNS}})_{ik} 
\eeq
\beq
\Gamma(C)_{ima}^{R(e)} =
g \frac{m_{e_i}}{\sqrt{2} M_{\mathrm{W}} \cos \beta} 
(O_{-})_{2m} (O_{\tilde \nu})_{ai}  
\eeq
\beq
\Gamma(C)_{ima}^{L(e)} =
- g (O_{+})_{1m} (O_{\tilde \nu})_{ai} 
\eeq 
\beq
\Gamma(N)_{ima}^{R(u)} = 
-\frac{g}{\sqrt{2}} \left\{
\frac{m_{u_i}}{M_{\mathrm{W}} \sin \beta} 
(O_{N}^{\dagger})_{m4} (O_{\tilde u})_{a,i} 
-
\frac{4}{3} \tan \theta_{W} (O_{N}^{\dagger})_{m2} 
(O_{\tilde u})_{a,i+3} \right\} 
\nonumber\\
\eeq
\begin{displaymath}
\Gamma(N)_{ima}^{L(u)} =
-\frac{g}{\sqrt{2}} \times
\end{displaymath}
\beq
\left\{
\frac{m_{u_i}}{M_{\mathrm{W}} \sin \beta} 
(O_{N})_{4m} (O_{\tilde u})_{a,i+3} 
+
\left[(O_{N})_{2m} + 
\frac{1}{3}\tan \theta_{W} (O_{N})_{1m} \right] 
(O_{\tilde u})_{ai} \right\} 
\eeq
\beq
\Gamma(N)_{ima}^{R(d)} =
-\frac{g}{\sqrt{2}} \left\{
\frac{m_{d_i}}{M_{\mathrm{W}} \cos \beta} 
(O_{N}^{\dagger})_{m3} (O_{\tilde d})_{ak} (V_{CKM})_{ki} 
+
\frac{2}{3} \tan \theta_{W} (O_{N}^{\dagger})_{m1} 
(O_{\tilde d})_{a,i+3} \right\}
\nonumber\\ 
\eeq 
\begin{displaymath}
\Gamma(N)_{ima}^{L(d)} = 
 \frac{g}{\sqrt{2}} \times 
\nonumber
\end{displaymath}
\beq 
\left\{
- \frac{m_{d_k}}{M_{\mathrm{W}} \cos \beta} 
(O_{N})_{3m} (O_{\tilde d})_{a,i+3} 
+
\left[(O_{N})_{2m} -
\frac{1}{3} \tan \theta_{W} (O_{N})_{1m} \right]
(O_{\tilde d})_{ak} (V_{CKM})_{ki}\right\}
\eeq
\beq
\Gamma(N)_{ima}^{L(\nu)} =
-\frac{g}{\sqrt{2}} 
\left[(O_{N})_{2m} - 
\tan \theta_{W} (O_{N})_{1m} \right] 
(O_{\tilde \nu})_{a,k}
(V_{\mathrm{PMNS}})_{ki} 
\eeq 
\beq
\Gamma(N)_{ima}^{R(e)} =
- g \sqrt{2} \left\{
\frac{m_{e_i}}{2 M_{\mathrm{W}} \cos \beta} 
(O_{N}^{\dagger})_{m3} (O_{\tilde e})_{ai}
+ \tan \theta_{W} (O_{N}^{\dagger})_{m1} 
(O_{\tilde e})_{a,i+3} \right\} 
\eeq 
\begin{displaymath}
\Gamma(N)_{ima}^{L(e)} =
g \sqrt{2} \times 
\end{displaymath}
\beq 
\left\{
- \frac{m_{e_i}}{2 M_{\mathrm{W}} \cos \beta} 
(O_{N})_{3m} (O_{\tilde e})_{a,i+3} 
+
\left[\frac{1}{2}(O_{N})_{2m} +
\frac{1}{2} \tan \theta_{W} (O_{N})_{1m} \right]
(O_{\tilde e})_{ai} \right\}
\eeq
where the squark, slepton mass-squared matrix $M_{\widetilde{f}}^2$,  
chargino and neutralino mass matrices $M_C$ and $M_N$ are diagonalized 
by the unitary matrices $O_{\widetilde f}$, $O_{-}$, $O_{+}$ and $O_N$, 
respectively.  
\beq
O_{\widetilde f} \ M_{\widetilde{f}}^2 \ O_{\widetilde f}^\dagger 
=
(M_{\widetilde{f}}^2)^{diag}
\eeq

\beq
O_{-}^{\dagger} \ M_C \ O_+ 
=
(M_C)^{diag}
\eeq

\beq
O_N^* \ M_N \ O_N^{\dagger} 
=
(M_N)^{diag}
\eeq

\section{Sparticle spectrum and renormalization}
In this appendix we exhibit the sparticle mass matrices 
that enter in the analysis of the dressings of the 
dimension five operators. The matrices are given at the 
electroweak scale, and they are the most general ones including CP phases. 
We list all relevant renormalization group equations at the 
one-loop level for the soft parameters in the MSSM. As discussed already
in Sec.(4.2), in MSSM the superpotential is given by
\beq
W = \hat{U}^C Y_u \hat{Q} \hat{H}_u + \hat{D}^C Y_d \hat{Q} \hat{H}_d + \hat{E}^C Y_e \hat{L} \hat{H}_d + \mu \hat{H}_u \hat{H}_d
\eeq
where $Y_{u,d,e}$ are matrices in family space. The soft SUSY-breaking 
Lagrangian contains scalar couplings
\beq
{\cal L}_{soft} \owns \tilde{u}^C h_u \tilde{Q} H_u + \tilde{d}^C h_d \tilde{Q} H_d 
+ \tilde{e}^C h_e \tilde{L} H_d + B H_u H_d + h.c. 
\eeq
where $h_{u,d,e}$ are $3 \times 3$ matrices. There are also scalar masses
\beq
{\cal L}_{soft} \owns m_{H_u}^2 H_u^{\dagger} H_u + m_{H_d}^2 H_d^{\dagger} H_d 
+ \tilde{Q}^{\dagger} M_{\tilde Q}^2 \tilde{Q} 
+ \tilde{L}^{\dagger} M_{\tilde L}^2 \tilde{L} \nonumber\\
\eeq
\beq 
+ \tilde{u}^{C \dagger} m_{\tilde u}^2 \ \tilde{u}^C + 
\tilde{d}^{C \dagger} m_{\tilde d}^2 \ \tilde{d}^C + 
\tilde{e}^{C \dagger} m_{\tilde e}^2 \ \tilde{e}^C
\eeq 

Here again $M_{\tilde Q}^2$, $M_{\tilde L}^2$, $m_{\tilde u}^2$, 
$m_{\tilde d}^2$, and $m_{\tilde e}^2$ are $3 \times 3$ matrices 
in family space.
The renormalization group equations for the gauge couplings are:

\beq
\frac{d g_a}{dt}=\frac{g_a^3}{16 \pi^2} B_a^{(1)} + \nonumber\\ 
\frac{g_a^3}{(16 \pi^2)^2} \left(  \sum_{b=1}^3 B_{ab}^{2} g_b^2 - \sum_{x=u,d,e} C_a^x Tr (Y_x^{\dagger} Y_x) \right)
\eeq
with $B_a^{(1)}= (33/5, 1, -3)$ for $U(1)_Y$, $SU(2)_L$ 
and $SU(3)_C$, respectively.

\begin{equation}
B_{ab}^{(2)}=\left(
\begin{array}{ccc}
  \frac{199}{25} & \frac{27}{5} & \frac{88}{5} \\ \\
  \frac{9}{5} & 25 & 24 \\ \\
  \frac{11}{5} & 9 & 14 \\ \\
\end{array}\right)
\end{equation}

\begin{equation}
C_{a}^{u,d,e}=\left(
\begin{array}{ccc}
  \frac{26}{5} & \frac{14}{5} & \frac{18}{5} \\ \\
  6 & 6 & 2 \\ \\
  4 & 4 & 0 \\ \\
\end{array}\right)
\end{equation}

The one-loop renormalization group equations 
for the three gaugino mass parameters are~\cite{Martin:1993zk}:

\beq
\frac{d M_a}{dt}= \frac{2 g_a^2}{16 \pi^2} B_a^{(1)} M_a
\eeq
while for the $\mu$ term, and the Yukawa couplings one has:
\beq
\frac{d \mu}{dt}= \frac{\beta_{\mu}^{(1)}}{16 \pi^2}
\eeq
\beq
\frac{d Y_{u,d,e}}{dt} = \frac{\beta_{Y_{u,d,e}}^{(1)}}{16 \pi^2}
\eeq
where
\beq
\beta_{\mu}^{(1)}=\mu \left(  Tr (3 Y_u Y_u^{\dagger} + 3 Y_d Y_d^{\dagger} + Y_e Y_e^{\dagger}) -3 g_2^2 - \frac{3}{5} g_1^2 \right)
\eeq
\beq
\beta_{Y_u}^{(1)}= Y_u \left(  3 Tr(Y_u Y_u^{\dagger}) + 3 Y_u^{\dagger} Y_u + Y_d^{\dagger} Y_d - 
\frac{16}{3} g_3^2 - 3 g_2^2 - \frac{13}{15} g_1^2 \right)
\eeq
\beq
\beta_{Y_d}^{(1)}= Y_d \left( Tr(3 Y_d Y_d^{\dagger} + Y_e Y_e^{\dagger}) + 
3 Y_d^{\dagger} Y_d + Y_u^{\dagger} Y_u - \frac{16}{3}g_3^2 -3 g_2^2 - \frac{7}{15} g_1^2 \right)
\eeq
\beq
\beta_{Y_e}^{(1)}= Y_e \left(  Tr( 3 Y_d Y_d^{\dagger} + Y_e Y_e^{\dagger}) + 
3 Y_e^{\dagger} Y_e - 3 g_2^2 - \frac{9}{5} g_1^2 \right)
\eeq
For the trilinear terms the RG equations are
\beq
\frac{d h_{u,d,e}}{dt}= \frac{{\beta_{h_{u,d,e}}}^{(1)}}{16 \pi^2}
\eeq
where:
\begin{displaymath}
\beta_{h_{u}}^{(1)}= h_u \left(  3 Tr (Y_u Y_u^{\dagger}) + 5 Y_u^{\dagger} Y_u + Y_d^{\dagger} Y_d 
- \frac{16}{3} g_3^2 - 3 g_2^2 - \frac{13}{15} g_1^2 \right) + \nonumber
\end{displaymath}
\beq
+ Y_u \left( 6 Tr(h_u Y_u^{\dagger}) + 4 Y_u^{\dagger} h_u + 2 Y_d^{\dagger}h_d + 
\frac{32}{3} g_3^2 M_3 + 6 g_2^2 M_2 + \frac{26}{15} g_1^2 M_1   \right)
\eeq
\begin{displaymath}
\beta_{h_{d}}^{(1)}= h_d \left( Tr( 3 Y_d Y_d^{\dagger} + Y_e Y_e^{\dagger}) 
+ 5 Y_d^{\dagger} Y_d + Y_u^{\dagger} Y_u - \frac{16}{3}g_3^2 -3 g_2^2 - \frac{7}{15} g_1^2 \right) + \nonumber
\end{displaymath}
\beq
Y_d \left( Tr( 6 h_d Y_d^{\dagger} + 2 h_e Y_e^{\dagger})+ 4 Y_d^{\dagger} h_d 
+ 2 Y_u^{\dagger} h_u + \frac{32}{3} g_3^2 M_3 + 6 g_2^2 M_2 + \frac{14}{15}g_1^2 M_1 \right)
\eeq
\begin{displaymath}
\beta_{h_{e}}^{(1)}= h_e \left( Tr( 3 Y_d Y_d^{\dagger} + Y_e Y_e^{\dagger}) + 5 Y_e^{\dagger} Y_e - 3 g_2^2 
- \frac{9}{5}g_1^2 \right) +
\end{displaymath}
\beq
Y_e \left( Tr( 6 h_d Y_d^{\dagger} + 2 h_e Y_e^{\dagger}) + 4 Y_e^{\dagger} h_e + 6 g_2^2 M_2 + \frac{18}{5} g_1^2 M_1 \right)
\eeq
The renormalization group equation for the B-term is given by

\beq
\frac{d B}{dt}= \frac{\beta_{B}^{(1)}}{16 \pi^2}
\eeq
where 
\begin{displaymath}
\beta_B^{(1)}=B \left( Tr(3 Y_u Y_u^{\dagger} + 3 Y_d Y_d^{\dagger} + Y_e Y_e^{\dagger})- 3 g_2^2 - \frac{3}{5}g_1^2\right)+
\end{displaymath}
\beq
\mu \left( Tr( 6 h_u Y_u^{\dagger} + 6 h_d Y_d^{\dagger} + 2 h_e Y_e^{\dagger})+ 6 g_2^2 M_2 + \frac{6}{5} g_1^2 M_1 \right)
\eeq
while the RG equations for the soft masses are
\beq
\frac{d}{dt}m^2 = \frac{\beta_{m^2}^{(1)}}{16 \pi^2}
\eeq
where 
\begin{displaymath}
\beta_{m_{H_u}^2}^{(1)}= 6 Tr\left( (m_{H_u}^2 + M_{\tilde Q}^2 ) Y_u^{\dagger} Y_u 
+ Y_u^{\dagger} m_{\tilde u}^2 Y_u + h_u^{\dagger} h_u \right)
- 6 g_2^2 |M_2|^2 - \frac{6}{5} g_1^2 |M_1|^2 
\end{displaymath}
\beq
+ \frac{3}{5} g_1^2 S
\eeq
\begin{displaymath}
\beta_{m_{H_d}^2}^{(1)}= Tr ( 6 (m_{H_d}^2 + M_{\tilde Q}^2 ) Y_d^{\dagger} Y_d 
+ 6 Y_d^{\dagger} m_{\tilde d}^2 Y_d + 
2 (m_{H_d}^2 + M_{\tilde L}^2) Y_e^{\dagger} Y_e + 2 Y_e^{\dagger} m_{\tilde e}^2 Y_e 
\end{displaymath}
\beq
+ 6 h_d^{\dagger} h_d + 2 h_e^{\dagger} h_e ) - 6 g_2^2 |M_2|^2 - \frac{6}{5} |M_1|^2 - \frac{3}{5}g_1^2 S
\eeq
\beq
\beta_{M_{\tilde Q}^2}^{(1)}=(M_{\tilde Q}^2 + 2 m_{H_u}^2) Y_u^{\dagger} Y_u + 
(M_{\tilde Q}^2 + 2 m_{H_d}^2)Y_d^{\dagger} Y_d + 
(Y_u^{\dagger} Y_u + Y_d^{\dagger} Y_d) M_{\tilde Q}^2  + 2 Y_u^{\dagger} m_{\tilde u}^2 Y_u
\eeq
\begin{displaymath}
\beta_{M_{\tilde L}^2}^{(1)}=(M_{\tilde L}^2 + 2 m_{H_d}^2) Y_e^{\dagger} Y_e + 
2 Y_e^{\dagger} m_{\tilde e}^2 Y_e + Y_e^{\dagger} Y_e M_{\tilde L}^2 
+2 h_e^{\dagger} h_e -
\end{displaymath}
\beq
- 6 g_2^2 |M_2|^2 -\frac{6}{5} g_1^2 |M_1|^2 - \frac{3}{5} g_1^2 S
\eeq
\begin{displaymath}
\beta_{m_{\tilde u}^2}^{(1)}=( 2 m_{\tilde u}^2 + 4 m_{H_u}^2) Y_u Y_u^{\dagger} 
+ 4 Y_u M_{\tilde Q}^2 Y_u^{\dagger} + 2 Y_u Y_u^{\dagger} m_{\tilde u}^2 + 
\end{displaymath}
\beq
+ 4 h_u h_u^{\dagger} - \frac{32}{3} g_3^2 |M_3|^2 -\frac{32}{15} g_1^2 |M_1|^2 -\frac{4}{5}g_1^2 S 
\eeq
\begin{displaymath}
\beta_{m_{\tilde d}^2}^{(1)}= ( 2 m_{\tilde d}^2 + 4 m_{H_d}^2) Y_d Y_d^{\dagger} 
+ 4 Y_d M_{\tilde Q}^2 Y_d^{\dagger} + 2 Y_d Y_d^{\dagger} m_{\tilde d}^2 + 4 h_d h_d^{\dagger}-
\end{displaymath}
\beq
-\frac{32}{3} g_3^2 |M_3|^2 - \frac{8}{15} g_1^2 |M_1|^2 + \frac{2}{5} g_1^2 S
\eeq
\begin{displaymath}
\beta_{m_{\tilde e}^2}^{(1)}= ( 2 m_{\tilde e}^2 + 4 m_{H_d}^2) Y_e Y_e^{\dagger} 
+ 4 Y_e M_{\tilde L}^2 Y_e^{\dagger} + 2 Y_e Y_e^{\dagger} m_{\tilde e}^2 
+ 4 h_e h_e^{\dagger}-
\end{displaymath}
\beq
-\frac{24}{5} g_1^2 |M_1|^2 + \frac{6}{5} g_1^2 S
\eeq
and where  
\beq
S= m_{H_u}^2 - m_{H_d}^2 + Tr \left( M_{\tilde Q}^2 - M_{\tilde L}^2 - 2 m_{\tilde u}^2 
+ m_{\tilde d}^2 + m_{\tilde e}^2 \right)
\eeq
The full two loop RG equations can be founds in Refs.\cite{Machacek:1983tz,Martin:1993zk}.
Now, let us list the mass matrices for the sparticles in the MSSM.
The chargino mass matrix was already given in Eq.(\ref{chmatrix}). 
For the  neutralino mass matrix one has 

\begin{displaymath}
M_{{\tilde \chi}^0}=\left(\matrix{M_1 & 0 & -M_Z s_W  c_\beta    & M_Z    s_W  s_\beta \cr
                 0  & M_2 & M_Z c_W   c_\beta & -M_Z c_W  s_\beta \cr
		 -M_Z s_W  c_\beta &  M_Z c_W c_\beta & 0 & -\mu \cr
		 M_Z s_W s_\beta  & -M_Z  c_W s_\beta & -\mu & 0}
			\right) \nonumber 
\end{displaymath}
\beq
\eeq
where $\theta_W$ is the weak angle, $s_W=\sin \theta_W$, $s_\beta= \sin\beta$, 
$c_\beta = \cos \beta$, and $s_\beta=\sin \beta$. 
The squark $(mass)^2$  matrix 
for $\tilde{u}$  at the electroweak scale is given by

\begin{displaymath}
M_{\tilde{u}}^2=\left(\matrix{M_{\tilde{Q}}^2+m{_u}^2 + M_{Z}^2(\frac{1}{2}-Q_u
s^2_W)\cos2\beta & m_u(A_{u}^{*} - \mu \cot\beta) \cr
   	          	m_u(A_{u} - \mu^{*} \cot\beta) & 
m_{\tilde{u}}^2+m{_u}^2 + M_{Z}^2 Q_u s^2_W \cos2\beta}
		\right) \nonumber
\end{displaymath}
\beq
\eeq
where $Q_u=\frac{2}{3}$,  and  the squark $(mass)^2$  matrix 
for $\tilde{d}$   is given by

\begin{displaymath}
M_{\tilde{d}}^2=\left(\matrix{M_{\tilde{Q}}^2+m{_d}^2-M_{Z}^2(\frac{1}{2}+Q_d
s^2_W)\cos2\beta & m_d(A_{d}^{*} - \mu \tan\beta) \cr
                        m_d(A_{d} - \mu^{*} \tan\beta) & m_{\tilde{d}
}^2+m{_d}^2+M_{Z}^2 Q_d s^2_W \cos2\beta}
                \right) \nonumber
\end{displaymath}
\beq
\eeq
We note that here we are using the relations $h_{u,d,e}=Y_{u,d,e} A_{u,d,e}$ 
for the trilinear terms. $Q_d=-\frac{1}{3}$. 
Finally, the slepton mass matrix is given by 

\begin{displaymath}
M_{\tilde {\it l}}^2=\left(\matrix{ M_{\tilde L}^2+m_{e}^2 -M_Z^2(\frac{1}{2}
- s^2_W)\cos 2\beta & m_{e}(A_{e}^* - \mu\tan\beta)\cr
m_{e}(A_{e} - \mu^*\tan\beta) &  m_{\tilde e}^2 + m_{e}^2 -M_Z^2 s^2_W
	\cos 2\beta  }
            \right)\nonumber
\end{displaymath}
\beq
\eeq
Further details of supersymmetry phenomenology can be found in 
Refs.~\cite{applied,Nilles:1983ge,Haber:1984rc,recent_susy}.
\section{Renormalization of the $d=5$ and $d=6$ operators} 
In this appendix we  discuss the renormalization effects for the 
$d=5$ and $d=6$ operators for proton decay. Typically
the $d=5$ effective operators are obtained at the GUT scale, once we integrate out 
the colored triplets. Before we dress those operators at the electroweak 
scale to obtain the $d=6$ effective operators, we have 
to run them from the GUT scale to the electroweak scale. After the dressing, 
we have to compute their coefficients at the proton decay scale $1$ GeV, 
and then use the Chiral Lagrangian technique to compute the lifetime for the 
different decay channels.    

The superpotential for the $d=5$ operators is 
\begin{eqnarray}
W_5 &=& \frac{1}{M_T} C_{L}^{ijkl} 
\ \epsilon_{abc} \ \epsilon_{\alpha \beta} \ \epsilon_{\gamma \delta} 
\ \hat{Q}_k^{a \alpha} \ \hat{Q}_l^{b \beta} \ 
\hat{Q}_i^{c \gamma} \ \hat{L}_{j}^{\delta} \ \nonumber\\
&+& \
\frac{1}{M_T} C_{R}^{ijkl} \ \epsilon_{abc} \ \hat{E}^C_k \ 
\hat{U}^C_{la} \ \hat{U}^C_{ib} \ \hat{D}^C_{jc} 
\end{eqnarray}   
where a, b, and c are the color indices. The coefficients $C_{L}$ 
and $C_{R}$ are functions of the Yukawa couplings and fermionic 
mixings at the GUT scale. Therefore in each model we have to find 
their expressions and values at the GUT scale and carry out the  RG 
evolution from the GUT scale down to the SUSY breaking scale~\cite{Goto:1998qg}. 

\begin{eqnarray}
(4 \pi)^2 \mu \frac{d}{d\mu} C_{L}^{ijkl} &=& 
(-8 g_3^2 -6 g_2^2 - \frac{2}{3} g_1^2) C_{L}^{ijkl} \nonumber\\
&+& C_{L}^{mjkl} (Y_D Y_D^{\dagger} + Y_U Y_U^{\dagger})^i_m \ + \
C_{L}^{imkl} (Y_L^{\dagger} Y_L)^j_m \nonumber\\
&+& C_{L}^{ijml} ( Y_D Y_D^{\dagger} + Y_U Y_U^{\dagger})^k_m 
\nonumber\\
&+& C_{L}^{ijkm} (Y_D Y_D^{\dagger} + Y_U Y_U^{\dagger})^l_m
\end{eqnarray}
and
\begin{eqnarray}
(4 \pi)^2 \mu \frac{d}{d\mu} C_{R}^{ijkl} &=& (-8 g_3^2 - 4 g_1^2) 
C_{R}^{ijkl} \ +\ C_{R}^{mjkl} ( 2 Y_U^{\dagger} Y_U)^i_m \ + \ \nonumber\\
&+&  C_{R}^{imkl} (2 Y_D^{\dagger} Y_D)^j_m \ + \ C_{R}^{ijml} 
(2 Y_L Y_L^{\dagger})^k_m \nonumber\\
&+& C_{R}^{ijkm} (2 Y_U^{\dagger} Y_U)^l_m
\end{eqnarray}
where $\mu$ is the renormalization scale, $Y_i$, and $g_i$ are the 
Yukawa matrices and gauge couplings.

Once we know  $C_{L}$ and $C_{R}$ at the electroweak scale, we can 
dress the $d=5$ operators. In order to estimate 
the value of the effective operators at the proton decay scale, we have 
to consider the long-range renormalization factor due to the QCD 
interaction between the SUSY scale ($m_{SUSY} \approx m_Z$) 
and the proton decay scale of 1 GeV. This factor is given by~\cite{Hisano:2000dg}:

\begin{equation}
A_L = \left(\frac{\alpha_s ( \mu_{had})}{ \alpha_s (m_b)}\right)^{6/25} 
\times 
\ \left(\frac{\alpha_s (m_b)}{\alpha_s (m_Z)}\right)^{6/23} \approx 1.4 
\end{equation} 
\\
\\
Long range effects  also  receive important  two loop QCD corrections 
which have  been computed  in ~\cite{Nihei:1994tx}. The reader is 
referred to this work for further details. 
In Sec.(3) we discussed  the most generic predictions 
for nucleon decay from the gauge $d=6$ operators. In this case 
proton decay is mediated by superheavy gauge bosons with mass $M_V$. 
Those effective operators are obtained at the GUT scale 
once the gauge bosons are integrated out. Since we have to 
compute the lifetime of the proton at 1 GeV, 
we have to carry out the RG evolution of these operators from the GUT scale 
to the electroweak scale and from the $M_Z$ scale to 1 GeV. In this case 
the effective $d=6$ operator will be multiply by a factor 
$A_R = A_R^{SD} A_L$, where the coefficient $A_{R}^{SD}$ is the 
short-distance renormalization factor which at one-loop 
(neglecting the flavour dependence of those operators) 
is given by~\cite{Hisano:2000dg}: 

\begin{equation}
A_R^{SD} = \left(\frac{\alpha_3 (m_Z)}{\alpha_{GUT}}\right)^{\frac{4}{3b_3}} \times
\left( \frac{\alpha_2 (m_Z)}{\alpha_{GUT}}\right)^{\frac{3}{2 b_2}} 
\end{equation}
where $b_3 = 9-2 n_g$ and $b_2 = 5- 2 n_g$ with $n_g$ is the number of 
families. $A_R^{SD} \approx 2.0$ if SUSY is the low energy effective 
theory below the GUT scale.

\section{Effective lagrangian for nucleon decay}

As discussed already in supersymmetric theories \bl ~violating 
dimension five operators must be  dressed by gluino, chargino 
and neutralino exchanges to produce dimension six operators. 
These operators are typically of two types: 
$\Delta S=0$ and $\Delta S=1$. 
The \bl violating  $\Delta S=0$ operators are:

\beqn
O_{RL}^{\nu_i}=\epsilon_{abc} \overline{(d_{Ra})^C} u_{Rb} \overline{(d_{Lc})^C} \nu_{iL} \nonumber\\
O_{LL}^{\nu_i}=\epsilon_{abc} \overline{(d_{La})^C} u_{Lb} \overline{(d_{Lc})^C} \nu_{iL} \nonumber\\
O_{RL}^{e_i}=\epsilon_{abc} \overline{(d_{Ra})^C} u_{Rb} \overline{(u_{Lc})^C}  e_{iL} \nonumber\\
O_{LR}^{e_i}=\epsilon_{abc} \overline{(d_{La})^C} u_{Lb} \overline{(u_{Rc})^C}  e_{iR} \nonumber\\
O_{LL}^{e_i}=\epsilon_{abc} \overline{(d_{La})^C} u_{Lb} \overline{(u_{Lc})^C}  e_{iL} \nonumber\\
O_{RR}^{e_i}=\epsilon_{abc} \overline{(d_{Ra})^C} u_{Rb} \overline{(u_{Rc})^C}  e_{iR} 
\label{sz}
\eeqn
In the above a,b,c =1,2,3 are the color indices, and i is the generation index. 
For the $|\Delta S|=1$ the \bl violating dimension six operators are:

\beqn
\tilde O_{RL1}^{\nu_i}=\epsilon_{abc} \overline{(s_{Ra})^C} u_{Rb} \overline{(d_{Lc})^C} \nu_{iL} \nonumber\\
\tilde O_{LL1}^{\nu_i}=\epsilon_{abc} \overline{(s_{La})^C} u_{Lb} \overline{(d_{Lc})^C} \nu_{iL} \nonumber\\
\tilde O_{RL}^{e_i}=\epsilon_{abc} \overline{(s_{Ra})^C} u_{Rb} \overline{(u_{Lc})^C}  e_{iL} \nonumber\\
\tilde O_{LR}^{e_i}=\epsilon_{abc} \overline{(s_{La})^C} u_{Lb} \overline{(u_{Rc})^C}  e_{iR} \nonumber\\
\tilde O_{LL}^{e_i}=\epsilon_{abc} \overline{(s_{La})^C} u_{Lb} \overline{(u_{Lc})^C}  e_{iL} \nonumber\\
\tilde O_{RR}^{e_i}=\epsilon_{abc} \overline{(s_{Ra})^C} u_{Rb} \overline{(u_{Rc})^C}  e_{iR}\nonumber\\ 
\tilde O_{RL2}^{\nu_i}=\epsilon_{abc} \overline{(d_{Ra})^C} u_{Rb} \overline{(s_{Lc})^C} \nu_{iL} \nonumber\\
\tilde O_{LL2}^{\nu_i}=\epsilon_{abc} \overline{(d_{La})^C} u_{Lb} \overline{(s_{Lc})^C} \nu_{iL} 
\label{s1}
\eeqn
 Eq.(\ref{sz}) and Eq.(\ref{s1}) contain all the possible type of dimension six operators, i.e., 
 of  chirality types  RRLL,  LLLL, LLRR, and  RRRR. 

In obtaining the above set of operators one uses a  Fierz reordering. This is best 
 accomplished by defining a set of 16 matrices as follows (see, e.g., ~\cite{Chattopadhyay:1998wb})
 \beq
\Gamma^A=\{1,\gamma^0,i\gamma^i,i\gamma^0\gamma_5,
\gamma^i\gamma_5,\gamma_5,i\sigma^{0i},\sigma^{ij}\}: ~~i,j=1-3
\eeq
which are normalized so that 
\beq
tr(\Gamma^A\Gamma^B)=4\delta^{AB}
\eeq
With the above definitions and normalizations,  the Fierz rearrangement formula 
takes on the  form
\beq
(\bar{u_1}\Gamma^Au_2)(\bar{u_3}\Gamma^Bu_4)=\sum_{C,D}F^{AB}_{CD}
(\bar{u_1}\Gamma^Cu_4)(\bar{u_3}\Gamma^Du_2)
\eeq
where   $u_j$ may be   Dirac or Majorana spinors and
\beq
F^{AB}_{CD}=
-(+)\frac{1}{16}tr(\Gamma^C\Gamma^A\Gamma^D\Gamma^B)
\eeq
In the above the  
+ve (-ve) sign is for commuting (anticommuting) u spinors.
 A -ve sign should be  chosen when  dealing with quantum Majorana 
 and Dirac fields in the Lagrangian. 
 
 The general Lagrangian with \bl ~violating dimension
 six operators will then have the form
 \begin{eqnarray}
 {\cal L_{BL}}&=& C_{RL}^{\nu_i} O_{RL}^{\nu_i} +  C_{LL}^{\nu_i}O_{LL}^{\nu_i} + C_{RL}^{e_i}O_{RL}^{e_i}
+ C_{LR}^{e_i} O_{LR}^{e_i} 
\nonumber\\
&+&  C_{LL}^{e_i}O_{LL}^{e_i}+ C_{RR}^{e_i} O_{RR}^{e_i} + \tilde C_{RL1}^{\nu_i} \tilde O_{RL1}^{\nu_i} 
\nonumber\\
&+& \tilde C_{LL1}^{\nu_i} \tilde O_{LL1}^{\nu_i} + \tilde C_{RL}^{e_i} \tilde O_{RL}^{e_i} + 
\tilde C_{LR}^{e_i} \tilde O_{LR}^{e_i} + \tilde C_{LL}^{e_i} \tilde O_{LL}^{e_i}+
\nonumber\\ 
&+& \tilde C_{RR}^{e_i} \tilde O_{RR}^{e_i} +\tilde C_{RL2}^{\nu_i} \tilde O_{RL2}^{\nu_i} + 
\tilde C_{LL2}^{\nu_i} \tilde O_{LL2}^{\nu_i} 
 \label{lagbl}
\end{eqnarray}
We note that in Eq.(\ref{lagbl}) the neutrinos $\nu_{i}$ are in mass diagonal state and hence are not
related by a simple $SU(2)_L$ symmetry to the corresponding operators with $e_{iL}$.
If we assume that the $\nu_i$ are flavor diagonal, then some of the co-efficients $C's$ and $\tilde C's$
can be related. Thus in this case $C_{RL}^{\nu_i}=-C_{RL}^{e_i}$, $C_{LL}^{\nu_i}=-C_{LL}^{e_i}$,
and $\tilde C_{RL1}^{\nu_i}=-\tilde C_{RL}^{e_i}$, $\tilde C_{LL1}^{\nu_i}=-\tilde C_{LL}^{e_i}$.
These reduce the number of independent couplings from six to four for the $\Delta S=0$ case
and from eight to six  for the $|\Delta S|=1$.
 The co-efficients $C_k^i, \tilde C_k^i$ are determined by the details of the underlying 
 GUT or string theory.  
 In trying to extract the physical implications of this interaction one uses the technique of 
 effective or  phenomenological Lagrangians~\cite{effec_lag}. Specifically 
 what one wishes to do is convert the above interaction which contains
 quarks and leptons into an interaction involving mesons, baryons and leptonic fields. 
 To this end it is useful to classify the operators according to their transformation properties
 under $SU(3)_L\times SU(3)_R$.  
While the analysis below follows closely the work of Ref.~\cite{Claudson:1981gh,Chadha:1983sj,Chadha:1983mh}
   it is more general.  First, we have not imposed any SU(2) symmetry on the operators in 
   Eq.(\ref{sz}) and Eq.(\ref{s1}) since one is below the electro-weak symmetry breaking 
scale where the  residual symmetry is only $SU(3)_C\times U(1)_{em}$. 
Secondly, in the analysis of Chadha and Daniel~\cite{Chadha:1983sj,Chadha:1983mh} only 
the chirality  LLLL type operators were considered in computing the decays. Specifically the
   LLRR and RRLL type operators were not fully included in the computation of proton decay rates.
   This was subsequently corrected in Ref.~\cite{Nath:1985ub}.
    In the following we will give  a full analysis including
   all four types  of operators, i.e., LLLL, RRLL, LLRR and RRRR (For a recent update 
see Ref.~\cite{Aoki:1999tw}). 
We give now the details of  the effective Lagrangian approach.  
  Noting that  $u_{La}, d_{La}$  transform like $3_L$, 
  $u_{Ra}, d_{Ra}$ transform like $3_R$, $3_L\times 3_L =3_L^* +6_L$, $3_L\times 3_L^*=8_L+ 1_L$ etc,
one finds the transformations of the operators  listed in Table (\ref{dim6trans}).  

 \begin{table}[h]
\begin{center}
\begin{tabular}{|r|r|r|}
\hline\hline
Dim 6 operator & Chirality type & Transformation \\
\hline
$ O_{RL}^{\nu_i}, O_{RL}^{e_i},
 \tilde O_{RL1}^{\nu_i}, \tilde O_{RL}^{e_i}, \tilde O_{RL2}^{\nu_i}$& RRLL &$ (3, 3^*)$\\
$ O_{LR}^{e_i}, \tilde O_{LR}^{e_i}$ &  LLRR & $(3^*, 3)$\\
$O_{LL}^{\nu_i}, O_{LL}^{e_i}, \tilde O_{LL1}^{\nu_i}, 
\tilde  O_{LL}^{e_i}, \tilde O_{LL2}^{e_i}$ & LLLL & (8,1)\\
$O_{RR}^{e_i}, \tilde O_{RR}^{e_i}$ & RRRR  & (1,8)\\
\hline
\hline
\end{tabular}
\end{center}
\caption{Properties  of the  dimension six operators  under $SU(3)_L\times SU(3)_R$.} 
\label{dim6trans}
\end{table}

To obtain the effective  Lagrangian for the operators  we  want to simulate  the transformations 
of Table (\ref{dim6trans}) using the baryon and  meson fields. For the baryons we introduce the
matrix
\begin{equation}
B=\sum_{a=1}^8\lambda_a B_a =\pmatrix{\offinterlineskip
 \frac{\Sigma^0}{\sqrt{2}}+\frac{\Lambda}{\sqrt{6}}&\Sigma^+&p\cr
\Sigma^-&     -\frac{\Sigma^0}{\sqrt{2}}+\frac{\Lambda}{\sqrt{6}}   &n\cr
\Xi^-& \Xi^0 & -\sqrt{\frac{2}{3}} \Lambda\cr}
\end{equation}
which transforms under $SU(3)_L\times SU(3)_R$ as  follows
\beqn
B'=UBU^{\dagger}
\eeqn
 while  the transformations of the pseudo-Goldstone  bosons are described as follows
 \beqn
 \xi'= L\xi U^{\dagger} =U\xi R^{\dagger}, ~~\xi =e^{iM/f}
 \eeqn
where
\beqn
M=    \sum_{a=1}^8 \lambda_a \phi_a=
\pmatrix{\offinterlineskip
 \frac{\pi^0}{\sqrt{2}}+\frac{\eta}{\sqrt{6}}&\pi^+&K^+\cr
\pi^-&     -\frac{\pi^0}{\sqrt{2}}+\frac{\eta}{\sqrt{6}}   &K^0\cr
K^-& \bar K^0 & -\sqrt{\frac{2}{3}} \eta\cr}
\eeqn 
Then under  $SU(3)_L\times SU(3)_R$ transformations we have:
\beqn
\xi B\xi \to L\xi B\xi R^{\dagger},~~
\xi^{\dagger}B\xi^{\dagger} \to R\xi^{\dagger} B \xi^{\dagger} L^{\dagger}\nonumber\\
\xi B \xi^{\dagger} \to L\xi B\xi^{\dagger} L^{\dagger},~~
\xi^{\dagger} B\xi \to R \xi^{\dagger} B\xi R^{\dagger}
\eeqn
The above transformations are of the type $(3,3^*)$, $(3^*,3)$, $(8,1)$, and $(1,8)$, 
respectively. However, we must use projection operators to precisely get 
the operators of type in Eq.(\ref{sz}) and Eq.(\ref{s1}). 
We can now  write the $O$  operators as follows~\cite{Claudson:1981gh,Chadha:1983sj}
\beqn
O_{RL}^{\nu_i}= \alpha\overline{(\nu_{iL})^C} Tr(P'\xi B_L \xi)\nonumber\\
O_{LL}^{\nu_i}= \beta\overline{(\nu_{iL})^C} Tr(P'\xi B_L \xi^{\dagger})\nonumber\\
O_{RL}^{e_i}= \alpha\overline{(e_{iL})^C} Tr(P\xi B_L \xi)\nonumber\\
O_{LR}^{e_i}=\alpha\overline{(e_{iR})^C} Tr(P\xi^{\dagger} B_R \xi^{\dagger})\nonumber\\
O_{LL}^{e_i}=\beta\overline{(e_{iL})^C} Tr(P\xi B_L \xi^{\dagger})\nonumber\\
O_{RR}^{e_i}=\beta\overline{(e_{iR})^C} Tr(P\xi^{\dagger} B_R \xi)
\eeqn
where
\beqn
P=\pmatrix{\offinterlineskip
 0 & 0&0\cr
 0 & 0& 0\cr
 1 & 0 & 0 \cr}, ~~
P'=\pmatrix{\offinterlineskip
 0 & 0&0\cr
 0 & 0& 0\cr
 0 & 1 & 0 \cr}
\eeqn 
and where $\alpha$ and $\beta$ are matrix elements of the three quark states between nucleon
and the vacuum state (see, e.g., Ref.~\cite{Aoki:1999tw})

\beqn
<0|\epsilon_{abc} \epsilon_{\alpha\beta}u^{\alpha}_{aR}d^{\beta}_{bR}u^{\gamma}_L|p>=\alpha  u^{\gamma}_L,\nonumber\\
<0|\epsilon_{abc} \epsilon_{\alpha\beta}u^{\alpha}_{aL}d^{\beta}_{bL}u^{\gamma}_L|p>=\beta u^{\gamma}_L
\label{matrixelements}
\eeqn 
$\alpha$ and $\beta$ are known to satisfy the constraint~\cite{Brodsky:1983st,Gavela:1988cp}
$|\alpha|=|\beta|$. 

Recent lattice QCD calculation of the proton decay matrix element by the CP-PACS and the
JLQCD Collaborations  gives~\cite{Tsutsui:2004qc}
\beqn
|\alpha|=0.0090(09)(^{+5}_{-19}) ~{\rm GeV}^3\nonumber\\
|\beta=0.0096(09)(^{+6}_{-20}) ~{\rm GeV}^3
\label{alphabeta}
\eeqn
where $\alpha$ and $\beta$ have a relatively opposite sign. In the above the first error  is statistical and the
second error systematic.   
The lattice analysis uses the quenched QCD calculation in the continuum limit  where the continuum operators
are defined in the naive dimensional regularization (NDR) with $\overline{MS}$ subtraction scheme.  
The evaluation of $\alpha$ and $\beta$ given above is at the scale $Q=2$ GeV.  
The result of Eq.(\ref{alphabeta}) is smaller the previous evaluation by the JLQCD Collaboration which 
gave~\cite{Aoki:1999tw} 
$|\alpha|=0.0151(1) ~{\rm GeV}^3$, 
$|\beta=0.014(1) ~{\rm GeV}^3$ 
where again the analysis is done using NDR and the evaluations are scale $Q=2.30(4)$ GeV. 
Further, one may compare the result of Eq.(\ref{alphabeta}) with the 
preliminary results of the RBC Collaboration at the scale $Q=1.23(5)$ GeV which gives
$|\alpha|=0.0061(1) ~{\rm GeV}^3$, 
$|\beta=0.007(1) ~{\rm GeV}^3$ and are about 30\% smaller than those of  Eq.(\ref{alphabeta}). 
The above difference cannot be accounted for by the renormalization group effects in going 
from the scale Q=1.23 GeV to the scale Q=2 GeV which gives  about 3.5\%  effect~\cite{Tsutsui:2004qc}.
It should be noted that the analysis of Eq.(\ref{alphabeta}) is about a factor of 3 larger than than
the early QCD calculations of these  matrix ~\cite{Donoghue:1982jm}.   Returning to 
the analysis of  Ref.~\cite{Tsutsui:2004qc} the relative sign 
between $\alpha$ and $\beta$ is important as it can affect very significantly 
the proton decay rates. Similarly we can write the $\tilde O$ operators as  follows 
\beqn
\tilde O_{RL1}^{\nu_i}=\alpha\overline{(\nu_{iL})^C} Tr( \tilde P'\xi B_L \xi)\nonumber\\
\tilde O_{LL1}^{\nu_i}=\beta\overline{(\nu_{iL})^C} Tr(\tilde P'\xi B_L \xi^{\dagger})\nonumber\\
\tilde O_{RL}^{e_i}=\alpha\overline{(e_{iL})^C} Tr(\tilde P\xi B_L \xi)\nonumber\\
\tilde O_{LR}^{e_i}=\alpha\overline{(e_{iR})^C} Tr( \tilde P\xi^{\dagger} B_R \xi^{\dagger})\nonumber\\
\tilde O_{LL}^{e_i}=\beta\overline{(e_{iL})^C} Tr(\tilde P\xi B_L \xi^{\dagger})\nonumber\\
\tilde O_{RR}^{e_i}=\beta\overline{(e_{iR})^C} Tr(\tilde P\xi^{\dagger} B_R \xi)\nonumber\\
\tilde O_{RL2}^{\nu_i}=\alpha\overline{(\nu_{iL})^C} Tr( \tilde P''\xi B_L \xi)\nonumber\\
\tilde O_{LL2}^{\nu_i}=\beta\overline{(\nu_{iL})^C} Tr(\tilde P''\xi B_L \xi^{\dagger})
\eeqn
where
\beqn
\tilde P=\pmatrix{\offinterlineskip
 0 & 0&0\cr
 -1 & 0& 0\cr
 0 & 0 & 0 \cr}, ~~
\tilde P'=\pmatrix{\offinterlineskip
 0 & 0&0\cr
 0 & -1& 0\cr
 0 & 0 & 0 \cr}, ~~
 \tilde P''=\pmatrix{\offinterlineskip
 0 & 0&0\cr
 0 & 0& 0\cr
 0 & 0 & 1 \cr}
\eeqn 
In extracting the \bl violating parts from Eq.(\ref{lagbl}) we must compute both the quadratic and
a cubic term. The quadratic part is easily extracted. It is  
 \begin{eqnarray}
 {\cal L_{BL}}^{(2)} &=& (\alpha C_{RL}^{\nu_i}+\beta C_{LL}^{\nu_i})\overline{(\nu_{iL})^C} n_L + 
 (\alpha C_{RL}^{e_i} + \beta C_{LL}^{e_i})\overline{(e_{iL})^C} p_L + 
\nonumber\\
&& (\alpha C_{LR}^{e_i}+\beta C_{RR}^{e_i})\overline{(e_{iR})^C} p_R 
+ (\alpha \tilde C_{RL1}^{\nu_i}+\beta \tilde C_{LL1}^{\nu_i})\overline{(\nu_{iL})^C} 
(\frac{\Sigma^0_L }{\sqrt 2} -\frac{\Lambda^0_L}{\sqrt 6})
\nonumber\\
 &-& (\alpha\tilde C_{RL}^{e_i}+\beta\tilde C_{LL}^{e_i})\overline{(e_{iL})^C}\Sigma_L^+ 
\nonumber\\
 &-& (\alpha\tilde C_{LR}^{e_i}+\beta\tilde C_{RR}^{e_i})\overline{(e_{iR})^C} \Sigma_R^+
-(\alpha\tilde C_{RL2}^{\nu_i}
+ \beta\tilde C_{LL2}^{\nu_i}) \sqrt {\frac{2}{3}} \overline{(\nu_{iL})^C}\Lambda_L^0 
\nonumber\\
\end{eqnarray}
while the \bl violating cubic interaction is
 \begin{eqnarray}
 {\cal L_{BL}}^{(3)} &=& \frac{i}{f}  \{
     \alpha C_{RL}^{\nu_i} ( \overline{(\nu_{iL})^C} p_L\pi^- 
 -  \overline{(\nu_{iL})^C} n_L( \frac{\pi^0}{\sqrt 2} + \frac{\eta}{\sqrt 6}) )  
\nonumber\\ 
&+&\beta C_{LL}^{\nu_i} 
 ( \overline{(\nu_{iL})^C} p_L\pi^- 
 +  \overline{(\nu_{iL})^C} n_L( -\frac{\pi^0}{\sqrt 2} + \frac{3}{\sqrt 6}\eta) )  
 \nonumber\\
 &+& \alpha  C_{RL}^{e_i} 
 ( \overline{(e_{iL})^C} n_L\pi^+ 
 +  \overline{(e_{iL})^C} p_L( \frac{\pi^0}{\sqrt 2} - \frac{1}{\sqrt 6}\eta) ) 
\nonumber\\ 
&-& \alpha C_{LR}^{e_i} 
 ( \overline{(e_{iR})^C} n_R\pi^+ 
 +  \overline{(e_{iR})^C} p_R( \frac{\pi^0}{\sqrt 2} - \frac{1}{\sqrt 6}\eta) ) 
 \nonumber\\
 &+& \beta C_{LL}^{e_i} 
 ( \overline{(e_{iL})^C} n_L\pi^+ 
 +  \overline{(e_{iL})^C} p_L( \frac{\pi^0}{\sqrt 2} + \frac{3}{\sqrt 6}\eta)  )
\nonumber\\ 
&-& \beta C_{RR}^{e_i} 
 ( \overline{(e_{iR})^C} n_R\pi^+ 
 +  \overline{(e_{iR})^C} p_R( \frac{\pi^0}{\sqrt 2} + \frac{3}{\sqrt 6}\eta)  )
\nonumber\\
 &+&  (-\alpha \tilde C_{RL1}^{\nu_i} +\beta \tilde C_{LL1}^{\nu_i})  \overline{(\nu_{iL})^C} n_L \bar K^0 
+ ( -\alpha \tilde C_{RL}^{e_i} 
\nonumber\\
&+& \beta \tilde C_{LL}^{e_i})  \overline{(e_{iL})^C} p_L \bar K^0
  + ( \alpha \tilde C_{LR}^{e_i} - \beta \tilde C_{RR}^{e_i})  \overline{(e_{iR})^C}p_R \bar K^0
 \nonumber\\
 &+& (\alpha\tilde C_{RL2}^{\nu_i} +\beta \tilde C_{LL2}^{\nu_i}) 
 ( \overline{(\nu_{iL})^C} n_L \bar K^0  + \overline{(\nu_{iL})^C} p_L  K^-) 
   \} + h.c. \nonumber\\
\end{eqnarray}

In addition there are baryon number conserving interactions. 
The relevant terms are~\cite{Chadha:1983sj}:
\begin{eqnarray}
 {\cal L_{BC}}&=& \frac{1}{2i} (D-F) Tr[ \bar B \gamma^{\mu} \gamma_5 B
 \{\partial_{\mu} \xi \xi^{\dagger} 
\nonumber\\
&-& \partial_{\mu}\xi^{\dagger} \xi\}]
- \frac{1}{2i} (D+F) Tr[ \bar B \gamma^{\mu} \gamma_5 
 \{\xi \partial_{\mu}\xi^{\dagger} - \xi^{\dagger} \partial_{\mu}\xi \}B]
\nonumber\\
&+& b_1 Tr[ \bar B \gamma_5(\xi^{\dagger} m\xi^{\dagger} -\xi  m\xi) B]
+b_2 Tr[ \bar B \gamma_5 B(\xi^{\dagger} m\xi^{\dagger} -\xi  m\xi) ]
\nonumber\\
\end{eqnarray}
where m is quark mass matrix  such that $ m=diag (m_u, m_d, m_s)$. 
In the above the terms with co-efficients ($D\pm F$) are invariant under $SU(3)_L\times SU(3)_R$
while the terms with co-efficients $b_1, b_2$ transform like $(3,3^*)+(3^*,3)$ and break 
$SU(3)_L\times SU(3)_R$ down to $SU(3)_V$. The matrix $\bar B$ is defined so that

\begin{equation}
\bar B=\pmatrix{\offinterlineskip
 \frac{\bar \Sigma^0}{\sqrt{2}}+\frac{\bar \Lambda}{\sqrt{6}}&\bar \Sigma^-&- \bar\Xi^-\cr
\bar\Sigma^+&     -\frac{\bar\Sigma^0}{\sqrt{2}}+\frac{\bar\Lambda}{\sqrt{6}}   & \bar \Xi^0\cr
\bar p & \bar n & -\sqrt{\frac{2}{3}} \bar\Lambda\cr}
\end{equation}
The relevant part of  $ {\cal L_{BC}} $ is 
\beqn
{\cal L_{BC}} = \left(  \frac{D-F}{\sqrt 2 f} \bar \Sigma^0 \gamma^{\mu} \gamma_5 p 
-\frac{D+3F}{\sqrt 6 f}  \bar \Lambda^0 \gamma^{\mu} \gamma_5 p
+\frac{D-F}{\sqrt 2 f} \bar \Sigma^- \gamma^{\mu} \gamma_5 n\right) \partial_{\mu}K^- 
\nonumber\\
+\left(\frac{D-F}{ f} \bar \Sigma^+ \gamma^{\mu} \gamma_5 p 
-\frac{D-F}{\sqrt 2 f}  \bar \Sigma^0 \gamma^{\mu} \gamma_5 n
-\frac{D+3F}{\sqrt 6 f} \bar \Lambda^0 \gamma^{\mu} \gamma_5 n\right) \partial_{\mu}\bar K^0 
\nonumber\\
+\frac{i}{f} (m_u+m_s) \left( \sqrt{\frac{2}{3}} (2b_1-b_2) \bar \Lambda^0\gamma_5p 
-\sqrt 2 b_2 \bar \Sigma^0\gamma_5 p +2b_2 \bar \Sigma^-\gamma_5 n \right)K^- 
\nonumber\\
+
\frac{i}{f} (m_d+m_s) \left(
 - 2b_2  \bar \Sigma^+\gamma_5p 
+\sqrt{\frac{2}{3}} (2b_1-b_2)  \bar \Lambda^0\gamma_5 n +
\sqrt 2 b_2 \bar \Sigma^0\gamma_5 n \right) \bar K^0  + h.c. \nonumber\\
\eeqn
The contribution of the $m_u, m_s$ terms is typically small and as is 
conventional we neglect them from here on. For simplicity we introduce the notation
\beqn
C_{RL}^{'\nu_i,e_i }=\alpha C_{RL}^{\nu_i,e_i}, ~C_{LL}^{'\nu_i,e_i}=\beta C_{LL}^{\nu_i,e_i},
~C_{LR}^{'e_i}=\alpha C_{LR}^{e_i}, ~C_{RR}^{'e_i}=\beta C_{RR}^{e_i},\nonumber\\
\tilde C_{RL1}^{'\nu_i}=\alpha C_{RL1}^{\nu_i}, 
~\tilde C_{LL1}^{'\nu_i}=\beta C_{LL1}^{\nu_i},
\tilde C_{RL}^{'e_i}=\alpha C_{R}^{e_i},
~\tilde C_{LR}^{'e_i}=\alpha \tilde C_{LR}^{e_i},
\nonumber\\
~\tilde C_{LL}^{'e_i}=\beta C_{LL}^{e_i},
~\tilde C_{RR}^{'e_i}=\beta \tilde C_{RR}^{e_i},
\tilde C_{RL2}^{'\nu_i}=\alpha C_{RL2}^{\nu_i}, 
~\tilde C_{LL2}^{'\nu_i} =\beta C_{LL2}^{\nu_i},
\label{cprime}
\eeqn

\noindent
We discuss now the decay widths for the various decay modes.

 \noindent
 (i) $\underline{p \to \bar\nu_i K^+}$  decay 

\begin{eqnarray}
\Gamma(p\to \bar\nu_i K^+) &= & (32\pi f^2 m_N^3)^{-1} (m_N^2-m_K^2)^2 
|(\tilde C_{LL2}^{'\nu_i}+\tilde C_{RL2}^{'\nu_i}) 
\nonumber\\
&+& \frac{m_N}{2m_{\Sigma^0}} ( \tilde C_{LL1}^{'\nu_i}+ 
\tilde C_{RL1}^{'\nu_i}) (D-F)
\nonumber\\
&+& \frac{m_N}{6m_{\Lambda}}  \{ \tilde C_{LL1}^{'\nu_i}+   \tilde C_{RL1}^{'\nu_i}+2 
(\tilde C_{LL2}^{'\nu_i}+\tilde C_{RL2}^{'\nu_i}) \} (D+3F)|^2 \nonumber\\
\end{eqnarray}
In the above and in the following the $C's$  
and  $\tilde C's$ are as defined in Eq.(\ref{cprime}).\\

\noindent
(ii) $\underline{n \to \bar\nu_i K^0}$ decay 
\begin{eqnarray}
\Gamma(n\to \bar\nu_i K^0) &=& (32\pi f^2 m_N^3)^{-1} (m_N^2-m_K^2)^2 
|( -\tilde C_{RL1}^{'\nu_i}+
\tilde C_{LL1}^{'\nu_i} +
\tilde C_{RL2}^{'\nu_i}+\tilde C_{LL2}^{'\nu_i})  
\nonumber\\
&-& \frac{1}{2} \frac{m_N}{m_{\Sigma^0}}  (\tilde C_{RL1}^{'\nu_i}+\tilde C_{LL1}^{'\nu_i})(D-F) 
+\frac{m_N}{6m_{\Lambda} }(\tilde C_{RL1}^{'\nu_i}+\tilde C_{LL1}^{'\nu_i}  
\nonumber\\
&+& 2\tilde C_{RL2}^{'\nu_i}+2\tilde C_{LL2}^{'\nu_i}) (D+3F) |^2
\end{eqnarray}
 
\noindent
 (iii) $\underline{p \to  l_i^+ K^0}$ decay
\begin{eqnarray}
\Gamma(p \to l_i^+ K^0) &=& (32\pi f^2 m_N^3)^{-1} (m_N^2-m_K^2)^2 
\{\frac{1}{2} [-\tilde  C_{RL}^{'e_i}+\tilde C_{LL}^{'e_i}+\tilde C_{LR}^{'e_i}
\nonumber\\
&-& \tilde C_{RR}^{'e_i} - \frac{m_N}{m_{\Sigma}} 
(\tilde C_{RL}^{'e_i} +\tilde C_{LL}^{'e_i} -\tilde C_{LR}^{'e_i} 
-\tilde C_{RR}^{'e_i})(D-F)]^2+
\nonumber\\
&+& \frac{1}{2} [-\tilde  C_{RL}^{'e_i}+\tilde C_{LL}^{'e_i}-\tilde C_{LR}^{'e_i}+\tilde C_{RR}^{'e_i} 
\nonumber\\
&-& \frac{m_N}{m_{\Sigma}} (\tilde C_{RL}^{'e_i} +\tilde C_{LL}^{'e_i} +\tilde C_{LR}^{'e_i} 
+\tilde C_{RR}^{'e_i})(D-F)]^2\} \nonumber\\
\end{eqnarray}

\noindent
(iv) $\underline{ p \to  \bar \nu_i \pi^+}$ decay
\beqn
\Gamma(p \to \bar \nu_i \pi^+) =(32\pi f^2 m_N^3)^{-1} (m_p^2-m_{\pi^+}^2)^2 
|C_{RL}^{'\nu_i} +C_{LL}^{'\nu_i}|^2(1+D+F)^2 \nonumber\\
\eeqn

\noindent
 (v) $\underline{ n \to  \bar\nu_i \pi^0}$  decay
\beqn
\Gamma(n\to \bar\nu_i \pi^0) =(32\pi f^2 m_N^3)^{-1} (m_n^2-m_{\pi^0}^2)^2 
 \frac{1}{2}  |C_{RL}^{'\nu_i} +C_{LL}^{'\nu_i}|^2
 (1+D+F)^2 \nonumber\\
\eeqn
Neglecting the  mass differences  of p and n and  of $\pi^+$ and $\pi^0$ we get
\beqn
\Gamma(n\to \bar \nu_i \pi^0)\simeq 0.5 \Gamma(p\to \bar \nu_i  \pi^+)
\eeqn

\noindent
(vi) $\underline{ n \to  \bar\nu_i \eta^0}$ decay
\beqn
\Gamma(n\to \bar\nu_i \eta^0) =(32\pi f^2 m_N^3)^{-1} (m_n^2-m_{\eta^0}^2)^2 \times
\nonumber\\
 \frac{3}{2}  | C_{RL}^{'\nu_i}(-\frac{1}{3} -\frac{D}{3}+F) +  
 C_{LL}^{'\nu_i} (1-\frac{D}{3} +F)|^2
\eeqn

\noindent
(vii) $\underline{p \to  e_i^+ \pi^0}$  decay
\beqn
\Gamma(p\to e_i^+ \pi^0) =(32\pi f^2 m_N^3)^{-1} (m_n^2-m_{\pi^0}^2)^2 \times
\nonumber\\
 \frac{1}{2}  \left(  | C_{RL}^{'e_i} +C_{LL}^{'e_i}|^2 +|C_{LR}^{'e_i}+C_{RR}^{'e_i}|^2  \right) 
(1+D+F)^2 \nonumber\\
\eeqn

\noindent
(viii)  $\underline{p \to  e_i^+ \eta^0}$   decay
\beqn
\Gamma(p\to e_i^+ \eta^0) =(32\pi f^2 m_N^3)^{-1} (m_n^2-m_{\eta^0}^2)^2 
\times
\nonumber\\
 \frac{3}{2}
 \{ [ C_{LL}^{'e_i}(1-\frac{D}{3} +F) +
 C_{RL}^{'e_i}(-\frac{1}{3}-\frac{D}{3} +F)]^2 
 \nonumber\\
  + [ C_{RR}^{'e_i}(1-\frac{D}{3} +F) +
 C_{LR}^{'e_i}(-\frac{1}{3}-\frac{D}{3} +F)]^2 \}
\eeqn

\noindent
 (ix) $\underline{n \to  e_i^+ \pi^-}$ decay   
\beqn
\Gamma(n\to e_i^+ \pi^-) =(32\pi f^2 m_N^3)^{-1} (m_n^2-m_{\pi^-}^2)^2 
 \times 
\nonumber\\
\frac{3}{2}
 \left( |C_{RL}^{'e_i}+C_{LL}^{'e_i}|^2+|C_{LR}^{'e_i}+C_{RR}^{'e_i}|^2 \right)  (1+D+F)^2 
 \eeqn
 Currently the effective lagrangian approach is the most reliable 
approach to the computation of proton decay amplitudes. 
Numerically $f=139$ MeV, while F and D can be gotten from a recent analysis  of  hyperon decays
which gives\cite{Cabibbo:2003cu} 
\beqn
F+D=1.2670\pm 0.0030, ~~F-D=-0.341\pm 0.016
\eeqn
As we already discussed, in order to compute the lifetime of the proton we have to take into account the renormalization effects from the GUT scale to 1 GeV. 
In the previous appendix we already discussed those effects for the $d=5$ and 
$d=6$ operators.

The chiral Lagrangian approach is currently the best available technique for the 
analysis of proton decay lifetime starting with the fundamental  Lagrangian  in terms of 
quark fields. The approach~\cite{effec_lag} had considerable success 
in the past but it has certain limitations. The technique works best in the so called 
soft pion approximation. In the present context,  the mesons  in  proton decay 
may have energies as large as 500 MeV so a certain extrapolation is 
necessary.  However, the technique  still remains the state of the art in the 
computation of proton decay life times.

\section{Details of the analysis on testing GUTs }
In Sec.(5.4) we discussed the discussed  the tests on $SU(5)$ models
with symmetric up Yukawa couplings. In this appendix we expand on those tests
to include other groups.  These are discussed below.\\

\noindent
(i) $SO(10)$ models with symmetric Yukawa couplings\\
Next we investigate the predictions in realistic grand unified theories 
based on the gauge group $SO(10)$~\cite{SO(10)} with symmetric 
Yukawa couplings. This is the case of $SO(10)$ theories 
with two Higgses $10_H$ and $126_H$.  
In these theories with symmetric Yukawa couplings we get 
the following relations for the mixing matrices, 
$U_C = U K_u$, $D_C = D K_d$ and $E_C = E K_e$, where 
$K_d$ and $K_e$ are diagonal matrices containing 
three phases. In those cases $V_1 = K^*_u$, 
$V_2= K^*_e V^{\dagger}_{DE}$, $V_3=K^*_d V_{DE}$ and 
$V_4=K^*_d$. Using these relations the coefficients in Eqs.(~\ref{cec}-\ref{cnuc}) 
are given by~\cite{FileviezPerez:2004hn}:

\begin{eqnarray}
\label{cecs}
c(e^C_{\alpha}, d_{\beta})_{sym}&=& (K_u^*)^{11}(K_e^*)^{\alpha
\alpha}
[ \delta^{\beta i} + V^{1 \beta}_{CKM} K_2^{\beta \beta} (K_2^*)^{ii}
(V^{\dagger}_{CKM})^{i1}  ] (V^*_{DE})^{i \alpha} \nonumber \\ 
\end{eqnarray}
\begin{eqnarray}
\label{ces}
c(e_{\alpha}, d_{\beta}^C)_{sym} 
&=& (K^*_u)^{11}(K^*_d)^{\beta \beta} \times \nonumber \\
&\times& [k_1^2 \delta^{\beta i} + k_2^2 (K_2^*)^{\beta
\beta}(V^{\dagger}_{CKM})^{\beta 1} V^{1i}_{CKM} K_2^{ii} ](V^{i
\alpha}_{DE}) \nonumber \\ 
\end{eqnarray}

\begin{eqnarray}
\label{cnus} 
c(\nu_l, d_{\alpha}, d^C_{\beta})_{sym}&=& (K_u^*)^{11}K_1^{11} \times \nonumber \\
&\times& [ k_1^2  \delta^{\alpha i} \delta^{\beta j} +
k_2^2 \delta^{\alpha \beta} \delta^{ij} (K_d^*)^{\alpha \alpha}
K_d^{ii}](V_{CKM} K_2)^{1i} (K_d^* V_{DE} V_{EN})^{jl} \nonumber \\ 
\end{eqnarray}

\begin{eqnarray}
\label{cnucs}
c(\nu_l^C, d_{\alpha}, d^C_{\beta})_{sym}&=& (K^*_d)^{\beta
\beta}(K_1^*)^{11} \times \nonumber \\ 
&\times& [ (K^*_2)^{\beta \beta} (V^{\dagger}_{CKM})^{\beta
1} \delta^{\alpha i} + \delta^{\alpha \beta} (K_2^*)^{ii}
(V^{\dagger}_{CKM})^{i1}](U_{EN}^{\dagger} K_e^*
V_{DE}^{\dagger})^{li} \nonumber \\
\end{eqnarray}
with $\alpha = \beta \neq 2\nonumber$.
Notice all overall phases in the different coefficients. In order to
compute the decay rate into antineutrinos we need the following expression:

\begin{displaymath}
\sum_{l=1}^3 c(\nu_l, d_{\alpha}, d_{\beta})_{sym}^* c(\nu_l, d_{\gamma},
d_{\delta})_{sym} = [ k_1^2 \delta^{\alpha i} \delta^{\beta j} + k_2^2
\delta^{\alpha \beta} \delta^{ij} K_d^{\alpha \alpha}
(K_d^*)^{ii}] \times
\end{displaymath}
\begin{equation}
\times [ k_1^2 \delta^{\gamma i^{'}} \delta^{\delta j} + k_2^2 \delta^{\gamma
\delta} \delta^{i^{'} j}(K_d^*)^{\gamma \gamma} K_d^{i^{'} i^{'}}
](V^*_{CKM} K_2^*)^{1i} (V_{CKM} K_2)^{1i^{'}}
\end{equation}
Using the above expression we find that it
is possible to determine the factor $k_1= g_{GUT}/ \sqrt{2} M_{(X,Y)}$ 
so that~\cite{FileviezPerez:2004hn}:

\begin{eqnarray}
\label{k1}
k_1&=& \frac{Q_1^{1/4}}{[ \left|A_1\right|^2
\left|V_{CKM}^{11}\right|^2+ \left|A_2\right|^2
\left|V_{CKM}^{12}\right|^2]^{1/4} } 
\end{eqnarray} 
where:
\begin{eqnarray}
Q_1&=& \frac{8 \pi m_p^3 f^2_{\pi} \Gamma(p \to K^+ \bar{\nu})}
{(m_p^2 - m_K^2)^2 A_L^2 \left|\alpha \right|^2}\\
A_1&=& \frac{2 m_p }{3 m_B} D\\
A_2&=& 1 + \frac{m_p}{3 m_B}(D + 3F)
\end{eqnarray}
Here one finds that the amplitude for the decay $p \to K^+ \bar{\nu}$ 
is independent of all unknown mixing and phases, and 
only depends on the factor $k_1$. Thus it appears possible to test
any grand unified theory with symmetric Yukawa matrices through this decay
mode. Once $k_1$ is known $k_2$ can be gotten by solving the 
following equation~\cite{FileviezPerez:2004hn}:

\begin{eqnarray}
\label{k2}
k_2^4 + 2 k_2^2 k_1^2 \left|V_{CKM}^{11}\right|^2 + k_1^4
\left|V^{11}_{CKM}\right|^2 - \frac{8 \pi f^2_{\pi} \Gamma(p \to \pi^+ \bar{\nu})}
{m_p A_L^2 \left|\alpha \right|^2 (1+D+F)^2} &=&0 \nonumber \\
\end{eqnarray}
which gives
\begin{eqnarray}
\label{k2s}
k_2 &=&k_1 \left|V_{CKM}^{11}\right| \{ - 1 + \sqrt{Q_2} \}^{1/2} 
\end{eqnarray}
where
\begin{eqnarray}
Q_2&=& 1 + \frac{8 \pi f_{\pi}^2 \Gamma( p \to \pi^+ \bar{\nu})}{
k_1^4 \left|V_{CKM}^{11}\right|^4 m_p A^2_L \left|\alpha\right|^2 (1+D+F)^2}- \left|V_{CKM}^{11}\right|^{-2}
\end{eqnarray}
Using the condition $ Q_2 > 1$, we get the following relation
\begin{eqnarray}
\label{R1}
\frac{\tau(p \to K^+
\bar{\nu})}{\tau(p \to \pi^+ \bar{\nu})}&>&\frac{m_p^4 \left|V_{CKM}^{11}\right|^2 (1+D+F)^2}{(m_p^2 -m_K^2)^2 [\left|A_1\right|^2 \left|V_{CKM}^{11}\right|^2+ \left|A_2\right|^2 \left|V_{CKM}^{12}\right|^2 ]}
\end{eqnarray}
The  above is a clean prediction of a GUT model with symmetric Yukawa couplings. 
The relations among the nucleon decays read as follows:
\begin{eqnarray}
\label{R2}
\frac{\tau(n \to K^0 \bar{\nu})}{\tau(p \to K^+ \bar{\nu})}&=&
\frac{m_n^3 (m_p^2 -m_K^2)^2 [\left|A_1\right|^2
\left|V_{CKM}^{11}\right|^2+ \left|A_2\right|^2
\left|V_{CKM}^{12}\right|^2 ]}{m_p^3 (m_n^2 -m_K^2)^2 [\left|A_3
\right|^2 \left|V_{CKM}^{11}\right|^2+ \left|A_2\right|^2
\left|V_{CKM}^{12}\right|^2 ]} \nonumber \\ \\
\label{R3}
\frac{\tau(n \to \pi^0 \bar{\nu})}{\tau(p \to \pi^+ \bar{\nu})}&=&
\frac{2 m_p}{m_n}\\
\label{R4}
\frac{\tau(n \to \eta^0 \bar{\nu})}{\tau(p \to \pi^+ \bar{\nu})}&=&
\frac{6 m_p m_n^3 (1+D+F)^2}{(m_n^2-m_{\eta}^2)^2 (1-D-3F)^2}
\end{eqnarray}
with
\begin{eqnarray}
A_3= 1 + \frac{m_n}{3 m_B}(D-3F)
\end{eqnarray}
Thus using the expressions for $k_1$ and $k_2$ 
(Eqs.~\ref{k1} and~\ref{k2s}), and the relation among the different 
decay rates of the neutron and the proton into an antineutrino 
(Eqs.~\ref{R1}-~\ref{R4}), it is possible to make a clear test of 
a grand unified theory with symmetric Yukawa couplings.  

Next we look at the predictions for the proton decay into charged antileptons. 
To write the decay rate for these modes we need the following expression: 

\begin{eqnarray}
\label{CL}
\sum_{\alpha =1}^2 c(e^C_{\alpha}, d_{\beta})^*_{sym}
c(e^C_{\alpha}, d_{\gamma})_{sym}
&=& 
[ \delta^{\beta i} + V^{1 \beta}_{CKM} K_2^{\beta \beta} (K_2^*)^{ii}
(V^{\dagger}_{CKM})^{i1}  ] \nonumber\\
&&[ \delta^{\gamma j} + V^{1 \gamma}_{CKM} K_2^{\gamma \gamma} (K_2^*)^{jj}
(V^{\dagger}_{CKM})^{j1} ] \times \nonumber \\
&\times& \sum^2_{i=1} V^{i \alpha}_{DE} (V^{j \alpha}_{DE})^{*}
\end{eqnarray}
Thus the decay of the channels with charged antileptons 
always depend on the matrices $K_2$ and $V_{DE}$. In the theories 
with the $10_H$ and/or $126_H$ Higgses there is a specific 
expression for the matrix $V_{DE}$: 

\begin{eqnarray}
\label{VDE}
4 V^T_{UD} K^*_u Y^{diag}_U V_{UD}-(3 \tan \alpha_{10} + \tan
\alpha_{126}) K^*_d Y^{diag}_D&=& \nonumber \\ 
V^*_{DE}K^*_e Y^{diag}_E
V^{\dagger}_{DE} ( \tan \alpha_{10}- \tan \alpha_{126}) &&\nonumber \\
\end{eqnarray}
where $\tan \alpha_{10}= v^{U}_{10} / v^{D}_{10}$,
and $\tan \alpha_{126}= v^{U}_{126} / v^{D}_{126}$. Here 
we see explicitly the relation among the different factors
entering in the proton decay predictions. Thus in this case
 it is very difficult to get clean predictions from those 
channels. However, these relations are still very useful 
as they allow on to distinguish among different models for 
the fermion masses.\\  

\noindent
(ii) Renormalizable flipped $SU(5)$ models\\

As is well known the electric charge is a generator 
of conventional $SU(5)$. However, it is possible to embed 
the electric charge in such a manner that it is a linear 
combination of the generators operating in both $SU(5)$ and an
extra $U(1)$, and still reproduce the SM charge assignment. 
This is exactly what is done in  flipped 
$SU(5)$~\cite{DeRujula:1980qc,Barr:1981qv,Derendinger:1983aj,Antoniadis:1989zy}. 
The matter now unifies in a different manner, which can be obtained
from the $SU(5)$ assignment by a flip: $d^C \leftrightarrow u^C$,
$e^C \leftrightarrow \nu^C$, $u \leftrightarrow d$ and
$\nu \leftrightarrow e$. 
In the case of flipped $SU(5)$ the gauge bosons responsible for
proton decay are: $(X', Y')=({{\bf 3}},{\bf 2},-1/3)$. 
The electric charge of $Y'$ is $-2/3$, while
$X'$ has the same charge as $Y$. Since the gauge sector and the
matter unification differ from $SU(5)$ case, the proton decay
predictions are also different~\cite{Barr:1981qv}.

Flipped $SU(5)$ is well motivated from string theory scenarios, since
one does not need large representations to achieve the GUT symmetry
breaking \cite{Antoniadis:1989zy}.
Another nice feature of flipped $SU(5)$ is that the dangerous $d=5$
operators are suppressed due to an extremely economical missing
partner mechanism. 
In renormalizable flipped $SU(5)$ one has $Y_D=Y_D^T$, so 
$D_C = D K_d$. In this case the coefficients entering  the proton
decay predictions are~\cite{Dorsner:2004xx}:
\\

\begin{eqnarray}
\label{sumflippedSU(5)1} \sum_{l=1}^3 c(\nu_l, d_{\alpha},
d^C_{\beta})_{SU(5)'}^*  \ c(\nu_l, d_{\gamma},
d^C_{\delta})_{SU(5)'} & = & k_2^4 K_d^{\beta \beta} \delta^{\beta
\alpha} (K_d^*)^{\delta \delta} \delta^{\delta \gamma}
\\ \nonumber\\
\left|c(e_{\alpha}, d^C_{\beta})\right|^2=k_2^4 \left|V_{CKM}^{1
\beta}\right|^2 \left|(V_1 V_{UD} V^\dagger_4 V_3)^{1
\alpha}\right|^2 & = & k_2^4 \left|V_{CKM}^{1 \beta}\right|^2
\left|(U_C^\dagger E)^{1 \alpha}\right|^2 \nonumber \\
\end{eqnarray}

Using these equations one gets the following relations~\cite{Dorsner:2004xx}:

\begin{eqnarray}
\label{xxx5}
\Gamma(p \rightarrow \pi^+ \bar{\nu})&=& k_2^4 \ C_2
\\ \nonumber\\
\label{flippedpi0}
\Gamma(p \rightarrow \pi^0 e_{\alpha}^+)&=&\frac{1}{2} \ \Gamma(p
\rightarrow \pi^+\bar{\nu}) \ \left|V_{CKM}^{1 1}\right|^2
\left|(U_C^\dagger E)^{1 \alpha}\right|^2
\\ \nonumber\\
\label{flippedK0}
\frac{\Gamma (p \to K^0 e_{\alpha}^+)}
{\Gamma (p \to \pi^0 e_{\alpha}^+)}
        &=& 2 \frac{C_3}{C_2} \ \frac{\left|V_{CKM}^{12}\right|^2}
{\left|V_{CKM}^{11}\right|^2}
\end{eqnarray}

where:

\begin{equation}
C_3 = \frac{(m_p^2-m_K^2)^2}{8 \pi f_\pi^2 m_p^3}  A_L^2
        \left|\alpha\right|^2 \left[1+{\frac{m_p}{m_B}} (D-F)\right]^2
\end{equation}

We note that in this case, $\Gamma(p \to K^+\bar{\nu})= 0$, and $\Gamma(n
\to K^0 \bar{\nu}) =0$. In Eq.~(\ref{flippedK0}) we assume
$(U_C^\dagger E)^{1 \alpha} \neq 0$.
Thus the renormalizable flipped $SU(5)$ can
be verified by looking at the channel $p \to \pi^+ \bar{\nu}$, and
using the correlation stemming from Eq.~(\ref{flippedK0}). 
This is a nontrivial result and can help us to test 
this scenario, if proton decay 
is found in the next generation of experiments.
If this channel is measured, we can make the predictions 
for decays into charged leptons using Eq.~(\ref{flippedpi0}) 
for a given model for fermion masses.

Thus it is 
possible to differentiate among different fermion mass models.
We note the difference between Eqs.~(\ref{xxx2}) and (\ref{xxx5}); there 
appears a suppression factor for the channel $p \to \pi^+ {\overline{\nu}}$ 
in the case of $SU(5)$.   
Since the nucleon decays into $K$ mesons are absent in the case of flipped
$SU(5)$, this presents an independent way to distinguish this model from
$SU(5)$, where these decay modes are always present.
The discussion of this section demonstrates that an analysis of proton decay modes 
and specifically of proton decay into antineutrinos allows one to differentiate among
different grand unification scenarios.
\section{Detailed analysis of  upper bounds} 
In this appendix we give details of the analysis presented in Sec.(5.6).
As pointed out in that section  the 
minimization of the total decay rate represents a formidable task
since there are in principle 42 unknown parameters in 
equations~(\ref{cec} - \ref{cnuc}). 
One possibility is to look for solutions where the ``$SU(5)$ contributions'' and
the  ``flipped $SU(5)$ 
contributions'' are suppressed (minimized) independently~\cite{Dorsner:2004xa}. 
Since one expects that in general the associated gauge bosons 
and couplings have different values this is also the most natural 
way to look for the minimal decay rate. Moreover, 
the bounds obtained is such a manner will be independent 
of the underlying gauge symmetry.
As discussed in the previous sections the 
``flipped $SU(5)$ contributions'' are set to zero by the 
following two conditions:

\begin{eqnarray}
V_4^{\beta \alpha}= (D_C^{\dagger} D)^{\beta \alpha}&=&0, \
\alpha=1 \ \textrm{or} \ \beta=1, \ \ (\textrm{Condition I}) \nonumber\\
(U_C^\dagger E)^{1\alpha}&=&0. \ \ (\textrm{Condition II}) \nonumber
\end{eqnarray}
Therefore, in the presence of all gauge $d=6$ contributions, 
in the Majorana neutrino case, there only remain the contributions 
appearing in $SU(5)$ models. But, those can be significantly 
suppressed. There are two major scenarios to be considered that differ 
the way proton decays~\cite{Dorsner:2004xa}:\\

\noindent
{{(A) There are no decays into the meson-charged 
antilepton pairs}}\\

All contributions to the decay of the proton into charged
antileptons and a meson can be set to zero . Namely, after we
implement Conditions I and II, we can set to zero Eq.~(\ref{ce})
by choosing
\begin{equation}
V_1^{11}= (U_C^\dagger U)^{11}=0 \ \ (\textrm{Condition III})
\end{equation}
(This condition cannot be implemented in the case of symmetric
up-quark Yukawa couplings.) On the other hand, Eq.~(\ref{cec}) can
be set to zero only if we impose
\begin{equation}
(V_2 V_{UD}^\dagger)^{\alpha1 }= (E_C^\dagger U)^{\alpha 1}=0 \ \
(\textrm{Condition IV})
\end{equation}
Thus with conditions I--IV there are only decays into
antineutrinos and, in the Majorana neutrino case, the only
non-zero coefficients are:

\begin{equation}
\label{cnu1new} c(\nu_l, d_{\alpha}, d^C_{\beta})= k_1^2 \ ( V_1
V_{UD} )^{1 \alpha} ( V_3 V_{EN})^{\beta l}
\end{equation}
So, indeed, there exists a large class of models for fermion
masses where there are no decays into a meson and charged
antileptons.
Up to this point all conditions we impose are consistent
with the unitarity constraint and experimental data on fermion
mixing. (In the $SU(5)$ case we have to impose Conditions III and
IV only.) Let us see the decay channels with 
antineutrinos. From Eq.~(\ref{cnu1new}) we see that it is not
possible to set to zero all decays since the factor $( V_1 V_{UD}
)^{1 \alpha}$ can be set to zero for only one value of $\alpha$ in
order to satisfy the unitarity constraint. Therefore we have to
compare the following two cases:

\begin{enumerate}
\item {Case (a)} $( V_1 V_{UD} )^{1 1}=0$ (Condition V)

In this case:

\begin{equation}
\Gamma_a(p \to \pi^{+} \overline{\nu}_i)=0
\end{equation}

Using chiral langragian technique yields
\begin{eqnarray}
\Gamma_a(p \to K^{+} \bar{\nu})&=& C(p,K) \left[1 + \frac{m_p}{3
m_B} (D + 3F) \right]^2 \frac{s_{13}^{2}}
{s_{12}^{2} + c_{12}^2 s_{13}^{2}} \nonumber \\  
\end{eqnarray}
where:
\begin{equation}
C(a,b)= \frac{(m_a^2 - m_b^2)^2}{8 \pi m_a^3 f^2_{\pi}} \ A_L^2 \
|\alpha|^2 \ k_1^4
\end{equation}

\item {Case (b)} $( V_1 V_{UD} )^{1 2}=0$ (Condition VI).

All the decay channels into antineutrinos are non-zero in this
case. The associated decay rates are:

\begin{eqnarray}
\Gamma_b(p \to \pi^{+} \bar{\nu})&=& C(p,\pi) \left[1 + D +
F\right]^2 \frac{s_{13}^2}{c_{12}^2 + s_{12}^2 s_{13}^2 }\\
\Gamma_b(p \to K^{+} \bar{\nu})&=& C(p,K) \left[\frac{2 m_p}{3
m_B} D\right]^2  \frac{s_{13}^2}{c_{12}^2 + s_{12}^2 s_{13}^2 }
\end{eqnarray}

\end{enumerate}
We note that these results are independent of 
\textit{all}\/ phases including those
of $V_{CKM}$ and $V_l$ and any mixing angles beyond the $CKM$
ones (This is rather unexpected since there are in principle
42 different angles and phases that could \textit{a priori}\/
enter the analysis.).
 Also, in the limit $V_{CMK}^{13} \rightarrow
0$ all decay rates vanish as required in the case of three
generations of matter fields. Here they have used the 
so-called ``standard'' parametrization of $V_{CKM}$ that 
utilizes angles $\theta_{12}$, $\theta_{23}$, $\theta_{13}$, 
and a phase $\delta_{13}$ (For example, in that parametrization
$V_{CKM}^{13}=e^{-i \delta_{13}} s_{13}$.), where 
$c_{ij}=\cos \theta_{ij}$ and $s_{ij}=\sin \theta_{ij}$. 
Hence, all one needs to know are angles 
$\theta_{12}$ and $\theta_{13}$.
Clearly of the two cases studied, it is {Case (b)} that gives 
the lowest total decay rate in the Majorana neutrino case.\\  

\noindent
{{(B) There are no decays into the meson-antineutrino pair
in the Majorana neutrino case}}\\

We now show that it is also possible to set to zero all nucleon
decay channels into a meson and antineutrinos. After Conditions I
and II, it is possible to impose $(V_1 V_{UD})^{1 \alpha}=0$ (Condition
VII) instead of $V_1^{11}=0$. (Again, these two equalities are
exclusive in the case $V^{13}_{CKM} \neq 0$.) Therefore, in the
Majorana neutrino case, there are no decays into antineutrinos
(see Eq.~\ref{cnu}). In this case the property that the gauge
contributions vanish as $|V^{13}_{CKM}| \rightarrow 0$ is obvious
since $|V_1^{11}|=|V^{13}_{CKM}|$. We have to further investigate
all possible values of $V_2^{\beta \alpha}$ and $V_3^{\beta
\alpha}$. Now, it is possible to choose $V_2^{\beta \alpha}=0$ and
$V_3^{\beta \alpha}=0$, except for the case $\alpha=\beta=2$
(Condition VIII). In that case there are only decays into a
strange mesons and muons. Let us call this {Case c)}. To
understand which case gives us an upper bound on the total proton
decay lifetime in the Majorana neutrino case, we compare the
predictions coming from the {Case (b)} and {Case (c)}.
The ratio between the relevant decay rates is 
given by~\cite{Dorsner:2004xa}:

\begin{equation}
\frac{\Gamma_c(p \to K^0 \mu^+)}{\Gamma_b(p \to \pi^+ \bar{\nu})}=
2 (c_{12}^2 + s_{12}^2 s_{13}^2) \frac{(m_p^2 - m_K^2)^2}{(m_p^2 -
m_{\pi}^2)^2}\frac{[1 + \frac{m_p}{m_B}(D-F)]^2}{[1+D+F]^2}=0.33
\end{equation}

Thus, the upper bound on the proton lifetime in the case of
Majorana neutrinos indeed corresponds to the total lifetime of 
{Case (c)}. One finds~\cite{Dorsner:2004xa}:

\begin{equation}
\tau_p\leq 6.0^{+0.5}_{-0.3} \times 10^{39} \
\frac{(M_X/10^{16}\,\textrm{GeV})^4}{\alpha_{GUT}^2} \
(0.003\,\textrm{GeV}^3 / \alpha)^2\,\textrm{years}
\end{equation}
where the gauge boson mass is given in units of $10^{16}$\,GeV. 
It explicitly indicates the dependence of the results on the nucleon
decay matrix element. These bounds are applicable to any GUT
regardless whether the scenario is supersymmetric or not. If the
theory is based on $SU(5)$ the above bounds are obtained by
imposing Conditions VII and VIII. If the theory contains both
$SU(5)$ and flipped $SU(5)$ contributions, in addition to these,
one needs to impose Conditions I and II~\cite{Dorsner:2004xa}.
Thus following two observations are in order:
 (i) All three cases ({Case (a)}--{(c)}) yield
comparable lifetimes (within a factor of ten) even though they
significantly defer in decay pattern predictions;
(ii) Using the most stringent experimental limit on partial
proton lifetime as if it represents the limit on the total proton
lifetime. Even though this is not correct (see discussion
in~\cite{Eidelman:2004wy}) it certainly yields the most conservative bound
on $M_X$.
\section{Relating 4D parameters to parameters of M theory} 
The compactifications of an 11 dimensional theory to four dimensions allows one to
relate 4 dimensional parameters such as Newtons' constant $G_N$, the 
grand unification scale $M_G$ and the unified coupling constant $\alpha_G$
to parameters of the higher dimensional theory. In the  analysis  here we give  an 
abbreviated version of the work of Refs.~\cite{Friedmann:2002ty,hw}. We begin with
the gravity action in 11 dimensions which is 

\beqn
(2\kappa_{11}^2)^{-1} \int_{{\cal R}^4\times X} d^{11}x\sqrt g R
\label{aug27a}
\eeqn
Reduction of this action to four dimensions  gives 

\beqn
V_X(2\kappa_{11}^2)^{-1} \int_{{\cal R}^4} d^{4}x\sqrt g R
\label{aug27b}
\eeqn
where $V_X$ is the volume of the compact space $X$. The 4D action of general relativity is

\beqn
(16\pi G_N)^{-1}  \int_{{\cal R}^4} d^{4}x\sqrt g R
\label{aug27c}
\eeqn
This leads  to a determinations of $G_N$ in terms of the parameters  of eleven dimensions and
the volume of  compactification

\beqn
G_N= \kappa_{11}^2 (8\pi V_X)^{-1}
\label{aug27d}
\eeqn
Next we look at the Yang-Mills action on ${\cal R}^4\times Q$. For the case of 
Type IIA D6 branes, we can write the Yang-Mills action in the form

\beqn
(4(2\pi)^2g_s(\alpha')^{-3/2})^{-1}\int d^7x\sqrt g Tr(F_{\mu\nu} F^{\mu\nu}).
\label{aug27e}
\eeqn
Here $g_s$ is the string coupling and the trace is taken in the fundamental 
representation of $U(n)$. We can write Eq.(\ref{aug27e}) in the form  

\beqn
(8(2\pi)^4g_s(\alpha')^{3/2})^{-1}\int d^7x\sqrt g \sum_{\alpha} F_{\mu\nu}^a F^{\mu\nu a}).
\label{aug27f}
\eeqn
where we have expanded $F_{\mu\nu}=\sum_a F_{\mu\nu}^aQ_a$ and used 
$Tr(Q_aQ_b)=\frac{1}{2} \delta_{ab}$.
Comparing with the Yang-Mills action in 7D  which is
$(4g_7^2 )^{-1}\int d^7x\sqrt g \times$ \\
$ \sum_{a} F_{\mu\nu}^a F^{\mu\nu a}$ one finds 

\beqn
g_7^2= 2^{4/3}(2\pi)^{4/3}\kappa_{11}^{2/3} 
\label{aug27g}
\eeqn
A further reduction of Eq.(\ref{aug27f}) to 4 dimensions on ${\cal R}^4\times Q$, 
and comparison of the action with the 4D Yang-Mills gives

\beqn
\alpha_G V_Q= (4\pi)^{1/3} \kappa_{11}^{2/3}
\label{aug27h}
\eeqn
$V_Q^{-1/3}$  has approximately the meaning of $M_G$. To make this connection more precise 
one can consider the gauge  coupling evolution in the above theory.  Now if $g_M$  is the unified gauge coupling as deduced in the M-theory, then $SU(3)$, $SU(2)$ and $U(1)$ gauge coupling 
constants are given by $1/g_i^2=k_i/g_M^2$ where $(k_1,k_2,k_3)= (5/3,1,1)$. On inclusion of 
loop corrections including the Kaluza-Klein harmonics on the compact space, one finds the 
evolution [From the evolution equations one  notes that the prediction of  $\sin\theta_W$  
is essentially unaffected by the tower of Kaluza-Klein states.].  

\beqn
\frac{16\pi^2}{g_i^2(\mu)}= (\frac{16\pi^2}{g_M^2}+10 {\cal T}_{\omega})k_i 
+ b_i log(\frac{L^{2/3}}{\mu^2V_Q^{2/3}})
\eeqn
where 

\beqn
 L_Q=exp({\cal T}_{\omega} -{\cal T}_{O})
\label{aug27h_1}
\eeqn
 and where ${\cal T}_{\omega}$, ${\cal T}_O$ are the so called analytic torsions that are
computable and the combination $L_Q$ is the so called Ray-Singer torsion~\cite{ray,rs},
and $\mu$ is the renormalization group 
scale. One may compare this evolution with what one expects in a GUT theory. Here one has

\beqn
\frac{16\pi^2}{g_i^2(\mu)}= (\frac{16\pi^2}{g_G^2})k_i +b_i log(\frac{M_G^2}{\mu^2})
\label{aug27h_2}
\eeqn
A comparison of the M-theory and the GUT theory results give
\beqn
g_G^{-2}=g_M^{-2}+\frac{5}{8\pi^2} {\cal T}_{\omega}\nonumber\\
\label{aug27h_3}
\eeqn
 and

 \beqn
 M_G=L_Q^{1/3} V_Q^{-1/3}
 \label{aug27h_4}
 \eeqn
Here  Eq.(\ref{aug27h_3}) gives the connection between the couplings of 
the M theory and the grand  unified theory while Eq.(\ref{aug27h_4}) makes more 
precise  the  definition of the GUT scale  for M theory compactifications.
Eliminating  $V_Q$ in terms of $M_G$ and $L_Q$  gives  a  determination
of $\kappa_{11}$

\beqn
\kappa_{11}= \frac{\alpha_G^{3/2} L_Q^{3/2}}{(4\pi)^{1/2} M_G^{9/2}}
\label{aug27i}
\eeqn
Using the definition of the 11 dimensional Planck scale
$M_{11}$\cite{polchinski}:

\beqn
2\kappa_{11}^2= (2\pi)^8 M_{11}^{-9}
\label{aug27j}
\eeqn
one gets a relation between $M_G$ and $M_{11}$

\beqn
M_{G}= (2\pi)^{-1} \alpha_G^{1/3}L_Q^{1/3} M_{11}
\label{aug27k}
\eeqn
From Eqs.(\ref{aug27g}), (\ref{aug27i}) and (\ref{aug27k}) one finds

\beqn
g_{7}^2 M_{11} =8\pi^2 \alpha_G^{2/3}  L_Q^{2/3}  M_G^{-2}
\label{aug27ka}
\eeqn
Interesting is the fact that  $M_G$ is  scaled down by a factor
$\alpha_G^{1/3}$ from the eleven dimensional Planck scale.  One can estimate the
size of  $M_{11}$ from above. Thus using $M_G=2\times 10^{16}$ GeV, $\alpha_G=0.04$
and $L_Q=8$, one finds $M_{11}=1.8\times 10^{17}$ GeV.\\   

Next  we consider Type IIA superstring. The action of the  gauge fields on
 a $D_6$ brane is given by\cite{polchinski} 
 
\beqn
( 4g_{D6}^2)^{-1}\int d^7x\sqrt{g_7}TrF_{ij}F^{ij},
\eeqn
where $g_{D6}$ is the gauge coupling constant and 
$F_{ij}$ are  the Yang-Mills field strengths. Here   $Tr$ is the
trace in the fundamental representation of $U(N)$.  Next  assume that
the $D6$-brane worldvolume has the product  ${\bf R}^4\times Q$, where $Q$ is a
compact three-manifold of volume $V_Q$.  With this assumption  the action in four
dimensions is 
\beqn
V_Q  (8g_{D6}^2)^{-1}\int d^4x \sum_a F_{ij}^a F^{ija}.
\eeqn
where as before 
we have  expanded  $F_{ij}=\sum_aF_{ij}^aQ_a$ and used  $Tr(Q_aQ_b)={1\over 2}\delta_{ab}$.
Comparing it to the conventional action of GUT gauge fields $( 4g_{G}^2)^{-1}$  
$\int d^4x$ $\sum_a F_{ij}^a F^{ija}$
where $g_{G}$ is the GUT coupling constant one finds the relation
\beqn
g_{G}^2=\frac{2g_{D6}^2}{V_Q}
\eeqn
Next we use the following relation on the $D6$ brane gauge 
coupling constant\cite{polchinski} 
\beqn
g_{D6}^2=(2\pi)^4g_s{\alpha'}^{3/2}
\eeqn
and get 

\beqn
g_{G}^2V_Q=2(2\pi)^4g_s{\alpha'}^{3/2}.
\label{jan3}
\eeqn
Now it is argued~\cite{Klebanov:2003my} that the relation of Eq.( \ref{aug27h_4}) is valid also for Type IIA theory. 
Using Eq.(\ref{aug27h_4}) in Eq.(\ref{jan3}) gives

\beqn
\alpha'= \frac{\alpha_G^{{2}/{3}} L_Q^{{2}/{3}}} {4\pi^2 g_s^{{2}/{3}} M_G^2}
\label{aug24_12a}
\eeqn
\section{Gauge coupling unification in string models}
As  noted  already aside from proton stability, gauge coupling unification is an important
constraint on unified models of particle interactions. For unification of gauge 
coupling constants it is not necessary that the gauge couplings  arise  from 
a grand unification since the Standard  Model gauge  group can emerge directly
at the string scale. Here  one has an additional constraint, i.e., not only
the  gauge couplings unify but also that  the  gauge couplings  unify with
gravity. Thus one has~\cite{ginsparg}

	\begin{equation}
	g_i^2k_i=g_{string}^2
	\end{equation}
	where $k_i$ are the Kac-Moody levels of  the subgroups,
	and $\alpha'$  is the Regge slope. 
	Models of this type  will in general
	possess fractionally charged neutral states unless the SM
	gauge group arises from an unbroken SU(5) at the
	string scale, or unless $k>1$~\cite{schell}.  
	In models with fractionally
	charged states one must either confine them to produce bound states which carry 
	integral charges or find a mechanism to make them massive.

	The unification of the gauge couplings and of gravity is automatic
	in string models, but these constraints must be checked with LEP  
	data.  The renormalization group evolution implies  
	\begin{equation}
	\frac{16\pi^2}{g_i^2(M_Z)}=k_i \frac{16\pi^2}{g_{string}^2}
	+b_iln(\frac{M_{str}^2}{M_Z^2})+\Delta_i
	\end{equation}
	where $\Delta_i$ contains stringy and non-stringy effects. 
	Now it is known that with the MSSM spectrum there is a unification of
	gauge coupling constants at a scale of $M_G\sim 2\times 10^{16}$ GeV 
	with $\alpha_G\sim 1/24$~\cite{Giunti:1991ta}. 
	The scale $M_G$ is about two orders of magnitude
	below the scale where the unification of gauge couplings and of gravity
	can occur as can be seen roughly by extrapolating $G_NE^2$  which acts
	like the fine structure constant for gravity. This discrepancy is a 
	serious problem for any string unified model~\cite{Dienes:1996du}.
	Some of the possible
	avenues to resolve this conflict are as follows
	\begin{enumerate}
	\item
	Extra matter at a high scale which can modify the RG evolution of gauge
	couplings to remove the 
	discrepancy~\cite{Dienes:1995sv,Dienes:1995bx,Barger:2005qy}
	\item 
	Non-standard hypercharge normalizations within string models with higher
	level gauge symmetries~\cite{Dienes:1995sq}.
	\item
	An M theory solution~\cite{Witten:1996mz} to the gauge coupling/gravity 
	unification, where the
	gravity propagates in a higher dimensional  bulk while gauge and  matter
	fields reside on  four dimensional wall.  Below a  certain scale, both
	matter, gauge and gravity propagate in four dimensions while above this
	scale matter and gauge fields  propagate  in four  dimensions while gravity
	propagates in higher dimensions which allows $\alpha_{gr}$ to evolve much
	faster allowing for unification at the conventional scale of $M_G$. \\
\end{enumerate}
	It is also of interest to discuss the issue of gauge coupling
	unification in intersecting D brane  models. Here typically  
	the gauge coupling unification is less transparent due to the
 product nature of the group structure at the string scale.
 Thus it  is instructive to explore the conditions under which 
 the gauge coupling unification may occur.  We recall that the 
 crucial constraint  in unification of the three  couplings is the 
 condition  $\alpha_2(M_X)=\alpha_3(M_X) =\frac{5}{3} \alpha_Y$. 
  In brane models it is not at all a priori obvious how a relation of this 
  type  might emerge.  For concreteness one may consider torroidal orbifold
  compactifications of  ${\cal T}^6/Z_2\times Z_2$ with ${\cal T}^6$ a product
  of two-tori. The moduli sector of this compactification includes the Kahler moduli
  $T_i$ (i=1,2,3) which shall  be the focus of our attention. In type  IIB picture
  which is dual to Type IIA, the D brane intersection angles  are  replaced  by 
  fluxes on the internal world volumes so that $F_a^{m}=m_a^{m}/n_a^{m}$, where
  a labels  a stack of D branes and m stands  for the components of the two torus m,
  and where $m_a^{m}$ and $n_a^{m}$ are rational numbers. The satisfaction of  N=1
  supersymmetry in type IIB can be  written in the form 
   \beqn
  \sum_{m=1,2,3} \frac{F_a^{m}}{Re(T_m)} =\prod \frac{F_a^{m}}{Re(T_m)} 
  \label{susy}
  \eeqn
While the unification of gauge coupling constants on intersecting branes in not
automatic such unification is not excluded. Thus an interesting observation is that
 one may choose intersecting  brane
 configurations for which the following relation holds 
\beqn
\frac{1}{\alpha_Y} = \frac{2}{3} \frac{1}{\alpha_3} +\frac{1}{\alpha_2}
\label{bls}
\eeqn 
If we work in the above class of models then the additional condition
\beqn
\alpha_2(M_X)=\alpha_3 (M_X)
\label{g2g3}
\eeqn
  would automatically lead to the desired relation 
 $\alpha_2(M_X)=\alpha_3(M_X) =\frac{5}{3} \alpha_Y$.
 It is interesting then to investigate the conditions under which the 
 constraint of Eq.(\ref{g2g3}) arises. 
A closer scrutiny reveals~\cite{Kors:2003wf} that there are three distinct classes of 
constraints which we label as A, B, and C that allow for the satisfaction 
of Eq.~(\ref{susy}).
The class A constraints  arise when none of the fluxes $F_a^i$ vanish.
In this case the  $Re(T_i)$ are all uniquely determined
and  the satisfaction of the relation $\alpha_2=\alpha_3$ can only
be accidental. 
That is to say for most models satisfying 
Eq.~(\ref{susy}) the satisfaction of the relation Eq.~(\ref{g2g3})
 and hence the unification of gauge coupling condition 
  $\alpha_2(M_X)=\alpha_3(M_X) =\frac{5}{3} \alpha_Y$ can only be
  accidental. 
   The class B constraints
arise when one of the fluxes $F_a^i$ vanishes (for each $a$) but one
still has a determination of the ratios $Re(T_1): Re(T_2): Re(T_3)$
but not a determination of the overall size. In this case again
one has the same problem in unifying the gauge couplings as in 
case A, i.e., the gauge coupling unification will have to be accidental.
 Finally, in case C one of the fluxes $F_a^i$ vanishes (for each 
$a$) and this time one has a determination of only one ratio. 
Thus, for example, one may determine  
  $Re(T_j): Re(T_k)$ while  $Re(T_i)$ ($i\neq j\neq k$) 
is unconstrained. In this case one has the possibility of unifying gauge
coupling constants by utilizing the free parameter $Re(T_i)$.
There are no known examples of models of  class A. 
  An example of class B model is that of Ref~\cite{Cvetic:2003yd} where
the ratio $Re(T_1): Re(T_2): Re(T_3)$ is determined and the
gauge coupling unification does not occur while an example
of  class C model is that of Ref.~\cite{Blumenhagen:2003jy,Blumenhagen:2003qd}
where $Re(T_2):Re(T_3)$ is 
determined, $Re(T_1)$ is left unconstrained and one may
achieve gauge coupling unification by constraining $Re(T_1)$.

\end{appendix}

\input{bibl_sort.tex}
\end{document}

%% file: box
\begin{picture}(80,40)(40,20)
\SetWidth{0.8}
\ArrowLine(0,40)(20,40)
\ArrowLine(0,0)(20,0)
\ArrowLine(80,40)(60,40)
\ArrowLine(80,0)(60,0)
\DashLine(20,40)(60,40)3
\DashLine(20,0)(60,0)3
\ArrowLine(20,20)(20,40)
\ArrowLine(20,20)(20,0)
\Line(17,17)(23,23)
\Line(17,23)(23,17)
\ArrowLine(60,20)(60,40)
\ArrowLine(60,20)(60,0)
\Line(57,17)(63,23)
\Line(57,23)(63,17)
\Text(10,30)[]{\trgor}
\Text(10,10)[]{\trdol}
\Text(40,30)[]{\sfgor}
\Text(40,10)[]{\sfdol}
\Text(75,30)[]{\spgor}
\Text(75,10)[]{\spdol}
\Text(-10,40)[]{\iena}
\Text(-10,0)[]{\idva}
\Text(90,0)[]{\itri}
\Text(90,40)[]{\isti}
\end{picture}

%% file: triangle
\begin{picture}(80,40)(40,20)
\SetWidth{0.8}
\ArrowLine(0,40)(10,20)
\ArrowLine(0,0)(10,20)
\ArrowLine(80,40)(60,40)
\ArrowLine(80,0)(60,0)
\DashLine(10,20)(40,20)3
\DashLine(40,20)(60,40)3
\DashLine(40,20)(60,0)3
\ArrowLine(60,20)(60,40)
\ArrowLine(60,20)(60,0)
\Line(57,17)(63,23)
\Line(57,23)(63,17)
\Text(25,30)[]{\tr}
\Text(45,35)[]{\sfgor}
\Text(45,5)[]{\sfdol}
\Text(75,30)[]{\spgor}
\Text(75,10)[]{\spdol}
\Text(-10,40)[]{\iena}
\Text(-10,0)[]{\idva}
\Text(90,0)[]{\itri}
\Text(90,40)[]{\isti}
\end{picture}